\documentclass[%
superscriptaddress,
 amsmath,amssymb,
aps, 
floatfix
]{revtex4-2}

\usepackage{graphicx}
\usepackage{dcolumn}
\usepackage{bm}
\usepackage{hyperref}

\newcommand{\equalcontrib}{\thanks{These authors contributed equally to this work. Correspondence: \href{mailto:qzhou75@wisc.edu}{qzhou75@wisc.edu}. }}

\usepackage{bbm}
\usepackage{braket}
\usepackage{booktabs}
\usepackage{array}
\usepackage{siunitx}

\newcolumntype{C}{S[table-format=5.0]}
\newcolumntype{E}{S[table-format=5.0]}
\newcolumntype{L}{S[table-format=3.0]}

\makeatletter
\newenvironment{arxivindependentbibliography}[1]{%
  \begingroup
  \def\arxiv@bibcount{#1}%
  \let\arxiv@original@ref\ref
  \def\ref##1{%
    \def\arxiv@label@name{##1}%
    \def\arxiv@lastbibitem{LastBibItem}%
    \ifx\arxiv@label@name\arxiv@lastbibitem
      \arxiv@bibcount
    \else
      \arxiv@original@ref{##1}%
    \fi
  }%
  \let\arxiv@original@label\label
  \def\label##1{%
    \def\arxiv@label@name{##1}%
    \def\arxiv@lastbibitem{LastBibItem}%
    \ifx\arxiv@label@name\arxiv@lastbibitem
    \else
      \arxiv@original@label{##1}%
    \fi
  }%
}{%
  \endgroup
}
\makeatother

\begin{document}

\preprint{APS/123-QED}

\title{Quantum Nonlinearity for Optical Neural Computing}

\author{Qingyi Zhou}
\equalcontrib
\affiliation{Department of Electrical and Computer Engineering, University of Wisconsin-Madison, Madison, WI 53706, USA}

\author{Jungmin Kim}
\equalcontrib
\affiliation{Department of Electrical and Computer Engineering, University of Wisconsin-Madison, Madison, WI 53706, USA}

\author{Yutian Tao}
\equalcontrib
\affiliation{The Computer Sciences Department, University of Wisconsin-Madison, Madison, WI 53706, USA}

\author{Guoming Huang}
\affiliation{Department of Electrical and Computer Engineering, University of Wisconsin-Madison, Madison, WI 53706, USA}

\author{Ming Zhou}
\affiliation{Department of Electrical Engineering, Stanford University, Stanford, CA 94305, USA}

\author{Zewei Shao}
\affiliation{Department of Electrical and Computer Engineering, University of Wisconsin-Madison, Madison, WI 53706, USA}

\author{Zongfu Yu}
\affiliation{Department of Electrical and Computer Engineering, University of Wisconsin-Madison, Madison, WI 53706, USA}

\date{\today}

\begin{abstract} 
The rapid scaling of deep neural networks comes at the cost of unsustainable power consumption. 
While optical neural networks offer an alternative, their capabilities remain constrained by the lack of efficient optical nonlinearities. 
To address this, we propose an optical neural computing architecture by embedding quantum emitters in inverse-designed nanophotonic structures. 
Due to their saturability, quantum emitters exhibit exceptionally strong nonlinearity compared with conventional materials. 
Using physics-aware training, we numerically demonstrate that the proposed architecture can solve complex tasks, including nonlinear classification and reinforcement learning,  within all-optical neural networks.
To enable fair comparison across different platforms, we introduce a framework that quantitatively links nonlinearity to a network’s expressive power. 
Analysis shows that our quantum activation operates at $\text{nW}/\mu\text{m}^2$ intensity, which is seven orders of magnitude below the nonlinearity threshold of conventional optical materials.
Looking ahead to large language models, we estimate the nonlinearity-limited optical power, which scales sublinearly with model size.
Our results indicate that quantum nanophotonics may provide a route toward sustainable AI inference. 
\end{abstract}

\maketitle


\section{Introduction} 
In the past decade, the rapid advancement of deep learning has profoundly transformed science and technology. 
Deep neural networks have achieved state-of-the-art performance across diverse fields, ranging from computer vision \cite{krizhevsky2012alexnet} and game-playing \cite{mnih2015atari} to protein design \cite{jumper2021alphafold} and language processing \cite{vaswani2017transformer}. 
Such progress has been driven by the continuous scaling of the model size. 
However, this scaling trend imposes an energy cost toward unsustainable levels \cite{patterson2021}. 
A growing effort has been directed towards finding alternative computing paradigms. 
In particular, following the pioneering work of Shen et al. \cite{shen2017}, optical neural networks (ONNs) have emerged as a promising candidate \cite{lin2018diffractive}, inspired by the vision that a passive optical device can implement linear transformation with high speed \cite{shekhar2024roadmap} and low energy cost \cite{wetzstein2020review}. 
Specifically, recent works have demonstrated that linear optical matrix operations can be performed with sub-photon energy consumption \cite{wang2022photon}. 
Despite these advantages, linear operations are insufficient for deep learning. 
The expressive power of ONN is severely limited by the lack of efficient optical nonlinearity \cite{wetzstein2020review, shi2025review}. 
In conventional materials, optical nonlinearities are often perturbative \cite{boyd2008}. 
As a result, existing all-optical nonlinear activation units demand high optical power and large footprint \cite{feldmann2019, wu2022low, wu2025field, jha2020reconfigurable, yanagimoto2025programmable}, making it difficult to scale up. 
This nonlinearity bottleneck has remained a long-standing challenge for the optical computing community.
Existing works that rely on hybrid opto-electronic architectures \cite{shen2017, williamson2019, pour2020, hu2025computing} suffer from additional latency and substantial system complexity due to frequent optical-electrical-optical (O/E/O) conversions.
More recently, ``structural nonlinearity'' schemes have been proposed \cite{yildirim2024nonlinear, xia2024nonlinear, wanjura2024nonlinear}, in which the input is encoded not in the optical field but in tunable parameters of a linear structure. 
However, the connection between such systems and standard deep learning models is often unclear. 

To address the nonlinearity bottleneck, we first point out that nonlinear optical phenomena are not intrinsically restricted to high intensities. 
A single quantum emitter (including atom, quantum dot, or color center) can be saturated by the absorption of a few photons per lifetime, leading to extremely strong optical nonlinearity \cite{lodahl2015review}. 
There have been both theoretical proposals \cite{zhu2025quantum, canora2025engineering} and experimental demonstrations \cite{zuo2019atom, ryou2021atom} of using quantum emitter media as activation units, underscoring their potential for realizing strong optical nonlinearities. 
However, it remains unclear what quantitative benefit such quantum nonlinearity offers for ONNs. Furthermore, to fully utilize the nonlinear functionality of quantum emitters, it is essential to enhance the interaction between light and quantum emitters, which can be achieved using properly designed nanophotonic structures. 
Recent progress in quantum nanophotonics has enabled deterministic integration of individual emitters with on-chip photonic structures \cite{Ohno2012_APL, chen2019laser, day2023laser, yama2026, schroder2017}. 
Therefore, we believe it is the right time to address the above-mentioned questions and systematically investigate whether quantum technologies can provide the strong nonlinearity required by optical neural networks. 

In this work, we present a theoretical proposal for a low-power optical neural network architecture that exploits the strong nonlinearity of individual quantum emitters. 
Specifically, we introduce a quantum-enhanced activation unit by embedding quantum emitters into adjoint-optimized nanophotonic structures \cite{lalau2013adjoint}. 
A strong nonlinear response can be achieved at intensity level of $\text{nW}/\mu\text{m}^{2}$. 
With full-wave simulations, we verify the performance on nonlinear classification task as well as reinforcement learning tasks, demonstrating functionality beyond linear models. 
To enable a fair comparison across different physical nonlinearities, we develop a theoretical framework that quantifies the ``expressive power'' of an arbitrary activation unit. 
Unlike existing theoretical analyses that focus on purely linear optics \cite{kulce2021capacity, miller2023thickness, li2025spatial, onodera2025} or restricted classes of unitary transformations \cite{yu2025nonlinear}, our framework directly links a nonlinear input-output response to the depth-wise growth of ONN expressivity. 
This allows us to translate a targeted expressive power into the required light intensity, enabling a quantitative assessment of ONN's energy efficiency. 
Applying this framework, we show that conventional platforms based on Kerr effect or saturable absorption would require prohibitively high intensities to match the digital baseline. 
In contrast, the proposed quantum-enhanced activation reduces the required intensity by seven orders of magnitude. 
Finally, we look into the future by estimating the nonlinearity-limited optical power for optical large language models (LLMs).
Our analysis shows that this lower bound, set by the nonlinear activation alone, remains at the watt level for the proposed scheme and scales sublinearly with model size.
Taken together, our results indicate that large-scale optical neural computing, powered by quantum-enhanced nonlinearities, could reduce the energy footprint of AI inference, providing a path beyond the limits of electronic hardware.

\section{Results}

\begin{figure*}[t]
  \centering
  \includegraphics[width=0.85\linewidth]{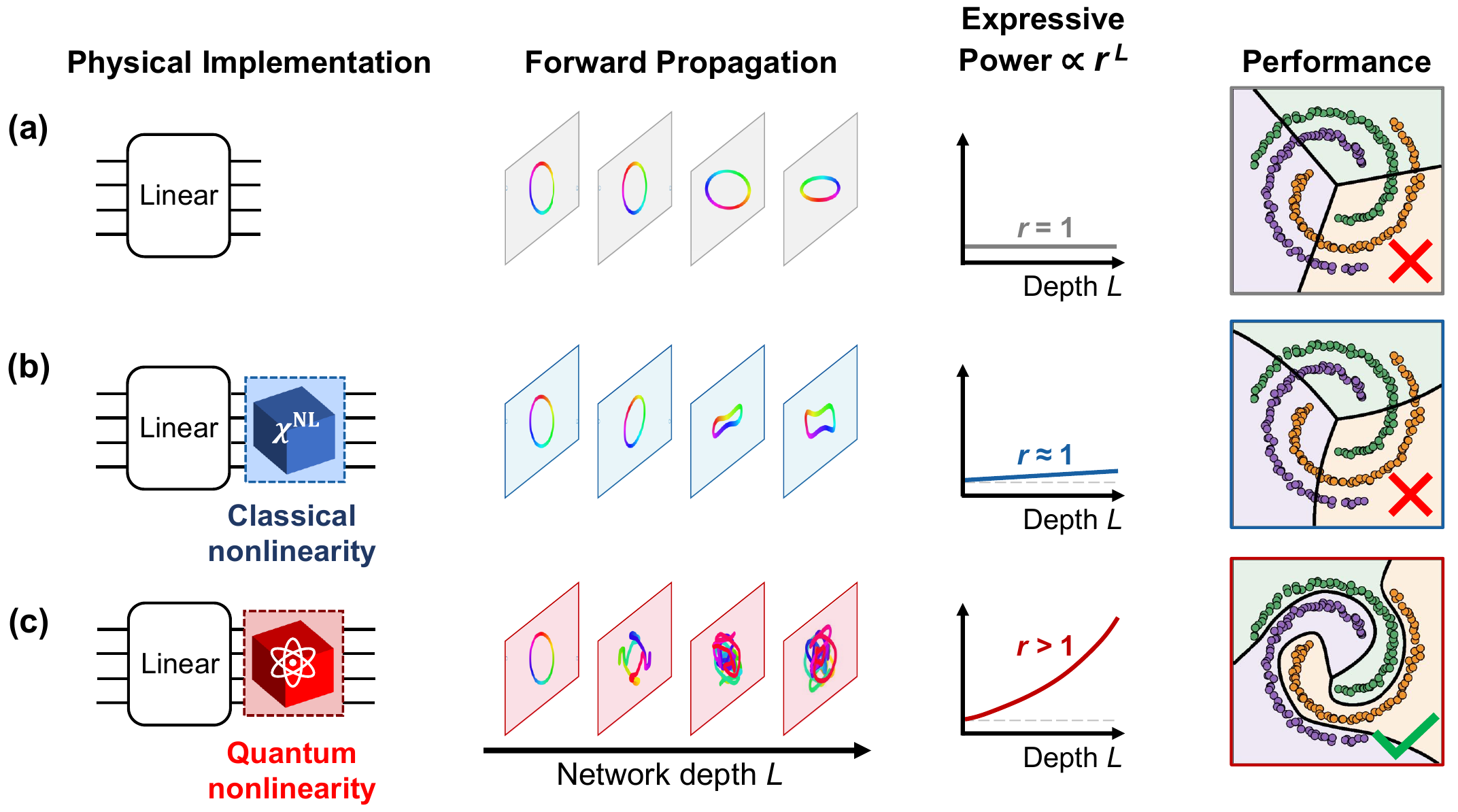}
  \caption{
  \textbf{Comparison of ONN architectures with different nonlinearities. }
  (a) A linear ONN, whose expressive power remains constant regardless of depth $L$. 
  (b) ONN with classical nonlinearity. With weak nonlinearity, the expressive power increases slightly with depth, yet is not enough for complex tasks. 
  (c) ONN with quantum nonlinearity. The expressive power $\propto r^{L}$ grows exponentially with $r>1$, and is able to handle complex task. }
  \label{fig:fig1}
\end{figure*}

\subsection{Overcoming nonlinearity bottleneck with quantum activations} 
We consider a generic multi-layer ONN architecture. 
Each layer consists of a linear transformation $\bm{y}=\bm{W}^{(i)}\bm{x}+\bm{b}^{(i)}$, comprising a weight matrix $\bm{W}^{(i)}$ and a bias $\bm{b}^{(i)}$, followed by a nonlinear activation $f(\cdot)$, mirroring the architecture of a typical multi-layer perceptron (MLP).
In machine learning theory, it is well established that the expressive power of a deep neural network grows exponentially with its depth \cite{montufar2014, poole2016, raghu2017}. 
A purely linear network cannot enjoy this benefit since a composition of linear transformations is still linear. 
As a result, the overall expressive power of an ONN is limited by its nonlinear activation units. 
We follow the framework developed in Refs.~\citenum{poole2016, raghu2017} and introduce a metric for quantifying expressive power. 
As illustrated in Fig.~\ref{fig:fig1}, a closed trajectory of data points serves as the input. 
The total curvature $K$ of this trajectory provides a robust measure of curve complexity, and is monitored as the curve propagates through successive layers. 
Intuitively, the total curvature measures the degree of ``folding'' applied to the data manifold, a capability that is fundamentally impossible with linear transformations. 
As shown in Fig.~\ref{fig:fig1}(a), for a purely linear network the total curvature remains constant. 
In contrast, in the presence of nonlinear activations, the curvature increases by a growth factor $r>1$ after each layer, leading to an exponential growth $\sim r^{L}$ with depth $L$ \cite{{poole2016, raghu2017}}. 
The growth factor $r$ is therefore used as a quantitative measure of expressive power (see Supplementary Note S10 for details). 

Conventional optical materials possess small nonlinear susceptibilities \cite{boyd2008}. 
At realistic light intensities, the response is only weakly nonlinear, which yields a growth factor $r \approx 1$. The expressive power, as shown in Fig.~\ref{fig:fig1}(b), shows little increase with depth. 
Existing all-optical activation units typically require mW$/\mu\text{m}^{2}$ laser intensities, together with large footprints \cite{feldmann2019, shi2022nonlinear, wu2022low, jha2020reconfigurable, li2023all} (see Supplementary Note S1 for a summary of representative designs obtained from literature).
Such requirements are incompatible with large-scale ONNs. 
On the other hand, it has long been recognized that low-dimensional systems exhibit much stronger optical nonlinearities than bulk media \cite{bao2009atomic, shi2017MoS2, liu2025Te}, owing to enhanced oscillator strength under quantum confinement \cite{hanamura1988}. 
In particular, zero-dimensional quantum emitters behave as two-level systems (TLSs) that can be saturated by the absorption of only a few photons per lifetime. 
This leads to extremely strong nonlinear scattering responses, which have been observed in various waveguide- and cavity-QED platforms \cite{javadi2015, volz2014, shomroni2014, hacker2016, lukin2020}. 
These observations suggest that emitter-based nonlinearities could be leveraged to overcome the nonlinearity bottleneck in ONNs (Fig.~\ref{fig:fig1}(c)). 

\subsection{Device design and verification on nonlinear classification} 
\begin{figure*}[t]
  \centering
  \includegraphics[width=1.0\linewidth]{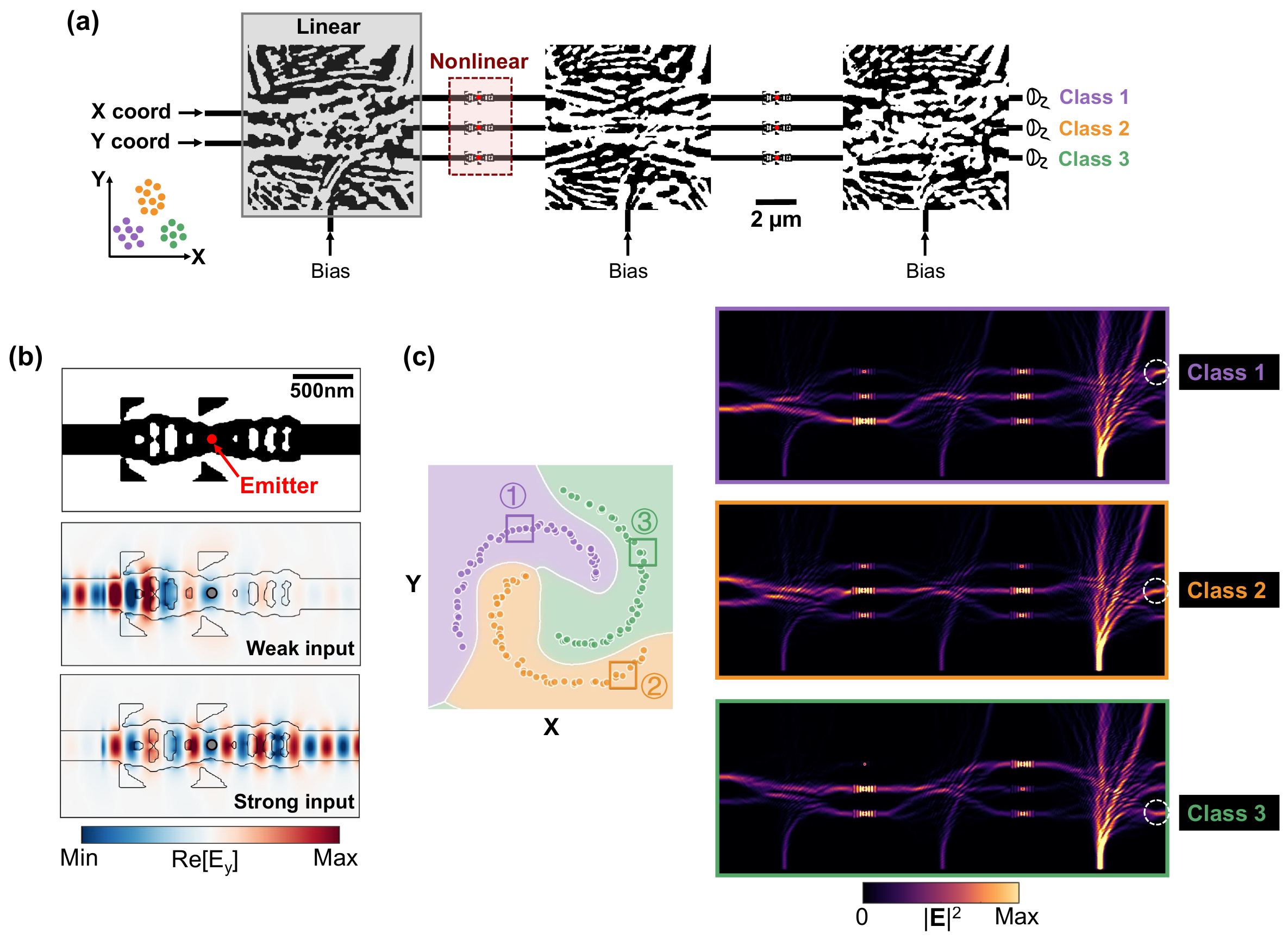}
  \caption{
  \textbf{Physics-aware training and verification on nonlinear classification.}
  (a) All-optical neural network design, constructed by stacking nonlinear activation units between linear blocks, which are also designed through adjoint optimization. 
  The 2D coordinates are encoded as input from the left ports, while a separate port at the bottom provides a constant optical bias. The detected intensities are interpreted as classification results.
  (b) The proposed quantum nonlinear activation unit. A single SiV$^{-}$ color center is embedded in an inverse-designed GaP-on-diamond structure. Simulated $\text{Re}[{E_{y}}]$ field distributions illustrate the transition from resonant scattering to saturation, resulting in a nonlinear response at low intensity.
  (c) Performance verification. Classification results for the ``spiral'' dataset are obtained via nonlinear FDFD simulations. 
  The $|E|^{2}$ intensity distributions of 3 representative inputs are visualized. Different intensity distributions at the output port correspond to different predictions. }
  \label{fig:fig2}
\end{figure*}

Motivated by the above considerations, we propose an all-optical activation unit, consisting of a single quantum emitter embedded inside an inverse-designed GaP-on-diamond nanophotonic structure (design region $1.5\times0.7~\mu\text{m}^{2}$), as shown in Fig.~\ref{fig:fig2}(a).
We utilized adjoint optimization to design a nanophotonic interface that maximizes light-matter interaction and minimizes loss (see Supplementary Note S4 for design details). 
The emitter is modeled as a TLS with a field-driven dipole moment ${d_{y}} \propto \frac{\Omega / \Gamma_{0}}{1 + 2(\Omega / \Gamma_{0})^{2}}$, where the Rabi frequency $\Omega$ is set by the local electric field \cite{zhou2024fdtd, wang2025lorentz} and $\Gamma_{0} = 2\pi\times 94$ MHz is the spontaneous emission rate (parameters are obtained from SiV$^{-}$ color centers, assuming lifetime-limited linewidth; see Supplementary Note S2 for details). 
The electric field distributions for two different input intensities are shown in Fig.~\ref{fig:fig2}(b), both obtained using full-wave three-dimensional nonlinear finite-difference frequency-domain (FDFD) simulations (see Supplementary Note~S3 for details). 
The nonlinear activation unit is constructed based on a realistic GaP-on-diamond platform ($200$~nm patterned GaP on a $160$~nm diamond film, with SiO$_{2}$ substrate). The passive linear weight blocks are designed based on the variational effective-index method~\cite{hammer2009eim, nikkhah2024_2D}. Note that the specific implementation of linear blocks is not central to this work and could be realized with other schemes such as Mach--Zehnder interferometers. 
The device is engineered to operate in two distinct regimes: in the weak-field limit the emitter acts as a linear scatterer, while in the strong-field limit it becomes nearly transparent. 
In this two-port geometry, the emitter is configured to induce a change in transmission coefficient $|\Delta t| = 1$ via interference, which results in a strong nonlinearity.
Using $N>1$ emitters could in principle achieve a larger transmission change of $|\Delta t| = 2$ (see Supplementary Note S4 for the analysis).
From the simulated input-output curve, we extract an effective nonlinear activation function with an ultra-low intensity threshold.
To evaluate the expressive power of the obtained activation function, we compute its growth factor $r(I)$, which reaches $r \approx 1.14$ at intensity $I=0.24~\text{nW}/\mu\text{m}^{2}$, exceeding the digital baseline.
In contrast, conventional silicon- and graphene-based nonlinearities remain near $r = 1$ under similar operating conditions (see Supplementary Note S11 and Supplementary Figure S22 for details). 
We have also analyzed the effect of low quantum efficiency and demonstrate that our method remains robust even when the TLS' quantum efficiency drops to $60\%$ (see Supplementary Note S5 for details). We also characterize the robustness of the activation unit against two different non-idealities: lateral position randomness and spectral disorder (see Supplementary Note S14 for details).
These results confirm our key physical intuition: by utilizing saturable quantum emitters embedded in nanophotonic structures, strong optical nonlinearity can be realized at ultra-low optical power. 

To demonstrate the practical advantage of the proposed activation, we benchmark our system on nonlinear classification task that is challenging for models that are weakly nonlinear. 
We adopt a physics-aware training approach to design our ONN in a physically consistent manner. 
We first characterize the nonlinear activation unit using nonlinear FDFD simulations to obtain its input-output transmission curve (see Supplementary Figure S8), which is then used as the activation function in a PyTorch-based ONN model. 
The linear weight matrices are represented as complex transmission matrices subject to energy-preserving constraints, ensuring that they can be implemented by passive structures. 
With this differentiable model, the ONN is trained in the digital domain via backpropagation. 
After training, we map the trained network to concrete photonic structures in a modular fashion. 
For each layer, the trained complex weight matrix is interpreted as a target transmission matrix between input and output ports. 
We then run a separate adjoint-based optimization to realize a compact block that implements this target matrix with high fidelity (see Supplementary Note S4 and S8 for quantitative metrics). 
The nonlinear layers are implemented by inserting the quantum activation units between these inverse-designed linear blocks, as illustrated in Fig.~\ref{fig:fig2}(a). 
By dissecting the network into multiple modules that can be designed individually, this approach avoids heavy full-wave simulations of the entire network during training (the training procedure is explained in Supplementary Note S7). 
We verify the final design using full-wave nonlinear FDFD simulations. 
The classification result for a three-class ``spiral'' dataset is shown in Fig.~\ref{fig:fig2}(c). 
The corresponding light intensities for three representative input points are also visualized, revealing how the network steers optical energy toward different output regions associated with different classes. 
We provide more examples in Supplementary Note S6 (performance on MNIST and FashionMNIST) and S8 (performance on nonlinear regression task). 
These results confirm that the physics-aware training yields physically realizable ONNs. 
Within realistic optical intensities well below $1~\text{mW}/\mu\text{m}^{2}$, conventional materials cannot provide sufficient expressive power. 
In stark contrast, quantum-enhanced nonlinearity can function below $1~\text{nW}/\mu\text{m}^{2}$, enabling the system to solve complex tasks. 

\begin{figure*}[t]
  \centering
  \includegraphics[width=1.0\linewidth]{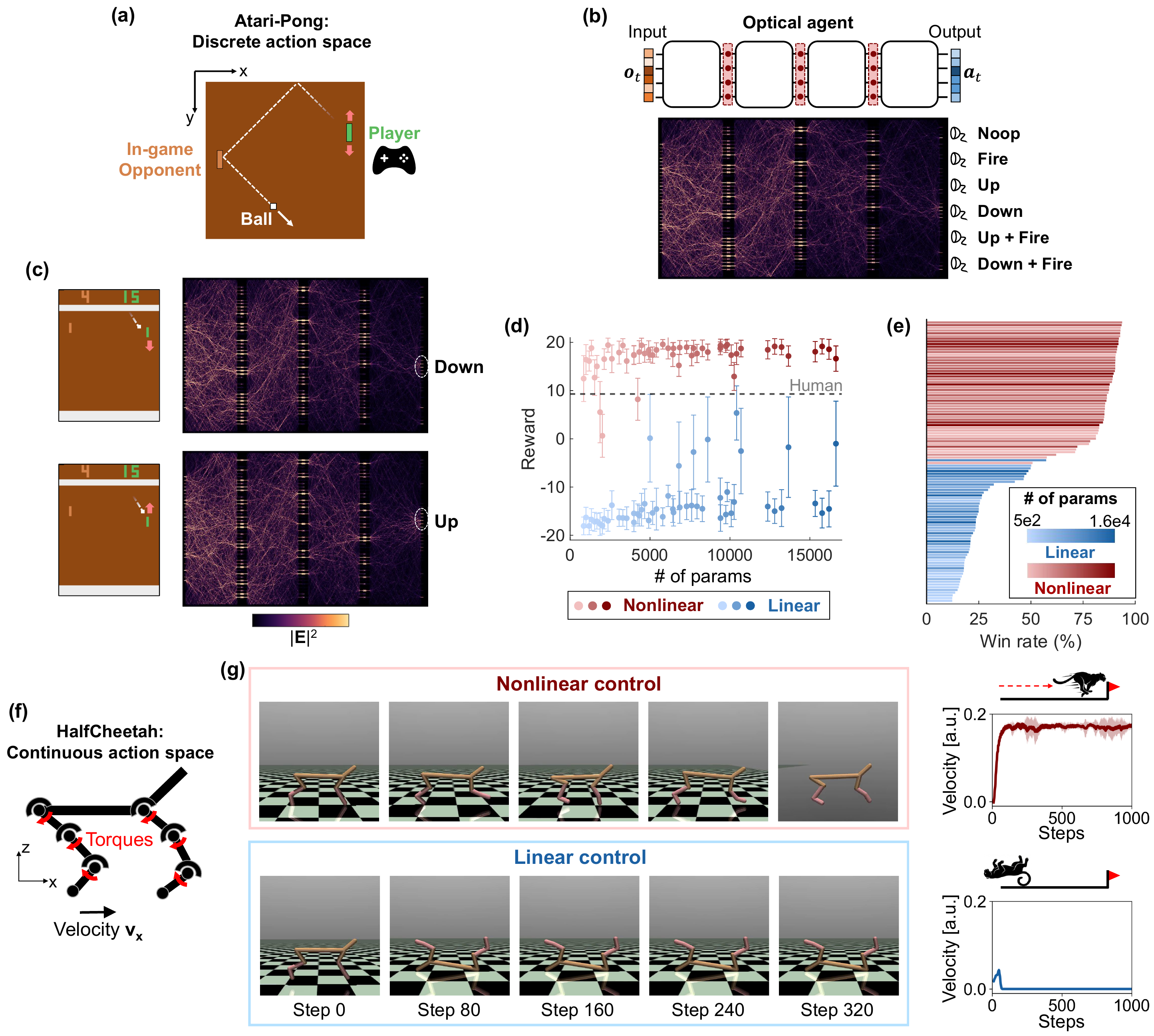}
  \caption{
  \textbf{Generalizing to reinforcement learning tasks.}
  (a) Schematic illustration of the Pong environment. 
  The player controls the green paddle and tries to block the ball (white) to win score. 
  (b) The structure of our optical agent, which functions as a policy network. 
  At each time step $t$, historical game frames are encoded into the input $\bm{o}_t$. The network outputs the logits for six discrete actions. 
  (c) Visualization of the learned policy. Light intensity distributions of two key frames are displayed. 
  (d) Final reward versus model size. 
  Linear models (blue) saturate at a low performance ceiling with high variance. 
  In contrast, nonlinear models (red) converge to near-perfect play very quickly. 
  Error bars denote standard deviations obtained over 30 episodes. 
  The human-level performance (reward$=9.3$) \cite{mnih2015atari} is visualized using gray dashed line.
  (e) Performance ranking. Best achieved rewards for 104 trained models (52 linear, 52 nonlinear) are sorted. 
  Nonlinear models consistently outperform linear models. 
  (f) Schematic illustration of the HalfCheetah control task. 
  (g) Snapshots obtained during testing. The nonlinear ONN runs stably, while the linear ONN falls down. 
  The insets plot the corresponding velocity curves, averaged over 10 episodes. The shaded areas visualize the standard deviations. }
  \label{fig:fig3}
\end{figure*}

\subsection{Generalizing to intelligent optical agents: Reinforcement learning} 
Having established that quantum-enhanced activations enable nonlinear classification, we next ask whether the same photonic building blocks can support more complex tasks. 
Reinforcement learning (RL), which has achieved impressive results on game-playing benchmarks \cite{silver2016alphago, mnih2015atari} and robotics \cite{hwangbo2019}, provides a natural testbed for our purpose. 
Note that RL based on photonic spiking network has recently been demonstrated on control benchmarks~\cite{xiang2025spiking_rl}. To demonstrate the generality of our approach, we evaluate it on two distinct tasks: a discrete game-playing task (``Atari Pong'') and a continuous control task (``HalfCheetah''), both provided by the Gymnasium library~\cite{towers2024gymnasium}. 

The Atari Pong environment is illustrated schematically in Fig.~\ref{fig:fig3}(a). 
In Pong, an agent controls the right paddle against the in-game opponent. An episode ends when a player reaches 21 points, and the reward is defined as the final score difference. 
As shown in Fig.~\ref{fig:fig3}(b), the ONN acts as a policy network: at each time step $t$, a stack of $F$ recent frames is encoded into an observation $\bm{o}_{t}$, which serves as the input (see Supplementary Figure S17 for details). 
The output intensities (divided into six regions) are interpreted as logits for six discrete actions. 
An action ${a}_{t}$ is sampled from this distribution and sent back to the environment, which then advances to the next time step. 
The policy parameters are trained using a standard proximal policy optimization (PPO) algorithm \cite{schulman2017ppo}, based on the same physics-aware framework described above (see Supplementary Note S9 for details). We note that while the nonlinear activation mechanism has been verified with 3D simulation in Fig.~2, the present Atari Pong demonstration relies on 2D simulations due to limited computing resource. 
The optical intensity distributions for two representative game frame are shown in Fig.~\ref{fig:fig3}(c). 
Different spatial configurations of the ball and paddles lead to distinct activation patterns and different intensity hotspots at the output ports. 
We then systematically benchmark the performance of linear versus nonlinear ONNs. 
We train 104 models, sweeping across network width $W$, number of hidden layers $L$, and the number of input frames $F$. All models are trained for identical number of iterations (see Supplementary Note S9 for details). 
The results are summarized in Fig.~\ref{fig:fig3}(d), where the final reward is plotted against number of parameters. 
Fig.~\ref{fig:fig3}(e) further ranks all the trained models based on their final rewards. 
Linear models saturate at a low performance ceiling regardless of model size, and exhibit high variance, indicating that the learned strategies cannot win reliably. 
In contrast, ONNs equipped with quantum activations achieve much higher rewards as they scale up, converging to near-perfect play. 

We further evaluate our ONN on the MuJoCo HalfCheetah control benchmark, as illustrated in Fig.~\ref{fig:fig3}(f). 
With a continuous action space, HalfCheetah is much more challenging than Pong. 
At each time step, the optical agent receives a 17-dimensional observation (joint positions and velocities, see Supplementary Figure S18) and outputs a 6-dimensional action vector that specifies torques applied at the six hinge joints. 
We adopt the similar ONN backbone as in Pong, with one key difference: the output uses balanced detection to support negative action values  (see Supplementary Note S9 for details). 
Training is performed using the standard soft actor-critic (SAC) algorithm \cite{haarnoja2018sac}.  
The corresponding snapshots collected during testing are shown in Fig.~\ref{fig:fig3}(g). 
With quantum activation the ONN learns to run smoothly, whereas a linear ONN fails to acquire a viable control policy and falls down early in the episode. 
The insets in Fig.~\ref{fig:fig3}(g) plot the averaged velocity $v_{x}$ over 10 episodes, highlighting the stability enabled by optical nonlinearity. 
The above results confirm that strong nonlinearity is essential for enabling complex capabilities in deep ONNs. 

\subsection{Nonlinearity-limited power requirements for large-scale networks} 
Having established the importance of strong nonlinearity at the device level, we now examine how the choice of nonlinearity constrains the optical power of large-scale systems. 
We consider only the optical power needed to drive the nonlinear activations, and ask what this nonlinearity-limited power would be for present-day LLMs.
As a quantitative baseline, we note that in standard digital MLPs, common activation functions typically increase the total curvature by $r_{\mathrm{digital}}\simeq 1.045\sim 1.095$ per layer (see Supplementary Note S10 for details). 
We therefore ask: for a given physical nonlinearity, what is the minimum optical intensity $I_{\min}$ required to match this digital baseline?
Using our established framework, we compute the expressive power $r(I)$ for three representative platforms: Kerr nonlinearity in a $50~\mu$m long silicon waveguide \cite{dinu2003third, dulkeith2006self}, saturable absorption in stacked graphene layers \cite{bao2009atomic, lau2022comparison} ($15$ nm total thickness), and our proposed quantum activation unit (see Supplementary Note S11 for details). 
We find that maintaining the target expressive power in silicon requires intensities exceeding $72.6\,\mathrm{W}/\mu\mathrm{m}^{2}$, while graphene requires approximately $0.02\,\mathrm{W}/\mu\mathrm{m}^{2}$.
In contrast, the quantum activation unit achieves the same baseline at merely $\sim {0.24}\,\mathrm{nW}/\mu\mathrm{m}^{2}$. 
This represents an efficiency improvement of roughly $8.3\times10^{7}$ times relative to graphene and $3.0\times10^{11}$ times relative to silicon.
This quantitative comparison reveals the inadequacy of conventional nonlinear materials, and shows that the quantum nonlinearity proposed here can overcome this limitation.

\begin{figure*}[t]
  \centering
  \includegraphics[width=1.0\linewidth]{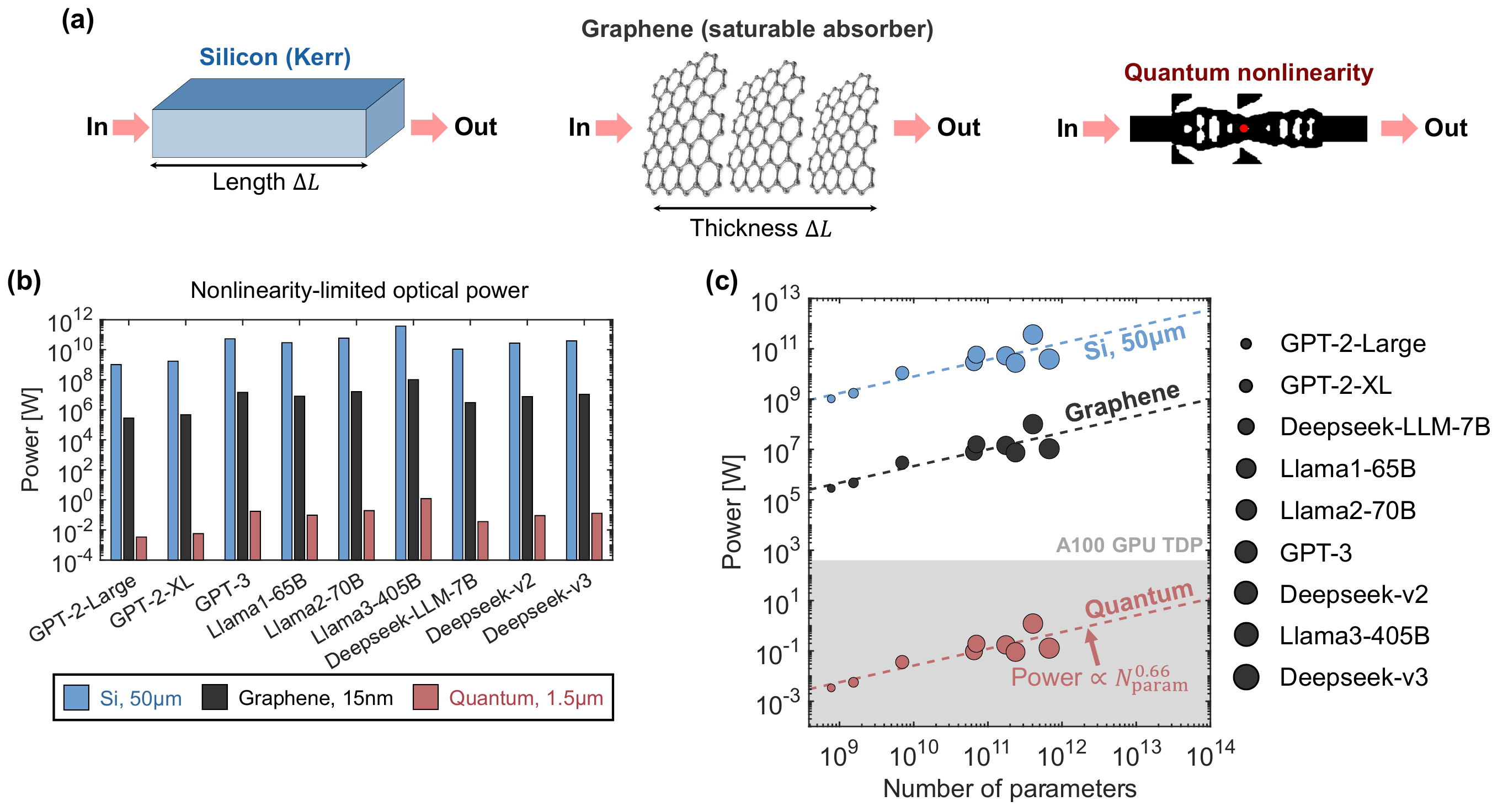}
  \caption{
  \textbf{Scalability of nonlinearity-limited power requirements. }
  (a) Schematic illustration of nonlinear platforms: conventional bulk material (silicon, $50~\mu\text{m}$ length), 2D saturable absorber (stacked graphene, $15~\text{nm}$ thickness), and the proposed quantum activation unit. 
  (b) Histogram showing the estimated nonlinearity-limited optical power of optical LLMs. 
  Conventional materials demand prohibitive power levels, while the proposed scheme is not restricted by nonlinearity.
  (c) Estimated nonlinearity-limited optical power versus model size $N_{\text{param}}$. The shaded area marks the thermal design power of a single NVIDIA A100 GPU ($\approx 400$W), shown only as a reference.
  ONNs follow a sublinear scaling $P \propto N_{\text{param}}^{0.66}$, indicating that optical computing has a growing advantage as models scale up. }
  \label{fig:fig4}
\end{figure*}
Given these intensity thresholds, we next estimate the nonlinearity-limited optical power for LLM-scale models. 
For a standard decoder-only transformer architecture with context length $L_{\mathrm{seq}}$, embedding dimension $d_{\mathrm{model}}$, and $L$ transformer layers \cite{vaswani2017transformer}, we estimate the optical input dimension per layer as $3 L_{\mathrm{seq}} d_{\mathrm{model}}$, accounting for the parallel projection of query, key, and value matrices.
Assuming each optical neuron occupies an effective cross-sectional area $A \approx 0.1~\mu\mathrm{m}^{2}$, corresponding to an on-chip waveguide mode,~\cite{shen2017} and is driven at $I_{\min}$, the nonlinearity-limited optical power is estimated as
\begin{equation}
  P \approx I_{\min} \, A \cdot \bigl( 3 L_{\mathrm{seq}} d_{\mathrm{model}} \bigr) \cdot L.
  \label{eq:LLM_power}
\end{equation}
This value should be understood as a lower bound set by the nonlinear activation alone and does not represent system-level power consumption.
Using the architectural parameters of representative LLMs ranging from GPT-2 to DeepSeek-V3 \cite{radford2019language, brown2020gpt3, touvron2023llama1, touvron2023llama2, dubey2024llama3, bi2024deepseek, liu2024deepseek-v2, liu2024deepseek-v3} (see Supplementary Note S12 for details), we evaluate Eq.~\eqref{eq:LLM_power} and summarize the results in Fig.~\ref{fig:fig4}(b). 
When using conventional nonlinearities based on silicon or graphene, the required optical power quickly reaches prohibitive levels (exceeding $10^{8}\,\mathrm{W}$ for the largest models). 
However, the proposed quantum architecture keeps this nonlinearity-limited power below $1.2~\mathrm{W}$ across all investigated models.
Finally, we analyze how this power requirement scales with model size. 
Fig.~\ref{fig:fig4}(c) plots the estimated optical power against the total number of trainable parameters, $N_{\text{param}}$.
We also indicate the thermal design power of a single high-end GPU (NVIDIA A100, $\approx 400$W) as a shaded area. We include this only as a rough reference for scale. We do not use it to claim that an optical system would consume less total power than a GPU.
For ONNs the data points follow a sublinear scaling law, $P \propto N_{\text{param}}^{0.66}$. 
This behavior stems from the geometric nature of the network: the parameter count grows with the ``volume'' of the network ($\sim L \cdot d_{\text{model}}^{2}$), whereas the required optical power scales with the number of inputs ($\sim L \cdot d_{\text{model}} \cdot L_\text{seq}$). 
Such distinction leads to a scaling exponent smaller than one, consistent with known results \cite{anderson2023optical}.
We would like to point out that the dimensionality of current integrated photonic circuits~\cite{lightelligence2025pace, ahmed2025upaia} remains far below LLM scale, so a near-term large-scale optical LLM is more plausibly realized on free-space platforms (e.g., diffractive networks). The same sublinear scaling applies to such platforms, since the exponent comes from the network geometry and is platform-independent.
In contrast, the power consumption of electronic processors typically scales linearly with $N_{\text{param}}$.
The nonlinearity-limited optical power therefore scales more favorably with model size, although translating this into a system-level energy advantage would require addressing the additional overheads.
Overall, these results suggest that quantum nonlinearity can substantially relax the optical-power bottleneck that conventional nonlinear materials impose at large scale.
By shifting to quantum nonlinearities, it should be possible to construct optical deep learning models within a feasible power budget.

\section{Discussion} 
In summary, we have presented a  theoretical framework to address the nonlinearity bottleneck in optical computing. 
At the device level, by integrating quantum emitters with inverse-designed nanophotonic structures, strong nonlinearity can be realized at intensities below $1~\text{nW}/\mu\text{m}^2$.
This enables complex functionalities ranging from nonlinear classification to reinforcement learning, presenting a clear performance gap over linear ONNs. 
Moreover, we have developed a general theoretical framework to quantify the expressive power of arbitrary nonlinear physical systems, which in turn allows us to determine the light intensity requirements. 
At the scale of large language models, we estimate the nonlinearity-limited optical power and find that it stays at the watt level, with a favorable sublinear scaling in model size.
Together, these results suggest that the lack of nonlinearity is not a fundamental limit, but rather an engineering challenge that could be overcome with quantum technologies. 

Despite these advances, transforming our theoretical proposal into large-scale hardware is still facing several practical challenges. 
A key trade-off exists between intensity threshold and operation bandwidth: the high sensitivity is inherently related to the emitter's long radiative lifetime. For a bare SiV$^-$ center, the intrinsic response bandwidth lies in the sub-GHz range, below the $10$--$50$~GHz modulation rates typically used in optical computing. We point out that this limit is not fundamental, since the response speed can be increased through Purcell enhancement in optimized photonic structures \cite{englund2005PhC}. Our inverse-designed activation units already exhibit an emergent Purcell factor $F_P \approx 2.74$ (Supplementary Note~S13),
and Purcell-enhanced linewidth as large as $2\pi \times 4.6$~GHz has been demonstrated in photonic crystal cavity \cite{evans2018}, offering a realistic route to GHz-scale operation.
Another challenge is the inhomogeneity of solid-state emitters. 
Since the activation requires each emitter to be resonant with the optical signal, the inhomogeneous broadening of transition frequencies becomes an obstacle as the system scales up.
To tackle this issue, solutions such as DC Stark tuning have been demonstrated to tune the resonance of individual emitters \cite{laucht2010stark}, which provides a route to align each emitter independently. 
Regarding integration, while deterministic placement of emitters remains difficult, recent advances in fabrication techniques offer promising solutions for large-scale integration \cite{chen2017laser, laferriere2022QD, yama2026}. 
Finally, solid-state quantum emitters often require cryogenic operation to suppress dephasing and to approach lifetime-limited linewidths \cite{Sipahigil2014, Rogers2014}, which introduces an additional power overhead for cooling. 

Looking forward, this work shows that in order to unlock the full potential of optical computing, we should exploit the strong nonlinearity provided by quantum emitters rather than conventional bulk materials. 
By combining inverse-designed nanophotonics with modern deep learning theory, we provide a path toward low-power optical neural computing, in which the lack of strong nonlinearity is no longer the limiting factor.
Realizing this vision will ultimately pave the way toward sustainable, next-generation artificial intelligence. 

\begin{acknowledgments}
The authors would like to thank Erfan Khoram, Zhicheng Wu, Prof. Jennifer. T. Choy, and Prof. E. Sifakis for insightful discussions. 
\end{acknowledgments}


\begin{arxivindependentbibliography}{84}
\end{arxivindependentbibliography}

\clearpage

\setcounter{section}{0}
\renewcommand{\thesection}{S\arabic{section}}
\renewcommand{\theHsection}{S\arabic{section}}
\setcounter{figure}{0}
\renewcommand{\thefigure}{S\arabic{figure}}
\renewcommand{\theHfigure}{S\arabic{figure}}
\setcounter{table}{0}
\renewcommand{\thetable}{S\arabic{table}}
\renewcommand{\theHtable}{S\arabic{table}}
\setcounter{equation}{0}
\renewcommand{\theequation}{S\arabic{equation}}
\renewcommand{\theHequation}{S\arabic{equation}}

\begin{center}
{\large\bfseries Supplementary Information for\par}
\vspace{0.5em}
{\LARGE\bfseries ``Quantum Nonlinearity for Optical Neural Computing''\par}
\vspace{1.5em}
Qingyi Zhou,$^{1,*}$ Jungmin Kim,$^{1,*}$ Yutian Tao,$^{2,*}$ Guoming Huang,$^1$\\
Ming Zhou,$^3$ Zewei Shao,$^1$ and Zongfu Yu$^1$\\[0.8em]
$^1$Department of Electrical and Computer Engineering, University of Wisconsin-Madison,
Madison, WI 53706, USA\\
$^2$The Computer Sciences Department, University of Wisconsin-Madison,
Madison, WI 53706, USA\\
$^3$Department of Electrical Engineering, Stanford University,
Stanford, CA 94305, USA\\[0.8em]
$^*$These authors contributed equally to this work.
Correspondence: \href{mailto:qzhou75@wisc.edu}{qzhou75@wisc.edu}.
\end{center}
\vspace{1em}

\section{Comparison with existing literature}
\begin{figure}[h]
\centering
\includegraphics[width=0.75\linewidth]{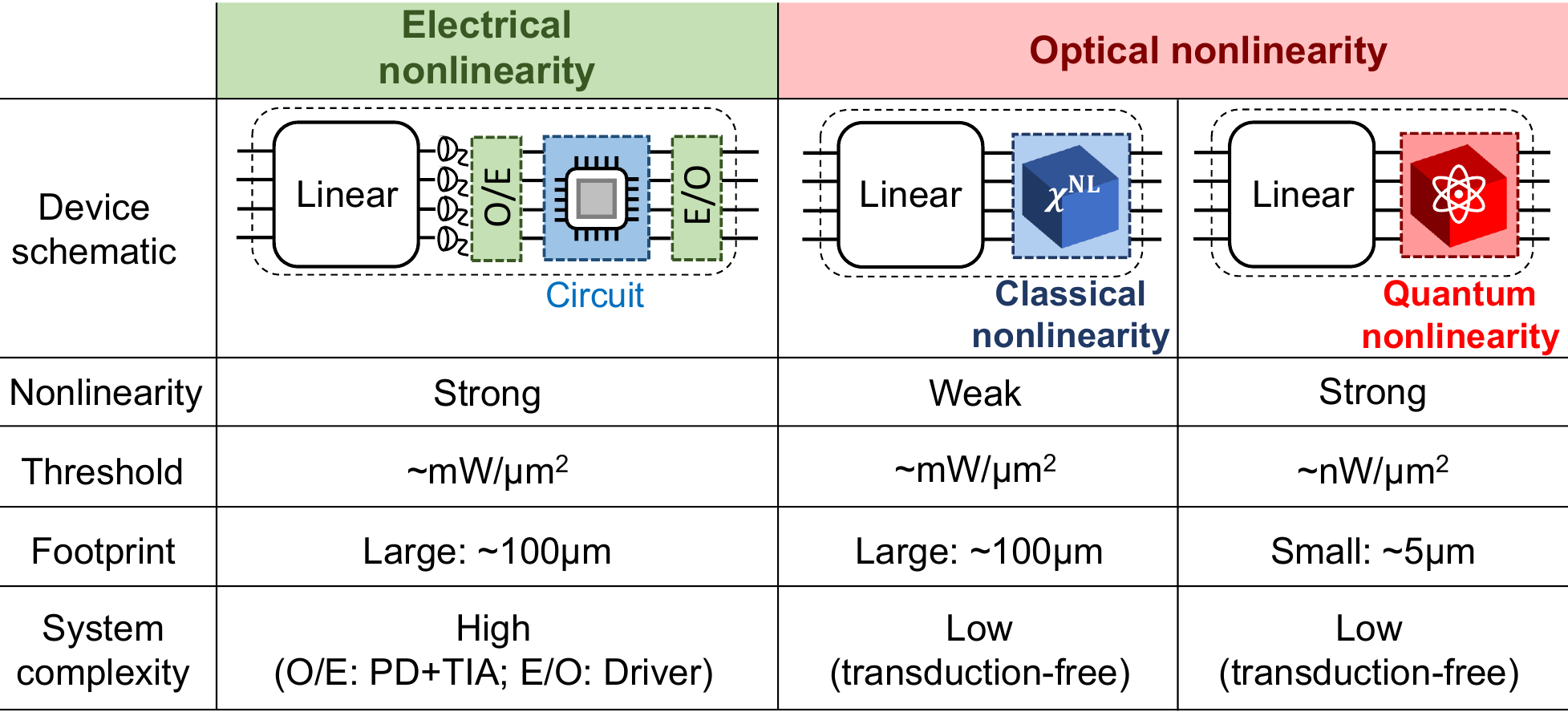}
\caption{
\textbf{Comparison of ONN architectures with different mechanisms for nonlinear activations. }
Carrying out nonlinear operations on electrical signal (the left column) requires frequent O/E/O conversions, which leads to extra overhead. 
Optical nonlinearities based on conventional materials (the middle column) are typically weak, thus requiring high laser power and large footprint. 
Our proposed quantum-enhanced nonlinearity (the right column) can function well under low laser power, with a much smaller footprint. }
\label{fig:schematic_compare}
\end{figure}

In this part, we compare the proposed quantum-enhanced nonlinear activation unit with other activation units proposed in existing literature. 
We first show in Fig.~\ref{fig:schematic_compare} a high-level comparison between different types of implementations. 
As shown in the left column, most existing optical neural network architectures choose to carry out activation function in electrical domain. 
In that sense, optical-electrical-optical (O/E/O) conversion is required for every single layer, making the system complicated.  
On the other hand, all-optical neural network architectures do not require O/E/O conversion for the intermediate layers. 
However, when relying on conventional materials, these ONNs often suffer from intrinsically small nonlinear susceptibilities. 
To compensate for such weak nonlinearity, high laser power and large footprint are often required for all-optical NNs. 
In order to address these challenges, we propose to construct an activation unit based on the strong nonlinearity of single quantum emitter. 
As shown in the right column of Fig.~\ref{fig:schematic_compare}, the proposed scheme reaches strong nonlinearity at low light intensity ($\sim \text{nW/}\mu\text{m}^{2}$), and has a very compact footprint ($\sim~5\mu\text{m}$) at the same time. 

\begin{figure}[h]
\centering
\includegraphics[width=0.65\linewidth]{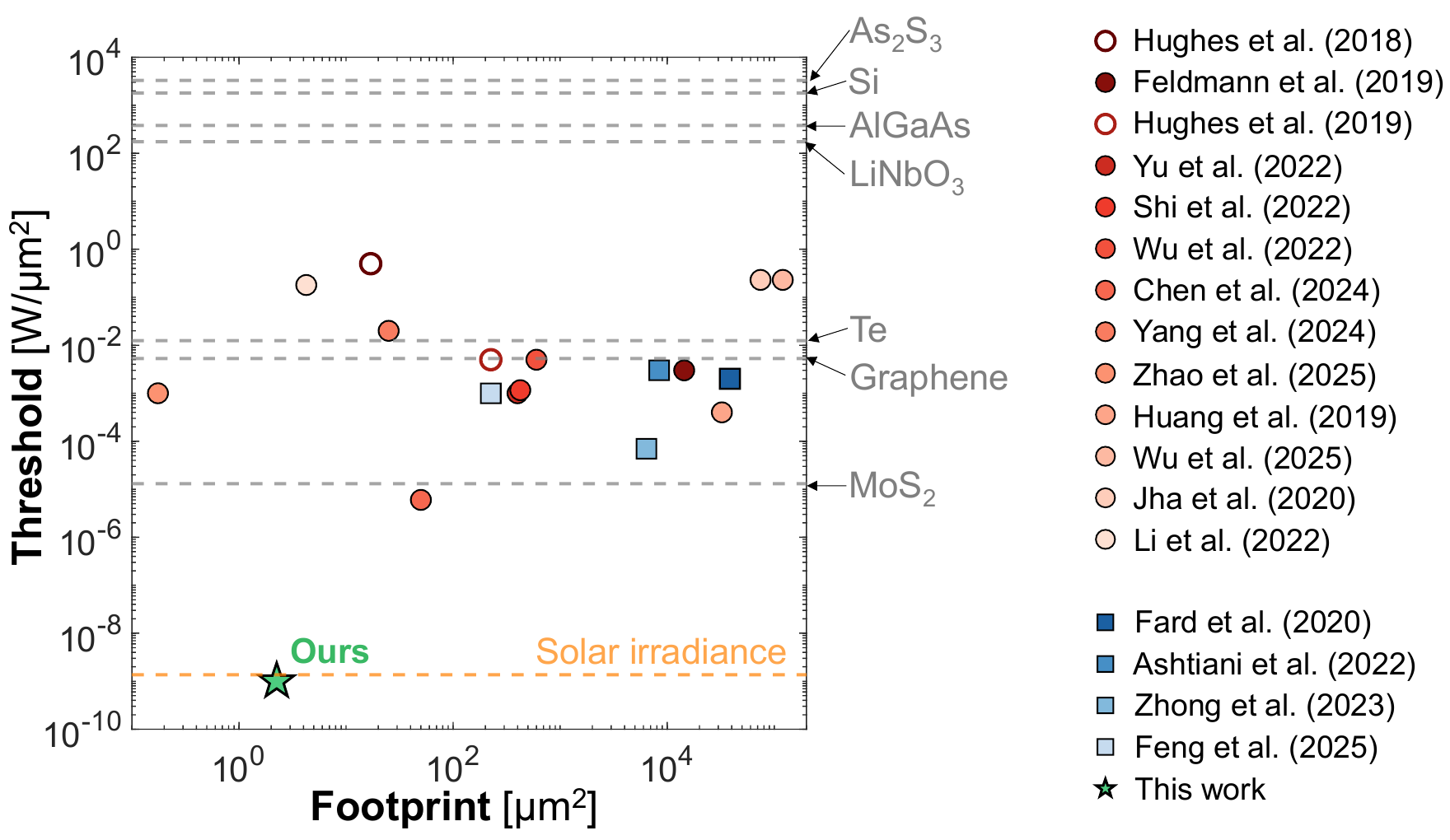}
\caption{\textbf{Detailed comparison. } 
We summarize the performance of different activation units proposed in literature \cite{si:hughes2018,si:feldmann2019,si:hughes2019,si:yu2022reconfigurable,si:shi2022nonlinear,si:wu2022low,si:chen2024ultra,si:yang2024inverse,si:zhao2025high,si:huang2019programmable,si:jha2020reconfigurable,si:wu2025field,si:li2023all,si:pour2020,si:ashtiani2022,si:zhong2023graphene,si:feng2025femtojoule}.
Unfilled markers indicate theoretical proposals. 
The nonlinearity thresholds of several materials have been visualized using gray dashed lines. Each material threshold corresponds to the input intensity at which the nonlinear index change reaches $\Delta n = 0.01$. The comparison is intended as an order-of-magnitude overview. }
\label{fig:literature_compare}
\end{figure}

Next, we display the comparison results by comparing our proposed activation unit with existing literature. 
We focus on two metrics: the footprint, defined as the 2D area occupied by the activation unit; the nonlinearity threshold, defined as the required light intensity to observe obvious nonlinear effect. 
The results are shown in Fig.~\ref{fig:literature_compare} as a scatter plot. 
Red circles correspond to all-optical implementations \cite{si:hughes2018,si:feldmann2019,si:hughes2019,si:yu2022reconfigurable,si:shi2022nonlinear,si:wu2022low,si:chen2024ultra,si:yang2024inverse,si:zhao2025high,si:huang2019programmable,si:jha2020reconfigurable,si:wu2025field,si:li2023all}, while blue squares correspond to opto-electronic implementations \cite{si:pour2020,si:ashtiani2022,si:zhong2023graphene,si:feng2025femtojoule}.
Unfilled circles indicate that the corresponding papers are theoretical proposals and include no experimental results. 
As a reference, we also visualize the nonlinearity thresholds for seven different materials, namely: \\
(1) silicon, with $\chi^{(3)}=2.45\times10^{-19}~\text{m}^{2}/\text{V}^{2}$ \cite{si:dulkeith2006self}; \\
(2) MoS$_\text{2}$, with $\chi^{(2)}=10^{-7}~\text{m}/\text{V}$ \cite{si:kumar2013second}; \\
(3) tellurene, with $\chi^{(2)}=3.26\times10^{-9}~\text{m}/\text{V}$ \cite{si:liu2025Te}; \\
(4) graphene, with intensity threshold $5.3\times 10^{-3}~\text{W}/\mu\text{m}^{2}$ \cite{si:bao2009atomic,si:bao2011monolayer}. \\
(5) As$_2$S$_3$, with $n_2 = 3\times10^{-18}~\text{m}^{2}/\text{W}$ \cite{si:lamont2008supercontinuum}; \\
(6) AlGaAs, with $n_2 = 2.6\times10^{-17}~\text{m}^{2}/\text{W}$ \cite{si:pu2016efficient}; \\
(7) LiNbO$_3$, with $\chi^{(2)} = 27~\text{pm/V}$ \cite{si:wang2017linbo3shg}; \\
The solar irradiance ($1360.8~\text{W}/\text{m}^{2}$) \cite{si:kopp2011solar} is also marked using the orange dashed line, which is much lower than the nonlinearity thresholds of existing activation units. 
The threshold of each reference material is obtained by requiring the nonlinear refractive-index change to reach $\Delta n = 0.01$. For a material with third-order nonlinearity, $\Delta n = n_2 I$, so the threshold intensity is $I_\text{th} = \Delta n / n_2$, with the nonlinear index obtained from the third-order susceptibility through $n_2 = 3\chi^{(3)} / (4 \varepsilon_0 c\, n_0^2)$. For a material with second-order nonlinearity, the field-induced index change is $\Delta n = \chi^{(2)} E$, giving a threshold intensity $I_\text{th} = \tfrac{1}{2}\varepsilon_0 c\,(\Delta n / \chi^{(2)})^2$. For the individual works compiled in Fig.~\ref{fig:literature_compare}, the operating intensity is estimated from the input condition reported in each paper. Note that this comparison is intended as an order-of-magnitude overview rather than a precisely standardized benchmark.
At the same time, our proposed activation unit shows a threshold that is orders of magnitude lower than all existing implementations. 

\newpage
\section{Dipole moment of two-level system} 
In this part we derive the dipole moment of a quantum emitter, under the illumination of a monochromatic incident wave. 
The quantum emitter is modeled as an ideal two-level system (TLS), with ground state $|g\rangle$ and excited state $|e\rangle$. 
The resonance frequency is denoted as $\omega_{0} = \frac{E_{e} - E_{g}}{\hbar}$. 
We denote the dipole moment matrix element as $\bm{d}_{0} = \langle e | \hat{\bm{d}} | g \rangle$, which is assumed to be a real quantity without loss of generality. 
Such a TLS has radiative decay rate $\Gamma_{0} = \frac{d_{0}^{2} \omega_{0}^{3}}{3\pi \hbar \epsilon_{0} c_{0}^{3}}$ in free space. 
The Hamiltonian of such a TLS driven by electromagnetic field can be written as
\begin{equation}
    H = H_0 + H_{\mathrm{int}}
  = \hbar\omega_0\ket{e}\bra{e}
  - \bm{d}_{0} \cdot \bm{E}(t)\,\bigl(\ket{g}\bra{e} + \ket{e}\bra{g}\bigr). 
\end{equation}
The time-evolution of the density matrix obeys the von-Neumann equation $\frac{d\rho}{dt} = -\frac{i}{\hbar}[H,\,\rho]$.
The operators can be resolved in the two-level basis:
\begin{equation}
    \rho = \begin{pmatrix}\rho_{gg}&\rho_{ge}\\[3pt]\rho_{eg}&\rho_{ee}\end{pmatrix},
    \,\,
    H_0 = \begin{pmatrix}0&0\\[3pt]0&\hbar\omega_0\end{pmatrix}, 
    \,\,
    H_{\mathrm{int}}(t) = \begin{pmatrix}
    0 & -\,\bm{d}_{0}\!\cdot\!\bm E(t) \\[3pt]
    -\,\bm{d}_{0}\!\cdot\!\bm E(t) & 0
    \end{pmatrix}.
\end{equation}
By calculating the commutators, the above equations give
\begin{equation}
    -\,\frac{i}{\hbar}\,[H_0,\rho]
    =
    \begin{pmatrix}
    0 & i\,\omega_0\,\rho_{ge} \\[4pt]
    -\,i\,\omega_0\,\rho_{eg} & 0
    \end{pmatrix},
\end{equation}
\begin{equation}
    -\,\frac{i}{\hbar}\,[H_{\mathrm{int}},\rho]
    = \frac{i\,(\bm{d}_{0}\!\cdot\!\bm E)}{\hbar}
    \begin{pmatrix}
    \rho_{ge}-\rho_{eg} & \rho_{ee}-\rho_{gg} \\[3pt]
    \rho_{gg}-\rho_{ee} & \rho_{eg}-\rho_{ge}
    \end{pmatrix}.
\end{equation}
The time-evolution of density matrix $\rho$ thus follows
\begin{equation}
\begin{cases}
\displaystyle
\frac{d\,\rho_{ee}}{dt}
= -\Gamma_{0}\,\rho_{ee} + \frac{2\,(\bm{d}_{0}\!\cdot\!\bm E)}{\hbar}\,\text{Im}\bigl[\rho_{eg}\bigr], \\[8pt]
\displaystyle
\frac{d\,\rho_{eg}}{dt}
= \Bigl(-\,i\omega_0 - \frac{\Gamma_{0}}{2} \Bigr)\rho_{eg}
+ \frac{i\,(\bm{d}_{0}\!\cdot\!\bm E)}{\hbar}\bigl(1 - 2\rho_{ee}\bigr).
\end{cases}
\label{eq:rho_evolve}
\end{equation}
Note that the radiative decay rate $\Gamma_{0}$ has been introduced phenomenologically. 
Next we assume that the polarization of the incident wave is aligned with the dipole moment $\bm{d}_{0}$, so that the inner product can be replaced by $d\cdot E(t)$. 
The monochromatic incident field oscillates at the laser frequency $\omega_{d}$, so that its time-profile can be written as $E(t) = E_{0} \cos(\omega_{d} t) = \frac{E_{0}}{2} (e^{i\omega_{d}t} + e^{-i\omega_{d}t} )$.
The corresponding Rabi frequency can be defined as $\Omega = \frac{d_{0} E_{0}}{\hbar}$.
We further define the detuning $\Delta = \omega_{d} - \omega_{0}$. Transforming into the frame rotating at the laser frequency, where the slowly-varying coherence is $\tilde{\rho}_{eg} = \rho_{eg} e^{i \omega_{d} t}$, the equations become
\begin{equation}
\begin{cases}
\displaystyle
\frac{d\,\rho_{ee}}{dt}
= \Omega\,\text{Im}\bigl[\tilde{\rho}_{eg}\bigr] - \Gamma_{0}\,\rho_{ee} , \\[8pt]
\displaystyle
\frac{d\,\tilde{\rho}_{eg}}{dt}
=  i \frac{\Omega}{2}\bigl(1 - 2\rho_{ee}\bigr) + \Bigl( i\Delta - \frac{\Gamma_{0}}{2} \Bigr) \tilde{\rho}_{eg}.
\end{cases}
\end{equation}
If we further assume that the laser is on resonance with the emitter, i.e.\ $\Delta = 0$, the above equations reduce to
\begin{equation}
\begin{cases}
\displaystyle
\frac{d\,\rho_{ee}}{dt}
= \Omega\,\text{Im}\bigl[\tilde{\rho}_{eg}\bigr] - \Gamma_{0}\,\rho_{ee} , \\[8pt]
\displaystyle
\frac{d\,\tilde{\rho}_{eg}}{dt}
=  i \frac{\Omega}{2}\bigl(1 - 2\rho_{ee}\bigr) - \frac{\Gamma_{0}}{2} \tilde{\rho}_{eg},
\end{cases}
\end{equation}
The steady state of the above equation can be derived by enforcing the time-derivatives to be zero.
The results can be summarized as
\begin{equation}
    \rho_{ee}(t\rightarrow \infty) = \frac{1}{2}\cdot \frac{\Omega^{2}}{\Omega^{2} + \Gamma_{0}^{2}/2},
\end{equation}
\begin{equation}
    \tilde{\rho}_{eg}(t\rightarrow \infty) = \frac{1}{2}\cdot \frac{i\Gamma_{0} \Omega}{\Omega^{2} + \Gamma_{0}^{2}/2}.
\end{equation}
At steady state, the TLS shows a dipole moment $\bm{d} = 2\bm{d}_{0} \cdot \tilde{\rho}_{eg}(t\rightarrow\infty) = 2i \bm{d}_{0} \cdot \frac{\Gamma_{0} \Omega}{\Gamma_{0}^{2} + 2 \Omega^{2}} $. 
The physical intuition behind these equations can be understood by checking two extreme cases: 
when the incident field is weak ($\Omega \ll \Gamma_{0}$), the TLS acts like a resonant dipole whose dipole moment $\bm{d} \propto \frac{\Omega}{\Gamma_{0}}$ is proportional to $E_{0}$; 
on the other hand, when the incident field is strong ($\Omega \gg \Gamma_{0}$), the TLS saturates with $\rho_{ee}\approx \frac{1}{2}$ and the dipole moment vanishes. 

The above derivations have assumed that the TLS is ideal. In reality, the presence of non-radiative decay rate $\Gamma_{\text{nrad}}$ leads to imperfect quantum efficiency $\eta_{Q} = \frac{\Gamma_{0}}{\Gamma_{0} + \Gamma_{\text{nrad}}} < 1$. By including this non-radiative decay, the time-evolution in eq.~\ref{eq:rho_evolve} becomes
\begin{equation}
\begin{cases}
\displaystyle
\frac{d\,\rho_{ee}}{dt}
= -(\Gamma_{0} + \Gamma_\text{nrad})\,\rho_{ee} + \frac{2\,(\bm{d}_{0}\!\cdot\!\bm E)}{\hbar}\,\text{Im}\bigl[\rho_{eg}\bigr], \\[8pt]
\displaystyle
\frac{d\,\rho_{eg}}{dt}
= \Bigl(-\,i\omega_0 - \frac{\Gamma_{0} + \Gamma_\text{nrad}}{2} \Bigr)\rho_{eg}
+ \frac{i\,(\bm{d}_{0}\!\cdot\!\bm E)}{\hbar}\bigl(1 - 2\rho_{ee}\bigr).
\end{cases}
\end{equation}
Note that we do not include pure dephasing rate $\gamma_{\phi}$ in this paper. The dipole moment now becomes $\bm{d} = 2i \bm{d}_{0} \cdot \frac{ (\Gamma_{0} + \Gamma_\text{nrad}) \cdot \Omega}{ ( \Gamma_{0}  + \Gamma_\text{nrad} )^{2} + 2 \Omega^{2}} $, which is weaker compared to the ideal case. 

\newpage
\section{Nonlinear FDFD simulation}
In this part, we explain in detail the formulation of our full-wave simulations. Our simulations are implemented by introducing a nonlinear dipole source into the standard finite-difference frequency domain (FDFD) approach. 
\subsection{FDFD with nonlinear dipole}
We start from Maxwell's equations in frequency domain:
\begin{equation}
    \nabla \times \bm{H} = \bm{J} - i\omega \epsilon_{r} \epsilon_{0} \bm{E}, ~~ 
    \nabla \times \bm{E} = i\omega \mu_{0} \bm{H},
\end{equation}
where $\bm{J}$ denotes the current source, and $\epsilon_{r}$ is the spatial distribution of relative permittivity. 
By eliminating $\bm{H}$ field the above equations can be simplified as
\begin{equation}
    -\nabla \times \nabla \times \bm{E} + \epsilon_{r} k_{0}^{2} \bm{E} = -i\omega \mu_{0} \bm{J},
    \label{eq:maxwell_E}
\end{equation}
where $k_{0}=\omega/c_{0}$ denotes the wave vector. 
In conventional electromagnetic simulations where the current source $\bm{J}$ is given, the $\bm{E}$ field can be obtained by treating eq.~\ref{eq:maxwell_E} as a linear equation $\bm{A} \bm{x} = \bm{b}$, where
\begin{equation}
    \bm{A}\sim -\nabla \times \nabla \times + \epsilon_{r} k_{0}^{2}, ~~ \bm{x} \sim \bm{E}, ~~ \bm{b} \sim -i\omega \mu_{0} \bm{J}. 
\end{equation}
In our case, however, the current source $\bm{J}$ is provided by the TLS and is related to the $\bm{E}$ field. 
Specifically, since we are working in frequency domain, the relationship between current density $\bm{J}$ and the TLS' dipole moment $\bm{d}$ is
\begin{equation}
    \bm{J}(\bm{r}) = -i\omega \bm{d}\cdot \delta(\bm{r} - \bm{r}_{0}),
\end{equation}
where $\delta(\bm{r} - \bm{r}_{0})$ is the Dirac $\delta$-function, indicating that the dipole is placed at $\bm{r} = \bm{r}_{0}$.

For a single TLS, its dipole moment has already been derived as
\begin{equation}
    \bm{d} = \frac{\Omega (-\Delta + i\frac{\Gamma_{0}}{2})}{\Delta^{2} + (\frac{\Gamma_{0}}{2})^{2} + \frac{\Omega^{2}}{2} } \cdot \bm{d}_{0}.
    \label{eq:dipole_moment}
\end{equation}
When discretized on a uniform grid (spatial resolution $\Delta x$), the above equation corresponds to a current source term at location $\bm{r}_{0}$: 
\begin{equation}
    \bm{b}_\text{TLS}(\bm{r}=\bm{r}_{0}) = -i\omega_{0}\mu_{0} \cdot \frac{-i\omega_{0} \bm{d}}{(\Delta x)^{3}} 
    = -\frac{\mu_{0} \omega_{0}^{2}}{(\Delta x)^{3}}\cdot \frac{\Omega (-\Delta + i\frac{\Gamma_{0}}{2})}{\Delta^{2} + (\frac{\Gamma_{0}}{2})^{2} + \frac{\Omega^{2}}{2} } \cdot \bm{d}_{0},
    \label{eq:b_TLS}
\end{equation}
where we have included $\frac{1}{(\Delta x)^{3}}$ as a discretization of the Dirac $\delta$-function. 
Note that $\bm{E}^\text{inc}(\bm{r}_{0})$ does not equal to the total field $\bm{E}(\bm{r}_{0})$ since the TLS cannot be driven by its primary radiation field \cite{si:zhou2024fdtd,si:wang2025lorentz}. 
In the next part we will explain how this can be handled numerically. 

\subsection{Discretized nonlinear equation}
In our FDFD framework, the simulation domain is discretized into $N$ grid points. 
The Maxwell's equations can be discretized as $\bm{A} \bm{x} = \bm{b}_{0} + \bm{b}_\text{TLS}$, 
where $\bm{A}$ corresponds to the operator $-\nabla\times\nabla\times + \epsilon_{r} k_{0}^{2}$ and is treated as an $3N\times 3N$ sparse matrix; 
$\bm{x}$ corresponds to $\bm{E}$ field distribution and is treated as an $3N\times 1$ vector; the source term $\bm{b}_{0}$ is determined by the incident waves, while the other source term $\bm{b}_\text{TLS}$ takes into account all TLSs. 

For the sake of convenience, we first define a $3N\times 3N$ sparse matrix $\bm{M}$, which serves as a ``mask'' and helps identify the locations of all TLSs: 
\begin{equation}
    M_{ij} = 
    \begin{cases}
    \displaystyle
    1, ~\text{if}~i=j~\text{and index}~(i-N)~\text{contains a TLS;} \\[8pt]
    \displaystyle
    0, ~\text{otherwise}. 
    \end{cases}
\end{equation}
Here we use the index $i-N$ to identify the $y$ component of the electric field, since the dipole moment of our quantum emitter is assumed to align with the $y$-axis. 
As we have mentioned, in eq.~\ref{eq:b_TLS} the dipole is driven by $\bm{E}^\text{inc}(\bm{r}_{0})$ instead of the total field $\bm{E}(\bm{r}_{0})$, since a field-driven dipole cannot be driven by its own primary radiation field. 
Therefore the primary radiation field should be excluded explicitly. 
Based on the above intuition, we further assume that the detune $\Delta = 0$. 
By defining the parameter $\alpha_\text{TLS} = -\frac{2i \mu_{0} \omega_{0}^{2} d_{0}^{2}}{\Gamma_{0}\hbar (\Delta x)^{3}}$, the discretized version of eq.~\ref{eq:b_TLS} can be written as
\begin{equation}
    \bm{b}_\text{TLS} = \alpha_\text{TLS} \bm{M} \cdot f\left(\bm{x} - \beta_\text{rad} \bm{b}_\text{TLS} \right),
    \label{eq:b_TLS_numerical}
\end{equation}
where the element-wise nonlinear function $f(\cdot)$ is defined as 
\begin{equation}
    f(x) = \frac{x}{1+2| \frac{d_{0} x }{\Gamma_{0} \hbar} |^{2}}. 
\end{equation}
The term $\beta_\text{rad} \bm{b}_\text{TLS}$ denotes the primary radiation field that TLSs produce at their own locations. 
Coefficient $\beta_\text{rad}$ is a constant which only depends on the background medium that the TLSs are embedded in. It is determined numerically through a small FDFD simulation, by simulating a point source in homogeneous background and extracting the Green's function at the source location. 
Based on eq.~\ref{eq:b_TLS_numerical}, we now proceed to determine the final form of our nonlinear FDFD. 
Maxwell's equations are discretized as $\bm{A} \bm{x} = \bm{b}_{0} + \bm{b}_\text{TLS}$, which gives $\bm{b}_\text{TLS} = \bm{A} \bm{x} - \bm{b}_{0}$. 
Thus, by eliminating $\bm{b}_\text{TLS}$ we arrive at a nonlinear equation w.r.t. electric field $\bm{x}$: 
\begin{equation}
    \bm{F}(\bm{x}) \triangleq \bm{A} \bm{x} - \bm{b}_{0} - \alpha_\text{TLS} \cdot \bm{M} \cdot f\left[ \bm{x} - \beta_\text{rad} (\bm{A} \bm{x} - \bm{b}_{0}) \right] = \bm{0}. 
    \label{eq:nonlinear_FDFD}
\end{equation}
The above nonlinear equation $\bm{F}(\bm{x})=0$ is what we aim to solve. 

\subsection{Newton-Raphson solver}
In order to solve for $\bm{F}(\bm{x})=0$ we apply the Newton-Raphson method \cite{si:nocedal2006numerical}. 
The vanilla Newton-Raphson algorithm can be understood as updating the solution $\bm{x}$ in an iterative manner: $\bm{x} \rightarrow \bm{x} - \bm{J}(\bm{x})^{-1} \bm{F}(\bm{x})$, where $\bm{J}(\bm{x}) \triangleq \frac{\partial \bm{F}(\bm{x})}{\partial \bm{x}}$ denotes the Jacobian matrix. 
In this part we derive the explicit form of $\bm{J}(\bm{x})$. 
For the sake of convenience, we define an intermediate variable $\bm{s} = \bm{x} - \beta_\text{rad} (\bm{A} \bm{x} - \bm{b}_{0})$. 
The change of variable leads to
\begin{equation}
    \bm{J}(\bm{x}) = \bm{A} - \alpha_\text{TLS} \cdot \bm{M} \cdot \frac{df(\bm{s})}{d\bm{s}} \cdot \frac{\partial \bm{s}}{\partial \bm{x}}. 
\end{equation}
Note that $\frac{\partial \bm{s}}{\partial \bm{x}} = \mathbbm{1} - \beta_\text{rad} \bm{A}$, where $\mathbbm{1}$ denotes the identity matrix. 
However, since $\bm{s}$ is complex, $\frac{df(\bm{s})}{d\bm{s}}$ should be treated carefully using Wirtinger derivatives \cite{si:wirtinger1927}. By re-writing $f(\bm{s})$ as
\begin{equation}
    f(\bm{s}, \bm{s}^{*}) = \frac{\bm{s}}{1+2\left(\frac{d_{0}}{\Gamma_{0} \hbar}\right)^{2}  \bm{s}^{*} \odot \bm{s}},
\end{equation}
the corresponding Wirtinger derivatives can be derived as
\begin{equation}
    \frac{\partial f}{\partial \bm{s}} = \frac{1}{\left( 1+2(\frac{d_{0}}{\Gamma_{0} \hbar})^2 |\bm{s}|^{2} \right) ^{2}},
    ~~~
    \frac{\partial f}{\partial \bm{s}^{*}} = -\frac{2 (\frac{d_{0}}{\Gamma_{0} \hbar})^2 \bm{s}^{2}}{\left( 1+2(\frac{d_{0}}{\Gamma_{0} \hbar})^2 |\bm{s}|^{2} \right) ^{2}}.
\end{equation}
Therefore, the Jacobian can be calculated as
\begin{equation}
\begin{aligned}
    \frac{\partial\bm{F}}{\partial \bm{x}} 
    &= \bm{A} - \alpha_\text{TLS} \bm{M} \cdot \left( \frac{\partial f(\bm{s}, \bm{s}^{*})}{\partial \bm{s}} \cdot \frac{\partial \bm{s}}{\partial \bm{x}} + \frac{\partial f(\bm{s}, \bm{s}^{*})}{\partial \bm{s}^{*}} \cdot \frac{\partial \bm{s}^{*}}{\partial \bm{x}} \right) \\
    &= \bm{A} - \alpha_\text{TLS} \bm{M} \cdot \text{diag}\left\{ \frac{1}{\left( 1+2(\frac{d_{0}}{\Gamma_{0} \hbar})^2 |\bm{s}|^{2} \right) ^{2}} \right\} \cdot (\mathbbm{1} - \beta_\text{rad} \bm{A}),
\end{aligned}
\end{equation}
\begin{equation}
\begin{aligned}
    \frac{\partial\bm{F}}{\partial \bm{x}^{*}} 
    &= - \alpha_\text{TLS} \bm{M} \cdot \left( \frac{\partial f(\bm{s}, \bm{s}^{*})}{\partial \bm{s}} \cdot \frac{\partial \bm{s}}{\partial \bm{x}^{*}} + \frac{\partial f(\bm{s}, \bm{s}^{*})}{\partial \bm{s}^{*}} \cdot \frac{\partial \bm{s}^{*}}{\partial \bm{x}^{*}} \right) \\
    &= - \alpha_\text{TLS} \bm{M} \cdot \text{diag}\left\{ -\frac{2 (\frac{d_{0}}{\Gamma_{0} \hbar})^2 \bm{s}^{2}}{\left( 1+2(\frac{d_{0}}{\Gamma_{0} \hbar})^2 |\bm{s}|^{2} \right) ^{2}} \right\} \cdot (\mathbbm{1} - \beta^{*}_\text{rad} \bm{A}^{*}).
\end{aligned}
\end{equation}
Based on the above results, we can finally arrive at the formulation of our Newton-Raphson solver. 
At each iteration, the solution is updated via $\bm{x} \rightarrow \bm{x} + \Delta \bm{x}$, where $\Delta\bm{x} = \text{Re}(\Delta\bm{x}) + i\text{Im}(\Delta\bm{x})$ can be solved using the following linear equation:
\begin{equation}
    \begin{bmatrix}
    \text{Re}(\frac{\partial\bm{F}}{\partial \bm{x}} ) +\text{Re}(\frac{\partial\bm{F}}{\partial \bm{x}^{*}} ) & -\text{Im}(\frac{\partial\bm{F}}{\partial \bm{x}} ) +\text{Im}(\frac{\partial\bm{F}}{\partial \bm{x}^{*}} ) \\[4pt]
    \text{Im}(\frac{\partial\bm{F}}{\partial \bm{x}} ) +\text{Im}(\frac{\partial\bm{F}}{\partial \bm{x}^{*}} ) & \text{Re}(\frac{\partial\bm{F}}{\partial \bm{x}} ) - \text{Re}(\frac{\partial\bm{F}}{\partial \bm{x}^{*}} )
    \end{bmatrix}
    \begin{bmatrix}
    \text{Re}(\Delta\bm{x}) \\[4pt]
    \text{Im}(\Delta\bm{x})
    \end{bmatrix}
    =
    -\begin{bmatrix}
    \text{Re}(\bm{F}(\bm{x})) \\[4pt]
    \text{Im}(\bm{F}(\bm{x}))
    \end{bmatrix}.
\end{equation}

\subsection{Benchmarking nonlinear FDFD}
\begin{figure}[h]
\centering
\includegraphics[width=1.0\linewidth]{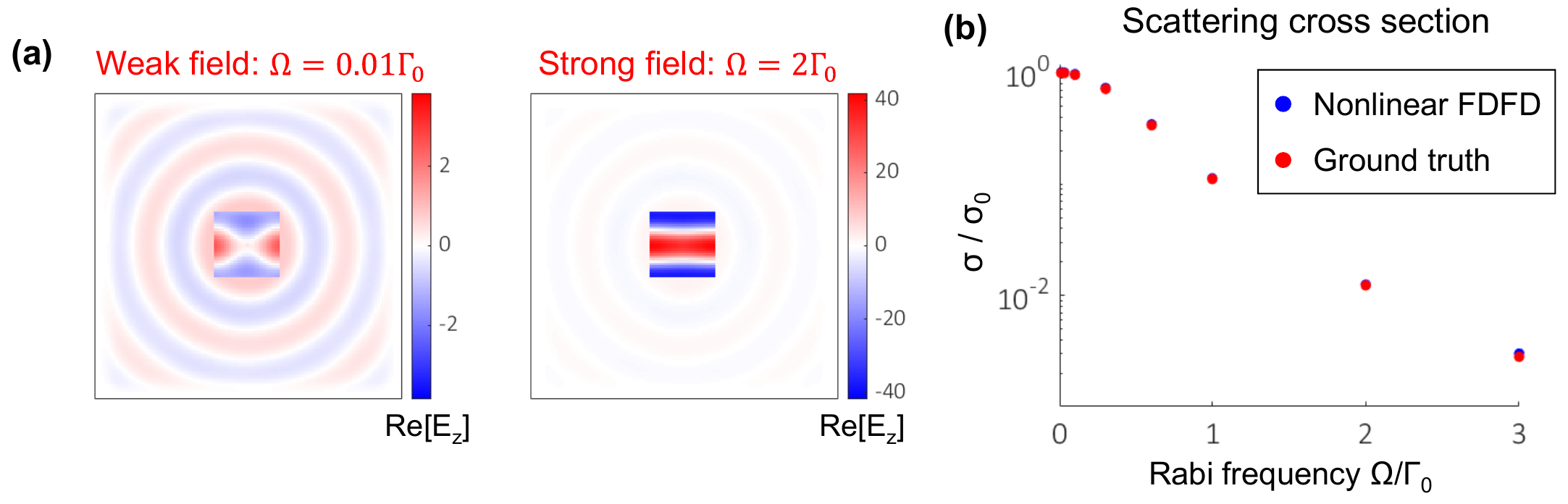}
\caption{
\textbf{Benchmark: scattering cross section of a single TLS. }
(a) The $\text{Re}[E_{z}]$ field distributions for weak incidence ($\Omega / \Gamma_{0} = 0.01$) and strong incidence ($\Omega / \Gamma_{0} = 2$). The square region in the middle marks the total-field region, introduced to provide plane-wave excitation. 
(b) Numerical results of $\sigma / \sigma_{0}$ under different Rabi frequencies. Our FDFD matches perfectly well with the ground truth. }
\label{fig:benchmark_scattering}
\end{figure}

We provide a simple example that serves as the benchmark for our nonlinear FDFD solver. 
Consider a single TLS placed inside vacuum, and a plane-wave serves as the incident wave. 
It is well-known that the scattering cross section of a resonant dipole cannot exceed $\sigma_{0}$, which is $\sigma_{0} = \frac{3\lambda_{0}^{2}}{2\pi}$ in 3D space. 
Here we benchmark using a 2D simulation, where the maximum scattering cross section is $\sigma_{0} = \frac{2\lambda_{0}}{\pi}$. 
We now consider the fact that this TLS can be saturated when the intensity of the incident wave is strong. 
Based on eq.~\ref{eq:dipole_moment}, when on resonance, the scattering cross section can be calculated analytically as
\begin{equation}
    \frac{\sigma}{\sigma_{0}} = \left( \frac{\Gamma_{0}^{2}}{\Gamma_{0}^{2} + 2\Omega^{2}} \right)^{2}, 
\end{equation}
where $\Omega$ denotes the Rabi frequency. 
We now vary $\Omega$ and check the scattering cross section obtained via our nonlinear FDFD solver. 
The simulation results are displayed in Fig.~\ref{fig:benchmark_scattering}. 
In Fig.~\ref{fig:benchmark_scattering}(a) we compare two different scenarios, namely $\Omega / \Gamma_{0} = 0.01$ and $\Omega / \Gamma_{0} = 2$. The TLS is placed at the middle. 
The scattered field distributions are different for these two cases, since for the second case the TLS already begins to saturate. 
Detailed comparison can be found in Fig.~\ref{fig:benchmark_scattering}(b). The cross sections calculated using our FDFD agrees perfectly with the analytical results, justifying the correctness of our numerical scheme. 

\newpage
\section{Adjoint-based optimization}

\subsection{Why optimization is required}
\begin{figure*}[h]
  \centering
  \includegraphics[width=1.0\linewidth]{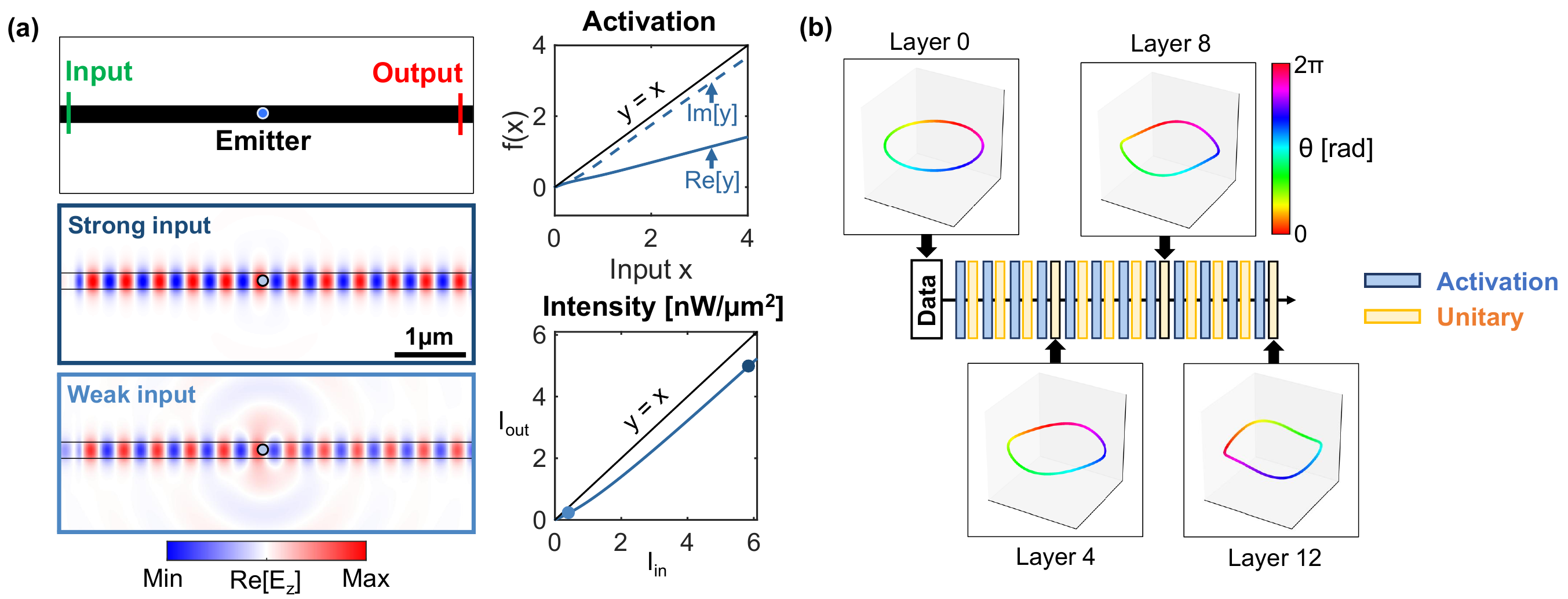}
  \caption{
  \textbf{Waveguide-based activation unit.} 
  (a) The input-output relationship shows weak nonlinearity. 
  (b) The evolution of data trajectory during forward-propagation. Since the activation function is only weakly nonlinear, the trajectory is only deformed slightly, indicating low expressive power. }
  \label{fig:waveguide_activation}
\end{figure*}

In this part we explain why we turn to adjoint optimization when designing the quantum-enhanced activation unit. 
We start from a simple implementation by embedding a quantum emitter inside a silicon waveguide ($250$ nm width). 
The quantum emitter has a radiative decay rate of $\Gamma_{0} = 2\pi \times 94$ MHz when placed inside homogeneous environment. 
The input-output relationship of such a waveguide QED system is displayed in Fig.~\ref{fig:waveguide_activation}(a). 
Here the input $x$ and the output $f(x)$ are calculated by multiplying the electric field amplitude with $\frac{d_{0}}{\Gamma_{0} \hbar}$. 
Although the intensity threshold is low ($\sim\text{nW}/\mu\text{m}^{2}$), the curves only show weak nonlinearity. 
Such behavior can be understood by inspecting the $E_{z}$ field distribution: at low intensity the TLS acts like a resonant dipole and scatters the incident field out of the waveguide. 
Na\"ively placing the emitter inside the silicon waveguide leads to a low coupling efficiency. 

We further test the effectiveness of this activation function, using the technique developed in Refs.~\citenum{si:poole2016,si:raghu2017}. 
We first construct a 24-layer ONN, plugging in this activation unit at each layer. 
All hidden layers contain $N_\text{dim} = 512$ neurons. 
The input data points form a circular trajectory, parameterized by $\theta\in [0, 2\pi]$:
\begin{equation}
    \bm{x}(\theta) = \sqrt{N} \cdot (\bm{u} \cos(\theta) + \bm{v} \sin(\theta)), 
\end{equation}
where $\bm{u}$ and $\bm{v}$ are two random unit vectors which form an orthonormal basis for a 2D subspace. The trajectory contains $10000$ sampling points, distributed uniformly on $\theta\in [0, 2\pi]$. 
We track the evolution of such a trajectory during the forward-propagation process, as shown in Fig.~\ref{fig:waveguide_activation}(b). For all the data trajectories we carry out a dimension reduction through principal component analysis (PCA), so that they can be visualized as curves in 3D space. 
It can be observed that the trajectory only deforms slightly. 
This is consistent with our understanding that using an activation function that is only weakly nonlinear leads to poor expressivity. 
Based on such intuition, we turn to adjoint optimization and see whether we can find a structure that increases the interaction between the incident light and the emitter, which hopefully results in a stronger nonlinearity. 

\subsection{Adjoint method}
In this part we explain how the conventional adjoint-based optimization can be generalized to our case, where we solve for nonlinear Maxwell's equations. 
Based on eq.~\ref{eq:nonlinear_FDFD}, the physical constraint is set by the following nonlinear equation: 
\begin{equation}
    \bm{A} \bm{x} - \bm{b}_{0} - \alpha_\text{TLS} \cdot \bm{M}\cdot f\left( \bm{x} - \beta_\text{rad} \bm{A} \bm{x} + \beta_\text{rad} \bm{b}_{0} \right) = \bm{0}, 
    \label{eq:nonlinear_FDFD_copy}
\end{equation}
where the matrix $\bm{A} = \bm{A}(\bm{p})$ depends on a set of design parameters $\bm{p}$. 
We would like to optimize for a given target function, denoted as $L(\bm{x})$. 
Without loss of generality, here we assume that $L$ does not depend explicitly on $\bm{p}$. 
Our aim is to calculate its gradient $\frac{dL}{dp_{i}}$ w.r.t. the design parameters. 
To achieve this, we first differentiate eq.~\ref{eq:nonlinear_FDFD_copy} w.r.t. parameter $p_{i}$: 
\begin{equation}
    \frac{\partial \bm{A}}{\partial p_{i}} \bm{x} + 
    \bm{A}\frac{d\bm{x} }{dp_{i}} 
    - \alpha_\text{TLS} \bm{M} \cdot \frac{df}{dp_{i}} = 0. 
\end{equation}
From the above equation, we can derive
\begin{equation}
    \frac{d\bm{x} }{dp_{i}} = -\left( \bm{A}^{'} \right)^{-1} \left[ \mathbbm{1} + \alpha_\text{TLS}\cdot \beta_\text{rad} \bm{M} \cdot \text{diag}\left\{ \frac{df}{d\bm{s}} \right\} \right] \cdot \frac{\partial \bm{A}}{\partial p_{i}} \bm{x},
\end{equation}
where we have defined $\bm{s} = \bm{x} - \beta_\text{rad} (\bm{A} \bm{x} - \bm{b}_{0})$, and $\bm{A}^{'} = \bm{A} - \alpha_\text{TLS} \bm{M} \cdot \text{diag}\left\{ \frac{df}{d\bm{s}} \right\} \cdot (\mathbbm{1} - \beta_\text{rad} \bm{A}) $. 
We now define the adjoint as
\begin{equation}
    \bm{x}_\text{adj} = -\left( \bm{A}^{'} \right)^{-T} \frac{\partial L}{\partial \bm{x}},
\end{equation}
which can be calculated by solving the following equation: 
\begin{equation}
    \left( \bm{A}^{'} \right)^{T} \bm{x}_\text{adj} = - \frac{\partial L}{\partial \bm{x}}. 
\end{equation}
Note that although the forward simulation involves solving a nonlinear equation, the adjoint simulation only involves solving a linear equation. 
Finally, we arrive at the gradient
\begin{equation}
    \frac{dL}{dp_{i}} = \left( \frac{\partial \bm{A}}{\partial p_{i}} \bm{x} \right)^{T} 
    \left[ \mathbbm{1} + \alpha_\text{TLS}\cdot \beta_\text{rad} \bm{M} \cdot \text{diag}\left\{ \frac{df}{d\bm{s}} \right\} \right] \cdot \bm{x}_\text{adj}. 
    \label{eq:adjoint}
\end{equation}
Based on the above eq.~\ref{eq:adjoint}, the calculated gradient can be used to update the design parameters $\bm{p}$.

\subsection{Material platform of the activation unit}
\begin{figure*}[h!]
  \centering
  \includegraphics[width=0.75\linewidth]{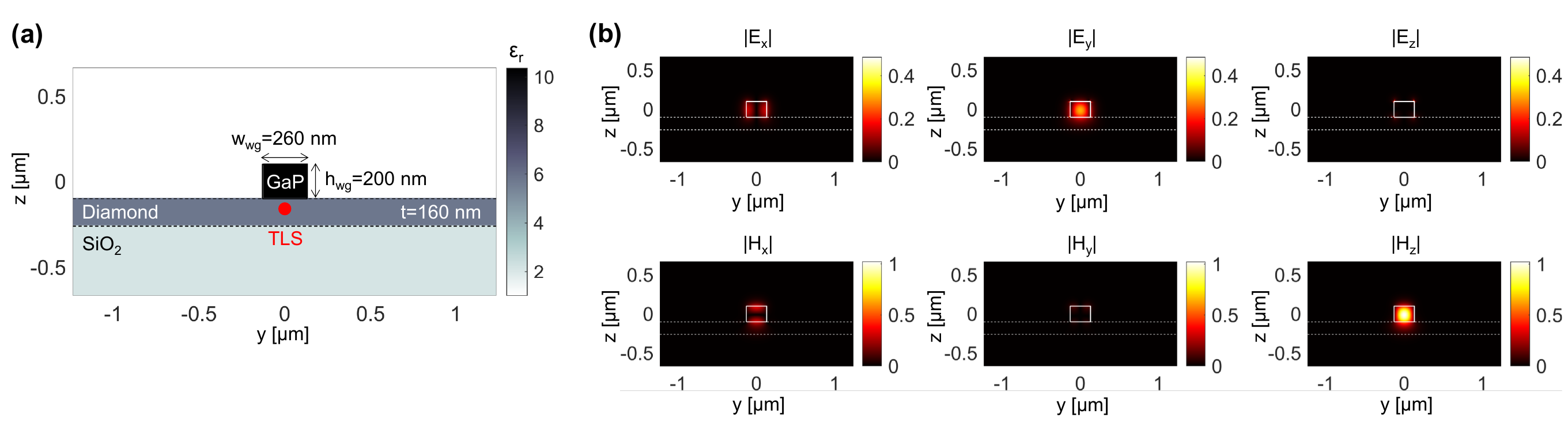}
  \caption{\textbf{Input waveguide of the 3D activation unit.} 
  (a) Cross-section of the GaP-on-diamond strip waveguide. 
  (b) Fundamental TE$_{0}$ guided-mode profile at $\lambda_{0} = 737.134$~nm, with effective index $n_\text{eff} = 2.646$.}
  \label{fig:waveguide_mode_3D}
\end{figure*}

In the revised proposal, the nonlinear activation unit is designed on a realistic GaP-on-diamond platform, and verified with 3D full-wave simulations. 
The device operates at the SiV$^{-}$ zero-phonon line $\lambda_{0} = 737.134$~nm \cite{si:hepp2014}.
It consists of a patterned top GaP layer of thickness $h_\text{wg} = 200$~nm on an unetched thin-film diamond layer ($t = 160$~nm) \cite{si:ding2024}, sitting on the SiO$_{2}$ substrate. 
The refractive indices of the involved materials are $n_\text{GaP} = 3.22$, $n_\text{diamond} = 2.40$, and $n_{\text{SiO}_2} = 1.45$. 
We choose GaP, not only because it is transparent at $737$~nm, but also because similar structures have been demonstrated in experiments \cite{si:chakravarthi2023,si:yama2026}.
Light is coupled into and out of the device through GaP strip waveguides ($w_\text{wg} = 260$~nm wide, $h_\text{wg} = 200$~nm thick), propagating along $x$ direction. 
We use the fundamental TE mode with effective index $n_\text{eff} = 2.646$. The waveguide cross-section and the corresponding guided-mode profile are shown in Fig.~\ref{fig:waveguide_mode_3D}.

The nonlinearity is provided by a single SiV$^{-}$ color center, modeled as a two-level system with its dipole moment oriented along $y$ (transverse to the propagation direction). 
The SiV$^{-}$ is positioned $30$~nm below the diamond surface. 
Such single emitters can be created deterministically by counted ion implantation, which has already been demonstrated experimentally \cite{si:schroder2017,si:titze2022}.
The emitter is assumed to operate at cryogenic temperature $T=4$K \cite{si:jahnke2015}.

\subsection{Optimization setup}
In this part we provide details related to optimizing the activation unit. 

\textbf{3D activation unit (GaP-on-diamond).} The design region is $1.5 \times 0.7~\mu\text{m}^{2}$, located in the top GaP layer of the platform described above, with the single SiV$^{-}$ emitter at its center. A conic filter with a radius of $100$~nm is applied to control the minimum feature size.

\textbf{2D study (silicon-on-insulator).} The 2D results presented in this paper were instead obtained on a silicon-based platform. 
The design region is $5~\mu\text{m} \times 1~\mu\text{m}$.
TLSs are placed at the middle part of the design region.
The spacing between two neighboring TLSs is fixed as $1~\mu$m.
A waveguide mode ($E_{z}\neq 0$; the effective refractive index is $n_\text{eff}=2.951$) serves as the input.

For both the 2D and 3D designs, the relative permittivity at each in-plane location $(x,y)$ inside the design region is parameterized as
\begin{equation}
    \epsilon_{r}(x, y) = \epsilon_\text{background} + (\epsilon_\text{structure} - \epsilon_\text{background}) \cdot \frac{1}{1+\exp\left[-\beta \cdot p(x,y)\right]},
    \label{eq:permittivity_parameterize}
\end{equation}
For the 2D design, the background and structural materials are SiO$_2$ and Si, respectively, such that $\epsilon_\text{background}=\epsilon_{\text{SiO}_2}=1.44^2$ and $\epsilon_\text{structure}=\epsilon_\text{Si}=3.47^2$. For the 3D design, they are vacuum and GaP, respectively, such that $\epsilon_\text{background}=\epsilon_\text{vac}=1$ and $\epsilon_\text{structure}=\epsilon_\text{GaP}=3.22^2$.
The parameter $\beta$ controls the binarization process and is gradually increased from $5.0$ to $150.0$ for the 2D design and from $5.0$ to $200.0$ for the 3D design throughout the optimization.

We now introduce the figure-of-merit (FoM) that we are trying to optimize. In order to achieve a strong nonlinearity, we consider two extreme cases: \\
(1) \textbf{case 1}: suppose the incident wave is so strong that all TLSs have saturated. The TLSs are transparent and can be ignored. \\
(2) \textbf{case 2}: suppose the incident wave is very weak, such that the TLSs can be treated as linear dipoles under the weak-excitation limit. \\
The above two cases are shown in Fig.~\ref{fig:fom_curves} schematically. 
The FoM consists of two parts. Specifically, for case $i$ ($i=1,2$), denote the amplitude measured at the output port as $a_{i}$, which is a complex scalar, calculated from the inner product between the electromagnetic field distribution and the desired waveguide mode. 
The input amplitude is normalized to be $1$. 
First, for case 1 the transmission should be maximized: 
\begin{equation}
    \text{FoM}_{1} = |a_{1}|^{2}.
\end{equation}
Second, in order to show strong nonlinearity, the difference between $a_{1}$ and $a_{2}$ should be as large as possible: 
\begin{equation}
    \text{FoM}_{2} = |a_{2} - a_{1}|^{2}.
\end{equation}
The final FoM we intend to optimize can be expressed as a weighted sum of these two terms:
\begin{equation}
    \text{FoM} = \lambda_{1} \cdot \text{FoM}_{1} + \lambda_{2} \cdot \text{FoM}_{2},
    \label{eq:fom}
\end{equation}
where we choose $\lambda_{1}=1.0$ and $\lambda_{2}=2.0$ for all the designs. 
Note that in practice, an extra penalty term is included to minimize the field intensity outside the design region. This term (neglected in the above equation for the sake of brevity) helps us prevent neighboring activation units from coupling to each other. 

We carry out gradient descent w.r.t. parameter $p(x,y)$ and iterate for $500$ steps to ensure convergence. 
Adam optimizer is used. 
The step size, initialized to be $0.02$, is decreased by a factor of $0.9958$ after each iteration.

\subsection{Understanding $N_\text{TLS}=2$ with coupled-mode theory}
\begin{figure*}[h]
  \centering
  \includegraphics[width=1.0\linewidth]{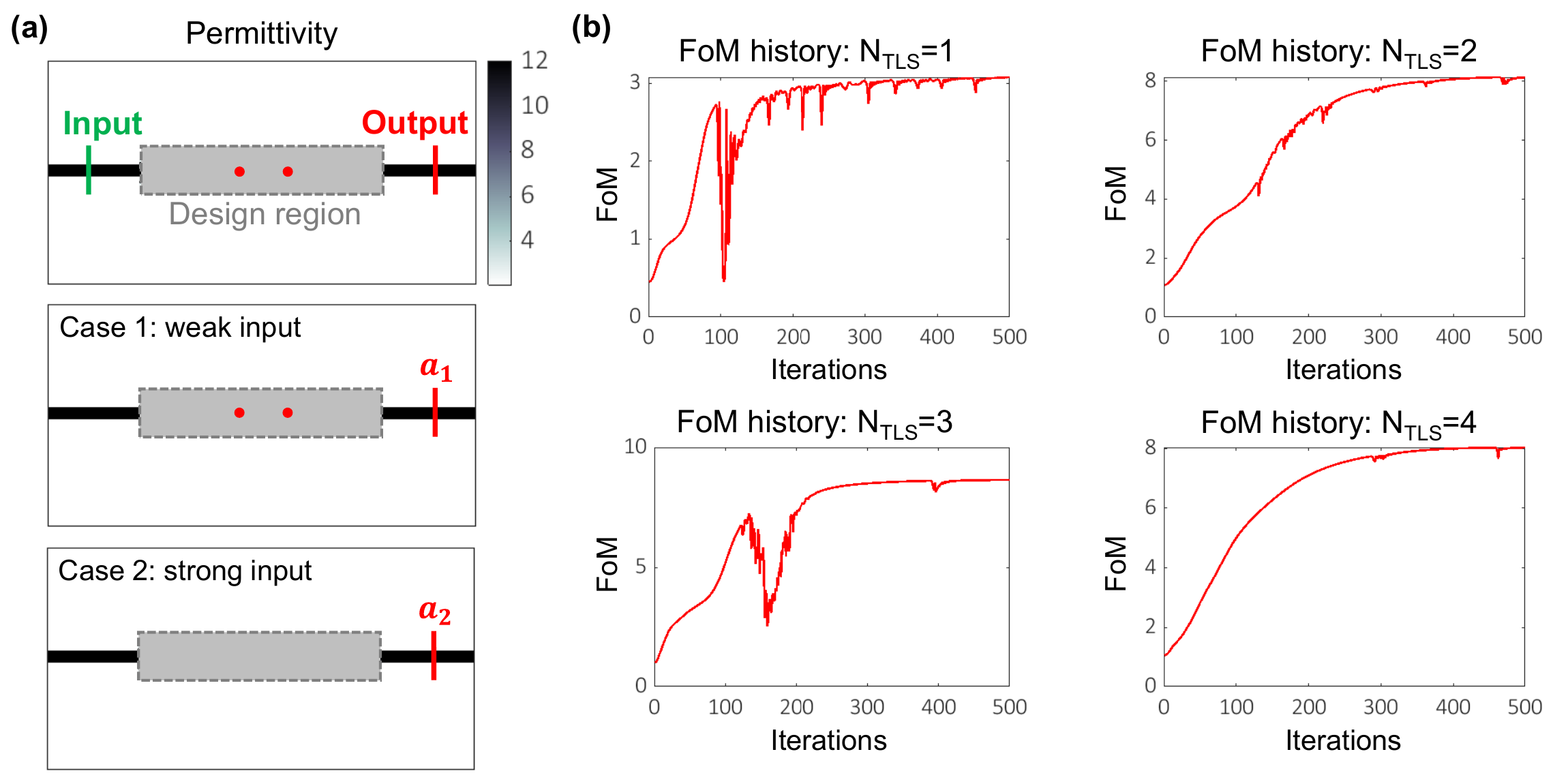}
  \caption{
  \textbf{Optimization of the activation unit (2D).}
  (a) Schematic illustration of the two extreme cases considered.
  (b) FoM curves obtained throughout the optimization. We vary the number of included emitters $N_\text{TLS}\in \{ 1,2,3,4\}$ and find that $N_\text{TLS}\geqslant 2$ results in much better performance. }
  \label{fig:fom_curves}
\end{figure*}

We recorded the FoM values obtained during the optimization of the 2D activation unit, varying the number of TLSs $N_\text{TLS}$ from $1$ to $4$, as visualized in Fig.~\ref{fig:fom_curves}. We observe that using $N_\text{TLS}>1$ leads to a much higher FoM. As shown below using coupled-mode theory, a single resonant dipole cannot modify the transmission coefficient from $-1$ to $1$, which limits the achievable nonlinear contrast. 
While using $N_\text{TLS}\geqslant 2$ improves the performance in this 2D study, deterministically placing more than one emitter inside the same compact device would significantly increase the experimental difficulty. For this reason, our final 3D proposal uses a single emitter ($N_\text{TLS}=1$), which still provides sufficient nonlinear contrast for the tasks demonstrated in this work.

\begin{figure*}[t]
  \centering
  \includegraphics[width=0.8\linewidth]{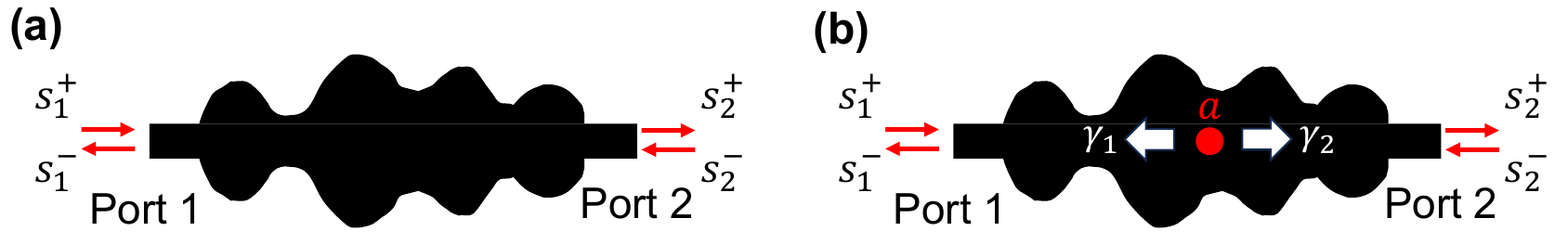}
  \caption{
  \textbf{Modeling the 2-port system using coupled-mode theory.} 
  (a) Without the emitter, the structure forms a background. 
  (b) Including $N_\text{TLS}=1$ emitter, which is treated as a single-mode resonator and couples to both ports. }
  \label{fig:cmt_illus}
\end{figure*}

In this part we use coupled-mode theory to prove why using $N_\text{TLS}=1$ leads to low FoM.
Consider a two-port system, as shown in Fig.~\ref{fig:cmt_illus}(a). 
We define the input vector as $| s^{+} \rangle = (s_{1}^{+}, s_{2}^{+})^{T}$, and the output vector as $| s^{-} \rangle = (s_{1}^{-}, s_{2}^{-})^{T}$. 
Without the resonant dipole, we denote the $S$-matrix of the background structure as a $2\times 2$ matrix $\bm{C}$:
\begin{equation}
    | s^{-} \rangle = \bm{C} | s^{+} \rangle. 
\end{equation}
The system is assumed to be passive. 
Now suppose a resonant dipole is included inside the structure, as shown in Fig.~\ref{fig:cmt_illus}(b). 
This resonant dipole is treated as a single-mode resonator, and the entire system can be modeled using temporal coupled-mode theory (TCMT):
\begin{equation}
    \frac{da}{dt} = (-i \omega_{0} - \gamma) a + \langle d | s^{+} \rangle, 
    \label{eq:tcmt}
\end{equation}
where $\gamma = \gamma_{1} + \gamma_{2}$ denotes the total decay rate, and $| d \rangle = (d_{1}, d_{2})^{T}$ depicts the coupling coefficients between the resonator and the two ports. 
These coefficients satisfy $\gamma_{1} = \frac{1}{2}|d_{1}|^{2}$ and $\gamma_{2} = \frac{1}{2}|d_{2}|^{2}$, which can be derived based on energy conservation constraint \cite{si:fan2003temporal}. 
The output now includes two contributions: 
\begin{equation}
    | s^{-} \rangle = \bm{C} | s^{+} \rangle + a | d \rangle. 
\end{equation}
Suppose we work under single frequency $\omega$. By applying the replacement $\frac{d}{dt} \rightarrow -i\omega$, eq.~\ref{eq:tcmt} can be solved to obtain
\begin{equation}
    a(\omega) = \frac{\langle d | s^{+} \rangle }{-i\Delta + \gamma},
\end{equation}
where detune $\Delta = \omega - \omega_{0}$. 
When on-resonance, the $S$-matrix of the entire system (defined as $|s^{-}\rangle = \bm{S}(\omega) |s^{+}\rangle$) can be derived as
\begin{equation}
    \bm{S}(\omega) = \bm{C}(\omega) + \frac{|d\rangle \langle d|}{-i\Delta + \gamma},
\end{equation}
which, when on resonance ($\Delta = 0$), can be further simplified as
\begin{equation}
    \bm{S}(\omega) = \bm{C}(\omega) + \frac{|d\rangle \langle d|}{ \gamma}.
\end{equation}
Without the resonant dipole, the transmission coefficient is $t=C_{21}$; with the dipole presented, the transmission coefficient becomes $t'=C_{21} + \frac{d_{1}^{*} d_{2}}{\gamma}$. 
The difference between $t'$ and $t$ can be simplified as
\begin{equation}
    |t' - t| = \left| \frac{d_{1}^{*} d_{2}}{\gamma} \right| = \frac{|d_{1}| \cdot |d_{2}|}{\gamma} = \frac{2\sqrt{\gamma_{1} \gamma_{2}}}{\gamma_{1} + \gamma_{2}} \leqslant 1.
\end{equation}
When using only $N_\text{TLS}=1$ emitter, we have proved that for the two extreme cases shown in Fig.~\ref{fig:fom_curves}, their transmission coefficients can differ by $|\Delta t| \triangleq |t' - t| \leqslant 1$. 
This helps explain why the FoM cannot exceed $3$ in Fig.~\ref{fig:fom_curves}(b): since $|a_{1}|\leqslant 1$ and $|a_{2} - a_{1}|\leqslant 1$, eq.~\ref{eq:fom} gives $\text{FoM} \leqslant \lambda_{1} + \lambda_{2} = 3$ given that $\lambda_{1}=1.0$ and $\lambda_{2}=2.0$. 
Using $N_\text{TLS}\geqslant 2$ emitters will break this restriction, boosting the performance to $\text{FoM} > 8$. 

\subsection{Final result}

\subsubsection{3D activation unit}
\begin{figure*}[h]
  \centering
  \includegraphics[width=0.7\linewidth]{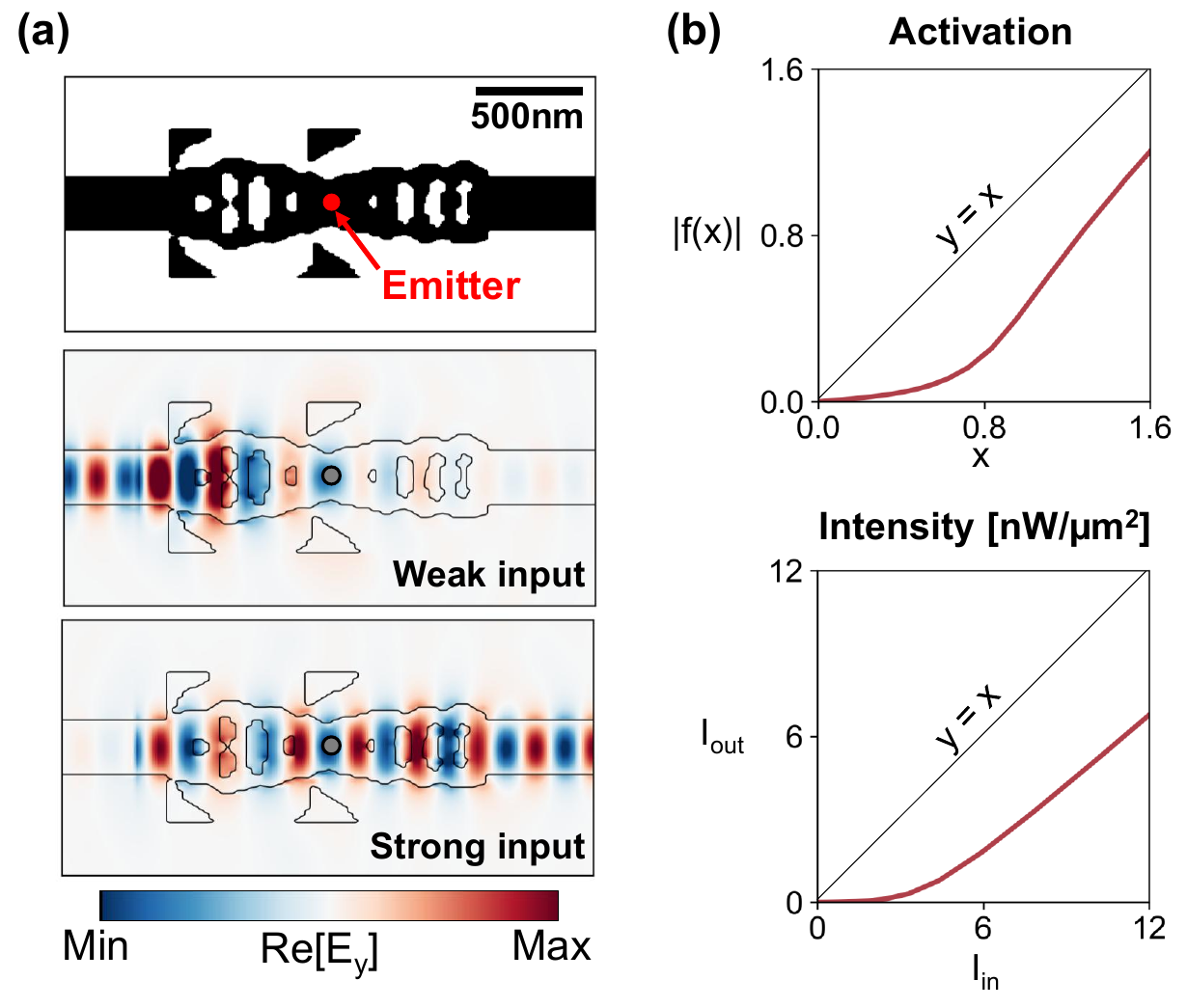}
  \caption{\textbf{3D inverse-designed activation unit.} 
  (a) Optimized binary GaP pattern in the $1.5 \times 0.7~\mu\text{m}^{2}$ design region (black: $200$~nm GaP; white: fully etched), with a single SiV$^{-}$ emitter placed $30$~nm below the diamond surface and dipole oriented along $y$. 
  The $\text{Re}[E_{y}]$ field distributions of two extreme cases are visualized. 
  (b) Nonlinear input-output activation function obtained from full-wave 3D nonlinear FDFD.}
  \label{fig:activation_curve_new}
\end{figure*}

We first present our main result, in which the nonlinear activation unit is designed on the GaP-on-diamond platform described above. 
As shown in Fig.~\ref{fig:activation_curve_new}(a), the inverse-designed structure is a binary GaP pattern, with a single SiV$^{-}$ emitter embedded $30$~nm below the diamond surface. 
The input-output relationship $S_{21}(x)$, obtained from the nonlinear FDFD simulation, is displayed in Fig.~\ref{fig:activation_curve_new}(b). 
The transmission $|S_{21}|^{2}$ rises from zero (weak-field limit) to $0.86$ (strong-field limit). 
Here we assume that the $S_{21}(x)$ curve takes a sigmoid-like form: 
\begin{equation}
    S_{21}(x) = b_\text{offset} + \frac{A e^{i\phi}}{1 + \exp\left[-\beta (\log_{10}(x) - x_{0}) \right]},
    \label{eq:S21_3D}
\end{equation}
Here $b_\text{offset}$ is a complex offset, $A$ and $\phi$ are the amplitude and phase of the sigmoid contribution, $\beta$ controls the sigmoid steepness, and $x_{0}$ sets the offset along the logarithmic input axis. 
Fitting the curve gives $b_\text{offset} = 0.0584 + 0.0603\,i$, $A = 0.8602$, $\phi = 1.5551$, $\beta = 7.972$, and $x_{0} = 0.0299$. 
The fitted activation is then implemented in PyTorch and used as the nonlinear activation when training the all-optical neural network. 
With this activation unit, the three-class spiral classification task is solved perfectly, as shown in Fig.~2 of the main text.

The resulting nonlinear threshold can be as low as a few $\text{nW}/\mu\text{m}^{2}$. 
This can be understood intuitively based on simple back-of-envelope calculation: for a two-level system whose dipole moment matrix element is $d_{0}$, its radiative decay rate is $\Gamma_{0}=\frac{n d_{0}^{2} \omega_{0}^{3}}{3\pi \hbar \epsilon_{0} c_{0}^{3}}$. Here $n$ denotes the refractive index of the emitter's environment. In order to observe saturation effect, the incident electric field $E_\text{inc}$ should satisfy $\frac{d_{0} E_\text{inc}}{\hbar} \sim \Gamma_{0}$. The physical intuition behind this equation is that the Rabi frequency should be comparable to its decay rate $\Gamma_{0}$. A quick estimation can be made by taking $n=2.4$ (corresponds to the refractive index of diamond at $737$ nm) and $\Gamma_{0}=2\pi\times 94$ MHz. The dipole moment can be estimated as $d_{0}\approx 17.7$ Debye. The electric field that's required can be estimated as $E_\text{inc}\approx 1.05\times 10^{3}~\text{V/m}$, corresponding to a light intensity of $I_\text{inc}=\frac{n E_\text{inc}^{2}}{2\eta_{0}} \approx 3.5~\text{nW/}\mu\text{m}^2$.

\subsubsection{2D activation unit}
We also show the performance of our 2D activation unit, which embeds two quantum emitters inside a silicon structure. This 2D unit serves two purposes: it provides the activation function used in the reinforcement learning demonstrations of Fig.~3 in the main text, and its trajectory evolution offers intuition on how the form of the activation function affects the expressive power.
The input-output relationship of this activation unit is displayed in Fig.~\ref{fig:inv_design_activation}(a). 
Contrary to the waveguide-based implementation, this time the curves show strong nonlinearity. 
By inspecting the $E_{z}$ field distribution we realize that the designed structure tries to focus the incident beam onto the two quantum emitters, thus forming two hot spots. 
This enhances the light-matter interaction, leading to a strongly nonlinear response. 

\begin{figure*}[h]
  \centering
  \includegraphics[width=0.45\linewidth]{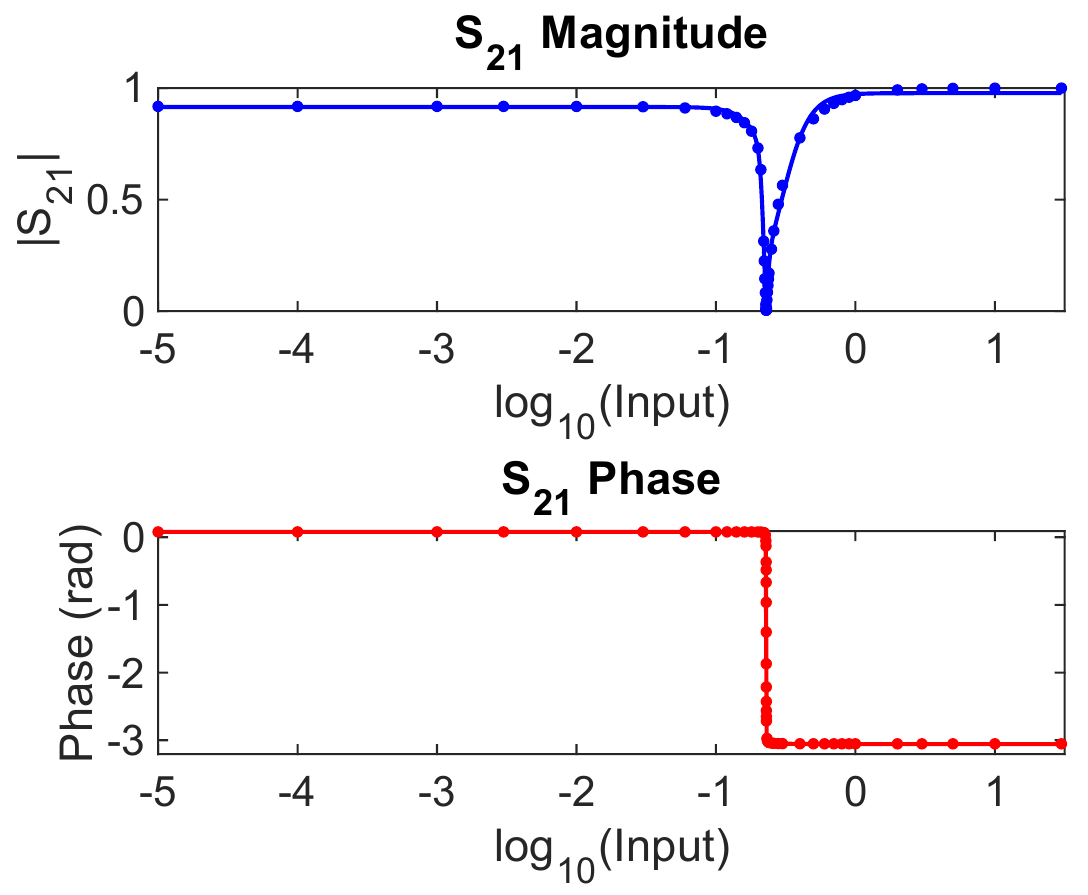}
  \caption{
  \textbf{Curve fitting for the nonlinear activation function.} 
  The upper panel displays the magnitude $|S_{21}|$, while the lower panel displays the phase $\angle S_{21}$.
  The circular markers show results obtained via nonlinear FDFD, while the solid curves represent the fitting results. }
  \label{fig:curve_fit}
\end{figure*}

\begin{figure*}[h]
  \centering
  \includegraphics[width=1.0\linewidth]{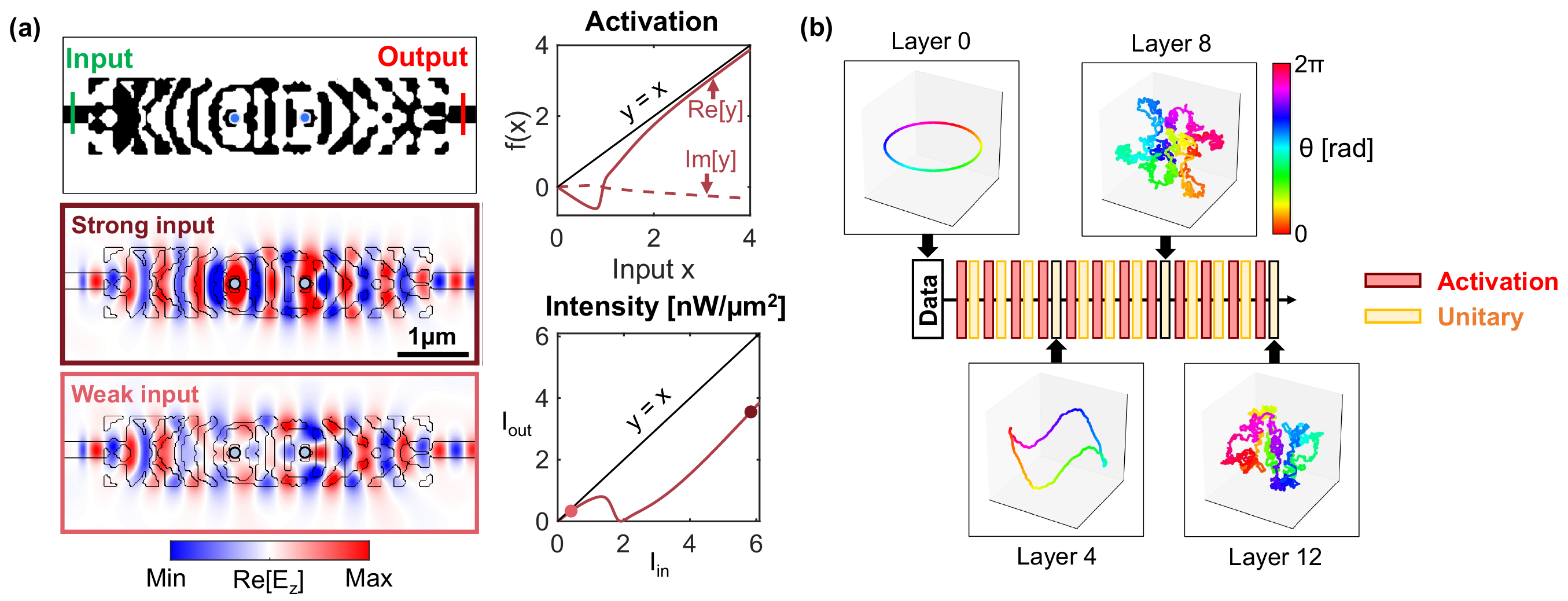}
  \caption{
  \textbf{Inverse-designed activation unit.} 
  (a) The input-output relationship shows strong nonlinearity. 
  (b) The evolution of data trajectory during forward-propagation. The trajectory is deformed strongly during the forward-propagation, indicating high expressive power. }
  \label{fig:inv_design_activation}
\end{figure*}

Again, we test the effectiveness of this activation function by tracking the evolution of a circular trajectory during the forward-propagation process. 
All the experimental settings remain the same as the previous case. 
We do a curve fitting based on the input-output relationship obtained from FDFD. Specifically, denote the input as $x$, the output $y = S_{21}(x)\cdot x$. The $S$-parameter, which now depends on $x$, is fitted using the following form:
\begin{equation}
    S_{21}(x) = b_\text{offset} + w_{1}(x) \cdot A_{1} e^{i\phi_{1}} + w_{2}(x) \cdot A_{2} e^{i\phi_{2}},
    \label{eq:S21}
\end{equation}
where the weights $w_{1}(x)$, $w_{2}(x)$ are sigmoid-like functions:
\begin{equation}
\begin{aligned}
    w_{1}(x) &= \frac{1}{1+\exp\left[-\beta_{1} ( \log_{10}(x) - x_{10}) \right] }, \\
    w_{2}(x) &= 1 - \frac{1}{1+\exp\left[-\beta_{2} ( \log_{10}(x) - x_{20}) \right]}.
\end{aligned}
\end{equation}
The curve fitting results are shown in Fig.~\ref{fig:curve_fit}. 
The fitted parameters are summarized as follows: $A_{1}=0.9250$, $\phi_{1}=-3.0528$, $A_{2}=0.9675$, $\phi_{2}=0.0747$, $b_\text{offset}=-0.0526$, $\beta_{1}=67.8915$, $x_{10}=-0.6538$, $\beta_{2}=10.1345$, $x_{20}=-0.5311$. 

The fitted curve is then implemented in PyTorch (version 2.4.1) and used as the activation function for training ONNs. 
As shown in Fig.~\ref{fig:inv_design_activation}(b), this time the trajectory changes very quickly as the network depth increases. 
The total curvature increases quickly, revealing that the designed ONN has strong expressive power. 

\subsection{Inverse design of linear blocks}
Finally, we provide details regarding how the linear blocks are designed via adjoint optimization.
The linear blocks are passive elements that route light to realize the trained weight matrices. 
We design them with a 2.5D variational effective-index method (vEIM)~\cite{si:hammer2009eim}, which is far cheaper than full 3D optimization. 
The blocks are GaP-on-diamond devices working at the same wavelength $\lambda_0 = 737.134$~nm. 
The top GaP layer is patterned by a $40$~nm shallow etch, switching between an unetched state ($200$~nm) and an etched state ($160$~nm). 
The vEIM projects the layered structure onto a single reference vertical mode, and collapses it into a 2D effective permittivity map $\epsilon_\text{eff}(x, y)$. 
The two states correspond to effective indices $n_\text{eff} = 2.89$ (unetched) and $n_\text{eff} = 2.76$ (etched). 
Such effective-index approach has been demonstrated experimentally for matrix-vector multiplication~\cite{si:nikkhah2024_2D}.

Within this 2.5D model, we run adjoint optimization based on FDFD for each layer, so that the device $S$-matrix reproduces the target weight matrix. 
For any given location inside the design region, the permittivity distribution is parameterized the same way as eq.~\ref{eq:permittivity_parameterize}, now interpolating between the etched and unetched effective permittivities. 
To ensure a fabricable structure, a conic filter with a radius of $200$~nm is applied, which helps control the minimum feature size. 
The parameter $\beta$, which controls the binarization process, is increased from $5.0$ to $200.0$ gradually throughout the optimization, which guarantees a fully binarized final design. 
We carry out gradient descent w.r.t. parameter $p(x,y)$ and iterate for $300$ steps to ensure convergence. 
Adam optimizer is used, with default parameter $\beta_{1}=0.9$ and $\beta_{2}=0.999$. 
The step size is decreased by a factor of $0.994$ after each iteration. 
With this procedure, each optimized block reproduces its target weight matrix to within $4$--$8\%$ relative error. 

\begin{figure*}[h!]
  \centering
  \includegraphics[width=1.0\linewidth]{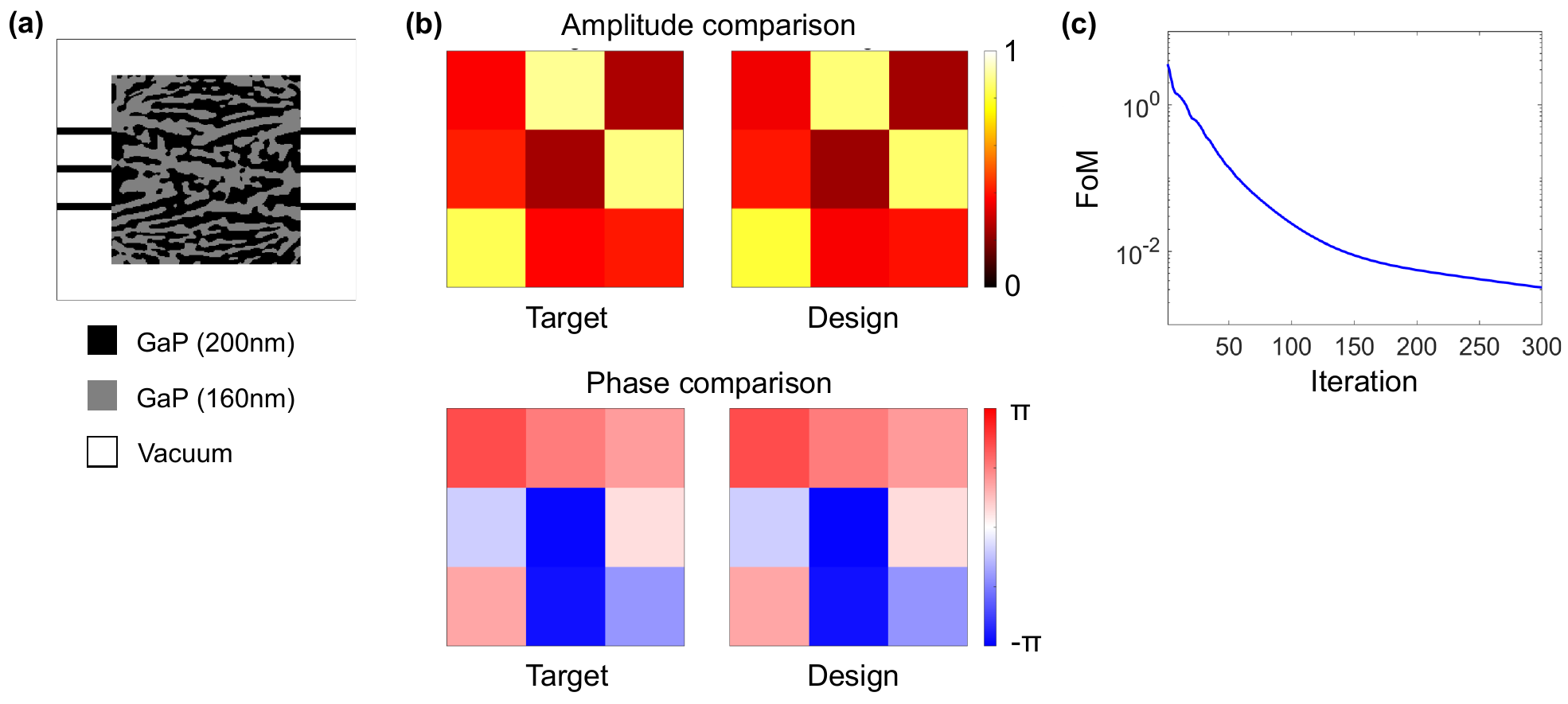}
  \caption{
  \textbf{Example of inverse-designed linear block.} 
  (a) The structure consists of etched GaP layer on top of thin-film diamond. 
  (b) Comparison between the $S$-matrix of the designed structure and the targeted unitary matrix. 
  (c) FoM curve obtained during the optimization. The FoM is defined as the Frobenius norm of the difference between the $S$-matrix and the target unitary matrix.
  }
  \label{fig:Smat_visualize}
\end{figure*}

\newpage
\section{Effect of low quantum efficiency}\label{si:eta_Q}
In this part we will show how a non-unity quantum efficiency will affect the design of our activation unit.
For the non-ideal case, denote the quantum efficiency as $\eta_{Q} = {\Gamma_{0}}/(\Gamma_{0} + \Gamma_\text{nrad} ) < 1$, where $\Gamma_\text{nrad}$ denotes the non-radiative decay rate. As derived in Supplementary Note~S2, a reduced $\eta_{Q}$ rescales the dressed dipole moment and weakens the coupling between the emitter and the optical field. It enters the nonlinear FDFD formalism through the replacement $f(x) \to f_{\eta_{Q}}(x) \triangleq f(\eta_{Q} x)$.

\begin{figure}[h]
\centering
\includegraphics[width=1.0\linewidth]{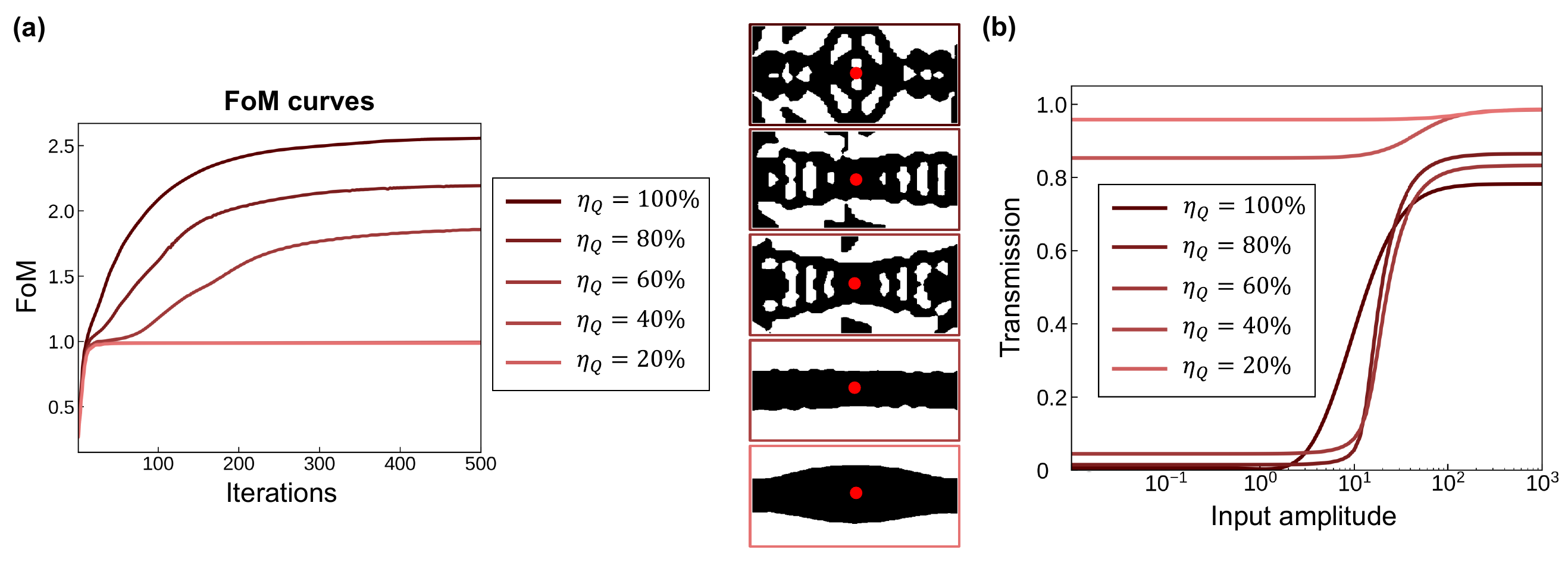}
\caption{
\textbf{The effect of non-unity quantum efficiency}.
(a) FoM curves obtained during the adjoint-based optimization.
(b) The resulting transmission curves $|S_{21}|^{2}$ for $\eta_{Q} \in \{20\%, 40\%, 60\%, 80\%, 100\%\}$. The nonlinear activation is retained for $\eta_{Q} \geqslant 60\%$, while the contrast collapses below $\eta_{Q} \approx 40\%$.
}
\label{fig:activations_eta}
\end{figure}

We fix the total decay rate as $\Gamma_{0} + \Gamma_\text{nrad} = 2\pi \times 94$ MHz, then repeat the 3D inverse design for different quantum efficiencies $\eta_{Q} \in \{ 20\%, 40\%, 60\%, 80\%, 100\% \}$. The results are displayed in Fig.~\ref{fig:activations_eta}.
Specifically, Fig.~\ref{fig:activations_eta}(a) shows that the best achievable FoM drops as the quantum efficiency decreases.
The resulting transmission curves are plotted in Fig.~\ref{fig:activations_eta}(b).
For $\eta_{Q} \geqslant 60\%$, the nonlinear activation retains a similar shape. 
Below $\eta_{Q} \approx 40\%$, however, the contrast collapses and the design becomes a transparent waveguide, because the weak-field response can no longer be suppressed once the radiative coupling becomes too weak. 
We'd like to point out that the quantum efficiency of SiV$^{-}$ centers has been measured to reach $\sim\!60\%$ \cite{si:bezard2024}.
Therefore we believe that our proposal remains valid with realistic emitters.

\newpage
\section{Performance on supervised learning tasks}
In this part we display the performance of our ONN on two supervised learning tasks: MNIST dataset \cite{si:lecun1998mnist} as well as FashionMNIST dataset \cite{si:xiao2017fashionmnist}. 
While these two datasets are simple, the results serve two purposes: 
first, by comparing the performance difference between nonlinear ONN and linear ONN, the effectiveness of our proposed activation function is verified; 
second, as we will show, a linear model can already solve the task reasonably well, which is consistent with existing literature. 
While the training is carried out in a physics-aware manner, all the models involved in this part have not been translated into concrete photonic devices, due to their high input dimensionality: designing linear blocks of this size by adjoint optimization is computationally expensive within our current pipeline. 
\subsection{MNIST}
The MNIST dataset consists of $28\times 28$ images of hand-written digits. 
These images are first resized to $14\times 14$, then flattened into a $196$-dim vector, which serves as the input of our ONN. 
The ONN consists of two hidden layers and one output layer, each containing $200$ neurons. 
We use the cross entropy loss to train our ONN, which is the standard approach for classification task. 
Adam optimizer is applied throughout the training process, with a fixed learning rate $\text{lr}=0.01$. 
The batch size is fixed as 100. 
Both models are trained for 30 epochs. The training curves are displayed in Fig.~\ref{fig:mnist}(a). 
The final classification accuracy on test dataset is $96.72\%$ (nonlinear) vs. $88.41\%$ (linear). The proposed activation unit leads to a $8.31\%$ improvement. 
\subsection{FashionMNIST}
\begin{figure*}[h]
  \centering
  \includegraphics[width=0.7\linewidth]{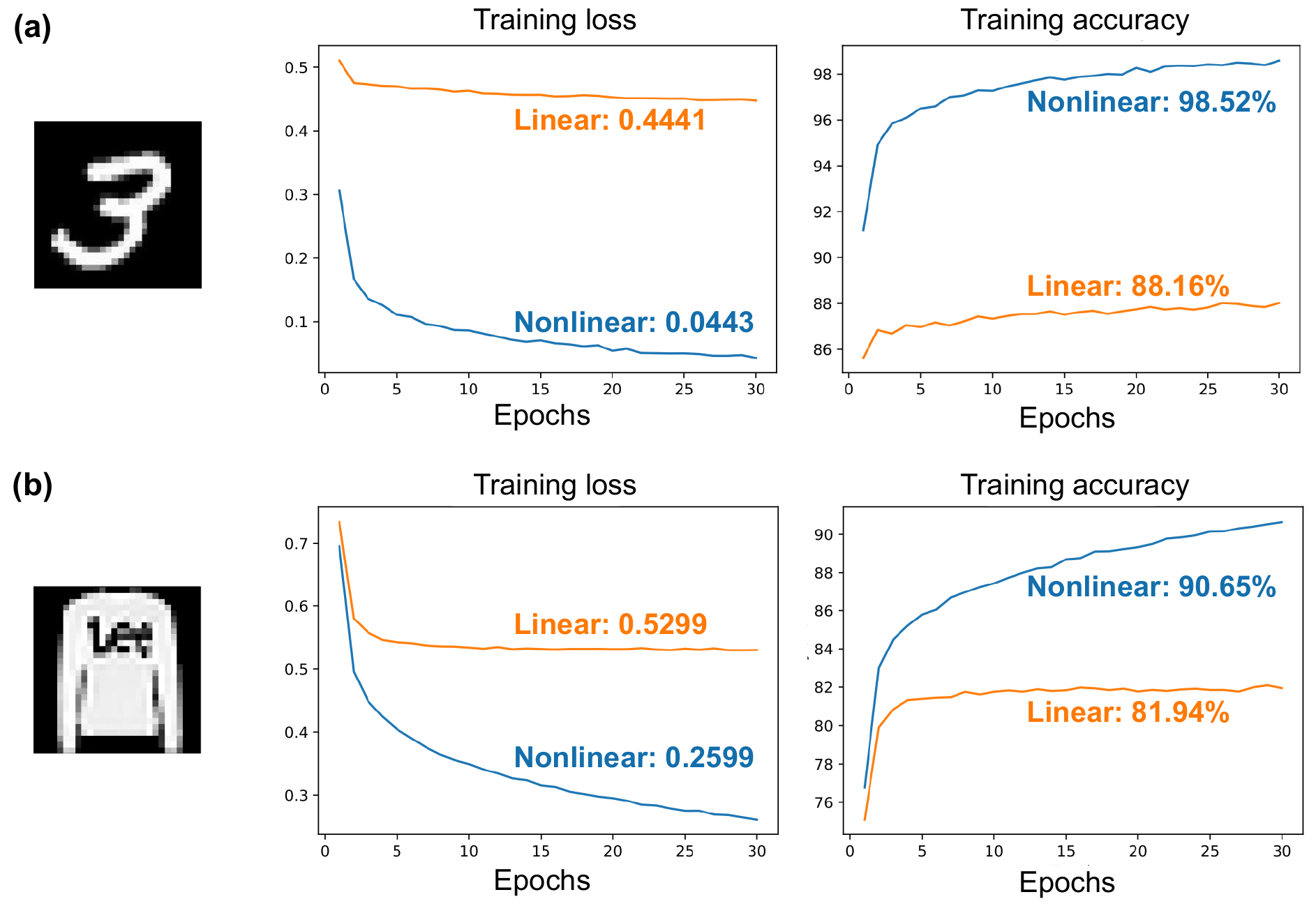}
  \caption{
  \textbf{Results for standard supervised learning tasks. } (a) The training curves for MNIST dataset. (b) The training curves for FashionMNIST dataset. }
  \label{fig:mnist}
\end{figure*}

The training setup for FashionMNIST are almost identical to that of MNIST. 
Specifically, the original $28\times 28$ images are resized to $14\times 14$, then flattened into a $196$-dim input vector. 
The ONN consists of two hidden layers and one output layer, each containing $200$ neurons. 
The cross entropy loss is used. 
Adam optimizer is applied, with a fixed learning rate $\text{lr}=0.001$. 
The batch size is fixed as 100. 
Both models are trained for 30 epochs. The training curves are displayed in Fig.~\ref{fig:mnist}(b). 
The final classification accuracy on test dataset is $87.87\%$ (nonlinear) vs. $80.23\%$ (linear). The proposed activation unit leads to a $7.64\%$ improvement. 

\newpage
\section{Physics-aware training procedure}
\begin{figure*}[h]
  \centering
  \includegraphics[width=1.0\linewidth]{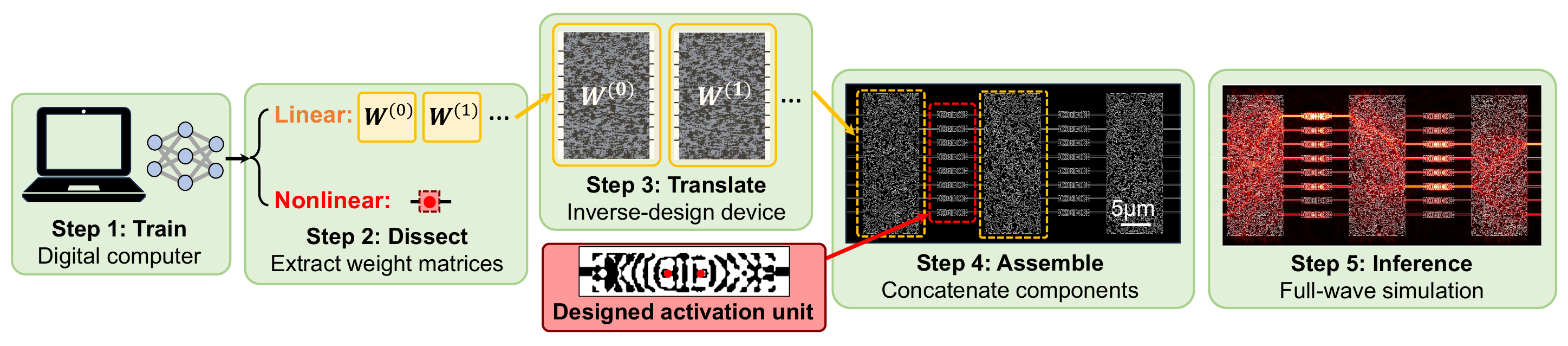}
  \caption{
  \textbf{Physics-aware training procedure. } 
  The ONN is first trained on a digital computer. The weight matrices are translated into linear blocks via adjoint optimization. After that, all the linear blocks are concatenated with the nonlinear activation units to form the ONN, which will be tested using full-wave simulations. }
  \label{fig:train_flow}
\end{figure*}

In this part we introduce the physics-aware training procedure in a detailed manner. 
As shown in Fig.~\ref{fig:train_flow}, the training procedure can be divided into five steps: \\
(1) \textbf{Train}: we first train a digital NN model with the help of PyTorch. Due to the energy-preserving constraint, we restrict the weight matrices $\bm{W}^{(i)}$ to be isometric, satisfying $ \left\lVert \bm{W}^{(i)} \bm{x} \right\rVert_{2}^{2} = \left\lVert \bm{x} \right\rVert_{2}^{2} $. The activation function used comes from eq.~\ref{eq:S21}. 
Beside these differences, this step is almost identical to training a normal multi-layer perceptron (MLP) in PyTorch. \\
(2) \textbf{Dissect}: all complex weight matrices $\bm{W}^{(i)}$ can now be extracted from the trained digital model. \\
(3) \textbf{Translate}: the weight matrix $\bm{W}^{(i)}$ is treated as a transmission matrix. Therefore it can be ``translated'' into a photonic device with the help of adjoint optimization. Note that other types of implementations (such as MZI mesh) can also be used: so long as they produce the same transmission matrix, the outcome should remain the same. \\
(4) \textbf{Assemble}: the optimized linear blocks are concatenated with the designed nonlinear activation units.  \\
(5) \textbf{Inference}: in order to test its performance, we apply full-wave FDFD simulations to obtain the electric field distributions for different input data. \\

In the main text, for the ``nonlinear classification'' task the presented results are obtained via full-wave simulation of the entire device. On the other hand, for the reinforcement learning task, the simulation domain is too large, making it expensive to simulate the entire device. As an alternative, we do a domain decomposition and simulate the output of each block individually. The output is then fed into the subsequent block as the input. Finally the $|\bm{E}|$ field distributions of different blocks are concatenated. The $|\bm{E}|^{2}$ distributions shown in main text Fig.~3 are obtained in this manner. Such alternative can be justified due to the fact that all our designed linear blocks show negligible reflection. Under such a condition, it is valid to simulate the light propagation in a section-by-section manner, which is a widely-accepted approach for designing diffractive neural networks \cite{si:lin2018diffractive,si:wu2019neuromorphic}. 

\newpage
\section{Nonlinear regression task} 
\begin{figure*}[h]
  \centering
  \includegraphics[width=0.8\linewidth]{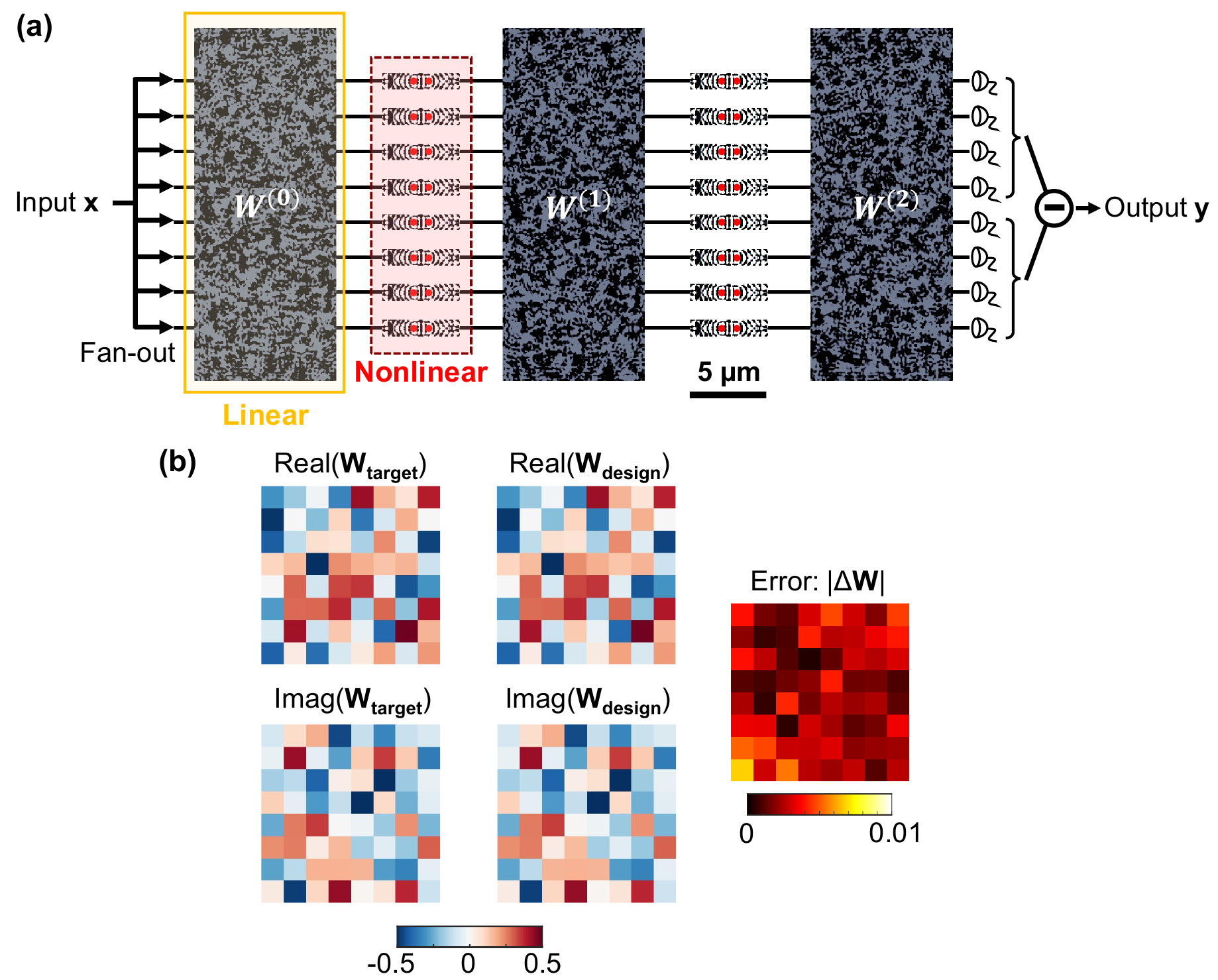}
  \caption{
  \textbf{Nonlinear regression task.} 
  (a) ONN architecture for nonlinear regression. The input $x\in [0, \pi]$ is duplicated before sent into input ports. Two hidden layers are included. At the output ports, differential readout is applied to support negative outputs. 
  (b) Transmission matrix design result for $\bm{W}^{(0)}$. Through comparison it can be concluded that the designed linear block can reproduce the desired transmission matrix with high fidelity. }
  \label{fig:sine_structure}
\end{figure*}

In this part, we display a nonlinear regression example. 
Specifically, we try to reproduce $y=\sin(2x)$ with the help of our proposed nonlinear activation unit. 
As shown in Fig.~\ref{fig:sine_structure}(a), the ONN's structure contains two hidden layers, each containing 8 neurons. 
The input $x$ is a scalar, satisfying $x\in [0, \pi]$. 
At the output ports, photodetectors are applied, which are modeled as $|\cdot|^{2}$ in our simulation. 
Considering that the output $y=\sin(2x)$ can be negative, differential readout is applied, so that the final result is produced by subtracting the total intensity obtained by two groups of detectors. 
In Fig.~\ref{fig:sine_structure}(b), we have displayed the transmission matrix of the first linear block, which corresponds to the matrix $\bm{W}^{(0)}$ in Fig.~\ref{fig:sine_structure}(a). 
The unitary matrix obtained from physics-aware training is denoted as $\bm{W}_\text{target}$, while the transmission matrix of the adjoint-optimized block is denoted as $\bm{W}_\text{design}$. 
Both the real parts and the imaginary parts of these two matrices match pretty well. 
The error $|\Delta \bm{W}| = |\bm{W}_\text{design} - \bm{W}_\text{target}|$ is also visualized, whose matrix elements are all well below $0.01$. 
Therefore we can safely conclude that through adjoint optimization, the linear blocks can reproduce the desired linear transformation with high fidelity. 

\begin{figure*}[h]
  \centering
  \includegraphics[width=0.9\linewidth]{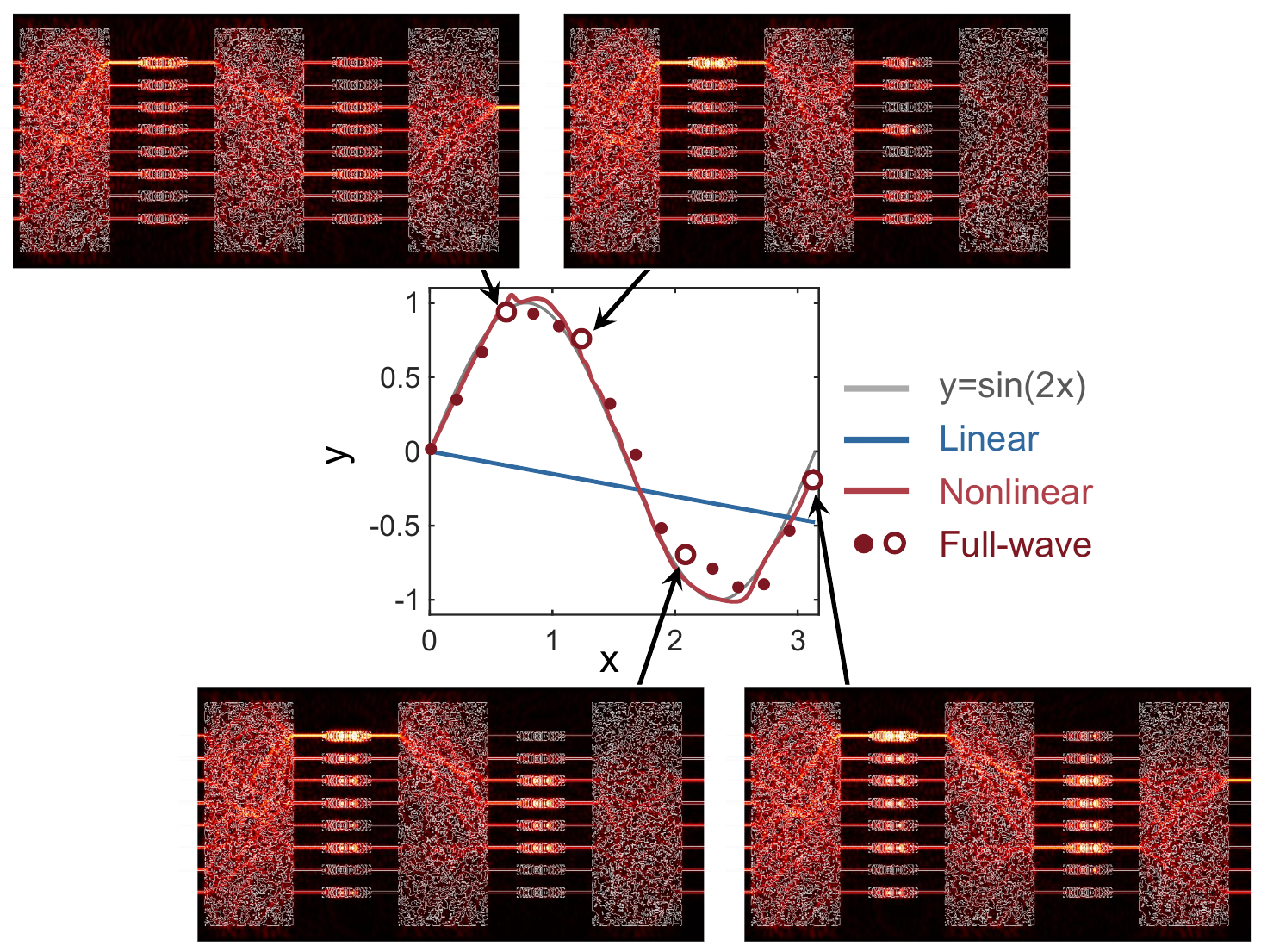}
  \caption{
  \textbf{Results for the nonlinear regression task.} The red curve indicates the curve fitting result of our trained nonlinear ONN, which reproduces the ground truth $y=\sin(2x)$ (gray curve) very well. 
  A linear model, on the other hand, fails completely, as shown by the blue curve. 
  The electric field distributions $|E_{z}|$ of four different data points are visualized. }
  \label{fig:sine_result}
\end{figure*}

The nonlinear regression results are displayed in Fig.~\ref{fig:sine_result}. 
The gray curve visualizes the ground truth $y=\sin(2x)$. 
The blue line corresponds to the results obtained through linear regression, which fails to capture the oscillatory behavior. 
On the other hand, with the quantum-enhanced activation unit, our nonlinear model can fit the $y=\sin(2x)$ curve perfectly (red curve). 
To verify that the physics-aware training leads to accurate result, we also carry out full-wave FDFD simulations for 16 data points, which produce consistent results. 
The electric field distributions ($|E_{z}|$) for 4 data points are also visualized in Fig.~\ref{fig:sine_result}. 
It can be observed that for different input intensities, the nonlinear activation units can steer the optical fields to different output ports, which then leads to different output results from the differential readout. 
While the nonlinear regression task is simple, such behavior (re-distribution of energy among different output ports when the input intensity increases) is impossible to achieve using a linear ONN, in which the output amplitude scales linearly with the input amplitude. 
Therefore the input-output relation is quadratic (considering the square-law detection), which cannot produce the oscillatory behavior of $y=\sin(2x)$. 

\newpage
\section{Reinforcement learning tasks: details}
In this part we discuss the details related to the training \& testing of reinforcement learning tasks. 
While the tasks we choose seem simple, both of them reveal clear performance gap between nonlinear ONN and linear ONN. This justifies the value of nonlinearity, since for complex tasks a nonlinear mapping is always beneficial. 
The Atari Pong results reported here are obtained using 2D FDFD, because a 3D verification for this case would be prohibitively expensive.

\subsection{Atari Pong}
\textbf{General setup:} the Atari Pong environment used to test our agent is provided by the Gymnasium library \cite{si:towers2024gymnasium} (version 1.1.1). 
Specifically, the game environment used is \texttt{ALE/Pong-ram-v5}. 
The observation type is set as \texttt{obs\_type = "ram"} (instead of using \texttt{obs\_type = "rgb"}), which means that for each frame the observation is 128-byte RAM, not the raw image. 
To obtain the input feature vector for our optical agent, we manually extract the $(x, y)$ coordinates of the player (denoted as $\bm{r}_{1}$, obtained from \texttt{ram[46], ram[51]}), the in-game AI (denoted as $\bm{r}_{2}$, obtained from \texttt{ram[45], ram[50]}), and the ball (denoted as $\bm{r}_{b}$, obtained from \texttt{ram[49], ram[54]}). 
We also calculate the ball's velocity $\bm{v}_{b} = (v_{x}, v_{y})$ by subtracting the ball's positions between two neighboring frames. 
Note that the positions are re-scaled, divided by \texttt{POSITION\_SCALE = 210}; the velocities are also re-scaled, divided by \texttt{VELOCITY\_SCALE = 8}. 
Each frame corresponds to a 8-dim feature vector, which concatenates the scaled $\bm{r}_{1}, \bm{r}_{2}, \bm{r}_{b}, \bm{v}_{b}$. 
Considering that historical information would be helpful for the agent, we concatenate the feature vectors obtained from the previous $F$ frames, which is a standard approach for tackling these Atari games \cite{si:mnih2015atari}. 
This is achieved by using the \texttt{VecFrameStack()} function. 
The final input vector has a dimension of $8F$. 

The optical agent can be viewed as a policy network, whose structure mimics a standard MLP. 
Specifically, the network contains $L$ hidden layers (each containing $H$ neurons) as well as an output layer. 
The input vector is first casted to \texttt{complex128} before sent into the neural network. 
Each layer consists of an isometric linear transformation and the proposed nonlinear activation function. 
Finally, the output vector is divided into six sections. 
For each section, the detected amplitudes are summed up to obtain the logits. 
\texttt{softmax()} is applied to the logits to obtain a 6-dim probability distribution, corresponding to the six discrete actions. 
During the training stage, the actual action is sampled randomly based on the probability distribution, while during inference the agent always takes the action with the highest probability. 

\begin{figure}[h]
\centering
\includegraphics[width=0.8\linewidth]{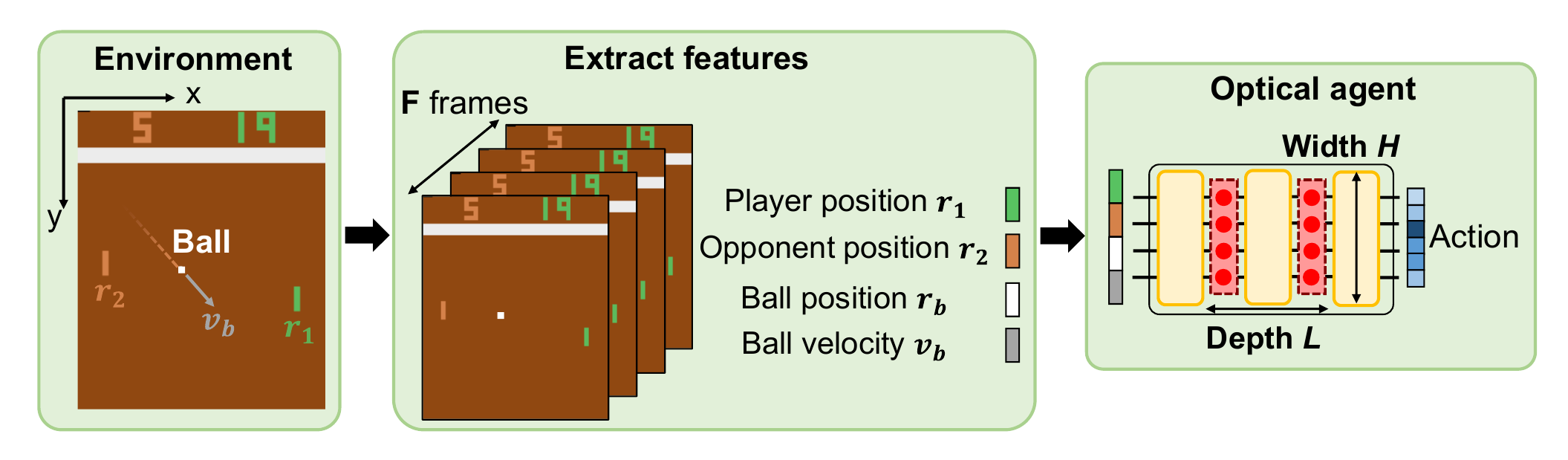}
\caption{
\textbf{Reinforcement learning setup for Atari Pong.} 
Hand-crafted features (including positions $\bm{r}_{1}$, $\bm{r}_{2}$, $\bm{r}_{b}$, and velocity $\bm{v}_{b}$) are extracted from the environment. 
The input vector is constructed by concatenating the feature vectors obtained from $F$ historical frames. }
\label{fig:RL_feature_pong}
\end{figure}

In order to obtain the results displayed in main text Fig.~3(d)(e), we change the model size by sweeping over $F$, $H$ and $L$ parameters. 
Specifically, we choose the number of frames $F\in \{1,2,3,4\}$, the network width $W\in\{18, 30, 42, 54\}$, and the network depth $L\in\{2,3,4,5\}$. 
Note that we add another restriction $W>8F$, because for an isometric linear transformation, the dimension of the input vector should be smaller than that of the output vector. 
This parameter sweeping provides us with $52$ different sets of parameter combinations. 
For each combination we train two models, one is linear model (with the activation function switched off), the other is nonlinear model (utilizing the proposed quantum-enhanced activation). 

\textbf{Training details:} in order to train our optical agent, we use the proximal policy optimization (PPO) algorithm \cite{si:schulman2017ppo}, provided by \texttt{stable-baselines3} \cite{si:raffin2021stable}. 
In order to accelerate the training process, we run multiple environments (setting \texttt{n\_env=8}) in parallel with the help of \texttt{SubprocVecEnv}. 
PPO optimizes a composite objective consisting of a clipped policy-gradient surrogate $L_{\mathrm{clip}}$, a value-function regression loss $L_{V}$, and an entropy bonus $\mathbb{E}_t\!\left[\mathcal{H}\right]$. 
Accordingly, \texttt{stable-baselines3} minimizes a loss function which consists a weighted sum of these three terms: 
\begin{equation}
\mathcal{L}=-L_{\mathrm{clip}}+\texttt{vf\_coef}\cdot L_V - \texttt{ent\_coef}\cdot \mathbb{E}_t\!\left[\mathcal{H}\right].
\end{equation}
In our experiments, we choose \texttt{vf\_coef = 0.5} and linearly change the entropy coefficient from \texttt{ent\_coef = 1e-2} to \texttt{1e-3} during the training process. 
Details of other PPO parameter settings are chosen as follows: \\
(1) the length of each rollout is set as \texttt{n\_steps = 2048}; \\
(2) batch size is fixed as $512$; \\
(3) each mini-batch is used to update the gradient descent for \texttt{n\_epochs = 10} times; \\
(4) when calculating the return, the discount factor is fixed as $\gamma=0.99$; \\
(5) we use the default optimizer, which is Adam. The learning rate is fixed as $2\times 10^{-4}$. \\
The Atari game environment can only be simulated on CPU. Running the optical agent on a GPU leads to frequent data transfer between CPU and GPU, which, in our case, slows down the training process since the optical agent is just a small MLP. 
Therefore, the training runs purely on CPU. 
All ONNs are trained using $5\times 10^{7}$ steps. 
During the training stage, we evaluate the performance every $20000$ steps. 
Specifically, the trained optical agent plays against the in-game AI for 30 episodes, and the mean/std reward is recorded. 
After the training stage, only the model with the best mean reward will be saved and tested. 

\subsection{HalfCheetah}
\textbf{General setup:} we evaluate our method on the \texttt{HalfCheetah-v5} benchmark provided by the Gymnasium library \cite{si:towers2024gymnasium}. The environment simulates a 2D cheetah-like robot using the MuJoCo physics engine. The agent controls the robot by applying torques to six hinge joints, resulting in a 6-dimensional continuous action space within the range $[-1, 1]$. Each episode lasts for 1000 steps. 
The raw observation is a 17-dimensional vector consisting of 8 positional values followed by 9 velocity values. 
To ensure numerical stability, we normalize the raw observation $\bm{o}_{t}$ by dividing each dimension by a fixed scaling constant. This normalization is implemented via a custom \texttt{ObservationWrapper} and is applied consistently across all training and testing phases. 
\begin{figure}[b]
\centering
\includegraphics[width=0.85\linewidth]{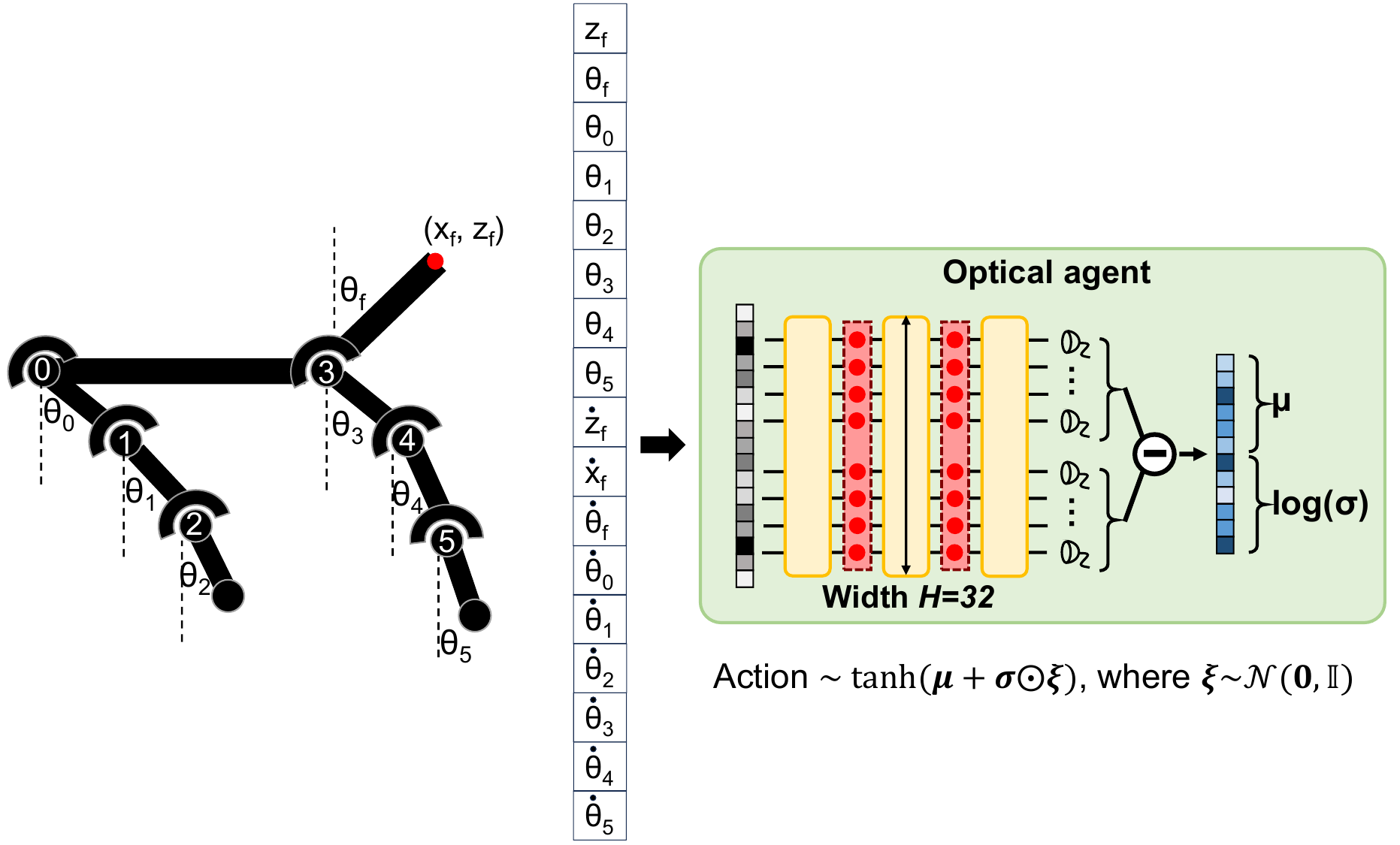}
\caption{
\textbf{Reinforcement learning setup for HalfCheetah control task.} 
The 17-dimensional input vector is shown explicitly. The variables with dots indicate time derivatives (velocities or angular velocities). 
The ONN outputs a 24-dimensional intensity vector, which is then converted into 12 signed values using balanced detection scheme. 
These values are split to parameterize a Gaussian distribution ($\bm{\mu}$ and $\log{\bm{\sigma}}$), from which the final action (six torques) is sampled. }
\label{fig:RL_feature_halfcheetah}
\end{figure}

The optical agent is implemented as a policy network, whose structure mimics a standard MLP (see Fig.~\ref{fig:RL_feature_halfcheetah}). 
Specifically, the policy network (actor) consists of $L=2$ hidden layers (each containing $H=32$ neurons) and an output layer.
The 17-dimensional input vector is first casted to \texttt{complex128} before sent into the neural network. 
Each layer consists of an isometric linear transformation and the proposed nonlinear activation function. 
Finally, to support the continuous control required by HalfCheetah, we implement a balanced detection mechanism. 
The ONN turns the input features into a 24-dimensional ($4 \times \text{action\_dim}$) output vector. 
These outputs are detected as non-negative intensities and split into two equal halves. The element-wise difference between these two halves leads to 12 signed values. 
The first six values serve as the mean $\bm{\mu}$, while the remaining six values serve as the log-standard deviation $\log{\bm{\sigma}}$ for a squashed Gaussian distribution. 
During the training stage, the continuous action is sampled randomly based on this squashed Gaussian distribution, while during inference the agent always takes the action based on the mean value. 


\textbf{Training \& evaluation details:} for the comparative experiments presented in the main text, we train two ONNs: a linear model (where the optical activation function is disabled) and the proposed nonlinear model (utilizing the quantum-enhanced activation). 
Both models share identical hyperparameters and training configurations, ensuring that the performance gap is caused by the nonlinearity. 
Training is performed using the soft actor-critic algorithm (SAC) \cite{si:haarnoja2018sac}, provided by \texttt{stable-baselines3} \cite{si:raffin2021stable}. 
We run multiple environments (\texttt{n\_env=8}) with the help of \texttt{SubprocVecEnv}. 
No temporal stacking is involved for the HalfCheetah task. 
Key hyperparameters are summarized as follows:
\\
(1) batch size is fixed as 256; \\
(2) the size of the replay buffer is \texttt{buffer\_size = 1000000}; \\
(3) the discount factor is $\gamma=0.99$; \\
(4) the soft update coefficient is $\tau=0.005$; \\
(5) we use the default optimizer, which is Adam. The learning rate is fixed as $3\times 10^{-4}$; \\
All ONNs are trained using $4\times 10^{6}$ steps. 
We run training on CPU, which is sufficient for the small optical MLPs used here. 
During training, we evaluate every $10^{4}$ steps using \texttt{EvalCallback}, running 20 evaluation episodes. The best-performing checkpoint is saved. 
During testing, the policy is evaluated for 10 episodes. 
To generate the velocity curves shown in main text Fig.~3(g), we record the horizontal position $x(t)$ (obtained from \texttt{qpos[0]}) at every timestep, and compute the forward velocity by discrete differencing
\begin{equation}
v_x(t) = \frac{x(t) - x(t-k)}{k},
\end{equation}
where $k=10$ is a sliding-window size. 
This windowed estimate helps suppress high-frequency fluctuations in the trajectory. 
We then plot $v_x(t)$ averaged over 10 episodes, with shaded areas indicating the standard deviation. 

\newpage
\section{Theory: quantifying expressive power}
In this part we introduce details related to quantifying the expressive power of any given activation function, regardless of whether it's digital or optical. 
We apply the method proposed in \cite{si:poole2016,si:raghu2017}, which is well-established in the machine learning theory community. 
Specifically, consider a trajectory of data points (can be understood as a closed curve inside high-dimensional feature space) that's been propagated layer-by-layer inside a neural network. 
The conclusion obtained by \cite{si:raghu2017} is summarized here. Suppose the network is wide enough. For each layer, the trajectory's total length $L$ increases by a factor of $\chi_{1}$, while the averaged local curvature $\kappa$ changes as
\begin{equation}
    \kappa^{2} \rightarrow \frac{\kappa^{2}}{\chi_{1}} + \frac{3\chi_{2}}{\chi_{1}^{2}},
\end{equation}
where $\chi_{1}$ can be calculated based on the 1st-order derivative of the activation function, while $\chi_{2}$ can be calculated based on its 2nd-order derivative. 
The total curvature $K$, which serves as a quantitative measure of the trajectory's complexity, is proportional to $\kappa^{2} \cdot L^{2}$ approximately. Therefore, after each layer, the increase of $K$ follows
\begin{equation}
    K^{2} \rightarrow K^{2} + \frac{3\chi_{2}}{\chi_{1}} L^{2}, 
\end{equation}
which only increases when the activation function is nonlinear ($\chi_{2}>0$). 
Intuitively, this explains that nonlinear activation function can help by injecting curvature into the data manifold. 
On the other hand, a purely linear transformation does not add complexity to the data manifold, and the total curvature remains constant $K=2\pi$. 
We then track how the total curvature of this trajectory evolves during its forward-propagation, since the total curvature $K$ serves as a good metric for evaluating the complexity of a trajectory. 
Based on the above expressions, it has been proved that with nonlinearity, the total curvature grows exponentially with network depth $L$, following $K\propto r^{L}$ \cite{si:raghu2017}. 
This helps us understand why deep neural networks have to be ``deep'': the expressive power of a neural network grows exponentially with its depth. 
\begin{figure}[h]
\centering
\includegraphics[width=0.6\linewidth]{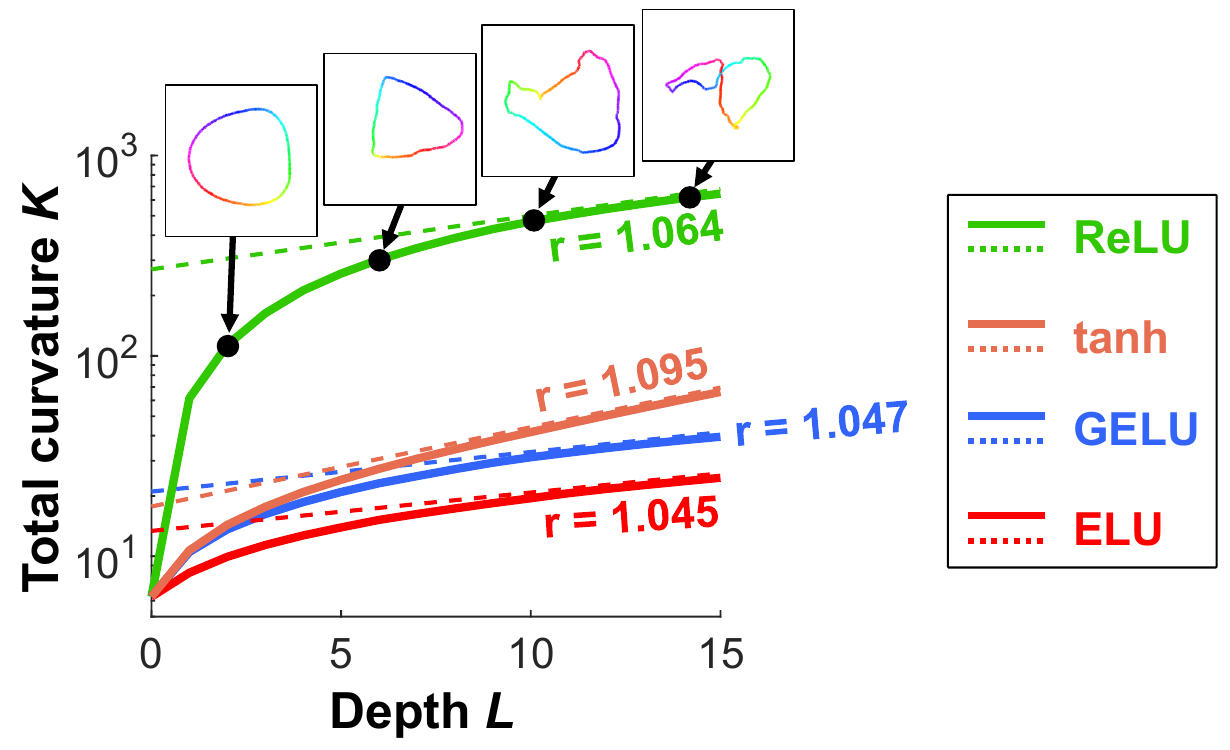}
\caption{ 
\textbf{Digital neural network baseline.} 
The expressive power of four different nonlinear activation functions are evaluated through curve fitting. }
\label{fig:DNN_baseline}
\end{figure}

Based on the above theory, the expressive power of any given nonlinear activation function can be quantified by ratio $r$. More specifically, by plugging the nonlinear activation into a neural network, the ratio $r$ can be evaluated numerically. To provide a baseline, we first evaluate $r$ for digital neural network (DNN), focusing on four popular activation functions: $\text{ReLU}$, $\text{tanh}$, $\text{GELU}$ and $\text{ELU}$. 
For each activation function, we construct a simple MLP with $L=15$ layers. Each hidden layer contains $N_\text{dim} = 512$ neurons. 
The input data points form a circular trajectory, parameterized by $\theta\in [0, 2\pi]$:
\begin{equation}
    \bm{x}(\theta) = \sqrt{N} \cdot (\bm{u} \cos(\theta) + \bm{v} \sin(\theta)), 
\end{equation}
where $\bm{u}$ and $\bm{v}$ are two random unit vectors which form an orthonormal basis for a 2D subspace. The trajectory contains $10000$ sampling points, distributed uniformly on $\theta\in [0, 2\pi]$.
We note that generally speaking, the growth factor $r$ depends not only on the activation function but also on the scale of the input data. In the digital baseline considered here this scale is fixed: the input trajectory is normalized by the $\sqrt{N}$ factor above, so that $r$ reduces to the single representative value reported below.
We track the evolution of such a trajectory during the forward-propagation process.
The results are shown in Fig.~\ref{fig:DNN_baseline}.
Four insets are included, which visualize the curves in a 2D subspace (dimension reduction done by PCA). 
Each data point corresponds to the average value obtained from $10$ Monte Carlo simulations. 
Finally, by doing a curve fitting $K\propto r^{L}$ using the data obtained from the last five NN layers, the value of ratio $r$ can be extracted. The fitted curves are displayed in Fig.~\ref{fig:DNN_baseline} using dashed lines. 
We arrive at the conclusion that, for a DNN, the data complexity increases exponentially with the number of layers $L$. 
Specifically, when using $\text{ELU}$, the data complexity increases by $4.5\%$ per layer. This value is used as the ``digital baseline'' when we evaluate the intensity requirement for optical nonlinearities. 

\newpage
\section{Expressive power of optical nonlinearities}
In this part we focus on understanding the expressive power of different optical nonlinear activations. We first introduce two conventional nonlinearities. 
Then we apply the proposed theoretical framework to quantify the expressive power of these activation functions, and compare with our proposed quantum-enhanced activation unit. 

\subsection{Optical nonlinearity of conventional materials}
In this part we introduce two different baselines, for realizing optical nonlinear activation function using conventional materials. 
The system we consider consists of one input port and one output port. At the input port, the electric field amplitude is denoted as $x$, while the amplitude at the output port is denoted as $y$. 
\subsubsection{Kerr effect of silicon}
Consider a silicon waveguide of length $\Delta L$. The phase difference induced by Kerr nonlinearity can be calculated as $\Delta \phi = n_{2} I \cdot k \Delta L$, where $I=\frac{n_{0}}{2\eta_{0}} |x|^{2}$ denotes the light intensity, $k=\frac{2\pi n_{0}}{\lambda_{0}}$ denotes the wave vector inside silicon. Here the wavelength is fixed as $\lambda_{0}=1.5~\mu$m. 
Based on existing experimental results \cite{si:dinu2003third,si:bristow2007two,si:dulkeith2006self}, the value of $n_{2}$ is chosen to be $5\times10^{-18}~\text{m}^{2}/\text{W}$. 
The refractive index $n_{0}\approx 3.5$. 
The input-output relationship of such a nonlinear activation unit can be formulated as
\begin{equation}
    y = x \cdot \exp(j\Delta \phi), 
\end{equation}
where the phase difference $\Delta \phi$ depends on input $x$. The phase difference $n_{0} k \Delta L$ has been ignored since it does not depend on intensity, thus providing no nonlinear effect. 
\subsubsection{Saturable absorption based on graphene}
Consider a saturable absorber, whose absorption coefficient depends on light intensity:
\begin{equation}
    \alpha = \frac{\alpha_{0}}{1 + I / I_\text{sat}},
\end{equation}
where $I_\text{sat}$ denotes the saturation intensity. 
Inside such a material, the evolution of light intensity follows
\begin{equation}
    \frac{dI}{dz}=-\alpha(I) \cdot I = -\frac{\alpha_{0} I}{1 + I / I_\text{sat}},
\end{equation}
which can be solved numerically to find the input-output relationship. 
In this paper we focus on graphene, with parameters $\alpha_{0}=6.8\times10^{7}~\text{m}^{-1}$. This coefficient is estimated based on experimental data: each graphene layer ($0.34$ nm thickness) leads to $2.3\%$ absorption \cite{si:lau2022comparison}. 
The saturation intensity is estimated as $I_\text{sat}=6.1\times10^{9}~\text{W/m}^{2}$ \cite{si:bao2009atomic}. It can be seen that for these conventional optical nonlinearities, the light intensity that's required to observe nonlinear effect is very strong. 

\subsection{Quantifying the expressive power}
\begin{figure}[h]
\centering
\includegraphics[width=0.75\linewidth]{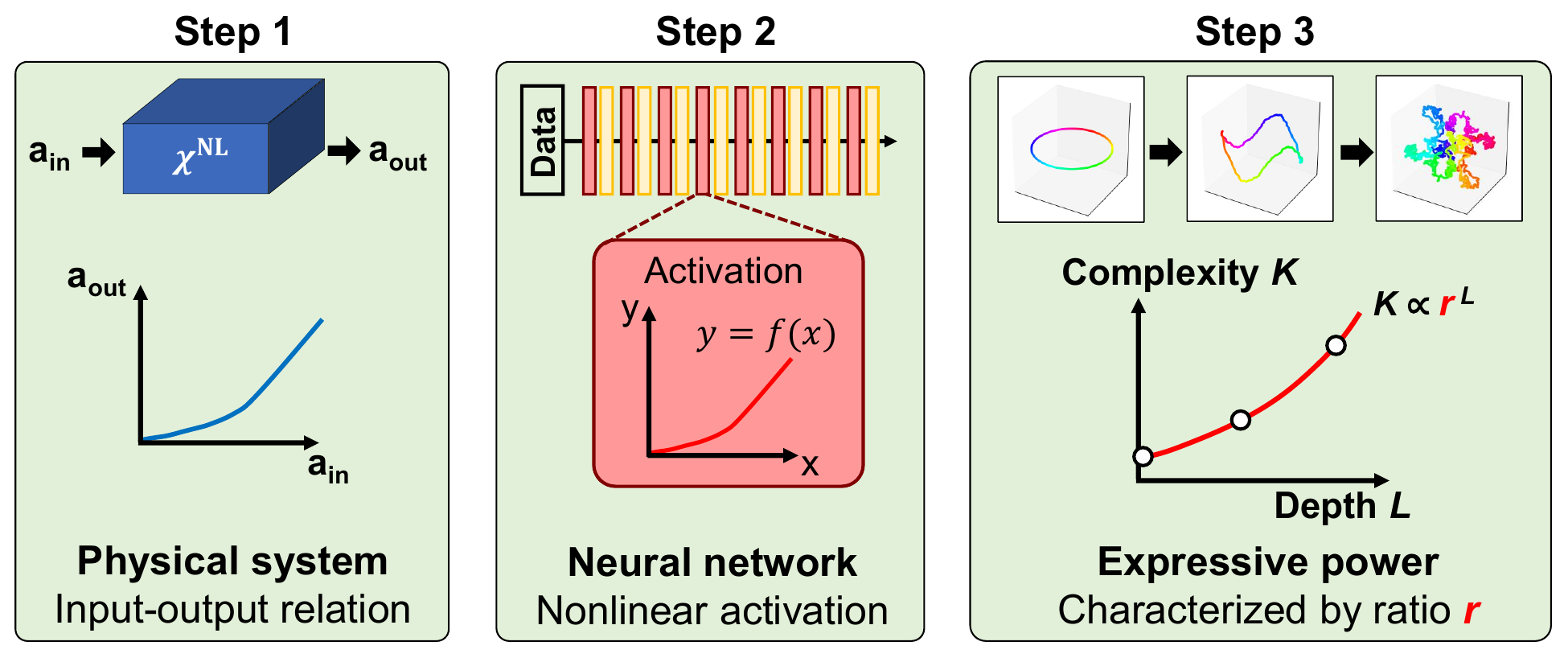}
\caption{
\textbf{Evaluating the expressive power of a physical system.} 
The 3-step procedure is general and can be applied to any nonlinear system. }
\label{fig:evaluate_r_schematics}
\end{figure}
\begin{figure}[h]
\centering
\includegraphics[width=0.7\linewidth]{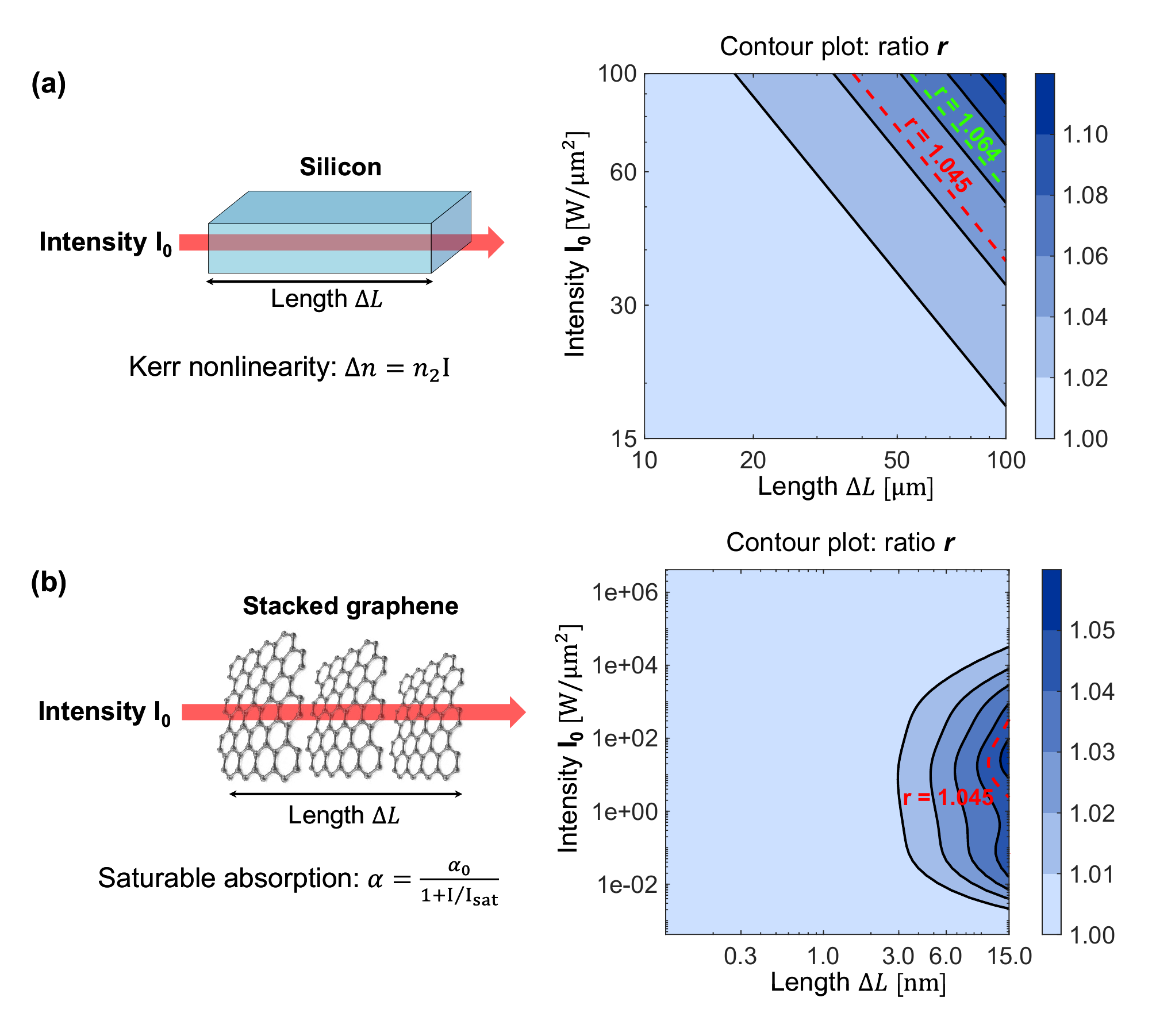}
\caption{\textbf{Expressive power of classical nonlinearities.} (a) Silicon waveguide with Kerr nonlinearity. (b) Stacked graphene layers as saturable absorbers. 
Here we do parameter sweeping w.r.t. length $\Delta L$ and light intensity $I_{0}$. For an optical nonlinearity, the ratio $r$ has to exceed $r_\text{min}=1.045$ (shown with red dashed contours) in order to be as powerful as its digital counterpart. }
\label{fig:contour_plot}
\end{figure}

In this part we aim to evaluate the expressive power for optical nonlinearities, by applying the same theoretical framework we used to understand digital neural networks. 
The procedure has been summarized as 3 steps, shown in Fig.~\ref{fig:evaluate_r_schematics}. For any nonlinear physical system, the first step involves obtaining its input-output relationship (shown as the blue curve). 
This curve is then interpreted as the nonlinear activation function, and is used to construct a deep neural network in the second step. 
Finally, the third step involves sending a circular data trajectory into the constructed NN. We then track how its total curvature $K$ increases, and extract the ratio $r$ through $K\propto r^{L}$ curve fitting, which quantifies the expressive power of this activation function. 
For any given optical nonlinearity, we apply the above 3-step procedure, then require that the corresponding activation function to have the same level of expressive power as the digital baseline. 
Specifically, here we require that ratio $r$ should be larger than $r_\text{min}=1.045$, which corresponds to the expressivity of $\text{ELU}$ activation function, evaluated in the previous part. 
Obviously, $r$ depends on the input light intensity $I_{0}$: very weak light leads to a linear mapping with $r=1$, thus its expressive power does not grow with depth. 
The above $r>r_\text{min}$ requirement posts constraint on the minimum light intensity. 
To obtain this intensity constraint, we have calculated the ratio $r$ for the two classical nonlinearities. The results are shown in Fig.~\ref{fig:contour_plot}. We sweep over two parameters, namely the length/thickness $\Delta L$ of the activation unit, and the input light intensity $I_{0}$. 
The $r=r_\text{min}$ threshold has also been displayed with red dashed lines in the contour plots. 
The minimum light intensity that's required to reach $r=r_\text{min}$ threshold can be extracted directly from these contour plots. 

\subsection{Comparison with quantum-enhanced activation}
\begin{figure}[h]
\centering
\includegraphics[width=0.75\linewidth]{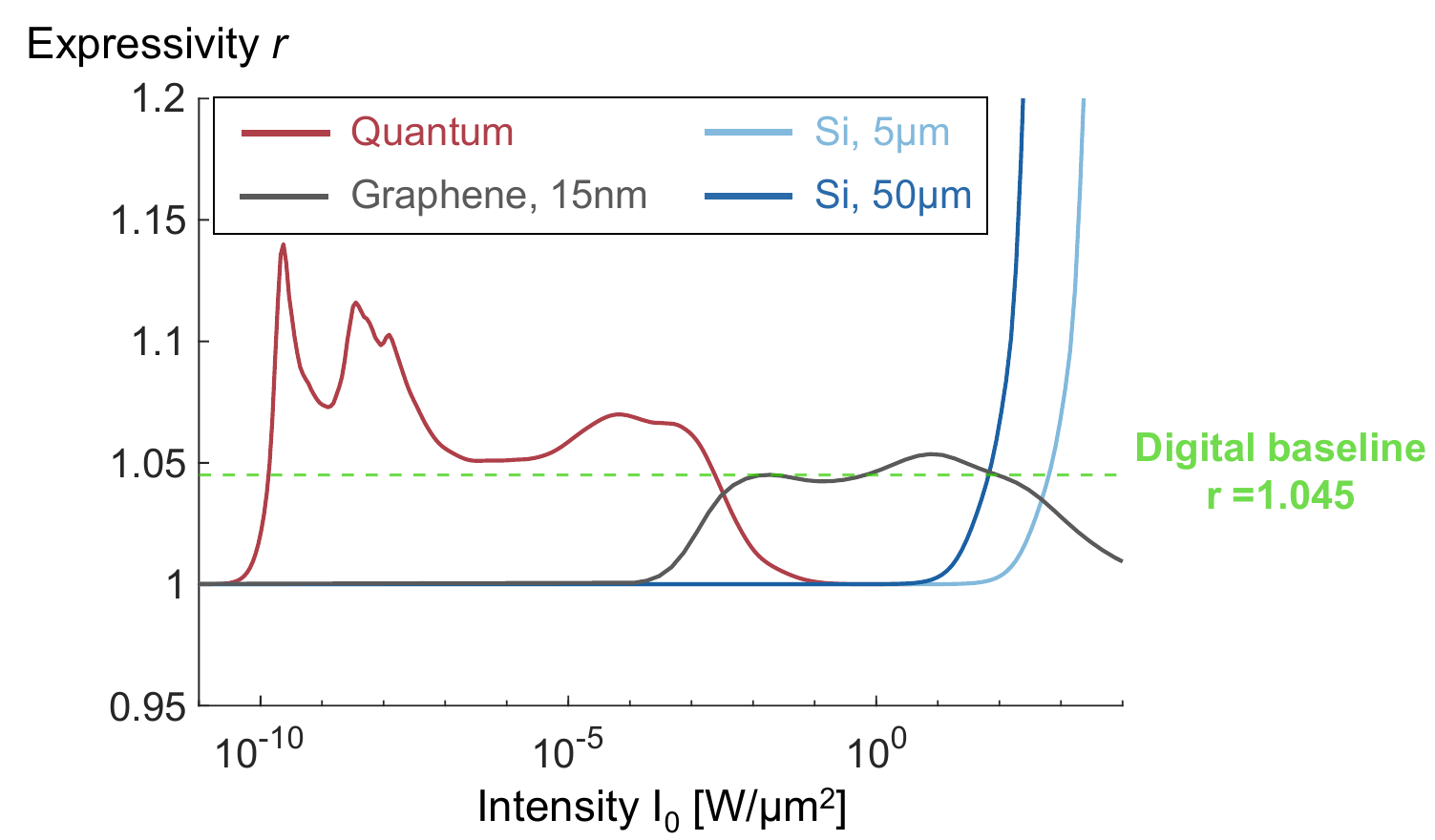}
\caption{
\textbf{Comparing the expressive power. }
For each optical nonlinearity, we visualize its corresponding ratio $r$ (which measures the expressive power) as a function of light intensity $I_{0}$. The proposed activation unit can provide enough nonlinearity at very low intensity.
 }
\label{fig:r_intensity_curve}
\end{figure}
To compare the expressive power of different types of nonlinearities in a more straight-forward manner, here we summarize the relationship between expressive power and light intensity in Fig.~\ref{fig:r_intensity_curve} . The two blue curves correspond to silicon waveguide, with length $\Delta L \in \{5~\mu\text{m}, 50~\mu\text{m}\} $. 
Even for a long waveguide with $\Delta L = 50~\mu\text{m}$, a very high intensity of $72.6~\text{W}/\mu\text{m}^{2}$ is required before the Kerr-based nonlinearity becomes useful. 
The gray dashed line corresponds to stacked graphene whose total thickness is $\Delta L = 15$ nm (this value is chosen such that the intensity does not decay below $1/e$, while keeping a high expressive power). Its threshold can be estimated as $0.02~\text{W}/\mu\text{m}^{2}$, which is lower than that of silicon. 
On the other hand, the proposed quantum nonlinearity, displayed with the red curve, shows strong nonlinearity when the input intensity is as low as $0.24~\text{nW}/\mu\text{m}^{2}$. 
The above intensity restrictions are used when estimating the nonlinearity-limited optical power of optical LLMs.

\newpage
\section{Power requirements of optical LLMs}
In this part we aim to estimate the nonlinearity-limited optical power of running popular LLMs with optical nonlinearities. 
Our calculation is based on the minimum light intensity constraint, obtained in the previous section. 
\subsection{Architecture of modern LLMs}
The Transformer architecture was proposed in \cite{si:vaswani2017transformer} and has now become the backbone of most popular LLMs. 
Many modern LLMs can be understood as stacked Transformer blocks. Suppose the input sequence contains $L_\text{seq}$ tokens (often referred to as the context length). After going through the embedding layer, the input matrix $X \in \mathbb{R}^{L_\text{seq}\times d_\text{model}}$, where $d_\text{model}$ denotes the dimension of embedding. 
The input is then multiplied with trained weight matrices to obtain the query matrix $Q$, the key matrix $K$, and the value matrix $V$:
\begin{equation}
    Q = X W_{Q}, 
    K = X W_{K}, 
    V = X W_{V}, 
\end{equation}
where both $Q, K, V \in \mathbb{R}^{L_\text{seq}\times d_\text{model}}$. 
Based on this architecture, considering that the weight matrices $W_{Q}$, $W_{K}$ and $W_{V}$ are fixed, only $X$ should be treated as optical input. 
Therefore when estimating the power consumption of optical LLM, we set the input dimension of each Transformer layer as $L_\text{seq}\times d_\text{model}\times 3$. The factor of $3$ comes from the fact that $3$ copies of $X$ are required to calculate $Q$, $K$ and $V$. 

In Table \ref{tab:LLM} we have summarized the architecture of several LLMs \cite{si:radford2019language,si:brown2020gpt3,si:touvron2023llama1,si:touvron2023llama2,si:dubey2024llama3,si:bi2024deepseek,si:liu2024deepseek-v2,si:liu2024deepseek-v3}. We will use these parameters later when evaluating the power consumption. Note that it's still unclear how self-attention and layer normalization can be implemented with optical computing. Therefore our analysis only serves as a crude estimation, by considering only the intensity requirement of nonlinear activation functions. 

\begin{table}[htbp]
  \centering
  \small
  \begin{tabular}{@{}l c C E L c@{}}
    \toprule
    {Name} & {Parameters~~} & {Context length $L_\text{seq}$~~} & {Embedding $d_\text{model}$~~} & {Layer $L$~~} & {Reference} \\
    \midrule
    GPT-2-Large       & 774M  & 1024 & 1280  & 36 & \cite{si:radford2019language} \\
    GPT-2-XL          & 1.54B  & 1024 & 1600  & 48 & \cite{si:radford2019language} \\
    GPT-3             & 175B & 2048 & 12288 & 96 & \cite{si:brown2020gpt3} \\
    Llama1-65B        & 65B & 2048 & 8192  & 80 & \cite{si:touvron2023llama1} \\
    Llama2-70B        & 70B & 4096 & 8192  & 80 & \cite{si:touvron2023llama2} \\
    Llama3-405B       & 405B & 8192 & 16384 & 126 & \cite{si:dubey2024llama3} \\
    Deepseek-LLM-7B   & 7B & 4096 & 4096  & 30 & \cite{si:bi2024deepseek} \\
    Deepseek-v2       & 236B & 4096 & 5120  & 60 & \cite{si:liu2024deepseek-v2} \\
    Deepseek-v3       & 671B & 4096 & 7168  & 61 & \cite{si:liu2024deepseek-v3} \\
    \bottomrule
  \end{tabular}
  \caption{\textbf{Decoder-only LLM architecture}. The parameters are used to estimate the optical power requirements of optical LLMs. }
  \label{tab:LLM}
\end{table}

\subsection{Estimating the nonlinearity-limited optical power}
\begin{figure}[h]
\centering
\includegraphics[width=1.0\linewidth]{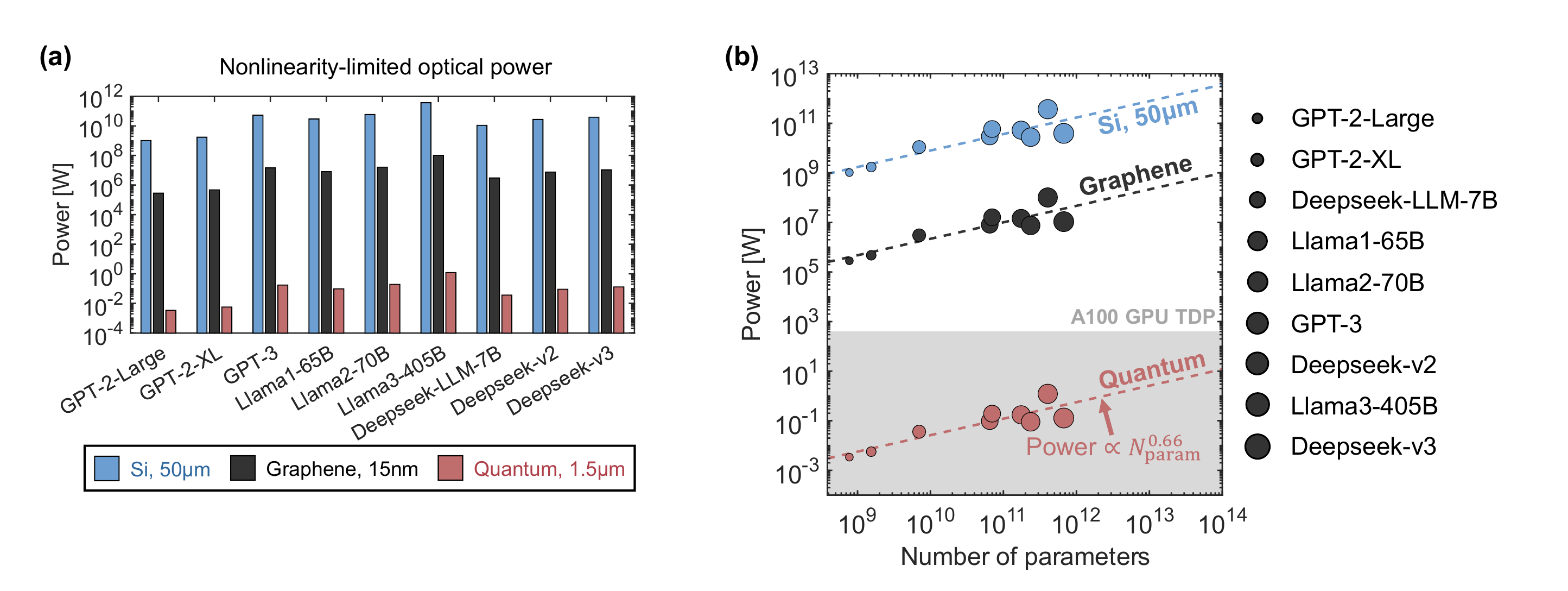}
\caption{
\textbf{Nonlinearity-limited optical power of LLMs}.
(a) Histogram showing the estimated nonlinearity-limited optical power of optical LLMs. Conventional materials demand prohibitive power levels, while the proposed scheme is not restricted by nonlinearity.
(b) Estimated nonlinearity-limited optical power versus model size $N_\text{param}$. ONNs follow a sublinear scaling $P\propto N_\text{param}^{0.66}$.
}
\label{fig:LLM_power}
\end{figure}

Intuitively, the nonlinearity-limited optical power of an optical LLM can be estimated based on the following formalism:
\begin{equation}
    \text{Power} = \text{Minimum intensity} \times \text{Cross section} \times \text{Input dimension} \times \text{Number of layers}.
\end{equation}
Here we have assumed that all inputs share the same level of light intensity, which is only limited by the strength of nonlinearity. Under such assumption, the optical power $P$ can be calculated using the following equation:
\begin{equation}
    P = I_\text{min}\cdot A \cdot (3 L_\text{seq} \cdot d_\text{model}) \cdot L,
\end{equation}
where $I_\text{min}$ stands for the minimum light intensity; $A$ denotes the physical cross section for each input dimension (fixed as $A\approx 0.1~\mu\text{m}^{2}$ \cite{si:shen2017}); $3 L_\text{seq} \cdot d_\text{model}$ gives the dimension of input, and $L$ stands for the number of layers contained in the LLM. 
Based on the above equation, we evaluate the nonlinearity-limited optical power of several open-source LLMs. We focus on the three types of optical nonlinearities mentioned previously, namely, silicon with Kerr nonlinearity (length $\Delta L = 50~\mu$m), graphene as saturable absorber (thickness $\Delta L = 15$ nm), and the proposed nonlinear unit based on quantum emitter. The results are displayed in Fig.~\ref{fig:LLM_power}, which presents the same results as Fig.~4(c) in the main text. 
In Fig.~\ref{fig:LLM_power}(a) the nonlinearity-limited optical power of different models are visualized using histogram. Fig.~\ref{fig:LLM_power}(b) visualizes the same set of data, while adding a curve fitting which shows that the optical power grows as $P\propto N_\text{param}^{0.66}$ empirically, consistent with \cite{si:anderson2023optical}. For electronic processors, the power consumption typically scales linearly with $N_\text{param}$. The nonlinearity-limited optical power therefore scales more favorably with model size, although a comparison at the system level would require accounting for additional overheads.

In conclusion, the proposed architecture based on quantum nonlinearity can run all LLMs with $<1.2$ W optical power. This would be $8.3\times10^{7}$ times more efficient than graphene, and $3.0\times 10^{11}$ times more efficient than silicon.
We emphasize that this value is a lower bound set by the nonlinear activation alone. It does not represent system-level power estimation since the power consumption related to fetching data, A/D (D/A) conversion, modulation, and cooling, are not included. We further note that the area $A \approx 0.1~\mu\text{m}^{2}$ assumed above corresponds to an on-chip waveguide mode. We would like to point out that the matrix sizes of current integrated photonic processors (of order $10^{2}$ ports~\cite{si:lightelligence2025pace,si:ahmed2025upaia}) remain far below the embedding dimensions of modern LLMs, so this estimate is, in the near term, more naturally realized on free-space platforms, where the per-neuron area is set by the pixel size. This will rescale the power estimation through the prefactor $I_{\min}A$ while leaving the sublinear scaling $N_\text{param}^{0.66}$ unchanged. 

\newpage
\section{Purcell factor of the activation unit}\label{si:purcell}

The response bandwidth of the activation unit is limited by the total decay rate of the emitter. 
As stated in the main text, this speed limit can be alleviated by engineering the photonic environment, which can modify its spontaneous emission rate through the Purcell effect.
We therefore quantify the modification of the Purcell factor $F_P$ at the emitter position in the designed activation unit \cite{si:novotny2012principles}. 

\begin{figure}[h!]
    \centering
    \includegraphics[width=0.85\linewidth]{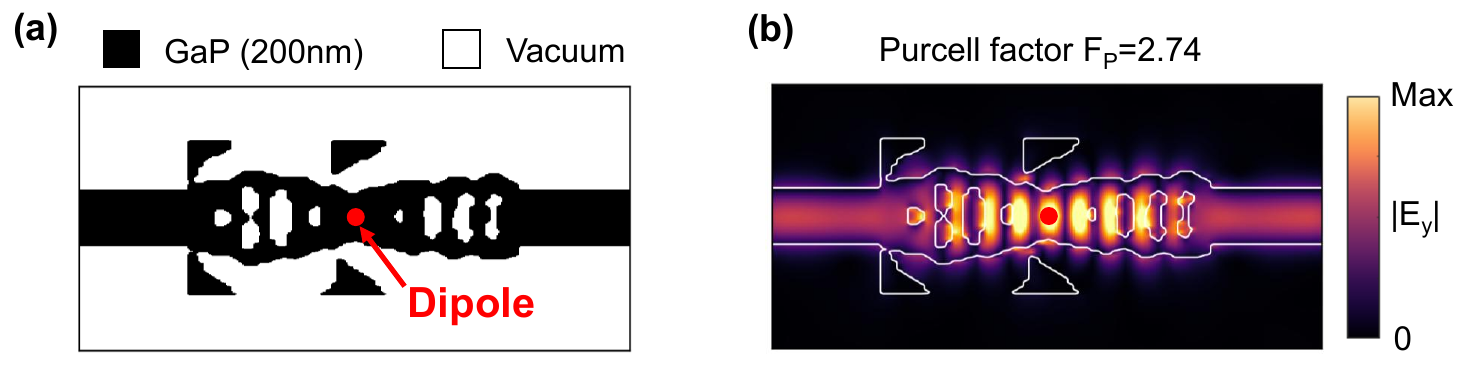}
    \caption{Purcell-factor analysis of the designed activation unit. (a) Schematic of the GaP-on-diamond device; the white marker indicates the position of the $y$-oriented SiV$^-$ emitter, located $30$~nm below the diamond surface at the center of the design region. (b) Simulated field distribution through the emitter. The emergent Purcell factor $F_P \approx 2.74$ is obtained as the ratio of the imaginary part of the dyadic Green's function at the emitter position in the device to that in homogeneous diamond.}
    \label{si:fig:purcell}
\end{figure}

The spontaneous emission rate is proportional to the imaginary part of the dyadic Green's function evaluated at the emitter position. 
We place a $y$-oriented point dipole at the emitter cell and perform two linear frequency-domain (3D FDFD) solves on an identical grid: 
one filled with the device permittivity $\varepsilon_\text{dev}$ and one filled with bulk-diamond $\varepsilon_\text{bulk}$. 
The Purcell factor is then
\begin{equation}
    F_P = \frac{\operatorname{Im} G_{yy}(\mathbf{r}_0, \mathbf{r}_0;\, \varepsilon_\text{dev})}{\operatorname{Im} G_{yy}(\mathbf{r}_0, \mathbf{r}_0;\, \varepsilon_\text{bulk})},
\end{equation}
where $\mathbf{r}_0$ is the emitter position and $G_{yy}$ is the $yy$-component of the dyadic Green's function. 
The device is a GaP-on-diamond structure (Fig.~\ref{si:fig:purcell}): a patterned GaP layer of thickness $200$~nm with an in-plane design region of $1.5 \times 0.7~\mu$m, on a $160$~nm-thick diamond layer. 
The SiV$^-$ emitter is located at the center of the design region, $30$~nm below the top diamond surface. 
For the designed activation unit we obtain $F_P \approx 2.74$. 
We emphasize that this enhancement is emergent since we didn't include the cavity quality factor as a optimization target explicitly. 
Using a larger photonic crystal cavity should be able to increase $F_P$ considerably, thus the response bandwidth should not be a fundamental limit.

\newpage
\section{Robustness against emitter non-idealities}\label{si:robustness}

In this part we characterize how the activation unit performs under two emitter non-idealities, namely the position randomness of the emitter and the spectral disorder.
The effect of a finite quantum efficiency has already been presented in Supplementary Note~S5. 
All the designs and analysis presented here are based on the 3D GaP-on-diamond platform. 
Material parameters, geometry, the FoM, and the optimization hyperparameters remain the same as explained in Supplementary Note~S4.

\subsection{Robustness to emitter position randomness}
\begin{figure}[h!]
    \centering
    \includegraphics[width=1.0\linewidth]{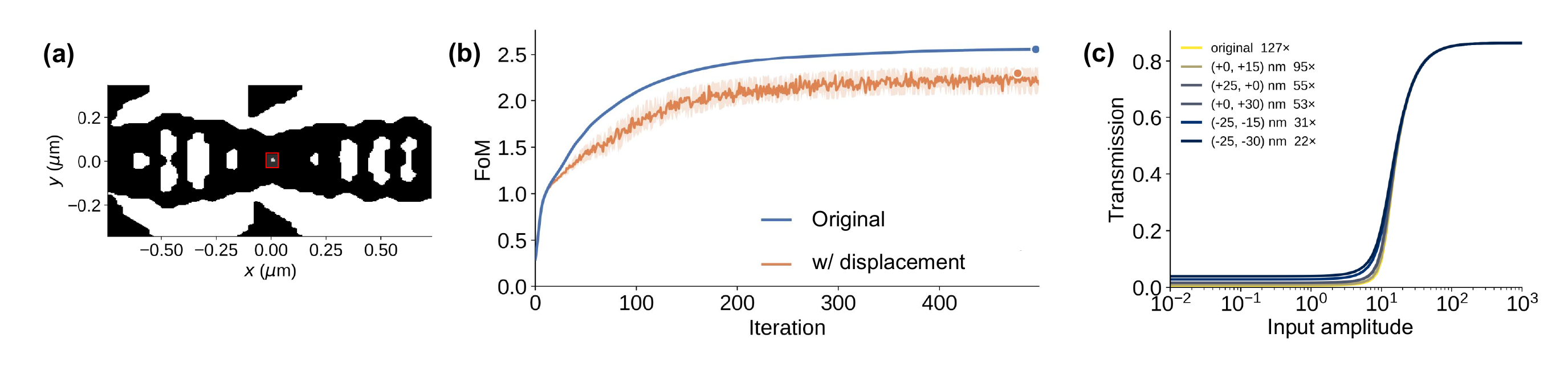}
    \caption{\textbf{Robustness to emitter position randomness.}
    {(a)} Setup: the inverse-designed GaP pattern. 
    The emitter's lateral position is sampled from a $50 \text{nm}\times 60\text{nm}$ region, marked by the red rectangle.
    {(b)} FoM curves: the original training curve (emitter position fixed at the middle) and the position-randomized training curve. 
    The shaded area marks the batch minimum/maximum at each iteration. 
    {(c)} The obtained transmission curve $|S_{21}|^2$ for several different emitter positions, showing that the activation curve is preserved.}
    \label{si:fig:robust_position}
\end{figure}

While placement of single SiV$^{-}$ can be realized by focused ion beam implantation, the position of the obtained emitter is random and often spans a few tens of nm \cite{si:schroder2017,si:titze2022}. 
To assess whether the activation unit tolerates such uncertainty, we randomize the lateral position of the SiV$^-$ emitter over a $50\text{nm} \times 60\text{nm}$ region, as shown in Fig.~\ref{si:fig:robust_position}(a).
The emitter depth is fixed as 30 nm. 
At each optimization iteration, six positions are drawn from this region and the FoM is averaged over the batch, 
so the inverse design should lead to a structure which is insensitive to the emitter's position. 

As shown in Fig.~\ref{si:fig:robust_position}(b), the optimization converges to $\text{FoM} = 2.295$ (the original design leads to $2.557$). 
We then verify the converged design on 5 different positions drawn from the same region.
The resulting transmission curves remain sigmoidal, as shown in Fig.~\ref{si:fig:robust_position}(c).
Across all tested positions, the transmission contrast $|S_{21}|^2$ between the strong- and weak-signal limits remains above $22\times$, well above the level required by the classification task.
Therefore, we conclude that with the help of inverse design, the activation unit can become pretty robust to realistic random placement.

\subsection{Robustness to spectral disorder}
\begin{figure}[h!]
    \centering
    \includegraphics[width=1.0\linewidth]{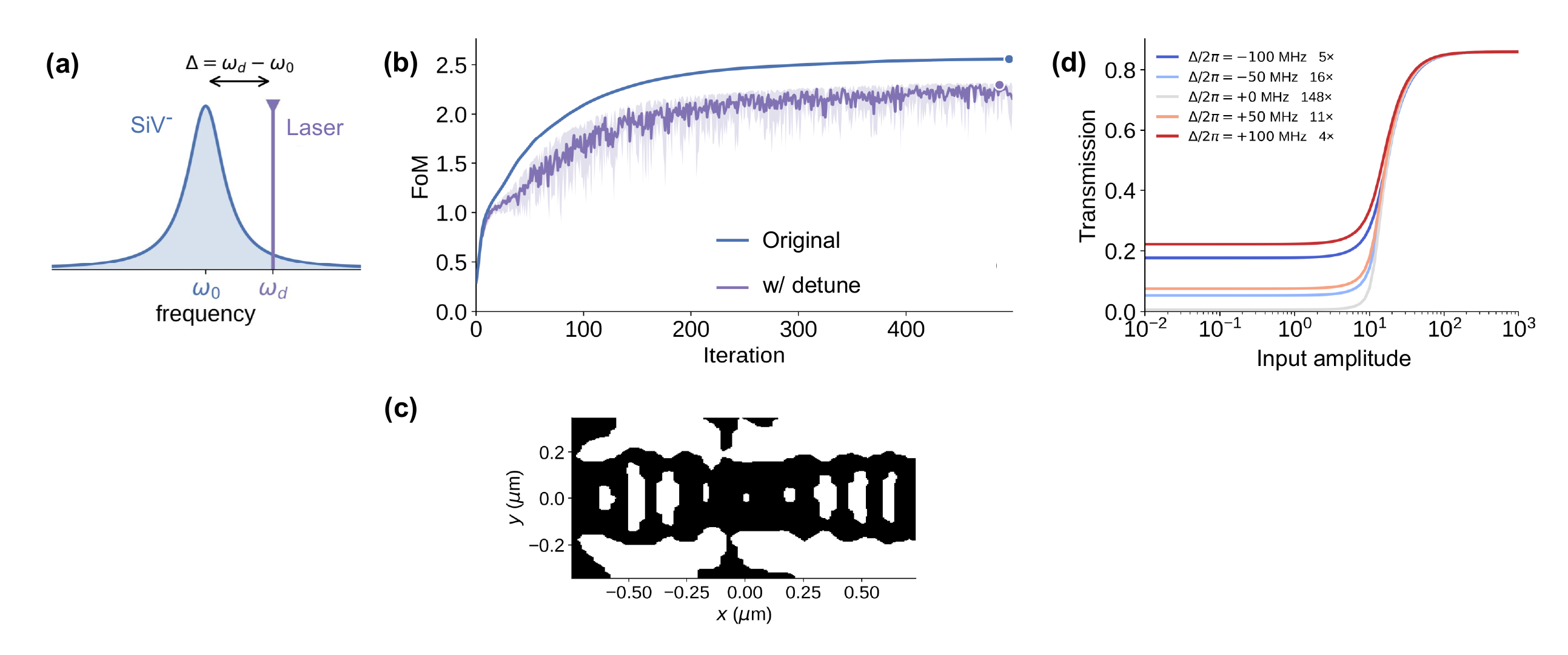}
    \caption{\textbf{Robustness to spectral disorder.}
    {(a)} Setup: schematic of the detuning $\Delta = \omega_d - \omega_0$. 
    {(b)} FoM curves: the original training curve ($\Delta = 0$) and the detuning-aware training curve. 
    The shaded area marks the batch minimum/maximum at each iteration.
    {(c)} The designed pattern which is robust against spectral disorder.
    {(d)} The obtained transmission curve $|S_{21}|^2$ for several detunings $\Delta/2\pi = 0, \pm 50, \pm 100$~MHz, showing how the contrast degrades with larger detune.}
    \label{si:fig:robust_detuning}
\end{figure}

Due to inhomogeneous broadening, the transition frequency of SiV$^-$ is not a fixed value. 
This leads to a detuning $\Delta$ between the emitter and the optical signal (see Fig.~\ref{si:fig:robust_detuning}(a)).
Since the activation relies on the emitter being resonant with the input signal, such detuning weakens the nonlinear response.
To understand how inhomogeneity affects the performance, we sample the detuning from Gaussian distribution $\Delta \sim \mathcal{N}(0, \sigma^2)$ with $\sigma/2\pi = 50$~MHz. 
During the optimization the FoM is averaged over six detunings, similar to the previous part. 

As shown in Fig.~\ref{si:fig:robust_detuning}(b), the optimization converges to $\text{FoM} = 2.297$.
The resulting design, shown in Fig.~\ref{si:fig:robust_detuning}(c), is then verified with 5 different detunings [Fig.~\ref{si:fig:robust_detuning}(d)].
As can be seen, while the sigmoidal shape is retained, the transmission contrast drops as the detuning goes to $100$ MHz.
We emphasize that the realistic inhomogeneous linewidth of SiV$^-$ is of order GHz, which is much higher than what we can handle.
Spectral disorder therefore remains a limitation. 
In principle, it can be mitigated by tuning the resonance of each emitter individually, for example via the DC Stark shift \cite{si:laucht2010stark}, as we have noted in the main text. 

\clearpage
\begin{arxivindependentbibliography}{70}
\end{arxivindependentbibliography}

\makeatletter
\global\let\auto@bib@innerbib\@empty
\makeatother


\begin{thebibliography}{84}%
\makeatletter
\providecommand \@ifxundefined [1]{%
 \@ifx{#1\undefined}
}%
\providecommand \@ifnum [1]{%
 \ifnum #1\expandafter \@firstoftwo
 \else \expandafter \@secondoftwo
 \fi
}%
\providecommand \@ifx [1]{%
 \ifx #1\expandafter \@firstoftwo
 \else \expandafter \@secondoftwo
 \fi
}%
\providecommand \natexlab [1]{#1}%
\providecommand \enquote  [1]{``#1''}%
\providecommand \bibnamefont  [1]{#1}%
\providecommand \bibfnamefont [1]{#1}%
\providecommand \citenamefont [1]{#1}%
\providecommand \href@noop [0]{\@secondoftwo}%
\providecommand \href [0]{\begingroup \@sanitize@url \@href}%
\providecommand \@href[1]{\@@startlink{#1}\@@href}%
\providecommand \@@href[1]{\endgroup#1\@@endlink}%
\providecommand \@sanitize@url [0]{\catcode `\\12\catcode `\$12\catcode
  `\&12\catcode `\#12\catcode `\^12\catcode `\_12\catcode `\%12\relax}%
\providecommand \@@startlink[1]{}%
\providecommand \@@endlink[0]{}%
\providecommand \url  [0]{\begingroup\@sanitize@url \@url }%
\providecommand \@url [1]{\endgroup\@href {#1}{\urlprefix }}%
\providecommand \urlprefix  [0]{URL }%
\providecommand \Eprint [0]{\href }%
\providecommand \doibase [0]{https://doi.org/}%
\providecommand \selectlanguage [0]{\@gobble}%
\providecommand \bibinfo  [0]{\@secondoftwo}%
\providecommand \bibfield  [0]{\@secondoftwo}%
\providecommand \translation [1]{[#1]}%
\providecommand \BibitemOpen [0]{}%
\providecommand \bibitemStop [0]{}%
\providecommand \bibitemNoStop [0]{.\EOS\space}%
\providecommand \EOS [0]{\spacefactor3000\relax}%
\providecommand \BibitemShut  [1]{\csname bibitem#1\endcsname}%
\let\auto@bib@innerbib\@empty
\bibitem [{\citenamefont {Krizhevsky}\ \emph {et~al.}(2012)\citenamefont
  {Krizhevsky}, \citenamefont {Sutskever},\ and\ \citenamefont
  {Hinton}}]{krizhevsky2012alexnet}%
  \BibitemOpen
  \bibfield  {author} {\bibinfo {author} {\bibfnamefont {A.}~\bibnamefont
  {Krizhevsky}}, \bibinfo {author} {\bibfnamefont {I.}~\bibnamefont
  {Sutskever}},\ and\ \bibinfo {author} {\bibfnamefont {G.~E.}\ \bibnamefont
  {Hinton}},\ }\bibfield  {title} {\bibinfo {title} {Imagenet classification
  with deep convolutional neural networks},\ }\href@noop {} {\bibfield
  {journal} {\bibinfo  {journal} {Advances in neural information processing
  systems}\ }\textbf {\bibinfo {volume} {25}} (\bibinfo {year}
  {2012})}\BibitemShut {NoStop}%
\bibitem [{\citenamefont {Mnih}\ \emph {et~al.}(2015)\citenamefont {Mnih},
  \citenamefont {Kavukcuoglu}, \citenamefont {Silver}, \citenamefont {Rusu},
  \citenamefont {Veness}, \citenamefont {Bellemare}, \citenamefont {Graves},
  \citenamefont {Riedmiller}, \citenamefont {Fidjeland}, \citenamefont
  {Ostrovski} \emph {et~al.}}]{mnih2015atari}%
  \BibitemOpen
  \bibfield  {author} {\bibinfo {author} {\bibfnamefont {V.}~\bibnamefont
  {Mnih}}, \bibinfo {author} {\bibfnamefont {K.}~\bibnamefont {Kavukcuoglu}},
  \bibinfo {author} {\bibfnamefont {D.}~\bibnamefont {Silver}}, \bibinfo
  {author} {\bibfnamefont {A.~A.}\ \bibnamefont {Rusu}}, \bibinfo {author}
  {\bibfnamefont {J.}~\bibnamefont {Veness}}, \bibinfo {author} {\bibfnamefont
  {M.~G.}\ \bibnamefont {Bellemare}}, \bibinfo {author} {\bibfnamefont
  {A.}~\bibnamefont {Graves}}, \bibinfo {author} {\bibfnamefont
  {M.}~\bibnamefont {Riedmiller}}, \bibinfo {author} {\bibfnamefont {A.~K.}\
  \bibnamefont {Fidjeland}}, \bibinfo {author} {\bibfnamefont {G.}~\bibnamefont
  {Ostrovski}}, \emph {et~al.},\ }\bibfield  {title} {\bibinfo {title}
  {Human-level control through deep reinforcement learning},\ }\href@noop {}
  {\bibfield  {journal} {\bibinfo  {journal} {nature}\ }\textbf {\bibinfo
  {volume} {518}},\ \bibinfo {pages} {529} (\bibinfo {year}
  {2015})}\BibitemShut {NoStop}%
\bibitem [{\citenamefont {Jumper}\ \emph {et~al.}(2021)\citenamefont {Jumper},
  \citenamefont {Evans}, \citenamefont {Pritzel}, \citenamefont {Green},
  \citenamefont {Figurnov}, \citenamefont {Ronneberger}, \citenamefont
  {Tunyasuvunakool}, \citenamefont {Bates}, \citenamefont {{\v{Z}}{\'\i}dek},
  \citenamefont {Potapenko} \emph {et~al.}}]{jumper2021alphafold}%
  \BibitemOpen
  \bibfield  {author} {\bibinfo {author} {\bibfnamefont {J.}~\bibnamefont
  {Jumper}}, \bibinfo {author} {\bibfnamefont {R.}~\bibnamefont {Evans}},
  \bibinfo {author} {\bibfnamefont {A.}~\bibnamefont {Pritzel}}, \bibinfo
  {author} {\bibfnamefont {T.}~\bibnamefont {Green}}, \bibinfo {author}
  {\bibfnamefont {M.}~\bibnamefont {Figurnov}}, \bibinfo {author}
  {\bibfnamefont {O.}~\bibnamefont {Ronneberger}}, \bibinfo {author}
  {\bibfnamefont {K.}~\bibnamefont {Tunyasuvunakool}}, \bibinfo {author}
  {\bibfnamefont {R.}~\bibnamefont {Bates}}, \bibinfo {author} {\bibfnamefont
  {A.}~\bibnamefont {{\v{Z}}{\'\i}dek}}, \bibinfo {author} {\bibfnamefont
  {A.}~\bibnamefont {Potapenko}}, \emph {et~al.},\ }\bibfield  {title}
  {\bibinfo {title} {Highly accurate protein structure prediction with
  alphafold},\ }\href@noop {} {\bibfield  {journal} {\bibinfo  {journal}
  {nature}\ }\textbf {\bibinfo {volume} {596}},\ \bibinfo {pages} {583}
  (\bibinfo {year} {2021})}\BibitemShut {NoStop}%
\bibitem [{\citenamefont {Vaswani}\ \emph {et~al.}(2017)\citenamefont
  {Vaswani}, \citenamefont {Shazeer}, \citenamefont {Parmar}, \citenamefont
  {Uszkoreit}, \citenamefont {Jones}, \citenamefont {Gomez}, \citenamefont
  {Kaiser},\ and\ \citenamefont {Polosukhin}}]{vaswani2017transformer}%
  \BibitemOpen
  \bibfield  {author} {\bibinfo {author} {\bibfnamefont {A.}~\bibnamefont
  {Vaswani}}, \bibinfo {author} {\bibfnamefont {N.}~\bibnamefont {Shazeer}},
  \bibinfo {author} {\bibfnamefont {N.}~\bibnamefont {Parmar}}, \bibinfo
  {author} {\bibfnamefont {J.}~\bibnamefont {Uszkoreit}}, \bibinfo {author}
  {\bibfnamefont {L.}~\bibnamefont {Jones}}, \bibinfo {author} {\bibfnamefont
  {A.~N.}\ \bibnamefont {Gomez}}, \bibinfo {author} {\bibfnamefont
  {{\L}.}~\bibnamefont {Kaiser}},\ and\ \bibinfo {author} {\bibfnamefont
  {I.}~\bibnamefont {Polosukhin}},\ }\bibfield  {title} {\bibinfo {title}
  {Attention is all you need},\ }\href@noop {} {\bibfield  {journal} {\bibinfo
  {journal} {Advances in neural information processing systems}\ }\textbf
  {\bibinfo {volume} {30}} (\bibinfo {year} {2017})}\BibitemShut {NoStop}%
\bibitem [{\citenamefont {Patterson}\ \emph {et~al.}(2021)\citenamefont
  {Patterson}, \citenamefont {Gonzalez}, \citenamefont {Le}, \citenamefont
  {Liang}, \citenamefont {Munguia}, \citenamefont {Rothchild}, \citenamefont
  {So}, \citenamefont {Texier},\ and\ \citenamefont {Dean}}]{patterson2021}%
  \BibitemOpen
  \bibfield  {author} {\bibinfo {author} {\bibfnamefont {D.}~\bibnamefont
  {Patterson}}, \bibinfo {author} {\bibfnamefont {J.}~\bibnamefont {Gonzalez}},
  \bibinfo {author} {\bibfnamefont {Q.}~\bibnamefont {Le}}, \bibinfo {author}
  {\bibfnamefont {C.}~\bibnamefont {Liang}}, \bibinfo {author} {\bibfnamefont
  {L.-M.}\ \bibnamefont {Munguia}}, \bibinfo {author} {\bibfnamefont
  {D.}~\bibnamefont {Rothchild}}, \bibinfo {author} {\bibfnamefont
  {D.}~\bibnamefont {So}}, \bibinfo {author} {\bibfnamefont {M.}~\bibnamefont
  {Texier}},\ and\ \bibinfo {author} {\bibfnamefont {J.}~\bibnamefont {Dean}},\
  }\bibfield  {title} {\bibinfo {title} {Carbon emissions and large neural
  network training},\ }\href@noop {} {\bibfield  {journal} {\bibinfo  {journal}
  {arXiv preprint arXiv:2104.10350}\ } (\bibinfo {year} {2021})}\BibitemShut
  {NoStop}%
\bibitem [{\citenamefont {Shen}\ \emph {et~al.}(2017)\citenamefont {Shen},
  \citenamefont {Harris}, \citenamefont {Skirlo}, \citenamefont {Prabhu},
  \citenamefont {Baehr-Jones}, \citenamefont {Hochberg}, \citenamefont {Sun},
  \citenamefont {Zhao}, \citenamefont {Larochelle}, \citenamefont {Englund}
  \emph {et~al.}}]{shen2017}%
  \BibitemOpen
  \bibfield  {author} {\bibinfo {author} {\bibfnamefont {Y.}~\bibnamefont
  {Shen}}, \bibinfo {author} {\bibfnamefont {N.~C.}\ \bibnamefont {Harris}},
  \bibinfo {author} {\bibfnamefont {S.}~\bibnamefont {Skirlo}}, \bibinfo
  {author} {\bibfnamefont {M.}~\bibnamefont {Prabhu}}, \bibinfo {author}
  {\bibfnamefont {T.}~\bibnamefont {Baehr-Jones}}, \bibinfo {author}
  {\bibfnamefont {M.}~\bibnamefont {Hochberg}}, \bibinfo {author}
  {\bibfnamefont {X.}~\bibnamefont {Sun}}, \bibinfo {author} {\bibfnamefont
  {S.}~\bibnamefont {Zhao}}, \bibinfo {author} {\bibfnamefont {H.}~\bibnamefont
  {Larochelle}}, \bibinfo {author} {\bibfnamefont {D.}~\bibnamefont {Englund}},
  \emph {et~al.},\ }\bibfield  {title} {\bibinfo {title} {Deep learning with
  coherent nanophotonic circuits},\ }\href@noop {} {\bibfield  {journal}
  {\bibinfo  {journal} {Nature photonics}\ }\textbf {\bibinfo {volume} {11}},\
  \bibinfo {pages} {441} (\bibinfo {year} {2017})}\BibitemShut {NoStop}%
\bibitem [{\citenamefont {Lin}\ \emph {et~al.}(2018)\citenamefont {Lin},
  \citenamefont {Rivenson}, \citenamefont {Yardimci}, \citenamefont {Veli},
  \citenamefont {Luo}, \citenamefont {Jarrahi},\ and\ \citenamefont
  {Ozcan}}]{lin2018diffractive}%
  \BibitemOpen
  \bibfield  {author} {\bibinfo {author} {\bibfnamefont {X.}~\bibnamefont
  {Lin}}, \bibinfo {author} {\bibfnamefont {Y.}~\bibnamefont {Rivenson}},
  \bibinfo {author} {\bibfnamefont {N.~T.}\ \bibnamefont {Yardimci}}, \bibinfo
  {author} {\bibfnamefont {M.}~\bibnamefont {Veli}}, \bibinfo {author}
  {\bibfnamefont {Y.}~\bibnamefont {Luo}}, \bibinfo {author} {\bibfnamefont
  {M.}~\bibnamefont {Jarrahi}},\ and\ \bibinfo {author} {\bibfnamefont
  {A.}~\bibnamefont {Ozcan}},\ }\bibfield  {title} {\bibinfo {title}
  {All-optical machine learning using diffractive deep neural networks},\
  }\href@noop {} {\bibfield  {journal} {\bibinfo  {journal} {Science}\ }\textbf
  {\bibinfo {volume} {361}},\ \bibinfo {pages} {1004} (\bibinfo {year}
  {2018})}\BibitemShut {NoStop}%
\bibitem [{\citenamefont {Shekhar}\ \emph {et~al.}(2024)\citenamefont
  {Shekhar}, \citenamefont {Bogaerts}, \citenamefont {Chrostowski},
  \citenamefont {Bowers}, \citenamefont {Hochberg}, \citenamefont {Soref},\
  and\ \citenamefont {Shastri}}]{shekhar2024roadmap}%
  \BibitemOpen
  \bibfield  {author} {\bibinfo {author} {\bibfnamefont {S.}~\bibnamefont
  {Shekhar}}, \bibinfo {author} {\bibfnamefont {W.}~\bibnamefont {Bogaerts}},
  \bibinfo {author} {\bibfnamefont {L.}~\bibnamefont {Chrostowski}}, \bibinfo
  {author} {\bibfnamefont {J.~E.}\ \bibnamefont {Bowers}}, \bibinfo {author}
  {\bibfnamefont {M.}~\bibnamefont {Hochberg}}, \bibinfo {author}
  {\bibfnamefont {R.}~\bibnamefont {Soref}},\ and\ \bibinfo {author}
  {\bibfnamefont {B.~J.}\ \bibnamefont {Shastri}},\ }\bibfield  {title}
  {\bibinfo {title} {Roadmapping the next generation of silicon photonics},\
  }\href@noop {} {\bibfield  {journal} {\bibinfo  {journal} {Nature
  Communications}\ }\textbf {\bibinfo {volume} {15}},\ \bibinfo {pages} {751}
  (\bibinfo {year} {2024})}\BibitemShut {NoStop}%
\bibitem [{\citenamefont {Wetzstein}\ \emph {et~al.}(2020)\citenamefont
  {Wetzstein}, \citenamefont {Ozcan}, \citenamefont {Gigan}, \citenamefont
  {Fan}, \citenamefont {Englund}, \citenamefont {Solja{\v{c}}i{\'c}},
  \citenamefont {Denz}, \citenamefont {Miller},\ and\ \citenamefont
  {Psaltis}}]{wetzstein2020review}%
  \BibitemOpen
  \bibfield  {author} {\bibinfo {author} {\bibfnamefont {G.}~\bibnamefont
  {Wetzstein}}, \bibinfo {author} {\bibfnamefont {A.}~\bibnamefont {Ozcan}},
  \bibinfo {author} {\bibfnamefont {S.}~\bibnamefont {Gigan}}, \bibinfo
  {author} {\bibfnamefont {S.}~\bibnamefont {Fan}}, \bibinfo {author}
  {\bibfnamefont {D.}~\bibnamefont {Englund}}, \bibinfo {author} {\bibfnamefont
  {M.}~\bibnamefont {Solja{\v{c}}i{\'c}}}, \bibinfo {author} {\bibfnamefont
  {C.}~\bibnamefont {Denz}}, \bibinfo {author} {\bibfnamefont {D.~A.}\
  \bibnamefont {Miller}},\ and\ \bibinfo {author} {\bibfnamefont
  {D.}~\bibnamefont {Psaltis}},\ }\bibfield  {title} {\bibinfo {title}
  {Inference in artificial intelligence with deep optics and photonics},\
  }\href@noop {} {\bibfield  {journal} {\bibinfo  {journal} {Nature}\ }\textbf
  {\bibinfo {volume} {588}},\ \bibinfo {pages} {39} (\bibinfo {year}
  {2020})}\BibitemShut {NoStop}%
\bibitem [{\citenamefont {Wang}\ \emph {et~al.}(2022)\citenamefont {Wang},
  \citenamefont {Ma}, \citenamefont {Wright}, \citenamefont {Onodera},
  \citenamefont {Richard},\ and\ \citenamefont {McMahon}}]{wang2022photon}%
  \BibitemOpen
  \bibfield  {author} {\bibinfo {author} {\bibfnamefont {T.}~\bibnamefont
  {Wang}}, \bibinfo {author} {\bibfnamefont {S.-Y.}\ \bibnamefont {Ma}},
  \bibinfo {author} {\bibfnamefont {L.~G.}\ \bibnamefont {Wright}}, \bibinfo
  {author} {\bibfnamefont {T.}~\bibnamefont {Onodera}}, \bibinfo {author}
  {\bibfnamefont {B.~C.}\ \bibnamefont {Richard}},\ and\ \bibinfo {author}
  {\bibfnamefont {P.~L.}\ \bibnamefont {McMahon}},\ }\bibfield  {title}
  {\bibinfo {title} {An optical neural network using less than 1 photon per
  multiplication},\ }\href@noop {} {\bibfield  {journal} {\bibinfo  {journal}
  {Nature Communications}\ }\textbf {\bibinfo {volume} {13}},\ \bibinfo {pages}
  {123} (\bibinfo {year} {2022})}\BibitemShut {NoStop}%
\bibitem [{\citenamefont {Shi}\ \emph {et~al.}(2025)\citenamefont {Shi},
  \citenamefont {Huang}, \citenamefont {Fu},\ and\ \citenamefont
  {Chen}}]{shi2025review}%
  \BibitemOpen
  \bibfield  {author} {\bibinfo {author} {\bibfnamefont {W.}~\bibnamefont
  {Shi}}, \bibinfo {author} {\bibfnamefont {Z.}~\bibnamefont {Huang}}, \bibinfo
  {author} {\bibfnamefont {T.}~\bibnamefont {Fu}},\ and\ \bibinfo {author}
  {\bibfnamefont {H.}~\bibnamefont {Chen}},\ }\bibfield  {title} {\bibinfo
  {title} {Review of nonlinear activation functions in optical neural
  networks},\ }\href@noop {} {\bibfield  {journal} {\bibinfo  {journal}
  {Advanced Photonics}\ }\textbf {\bibinfo {volume} {7}},\ \bibinfo {pages}
  {064004} (\bibinfo {year} {2025})}\BibitemShut {NoStop}%
\bibitem [{\citenamefont {Boyd}\ \emph {et~al.}(2008)\citenamefont {Boyd},
  \citenamefont {Gaeta},\ and\ \citenamefont {Giese}}]{boyd2008}%
  \BibitemOpen
  \bibfield  {author} {\bibinfo {author} {\bibfnamefont {R.~W.}\ \bibnamefont
  {Boyd}}, \bibinfo {author} {\bibfnamefont {A.~L.}\ \bibnamefont {Gaeta}},\
  and\ \bibinfo {author} {\bibfnamefont {E.}~\bibnamefont {Giese}},\ }\bibfield
   {title} {\bibinfo {title} {Nonlinear optics},\ }in\ \href@noop {} {\emph
  {\bibinfo {booktitle} {Springer Handbook of Atomic, Molecular, and Optical
  Physics}}}\ (\bibinfo  {publisher} {Springer},\ \bibinfo {year} {2008})\ pp.\
  \bibinfo {pages} {1097--1110}\BibitemShut {NoStop}%
\bibitem [{\citenamefont {Feldmann}\ \emph {et~al.}(2019)\citenamefont
  {Feldmann}, \citenamefont {Youngblood}, \citenamefont {Wright}, \citenamefont
  {Bhaskaran},\ and\ \citenamefont {Pernice}}]{feldmann2019}%
  \BibitemOpen
  \bibfield  {author} {\bibinfo {author} {\bibfnamefont {J.}~\bibnamefont
  {Feldmann}}, \bibinfo {author} {\bibfnamefont {N.}~\bibnamefont
  {Youngblood}}, \bibinfo {author} {\bibfnamefont {C.~D.}\ \bibnamefont
  {Wright}}, \bibinfo {author} {\bibfnamefont {H.}~\bibnamefont {Bhaskaran}},\
  and\ \bibinfo {author} {\bibfnamefont {W.~H.}\ \bibnamefont {Pernice}},\
  }\bibfield  {title} {\bibinfo {title} {All-optical spiking neurosynaptic
  networks with self-learning capabilities},\ }\href@noop {} {\bibfield
  {journal} {\bibinfo  {journal} {Nature}\ }\textbf {\bibinfo {volume} {569}},\
  \bibinfo {pages} {208} (\bibinfo {year} {2019})}\BibitemShut {NoStop}%
\bibitem [{\citenamefont {Wu}\ \emph {et~al.}(2022)\citenamefont {Wu},
  \citenamefont {Li}, \citenamefont {Tong}, \citenamefont {Dong},\ and\
  \citenamefont {Zhang}}]{wu2022low}%
  \BibitemOpen
  \bibfield  {author} {\bibinfo {author} {\bibfnamefont {B.}~\bibnamefont
  {Wu}}, \bibinfo {author} {\bibfnamefont {H.}~\bibnamefont {Li}}, \bibinfo
  {author} {\bibfnamefont {W.}~\bibnamefont {Tong}}, \bibinfo {author}
  {\bibfnamefont {J.}~\bibnamefont {Dong}},\ and\ \bibinfo {author}
  {\bibfnamefont {X.}~\bibnamefont {Zhang}},\ }\bibfield  {title} {\bibinfo
  {title} {Low-threshold all-optical nonlinear activation function based on a
  ge/si hybrid structure in a microring resonator},\ }\href@noop {} {\bibfield
  {journal} {\bibinfo  {journal} {Optical Materials Express}\ }\textbf
  {\bibinfo {volume} {12}},\ \bibinfo {pages} {970} (\bibinfo {year}
  {2022})}\BibitemShut {NoStop}%
\bibitem [{\citenamefont {Wu}\ \emph {et~al.}(2025)\citenamefont {Wu},
  \citenamefont {Li}, \citenamefont {Ge},\ and\ \citenamefont
  {Feng}}]{wu2025field}%
  \BibitemOpen
  \bibfield  {author} {\bibinfo {author} {\bibfnamefont {T.}~\bibnamefont
  {Wu}}, \bibinfo {author} {\bibfnamefont {Y.}~\bibnamefont {Li}}, \bibinfo
  {author} {\bibfnamefont {L.}~\bibnamefont {Ge}},\ and\ \bibinfo {author}
  {\bibfnamefont {L.}~\bibnamefont {Feng}},\ }\bibfield  {title} {\bibinfo
  {title} {Field-programmable photonic nonlinearity},\ }\href@noop {}
  {\bibfield  {journal} {\bibinfo  {journal} {Nature Photonics}\ }\textbf
  {\bibinfo {volume} {19}},\ \bibinfo {pages} {725} (\bibinfo {year}
  {2025})}\BibitemShut {NoStop}%
\bibitem [{\citenamefont {Jha}\ \emph {et~al.}(2020)\citenamefont {Jha},
  \citenamefont {Huang},\ and\ \citenamefont
  {Prucnal}}]{jha2020reconfigurable}%
  \BibitemOpen
  \bibfield  {author} {\bibinfo {author} {\bibfnamefont {A.}~\bibnamefont
  {Jha}}, \bibinfo {author} {\bibfnamefont {C.}~\bibnamefont {Huang}},\ and\
  \bibinfo {author} {\bibfnamefont {P.~R.}\ \bibnamefont {Prucnal}},\
  }\bibfield  {title} {\bibinfo {title} {Reconfigurable all-optical nonlinear
  activation functions for neuromorphic photonics},\ }\href@noop {} {\bibfield
  {journal} {\bibinfo  {journal} {Optics letters}\ }\textbf {\bibinfo {volume}
  {45}},\ \bibinfo {pages} {4819} (\bibinfo {year} {2020})}\BibitemShut
  {NoStop}%
\bibitem [{\citenamefont {Yanagimoto}\ \emph {et~al.}(2025)\citenamefont
  {Yanagimoto}, \citenamefont {Ash}, \citenamefont {Sohoni}, \citenamefont
  {Stein}, \citenamefont {Zhao}, \citenamefont {Presutti}, \citenamefont
  {Jankowski}, \citenamefont {Wright}, \citenamefont {Onodera},\ and\
  \citenamefont {McMahon}}]{yanagimoto2025programmable}%
  \BibitemOpen
  \bibfield  {author} {\bibinfo {author} {\bibfnamefont {R.}~\bibnamefont
  {Yanagimoto}}, \bibinfo {author} {\bibfnamefont {B.~A.}\ \bibnamefont {Ash}},
  \bibinfo {author} {\bibfnamefont {M.~M.}\ \bibnamefont {Sohoni}}, \bibinfo
  {author} {\bibfnamefont {M.~M.}\ \bibnamefont {Stein}}, \bibinfo {author}
  {\bibfnamefont {Y.}~\bibnamefont {Zhao}}, \bibinfo {author} {\bibfnamefont
  {F.}~\bibnamefont {Presutti}}, \bibinfo {author} {\bibfnamefont
  {M.}~\bibnamefont {Jankowski}}, \bibinfo {author} {\bibfnamefont {L.~G.}\
  \bibnamefont {Wright}}, \bibinfo {author} {\bibfnamefont {T.}~\bibnamefont
  {Onodera}},\ and\ \bibinfo {author} {\bibfnamefont {P.~L.}\ \bibnamefont
  {McMahon}},\ }\bibfield  {title} {\bibinfo {title} {Programmable on-chip
  nonlinear photonics},\ }\href@noop {} {\bibfield  {journal} {\bibinfo
  {journal} {Nature}\ ,\ \bibinfo {pages} {1}} (\bibinfo {year}
  {2025})}\BibitemShut {NoStop}%
\bibitem [{\citenamefont {Williamson}\ \emph {et~al.}(2019)\citenamefont
  {Williamson}, \citenamefont {Hughes}, \citenamefont {Minkov}, \citenamefont
  {Bartlett}, \citenamefont {Pai},\ and\ \citenamefont {Fan}}]{williamson2019}%
  \BibitemOpen
  \bibfield  {author} {\bibinfo {author} {\bibfnamefont {I.~A.}\ \bibnamefont
  {Williamson}}, \bibinfo {author} {\bibfnamefont {T.~W.}\ \bibnamefont
  {Hughes}}, \bibinfo {author} {\bibfnamefont {M.}~\bibnamefont {Minkov}},
  \bibinfo {author} {\bibfnamefont {B.}~\bibnamefont {Bartlett}}, \bibinfo
  {author} {\bibfnamefont {S.}~\bibnamefont {Pai}},\ and\ \bibinfo {author}
  {\bibfnamefont {S.}~\bibnamefont {Fan}},\ }\bibfield  {title} {\bibinfo
  {title} {Reprogrammable electro-optic nonlinear activation functions for
  optical neural networks},\ }\href@noop {} {\bibfield  {journal} {\bibinfo
  {journal} {IEEE Journal of Selected Topics in Quantum Electronics}\ }\textbf
  {\bibinfo {volume} {26}},\ \bibinfo {pages} {1} (\bibinfo {year}
  {2019})}\BibitemShut {NoStop}%
\bibitem [{\citenamefont {Pour~Fard}\ \emph {et~al.}(2020)\citenamefont
  {Pour~Fard}, \citenamefont {Williamson}, \citenamefont {Edwards},
  \citenamefont {Liu}, \citenamefont {Pai}, \citenamefont {Bartlett},
  \citenamefont {Minkov}, \citenamefont {Hughes}, \citenamefont {Fan},\ and\
  \citenamefont {Nguyen}}]{pour2020}%
  \BibitemOpen
  \bibfield  {author} {\bibinfo {author} {\bibfnamefont {M.~M.}\ \bibnamefont
  {Pour~Fard}}, \bibinfo {author} {\bibfnamefont {I.~A.}\ \bibnamefont
  {Williamson}}, \bibinfo {author} {\bibfnamefont {M.}~\bibnamefont {Edwards}},
  \bibinfo {author} {\bibfnamefont {K.}~\bibnamefont {Liu}}, \bibinfo {author}
  {\bibfnamefont {S.}~\bibnamefont {Pai}}, \bibinfo {author} {\bibfnamefont
  {B.}~\bibnamefont {Bartlett}}, \bibinfo {author} {\bibfnamefont
  {M.}~\bibnamefont {Minkov}}, \bibinfo {author} {\bibfnamefont {T.~W.}\
  \bibnamefont {Hughes}}, \bibinfo {author} {\bibfnamefont {S.}~\bibnamefont
  {Fan}},\ and\ \bibinfo {author} {\bibfnamefont {T.-A.}\ \bibnamefont
  {Nguyen}},\ }\bibfield  {title} {\bibinfo {title} {Experimental realization
  of arbitrary activation functions for optical neural networks},\ }\href@noop
  {} {\bibfield  {journal} {\bibinfo  {journal} {Optics Express}\ }\textbf
  {\bibinfo {volume} {28}},\ \bibinfo {pages} {12138} (\bibinfo {year}
  {2020})}\BibitemShut {NoStop}%
\bibitem [{\citenamefont {Hu}\ \emph {et~al.}(2025)\citenamefont {Hu},
  \citenamefont {Song}, \citenamefont {Zhu}, \citenamefont {Guo}, \citenamefont
  {Lu}, \citenamefont {Zhang}, \citenamefont {He}, \citenamefont {Franken},
  \citenamefont {Powell}, \citenamefont {Warner}, \citenamefont {Assumpcao},
  \citenamefont {Renaud}, \citenamefont {Wang}, \citenamefont {Magalh{\~a}es},
  \citenamefont {Rosborough}, \citenamefont {Shams-Ansari}, \citenamefont {Li},
  \citenamefont {Cheng}, \citenamefont {Luke}, \citenamefont {Yang},
  \citenamefont {Barbastathis}, \citenamefont {Zhang}, \citenamefont {Zhu},
  \citenamefont {Johansson}, \citenamefont {Beling}, \citenamefont {Sinclair},\
  and\ \citenamefont {Lon{\v c}ar}}]{hu2025computing}%
  \BibitemOpen
  \bibfield  {author} {\bibinfo {author} {\bibfnamefont {Y.}~\bibnamefont
  {Hu}}, \bibinfo {author} {\bibfnamefont {Y.}~\bibnamefont {Song}}, \bibinfo
  {author} {\bibfnamefont {X.}~\bibnamefont {Zhu}}, \bibinfo {author}
  {\bibfnamefont {X.}~\bibnamefont {Guo}}, \bibinfo {author} {\bibfnamefont
  {S.}~\bibnamefont {Lu}}, \bibinfo {author} {\bibfnamefont {Q.}~\bibnamefont
  {Zhang}}, \bibinfo {author} {\bibfnamefont {L.}~\bibnamefont {He}}, \bibinfo
  {author} {\bibfnamefont {C.~A.~A.}\ \bibnamefont {Franken}}, \bibinfo
  {author} {\bibfnamefont {K.}~\bibnamefont {Powell}}, \bibinfo {author}
  {\bibfnamefont {H.}~\bibnamefont {Warner}}, \bibinfo {author} {\bibfnamefont
  {D.}~\bibnamefont {Assumpcao}}, \bibinfo {author} {\bibfnamefont
  {D.}~\bibnamefont {Renaud}}, \bibinfo {author} {\bibfnamefont
  {Y.}~\bibnamefont {Wang}}, \bibinfo {author} {\bibfnamefont {L.}~\bibnamefont
  {Magalh{\~a}es}}, \bibinfo {author} {\bibfnamefont {V.}~\bibnamefont
  {Rosborough}}, \bibinfo {author} {\bibfnamefont {A.}~\bibnamefont
  {Shams-Ansari}}, \bibinfo {author} {\bibfnamefont {X.}~\bibnamefont {Li}},
  \bibinfo {author} {\bibfnamefont {R.}~\bibnamefont {Cheng}}, \bibinfo
  {author} {\bibfnamefont {K.}~\bibnamefont {Luke}}, \bibinfo {author}
  {\bibfnamefont {K.}~\bibnamefont {Yang}}, \bibinfo {author} {\bibfnamefont
  {G.}~\bibnamefont {Barbastathis}}, \bibinfo {author} {\bibfnamefont
  {M.}~\bibnamefont {Zhang}}, \bibinfo {author} {\bibfnamefont
  {D.}~\bibnamefont {Zhu}}, \bibinfo {author} {\bibfnamefont {L.}~\bibnamefont
  {Johansson}}, \bibinfo {author} {\bibfnamefont {A.}~\bibnamefont {Beling}},
  \bibinfo {author} {\bibfnamefont {N.}~\bibnamefont {Sinclair}},\ and\
  \bibinfo {author} {\bibfnamefont {M.}~\bibnamefont {Lon{\v c}ar}},\
  }\bibfield  {title} {\bibinfo {title} {Integrated lithium niobate photonic
  computing circuit based on efficient and high-speed electro-optic
  conversion},\ }\href {https://doi.org/10.1038/s41467-025-62635-8} {\bibfield
  {journal} {\bibinfo  {journal} {Nature Communications}\ }\textbf {\bibinfo
  {volume} {16}},\ \bibinfo {pages} {8178} (\bibinfo {year}
  {2025})}\BibitemShut {NoStop}%
\bibitem [{\citenamefont {Yildirim}\ \emph {et~al.}(2024)\citenamefont
  {Yildirim}, \citenamefont {Dinc}, \citenamefont {Oguz}, \citenamefont
  {Psaltis},\ and\ \citenamefont {Moser}}]{yildirim2024nonlinear}%
  \BibitemOpen
  \bibfield  {author} {\bibinfo {author} {\bibfnamefont {M.}~\bibnamefont
  {Yildirim}}, \bibinfo {author} {\bibfnamefont {N.~U.}\ \bibnamefont {Dinc}},
  \bibinfo {author} {\bibfnamefont {I.}~\bibnamefont {Oguz}}, \bibinfo {author}
  {\bibfnamefont {D.}~\bibnamefont {Psaltis}},\ and\ \bibinfo {author}
  {\bibfnamefont {C.}~\bibnamefont {Moser}},\ }\bibfield  {title} {\bibinfo
  {title} {Nonlinear processing with linear optics},\ }\href@noop {} {\bibfield
   {journal} {\bibinfo  {journal} {Nature Photonics}\ }\textbf {\bibinfo
  {volume} {18}},\ \bibinfo {pages} {1076} (\bibinfo {year}
  {2024})}\BibitemShut {NoStop}%
\bibitem [{\citenamefont {Xia}\ \emph {et~al.}(2024)\citenamefont {Xia},
  \citenamefont {Kim}, \citenamefont {Eliezer}, \citenamefont {Han},
  \citenamefont {Shaughnessy}, \citenamefont {Gigan},\ and\ \citenamefont
  {Cao}}]{xia2024nonlinear}%
  \BibitemOpen
  \bibfield  {author} {\bibinfo {author} {\bibfnamefont {F.}~\bibnamefont
  {Xia}}, \bibinfo {author} {\bibfnamefont {K.}~\bibnamefont {Kim}}, \bibinfo
  {author} {\bibfnamefont {Y.}~\bibnamefont {Eliezer}}, \bibinfo {author}
  {\bibfnamefont {S.}~\bibnamefont {Han}}, \bibinfo {author} {\bibfnamefont
  {L.}~\bibnamefont {Shaughnessy}}, \bibinfo {author} {\bibfnamefont
  {S.}~\bibnamefont {Gigan}},\ and\ \bibinfo {author} {\bibfnamefont
  {H.}~\bibnamefont {Cao}},\ }\bibfield  {title} {\bibinfo {title} {Nonlinear
  optical encoding enabled by recurrent linear scattering},\ }\href@noop {}
  {\bibfield  {journal} {\bibinfo  {journal} {Nature Photonics}\ }\textbf
  {\bibinfo {volume} {18}},\ \bibinfo {pages} {1067} (\bibinfo {year}
  {2024})}\BibitemShut {NoStop}%
\bibitem [{\citenamefont {Wanjura}\ and\ \citenamefont
  {Marquardt}(2024)}]{wanjura2024nonlinear}%
  \BibitemOpen
  \bibfield  {author} {\bibinfo {author} {\bibfnamefont {C.~C.}\ \bibnamefont
  {Wanjura}}\ and\ \bibinfo {author} {\bibfnamefont {F.}~\bibnamefont
  {Marquardt}},\ }\bibfield  {title} {\bibinfo {title} {Fully nonlinear
  neuromorphic computing with linear wave scattering},\ }\href@noop {}
  {\bibfield  {journal} {\bibinfo  {journal} {Nature Physics}\ }\textbf
  {\bibinfo {volume} {20}},\ \bibinfo {pages} {1434} (\bibinfo {year}
  {2024})}\BibitemShut {NoStop}%
\bibitem [{\citenamefont {Lodahl}\ \emph {et~al.}(2015)\citenamefont {Lodahl},
  \citenamefont {Mahmoodian},\ and\ \citenamefont {Stobbe}}]{lodahl2015review}%
  \BibitemOpen
  \bibfield  {author} {\bibinfo {author} {\bibfnamefont {P.}~\bibnamefont
  {Lodahl}}, \bibinfo {author} {\bibfnamefont {S.}~\bibnamefont {Mahmoodian}},\
  and\ \bibinfo {author} {\bibfnamefont {S.}~\bibnamefont {Stobbe}},\
  }\bibfield  {title} {\bibinfo {title} {Interfacing single photons and single
  quantum dots with photonic nanostructures},\ }\href@noop {} {\bibfield
  {journal} {\bibinfo  {journal} {Reviews of Modern Physics}\ }\textbf
  {\bibinfo {volume} {87}},\ \bibinfo {pages} {347} (\bibinfo {year}
  {2015})}\BibitemShut {NoStop}%
\bibitem [{\citenamefont {Zhu}\ \emph {et~al.}(2025)\citenamefont {Zhu},
  \citenamefont {Wang}, \citenamefont {McMahon},\ and\ \citenamefont
  {Soh}}]{zhu2025quantum}%
  \BibitemOpen
  \bibfield  {author} {\bibinfo {author} {\bibfnamefont {C.}~\bibnamefont
  {Zhu}}, \bibinfo {author} {\bibfnamefont {T.}~\bibnamefont {Wang}}, \bibinfo
  {author} {\bibfnamefont {P.~L.}\ \bibnamefont {McMahon}},\ and\ \bibinfo
  {author} {\bibfnamefont {D.}~\bibnamefont {Soh}},\ }\bibfield  {title}
  {\bibinfo {title} {Quantum optical neural networks using atom-cavity
  interactions to provide all-optical nonlinearity},\ }\href@noop {} {\bibfield
   {journal} {\bibinfo  {journal} {arXiv preprint arXiv:2511.06167}\ }
  (\bibinfo {year} {2025})}\BibitemShut {NoStop}%
\bibitem [{\citenamefont {Canora}\ \emph {et~al.}(2025)\citenamefont {Canora},
  \citenamefont {Xu}, \citenamefont {Niu}, \citenamefont {Alaeian},\ and\
  \citenamefont {Du}}]{canora2025engineering}%
  \BibitemOpen
  \bibfield  {author} {\bibinfo {author} {\bibfnamefont {R.}~\bibnamefont
  {Canora}}, \bibinfo {author} {\bibfnamefont {X.}~\bibnamefont {Xu}}, \bibinfo
  {author} {\bibfnamefont {Z.}~\bibnamefont {Niu}}, \bibinfo {author}
  {\bibfnamefont {H.}~\bibnamefont {Alaeian}},\ and\ \bibinfo {author}
  {\bibfnamefont {S.}~\bibnamefont {Du}},\ }\bibfield  {title} {\bibinfo
  {title} {Engineering nonlinear activation functions for all-optical neural
  networks via quantum interference},\ }\href@noop {} {\bibfield  {journal}
  {\bibinfo  {journal} {arXiv preprint arXiv:2504.04009}\ } (\bibinfo {year}
  {2025})}\BibitemShut {NoStop}%
\bibitem [{\citenamefont {Zuo}\ \emph {et~al.}(2019)\citenamefont {Zuo},
  \citenamefont {Li}, \citenamefont {Zhao}, \citenamefont {Jiang},
  \citenamefont {Chen}, \citenamefont {Chen}, \citenamefont {Jo}, \citenamefont
  {Liu},\ and\ \citenamefont {Du}}]{zuo2019atom}%
  \BibitemOpen
  \bibfield  {author} {\bibinfo {author} {\bibfnamefont {Y.}~\bibnamefont
  {Zuo}}, \bibinfo {author} {\bibfnamefont {B.}~\bibnamefont {Li}}, \bibinfo
  {author} {\bibfnamefont {Y.}~\bibnamefont {Zhao}}, \bibinfo {author}
  {\bibfnamefont {Y.}~\bibnamefont {Jiang}}, \bibinfo {author} {\bibfnamefont
  {Y.-C.}\ \bibnamefont {Chen}}, \bibinfo {author} {\bibfnamefont
  {P.}~\bibnamefont {Chen}}, \bibinfo {author} {\bibfnamefont {G.-B.}\
  \bibnamefont {Jo}}, \bibinfo {author} {\bibfnamefont {J.}~\bibnamefont
  {Liu}},\ and\ \bibinfo {author} {\bibfnamefont {S.}~\bibnamefont {Du}},\
  }\bibfield  {title} {\bibinfo {title} {All-optical neural network with
  nonlinear activation functions},\ }\href@noop {} {\bibfield  {journal}
  {\bibinfo  {journal} {Optica}\ }\textbf {\bibinfo {volume} {6}},\ \bibinfo
  {pages} {1132} (\bibinfo {year} {2019})}\BibitemShut {NoStop}%
\bibitem [{\citenamefont {Ryou}\ \emph {et~al.}(2021)\citenamefont {Ryou},
  \citenamefont {Whitehead}, \citenamefont {Zhelyeznyakov}, \citenamefont
  {Anderson}, \citenamefont {Keskin}, \citenamefont {Bajcsy},\ and\
  \citenamefont {Majumdar}}]{ryou2021atom}%
  \BibitemOpen
  \bibfield  {author} {\bibinfo {author} {\bibfnamefont {A.}~\bibnamefont
  {Ryou}}, \bibinfo {author} {\bibfnamefont {J.}~\bibnamefont {Whitehead}},
  \bibinfo {author} {\bibfnamefont {M.}~\bibnamefont {Zhelyeznyakov}}, \bibinfo
  {author} {\bibfnamefont {P.}~\bibnamefont {Anderson}}, \bibinfo {author}
  {\bibfnamefont {C.}~\bibnamefont {Keskin}}, \bibinfo {author} {\bibfnamefont
  {M.}~\bibnamefont {Bajcsy}},\ and\ \bibinfo {author} {\bibfnamefont
  {A.}~\bibnamefont {Majumdar}},\ }\bibfield  {title} {\bibinfo {title}
  {Free-space optical neural network based on thermal atomic nonlinearity},\
  }\href@noop {} {\bibfield  {journal} {\bibinfo  {journal} {Photonics
  Research}\ }\textbf {\bibinfo {volume} {9}},\ \bibinfo {pages} {B128}
  (\bibinfo {year} {2021})}\BibitemShut {NoStop}%
\bibitem [{\citenamefont {Ohno}\ \emph {et~al.}(2012)\citenamefont {Ohno},
  \citenamefont {Heremans}, \citenamefont {Bassett}, \citenamefont {Myers},
  \citenamefont {Toyli}, \citenamefont {Jayich}, \citenamefont {Palmstr{\o}m},\
  and\ \citenamefont {Awschalom}}]{Ohno2012_APL}%
  \BibitemOpen
  \bibfield  {author} {\bibinfo {author} {\bibfnamefont {K.}~\bibnamefont
  {Ohno}}, \bibinfo {author} {\bibfnamefont {F.~J.}\ \bibnamefont {Heremans}},
  \bibinfo {author} {\bibfnamefont {L.~C.}\ \bibnamefont {Bassett}}, \bibinfo
  {author} {\bibfnamefont {B.~A.}\ \bibnamefont {Myers}}, \bibinfo {author}
  {\bibfnamefont {D.~M.}\ \bibnamefont {Toyli}}, \bibinfo {author}
  {\bibfnamefont {A.~C.~B.}\ \bibnamefont {Jayich}}, \bibinfo {author}
  {\bibfnamefont {C.~J.}\ \bibnamefont {Palmstr{\o}m}},\ and\ \bibinfo {author}
  {\bibfnamefont {D.~D.}\ \bibnamefont {Awschalom}},\ }\bibfield  {title}
  {\bibinfo {title} {Engineering shallow spins in diamond with nitrogen
  delta-doping},\ }\href {https://doi.org/10.1063/1.4748280} {\bibfield
  {journal} {\bibinfo  {journal} {Applied Physics Letters}\ }\textbf {\bibinfo
  {volume} {101}},\ \bibinfo {pages} {082413} (\bibinfo {year}
  {2012})}\BibitemShut {NoStop}%
\bibitem [{\citenamefont {Chen}\ \emph {et~al.}(2019)\citenamefont {Chen},
  \citenamefont {Griffiths}, \citenamefont {Weng}, \citenamefont {Nicley},
  \citenamefont {Ishmael}, \citenamefont {Lekhai}, \citenamefont {Johnson},
  \citenamefont {Stephen}, \citenamefont {Green}, \citenamefont {Morley} \emph
  {et~al.}}]{chen2019laser}%
  \BibitemOpen
  \bibfield  {author} {\bibinfo {author} {\bibfnamefont {Y.-C.}\ \bibnamefont
  {Chen}}, \bibinfo {author} {\bibfnamefont {B.}~\bibnamefont {Griffiths}},
  \bibinfo {author} {\bibfnamefont {L.}~\bibnamefont {Weng}}, \bibinfo {author}
  {\bibfnamefont {S.~S.}\ \bibnamefont {Nicley}}, \bibinfo {author}
  {\bibfnamefont {S.~N.}\ \bibnamefont {Ishmael}}, \bibinfo {author}
  {\bibfnamefont {Y.}~\bibnamefont {Lekhai}}, \bibinfo {author} {\bibfnamefont
  {S.}~\bibnamefont {Johnson}}, \bibinfo {author} {\bibfnamefont {C.~J.}\
  \bibnamefont {Stephen}}, \bibinfo {author} {\bibfnamefont {B.~L.}\
  \bibnamefont {Green}}, \bibinfo {author} {\bibfnamefont {G.~W.}\ \bibnamefont
  {Morley}}, \emph {et~al.},\ }\bibfield  {title} {\bibinfo {title} {Laser
  writing of individual nitrogen-vacancy defects in diamond with near-unity
  yield},\ }\href@noop {} {\bibfield  {journal} {\bibinfo  {journal} {Optica}\
  }\textbf {\bibinfo {volume} {6}},\ \bibinfo {pages} {662} (\bibinfo {year}
  {2019})}\BibitemShut {NoStop}%
\bibitem [{\citenamefont {Day}\ \emph {et~al.}(2023)\citenamefont {Day},
  \citenamefont {Dietz}, \citenamefont {Sutula}, \citenamefont {Yeh},\ and\
  \citenamefont {Hu}}]{day2023laser}%
  \BibitemOpen
  \bibfield  {author} {\bibinfo {author} {\bibfnamefont {A.~M.}\ \bibnamefont
  {Day}}, \bibinfo {author} {\bibfnamefont {J.~R.}\ \bibnamefont {Dietz}},
  \bibinfo {author} {\bibfnamefont {M.}~\bibnamefont {Sutula}}, \bibinfo
  {author} {\bibfnamefont {M.}~\bibnamefont {Yeh}},\ and\ \bibinfo {author}
  {\bibfnamefont {E.~L.}\ \bibnamefont {Hu}},\ }\bibfield  {title} {\bibinfo
  {title} {Laser writing of spin defects in nanophotonic cavities},\
  }\href@noop {} {\bibfield  {journal} {\bibinfo  {journal} {Nature Materials}\
  }\textbf {\bibinfo {volume} {22}},\ \bibinfo {pages} {696} (\bibinfo {year}
  {2023})}\BibitemShut {NoStop}%
\bibitem [{\citenamefont {Yama}\ \emph {et~al.}(2026)\citenamefont {Yama},
  \citenamefont {Wu}, \citenamefont {Hatami},\ and\ \citenamefont
  {Fu}}]{yama2026}%
  \BibitemOpen
  \bibfield  {author} {\bibinfo {author} {\bibfnamefont {N.~S.}\ \bibnamefont
  {Yama}}, \bibinfo {author} {\bibfnamefont {C.-C.}\ \bibnamefont {Wu}},
  \bibinfo {author} {\bibfnamefont {F.}~\bibnamefont {Hatami}},\ and\ \bibinfo
  {author} {\bibfnamefont {K.-M.~C.}\ \bibnamefont {Fu}},\ }\bibfield  {title}
  {\bibinfo {title} {A scalable gallium-phosphide-on-diamond spin-photon
  interface},\ }\href@noop {} {\bibfield  {journal} {\bibinfo  {journal} {arXiv
  preprint arXiv:2601.04733}\ } (\bibinfo {year} {2026})}\BibitemShut {NoStop}%
\bibitem [{\citenamefont {Schr{\"o}der}\ \emph {et~al.}(2017)\citenamefont
  {Schr{\"o}der}, \citenamefont {Trusheim}, \citenamefont {Walsh},
  \citenamefont {Li}, \citenamefont {Zheng}, \citenamefont {Schukraft},
  \citenamefont {Sipahigil}, \citenamefont {Evans}, \citenamefont {Sukachev},
  \citenamefont {Nguyen}, \citenamefont {Pacheco}, \citenamefont {Camacho},
  \citenamefont {Bielejec}, \citenamefont {Lukin},\ and\ \citenamefont
  {Englund}}]{schroder2017}%
  \BibitemOpen
  \bibfield  {author} {\bibinfo {author} {\bibfnamefont {T.}~\bibnamefont
  {Schr{\"o}der}}, \bibinfo {author} {\bibfnamefont {M.~E.}\ \bibnamefont
  {Trusheim}}, \bibinfo {author} {\bibfnamefont {M.}~\bibnamefont {Walsh}},
  \bibinfo {author} {\bibfnamefont {L.}~\bibnamefont {Li}}, \bibinfo {author}
  {\bibfnamefont {J.}~\bibnamefont {Zheng}}, \bibinfo {author} {\bibfnamefont
  {M.}~\bibnamefont {Schukraft}}, \bibinfo {author} {\bibfnamefont
  {A.}~\bibnamefont {Sipahigil}}, \bibinfo {author} {\bibfnamefont {R.~E.}\
  \bibnamefont {Evans}}, \bibinfo {author} {\bibfnamefont {D.~D.}\ \bibnamefont
  {Sukachev}}, \bibinfo {author} {\bibfnamefont {C.~T.}\ \bibnamefont
  {Nguyen}}, \bibinfo {author} {\bibfnamefont {J.~L.}\ \bibnamefont {Pacheco}},
  \bibinfo {author} {\bibfnamefont {R.~M.}\ \bibnamefont {Camacho}}, \bibinfo
  {author} {\bibfnamefont {E.~S.}\ \bibnamefont {Bielejec}}, \bibinfo {author}
  {\bibfnamefont {M.~D.}\ \bibnamefont {Lukin}},\ and\ \bibinfo {author}
  {\bibfnamefont {D.}~\bibnamefont {Englund}},\ }\bibfield  {title} {\bibinfo
  {title} {Scalable focused ion beam creation of nearly lifetime-limited single
  quantum emitters in diamond nanostructures},\ }\href
  {https://doi.org/10.1038/ncomms15376} {\bibfield  {journal} {\bibinfo
  {journal} {Nature Communications}\ }\textbf {\bibinfo {volume} {8}},\
  \bibinfo {pages} {15376} (\bibinfo {year} {2017})}\BibitemShut {NoStop}%
\bibitem [{\citenamefont {Lalau-Keraly}\ \emph {et~al.}(2013)\citenamefont
  {Lalau-Keraly}, \citenamefont {Bhargava}, \citenamefont {Miller},\ and\
  \citenamefont {Yablonovitch}}]{lalau2013adjoint}%
  \BibitemOpen
  \bibfield  {author} {\bibinfo {author} {\bibfnamefont {C.~M.}\ \bibnamefont
  {Lalau-Keraly}}, \bibinfo {author} {\bibfnamefont {S.}~\bibnamefont
  {Bhargava}}, \bibinfo {author} {\bibfnamefont {O.~D.}\ \bibnamefont
  {Miller}},\ and\ \bibinfo {author} {\bibfnamefont {E.}~\bibnamefont
  {Yablonovitch}},\ }\bibfield  {title} {\bibinfo {title} {Adjoint shape
  optimization applied to electromagnetic design},\ }\href@noop {} {\bibfield
  {journal} {\bibinfo  {journal} {Optics express}\ }\textbf {\bibinfo {volume}
  {21}},\ \bibinfo {pages} {21693} (\bibinfo {year} {2013})}\BibitemShut
  {NoStop}%
\bibitem [{\citenamefont {Kulce}\ \emph {et~al.}(2021)\citenamefont {Kulce},
  \citenamefont {Mengu}, \citenamefont {Rivenson},\ and\ \citenamefont
  {Ozcan}}]{kulce2021capacity}%
  \BibitemOpen
  \bibfield  {author} {\bibinfo {author} {\bibfnamefont {O.}~\bibnamefont
  {Kulce}}, \bibinfo {author} {\bibfnamefont {D.}~\bibnamefont {Mengu}},
  \bibinfo {author} {\bibfnamefont {Y.}~\bibnamefont {Rivenson}},\ and\
  \bibinfo {author} {\bibfnamefont {A.}~\bibnamefont {Ozcan}},\ }\bibfield
  {title} {\bibinfo {title} {All-optical information-processing capacity of
  diffractive surfaces},\ }\href@noop {} {\bibfield  {journal} {\bibinfo
  {journal} {Light: Science \& Applications}\ }\textbf {\bibinfo {volume}
  {10}},\ \bibinfo {pages} {25} (\bibinfo {year} {2021})}\BibitemShut {NoStop}%
\bibitem [{\citenamefont {Miller}(2023)}]{miller2023thickness}%
  \BibitemOpen
  \bibfield  {author} {\bibinfo {author} {\bibfnamefont {D.~A.}\ \bibnamefont
  {Miller}},\ }\bibfield  {title} {\bibinfo {title} {Why optics needs
  thickness},\ }\href@noop {} {\bibfield  {journal} {\bibinfo  {journal}
  {Science}\ }\textbf {\bibinfo {volume} {379}},\ \bibinfo {pages} {41}
  (\bibinfo {year} {2023})}\BibitemShut {NoStop}%
\bibitem [{\citenamefont {Li}\ and\ \citenamefont
  {Monticone}(2025)}]{li2025spatial}%
  \BibitemOpen
  \bibfield  {author} {\bibinfo {author} {\bibfnamefont {Y.}~\bibnamefont
  {Li}}\ and\ \bibinfo {author} {\bibfnamefont {F.}~\bibnamefont {Monticone}},\
  }\bibfield  {title} {\bibinfo {title} {The spatial complexity of optical
  computing: toward space-efficient design},\ }\href@noop {} {\bibfield
  {journal} {\bibinfo  {journal} {Nature Communications}\ }\textbf {\bibinfo
  {volume} {16}},\ \bibinfo {pages} {8588} (\bibinfo {year}
  {2025})}\BibitemShut {NoStop}%
\bibitem [{\citenamefont {Onodera}\ \emph {et~al.}(2025)\citenamefont
  {Onodera}, \citenamefont {Stein}, \citenamefont {Ash}, \citenamefont
  {Sohoni}, \citenamefont {Bosch}, \citenamefont {Yanagimoto}, \citenamefont
  {Jankowski}, \citenamefont {McKenna}, \citenamefont {Wang}, \citenamefont
  {Shvets} \emph {et~al.}}]{onodera2025}%
  \BibitemOpen
  \bibfield  {author} {\bibinfo {author} {\bibfnamefont {T.}~\bibnamefont
  {Onodera}}, \bibinfo {author} {\bibfnamefont {M.~M.}\ \bibnamefont {Stein}},
  \bibinfo {author} {\bibfnamefont {B.~A.}\ \bibnamefont {Ash}}, \bibinfo
  {author} {\bibfnamefont {M.~M.}\ \bibnamefont {Sohoni}}, \bibinfo {author}
  {\bibfnamefont {M.}~\bibnamefont {Bosch}}, \bibinfo {author} {\bibfnamefont
  {R.}~\bibnamefont {Yanagimoto}}, \bibinfo {author} {\bibfnamefont
  {M.}~\bibnamefont {Jankowski}}, \bibinfo {author} {\bibfnamefont {T.~P.}\
  \bibnamefont {McKenna}}, \bibinfo {author} {\bibfnamefont {T.}~\bibnamefont
  {Wang}}, \bibinfo {author} {\bibfnamefont {G.}~\bibnamefont {Shvets}}, \emph
  {et~al.},\ }\bibfield  {title} {\bibinfo {title} {Arbitrary control over
  multimode wave propagation for machine learning},\ }\href@noop {} {\bibfield
  {journal} {\bibinfo  {journal} {Nature Physics}\ ,\ \bibinfo {pages} {1}}
  (\bibinfo {year} {2025})}\BibitemShut {NoStop}%
\bibitem [{\citenamefont {Yu}\ \emph {et~al.}(2025)\citenamefont {Yu},
  \citenamefont {Piao},\ and\ \citenamefont {Park}}]{yu2025nonlinear}%
  \BibitemOpen
  \bibfield  {author} {\bibinfo {author} {\bibfnamefont {S.}~\bibnamefont
  {Yu}}, \bibinfo {author} {\bibfnamefont {X.}~\bibnamefont {Piao}},\ and\
  \bibinfo {author} {\bibfnamefont {N.}~\bibnamefont {Park}},\ }\bibfield
  {title} {\bibinfo {title} {Nonlinear unitary circuits for photonic neural
  networks},\ }\href@noop {} {\bibfield  {journal} {\bibinfo  {journal} {ACS
  Photonics}\ } (\bibinfo {year} {2025})}\BibitemShut {NoStop}%
\bibitem [{\citenamefont {Mont{\'u}far}\ \emph {et~al.}(2014)\citenamefont
  {Mont{\'u}far}, \citenamefont {Pascanu}, \citenamefont {Cho},\ and\
  \citenamefont {Bengio}}]{montufar2014}%
  \BibitemOpen
  \bibfield  {author} {\bibinfo {author} {\bibfnamefont {G.}~\bibnamefont
  {Mont{\'u}far}}, \bibinfo {author} {\bibfnamefont {R.}~\bibnamefont
  {Pascanu}}, \bibinfo {author} {\bibfnamefont {K.}~\bibnamefont {Cho}},\ and\
  \bibinfo {author} {\bibfnamefont {Y.}~\bibnamefont {Bengio}},\ }\bibfield
  {title} {\bibinfo {title} {On the number of linear regions of deep neural
  networks},\ }\href@noop {} {\bibfield  {journal} {\bibinfo  {journal}
  {Advances in neural information processing systems}\ }\textbf {\bibinfo
  {volume} {27}} (\bibinfo {year} {2014})}\BibitemShut {NoStop}%
\bibitem [{\citenamefont {Poole}\ \emph {et~al.}(2016)\citenamefont {Poole},
  \citenamefont {Lahiri}, \citenamefont {Raghu}, \citenamefont
  {Sohl-Dickstein},\ and\ \citenamefont {Ganguli}}]{poole2016}%
  \BibitemOpen
  \bibfield  {author} {\bibinfo {author} {\bibfnamefont {B.}~\bibnamefont
  {Poole}}, \bibinfo {author} {\bibfnamefont {S.}~\bibnamefont {Lahiri}},
  \bibinfo {author} {\bibfnamefont {M.}~\bibnamefont {Raghu}}, \bibinfo
  {author} {\bibfnamefont {J.}~\bibnamefont {Sohl-Dickstein}},\ and\ \bibinfo
  {author} {\bibfnamefont {S.}~\bibnamefont {Ganguli}},\ }\bibfield  {title}
  {\bibinfo {title} {Exponential expressivity in deep neural networks through
  transient chaos},\ }\href@noop {} {\bibfield  {journal} {\bibinfo  {journal}
  {Advances in neural information processing systems}\ }\textbf {\bibinfo
  {volume} {29}} (\bibinfo {year} {2016})}\BibitemShut {NoStop}%
\bibitem [{\citenamefont {Raghu}\ \emph {et~al.}(2017)\citenamefont {Raghu},
  \citenamefont {Poole}, \citenamefont {Kleinberg}, \citenamefont {Ganguli},\
  and\ \citenamefont {Sohl-Dickstein}}]{raghu2017}%
  \BibitemOpen
  \bibfield  {author} {\bibinfo {author} {\bibfnamefont {M.}~\bibnamefont
  {Raghu}}, \bibinfo {author} {\bibfnamefont {B.}~\bibnamefont {Poole}},
  \bibinfo {author} {\bibfnamefont {J.}~\bibnamefont {Kleinberg}}, \bibinfo
  {author} {\bibfnamefont {S.}~\bibnamefont {Ganguli}},\ and\ \bibinfo {author}
  {\bibfnamefont {J.}~\bibnamefont {Sohl-Dickstein}},\ }\bibfield  {title}
  {\bibinfo {title} {On the expressive power of deep neural networks},\ }in\
  \href@noop {} {\emph {\bibinfo {booktitle} {international conference on
  machine learning}}}\ (\bibinfo {organization} {PMLR},\ \bibinfo {year}
  {2017})\ pp.\ \bibinfo {pages} {2847--2854}\BibitemShut {NoStop}%
\bibitem [{\citenamefont {Shi}\ \emph {et~al.}(2022)\citenamefont {Shi},
  \citenamefont {Ren}, \citenamefont {Chen}, \citenamefont {Liu}, \citenamefont
  {Jin}, \citenamefont {Guo}, \citenamefont {Yu},\ and\ \citenamefont
  {Zhang}}]{shi2022nonlinear}%
  \BibitemOpen
  \bibfield  {author} {\bibinfo {author} {\bibfnamefont {Y.}~\bibnamefont
  {Shi}}, \bibinfo {author} {\bibfnamefont {J.}~\bibnamefont {Ren}}, \bibinfo
  {author} {\bibfnamefont {G.}~\bibnamefont {Chen}}, \bibinfo {author}
  {\bibfnamefont {W.}~\bibnamefont {Liu}}, \bibinfo {author} {\bibfnamefont
  {C.}~\bibnamefont {Jin}}, \bibinfo {author} {\bibfnamefont {X.}~\bibnamefont
  {Guo}}, \bibinfo {author} {\bibfnamefont {Y.}~\bibnamefont {Yu}},\ and\
  \bibinfo {author} {\bibfnamefont {X.}~\bibnamefont {Zhang}},\ }\bibfield
  {title} {\bibinfo {title} {Nonlinear germanium-silicon photodiode for
  activation and monitoring in photonic neuromorphic networks},\ }\href@noop {}
  {\bibfield  {journal} {\bibinfo  {journal} {Nature Communications}\ }\textbf
  {\bibinfo {volume} {13}},\ \bibinfo {pages} {6048} (\bibinfo {year}
  {2022})}\BibitemShut {NoStop}%
\bibitem [{\citenamefont {Li}\ \emph {et~al.}(2023)\citenamefont {Li},
  \citenamefont {Sekine}, \citenamefont {Nehra}, \citenamefont {Gray},
  \citenamefont {Ledezma}, \citenamefont {Guo},\ and\ \citenamefont
  {Marandi}}]{li2023all}%
  \BibitemOpen
  \bibfield  {author} {\bibinfo {author} {\bibfnamefont {G.~H.}\ \bibnamefont
  {Li}}, \bibinfo {author} {\bibfnamefont {R.}~\bibnamefont {Sekine}}, \bibinfo
  {author} {\bibfnamefont {R.}~\bibnamefont {Nehra}}, \bibinfo {author}
  {\bibfnamefont {R.~M.}\ \bibnamefont {Gray}}, \bibinfo {author}
  {\bibfnamefont {L.}~\bibnamefont {Ledezma}}, \bibinfo {author} {\bibfnamefont
  {Q.}~\bibnamefont {Guo}},\ and\ \bibinfo {author} {\bibfnamefont
  {A.}~\bibnamefont {Marandi}},\ }\bibfield  {title} {\bibinfo {title}
  {All-optical ultrafast relu function for energy-efficient nanophotonic deep
  learning},\ }\href@noop {} {\bibfield  {journal} {\bibinfo  {journal}
  {Nanophotonics}\ }\textbf {\bibinfo {volume} {12}},\ \bibinfo {pages} {847}
  (\bibinfo {year} {2023})}\BibitemShut {NoStop}%
\bibitem [{\citenamefont {Bao}\ \emph {et~al.}(2009)\citenamefont {Bao},
  \citenamefont {Zhang}, \citenamefont {Wang}, \citenamefont {Ni},
  \citenamefont {Yan}, \citenamefont {Shen}, \citenamefont {Loh},\ and\
  \citenamefont {Tang}}]{bao2009atomic}%
  \BibitemOpen
  \bibfield  {author} {\bibinfo {author} {\bibfnamefont {Q.}~\bibnamefont
  {Bao}}, \bibinfo {author} {\bibfnamefont {H.}~\bibnamefont {Zhang}}, \bibinfo
  {author} {\bibfnamefont {Y.}~\bibnamefont {Wang}}, \bibinfo {author}
  {\bibfnamefont {Z.}~\bibnamefont {Ni}}, \bibinfo {author} {\bibfnamefont
  {Y.}~\bibnamefont {Yan}}, \bibinfo {author} {\bibfnamefont {Z.~X.}\
  \bibnamefont {Shen}}, \bibinfo {author} {\bibfnamefont {K.~P.}\ \bibnamefont
  {Loh}},\ and\ \bibinfo {author} {\bibfnamefont {D.~Y.}\ \bibnamefont
  {Tang}},\ }\bibfield  {title} {\bibinfo {title} {Atomic-layer graphene as a
  saturable absorber for ultrafast pulsed lasers},\ }\href@noop {} {\bibfield
  {journal} {\bibinfo  {journal} {Advanced Functional Materials}\ }\textbf
  {\bibinfo {volume} {19}},\ \bibinfo {pages} {3077} (\bibinfo {year}
  {2009})}\BibitemShut {NoStop}%
\bibitem [{\citenamefont {Shi}\ \emph {et~al.}(2017)\citenamefont {Shi},
  \citenamefont {Yu}, \citenamefont {Liu}, \citenamefont {He}, \citenamefont
  {Wang}, \citenamefont {Qin}, \citenamefont {Zhou}, \citenamefont {Li},
  \citenamefont {Zhou}, \citenamefont {Sui} \emph {et~al.}}]{shi2017MoS2}%
  \BibitemOpen
  \bibfield  {author} {\bibinfo {author} {\bibfnamefont {J.}~\bibnamefont
  {Shi}}, \bibinfo {author} {\bibfnamefont {P.}~\bibnamefont {Yu}}, \bibinfo
  {author} {\bibfnamefont {F.}~\bibnamefont {Liu}}, \bibinfo {author}
  {\bibfnamefont {P.}~\bibnamefont {He}}, \bibinfo {author} {\bibfnamefont
  {R.}~\bibnamefont {Wang}}, \bibinfo {author} {\bibfnamefont {L.}~\bibnamefont
  {Qin}}, \bibinfo {author} {\bibfnamefont {J.}~\bibnamefont {Zhou}}, \bibinfo
  {author} {\bibfnamefont {X.}~\bibnamefont {Li}}, \bibinfo {author}
  {\bibfnamefont {J.}~\bibnamefont {Zhou}}, \bibinfo {author} {\bibfnamefont
  {X.}~\bibnamefont {Sui}}, \emph {et~al.},\ }\bibfield  {title} {\bibinfo
  {title} {3r mos2 with broken inversion symmetry: a promising ultrathin
  nonlinear optical device},\ }\href@noop {} {\bibfield  {journal} {\bibinfo
  {journal} {Advanced Materials}\ }\textbf {\bibinfo {volume} {29}},\ \bibinfo
  {pages} {1701486} (\bibinfo {year} {2017})}\BibitemShut {NoStop}%
\bibitem [{\citenamefont {Liu}\ \emph {et~al.}(2025)\citenamefont {Liu},
  \citenamefont {Liang}, \citenamefont {Zhou}, \citenamefont {Khan},
  \citenamefont {Lu}, \citenamefont {Yildirim}, \citenamefont {Sun},
  \citenamefont {Rahman}, \citenamefont {Liu}, \citenamefont {Yu} \emph
  {et~al.}}]{liu2025Te}%
  \BibitemOpen
  \bibfield  {author} {\bibinfo {author} {\bibfnamefont {B.}~\bibnamefont
  {Liu}}, \bibinfo {author} {\bibfnamefont {K.}~\bibnamefont {Liang}}, \bibinfo
  {author} {\bibfnamefont {Q.}~\bibnamefont {Zhou}}, \bibinfo {author}
  {\bibfnamefont {A.~R.}\ \bibnamefont {Khan}}, \bibinfo {author}
  {\bibfnamefont {Z.}~\bibnamefont {Lu}}, \bibinfo {author} {\bibfnamefont
  {T.}~\bibnamefont {Yildirim}}, \bibinfo {author} {\bibfnamefont
  {X.}~\bibnamefont {Sun}}, \bibinfo {author} {\bibfnamefont {S.}~\bibnamefont
  {Rahman}}, \bibinfo {author} {\bibfnamefont {Y.}~\bibnamefont {Liu}},
  \bibinfo {author} {\bibfnamefont {Z.}~\bibnamefont {Yu}}, \emph {et~al.},\
  }\bibfield  {title} {\bibinfo {title} {Giant second harmonic generation in
  two-dimensional tellurene with synthesis and thickness engineering},\
  }\href@noop {} {\bibfield  {journal} {\bibinfo  {journal} {Applied physics
  reviews}\ }\textbf {\bibinfo {volume} {12}} (\bibinfo {year}
  {2025})}\BibitemShut {NoStop}%
\bibitem [{\citenamefont {Hanamura}(1988)}]{hanamura1988}%
  \BibitemOpen
  \bibfield  {author} {\bibinfo {author} {\bibfnamefont {E.}~\bibnamefont
  {Hanamura}},\ }\bibfield  {title} {\bibinfo {title} {Rapid radiative decay
  and enhanced optical nonlinearity of excitons in a quantum well},\
  }\href@noop {} {\bibfield  {journal} {\bibinfo  {journal} {Physical Review
  B}\ }\textbf {\bibinfo {volume} {38}},\ \bibinfo {pages} {1228} (\bibinfo
  {year} {1988})}\BibitemShut {NoStop}%
\bibitem [{\citenamefont {Javadi}\ \emph {et~al.}(2015)\citenamefont {Javadi},
  \citenamefont {S{\"o}llner}, \citenamefont {Arcari}, \citenamefont {Hansen},
  \citenamefont {Midolo}, \citenamefont {Mahmoodian}, \citenamefont
  {Kir{\v{s}}ansk{\.e}}, \citenamefont {Pregnolato}, \citenamefont {Lee},
  \citenamefont {Song} \emph {et~al.}}]{javadi2015}%
  \BibitemOpen
  \bibfield  {author} {\bibinfo {author} {\bibfnamefont {A.}~\bibnamefont
  {Javadi}}, \bibinfo {author} {\bibfnamefont {I.}~\bibnamefont {S{\"o}llner}},
  \bibinfo {author} {\bibfnamefont {M.}~\bibnamefont {Arcari}}, \bibinfo
  {author} {\bibfnamefont {S.~L.}\ \bibnamefont {Hansen}}, \bibinfo {author}
  {\bibfnamefont {L.}~\bibnamefont {Midolo}}, \bibinfo {author} {\bibfnamefont
  {S.}~\bibnamefont {Mahmoodian}}, \bibinfo {author} {\bibfnamefont
  {G.}~\bibnamefont {Kir{\v{s}}ansk{\.e}}}, \bibinfo {author} {\bibfnamefont
  {T.}~\bibnamefont {Pregnolato}}, \bibinfo {author} {\bibfnamefont
  {E.}~\bibnamefont {Lee}}, \bibinfo {author} {\bibfnamefont {J.}~\bibnamefont
  {Song}}, \emph {et~al.},\ }\bibfield  {title} {\bibinfo {title}
  {Single-photon non-linear optics with a quantum dot in a waveguide},\
  }\href@noop {} {\bibfield  {journal} {\bibinfo  {journal} {Nature
  communications}\ }\textbf {\bibinfo {volume} {6}},\ \bibinfo {pages} {8655}
  (\bibinfo {year} {2015})}\BibitemShut {NoStop}%
\bibitem [{\citenamefont {Volz}\ \emph {et~al.}(2014)\citenamefont {Volz},
  \citenamefont {Scheucher}, \citenamefont {Junge},\ and\ \citenamefont
  {Rauschenbeutel}}]{volz2014}%
  \BibitemOpen
  \bibfield  {author} {\bibinfo {author} {\bibfnamefont {J.}~\bibnamefont
  {Volz}}, \bibinfo {author} {\bibfnamefont {M.}~\bibnamefont {Scheucher}},
  \bibinfo {author} {\bibfnamefont {C.}~\bibnamefont {Junge}},\ and\ \bibinfo
  {author} {\bibfnamefont {A.}~\bibnamefont {Rauschenbeutel}},\ }\bibfield
  {title} {\bibinfo {title} {Nonlinear $\pi$ phase shift for single
  fibre-guided photons interacting with a single resonator-enhanced atom},\
  }\href@noop {} {\bibfield  {journal} {\bibinfo  {journal} {Nature Photonics}\
  }\textbf {\bibinfo {volume} {8}},\ \bibinfo {pages} {965} (\bibinfo {year}
  {2014})}\BibitemShut {NoStop}%
\bibitem [{\citenamefont {Shomroni}\ \emph {et~al.}(2014)\citenamefont
  {Shomroni}, \citenamefont {Rosenblum}, \citenamefont {Lovsky}, \citenamefont
  {Bechler}, \citenamefont {Guendelman},\ and\ \citenamefont
  {Dayan}}]{shomroni2014}%
  \BibitemOpen
  \bibfield  {author} {\bibinfo {author} {\bibfnamefont {I.}~\bibnamefont
  {Shomroni}}, \bibinfo {author} {\bibfnamefont {S.}~\bibnamefont {Rosenblum}},
  \bibinfo {author} {\bibfnamefont {Y.}~\bibnamefont {Lovsky}}, \bibinfo
  {author} {\bibfnamefont {O.}~\bibnamefont {Bechler}}, \bibinfo {author}
  {\bibfnamefont {G.}~\bibnamefont {Guendelman}},\ and\ \bibinfo {author}
  {\bibfnamefont {B.}~\bibnamefont {Dayan}},\ }\bibfield  {title} {\bibinfo
  {title} {All-optical routing of single photons by a one-atom switch
  controlled by a single photon},\ }\href@noop {} {\bibfield  {journal}
  {\bibinfo  {journal} {Science}\ }\textbf {\bibinfo {volume} {345}},\ \bibinfo
  {pages} {903} (\bibinfo {year} {2014})}\BibitemShut {NoStop}%
\bibitem [{\citenamefont {Hacker}\ \emph {et~al.}(2016)\citenamefont {Hacker},
  \citenamefont {Welte}, \citenamefont {Rempe},\ and\ \citenamefont
  {Ritter}}]{hacker2016}%
  \BibitemOpen
  \bibfield  {author} {\bibinfo {author} {\bibfnamefont {B.}~\bibnamefont
  {Hacker}}, \bibinfo {author} {\bibfnamefont {S.}~\bibnamefont {Welte}},
  \bibinfo {author} {\bibfnamefont {G.}~\bibnamefont {Rempe}},\ and\ \bibinfo
  {author} {\bibfnamefont {S.}~\bibnamefont {Ritter}},\ }\bibfield  {title}
  {\bibinfo {title} {A photon--photon quantum gate based on a single atom in an
  optical resonator},\ }\href@noop {} {\bibfield  {journal} {\bibinfo
  {journal} {Nature}\ }\textbf {\bibinfo {volume} {536}},\ \bibinfo {pages}
  {193} (\bibinfo {year} {2016})}\BibitemShut {NoStop}%
\bibitem [{\citenamefont {Lukin}\ \emph {et~al.}(2020)\citenamefont {Lukin},
  \citenamefont {Dory}, \citenamefont {Guidry}, \citenamefont {Yang},
  \citenamefont {Mishra}, \citenamefont {Trivedi}, \citenamefont {Radulaski},
  \citenamefont {Sun}, \citenamefont {Vercruysse}, \citenamefont {Ahn} \emph
  {et~al.}}]{lukin2020}%
  \BibitemOpen
  \bibfield  {author} {\bibinfo {author} {\bibfnamefont {D.~M.}\ \bibnamefont
  {Lukin}}, \bibinfo {author} {\bibfnamefont {C.}~\bibnamefont {Dory}},
  \bibinfo {author} {\bibfnamefont {M.~A.}\ \bibnamefont {Guidry}}, \bibinfo
  {author} {\bibfnamefont {K.~Y.}\ \bibnamefont {Yang}}, \bibinfo {author}
  {\bibfnamefont {S.~D.}\ \bibnamefont {Mishra}}, \bibinfo {author}
  {\bibfnamefont {R.}~\bibnamefont {Trivedi}}, \bibinfo {author} {\bibfnamefont
  {M.}~\bibnamefont {Radulaski}}, \bibinfo {author} {\bibfnamefont
  {S.}~\bibnamefont {Sun}}, \bibinfo {author} {\bibfnamefont {D.}~\bibnamefont
  {Vercruysse}}, \bibinfo {author} {\bibfnamefont {G.~H.}\ \bibnamefont {Ahn}},
  \emph {et~al.},\ }\bibfield  {title} {\bibinfo {title}
  {4h-silicon-carbide-on-insulator for integrated quantum and nonlinear
  photonics},\ }\href@noop {} {\bibfield  {journal} {\bibinfo  {journal}
  {Nature Photonics}\ }\textbf {\bibinfo {volume} {14}},\ \bibinfo {pages}
  {330} (\bibinfo {year} {2020})}\BibitemShut {NoStop}%
\bibitem [{\citenamefont {Zhou}\ \emph {et~al.}(2024)\citenamefont {Zhou},
  \citenamefont {Gangaraj}, \citenamefont {Zhou},\ and\ \citenamefont
  {Yu}}]{zhou2024fdtd}%
  \BibitemOpen
  \bibfield  {author} {\bibinfo {author} {\bibfnamefont {Q.}~\bibnamefont
  {Zhou}}, \bibinfo {author} {\bibfnamefont {S.}~\bibnamefont {Gangaraj}},
  \bibinfo {author} {\bibfnamefont {M.}~\bibnamefont {Zhou}},\ and\ \bibinfo
  {author} {\bibfnamefont {Z.}~\bibnamefont {Yu}},\ }\bibfield  {title}
  {\bibinfo {title} {Simulating quantum emitters in arbitrary photonic
  environments using fdtd: beyond the semi-classical regime},\ }\href@noop {}
  {\bibfield  {journal} {\bibinfo  {journal} {arXiv preprint arXiv:2410.16118}\
  } (\bibinfo {year} {2024})}\BibitemShut {NoStop}%
\bibitem [{\citenamefont {Wang}\ and\ \citenamefont
  {Fan}(2025)}]{wang2025lorentz}%
  \BibitemOpen
  \bibfield  {author} {\bibinfo {author} {\bibfnamefont {H.}~\bibnamefont
  {Wang}}\ and\ \bibinfo {author} {\bibfnamefont {S.}~\bibnamefont {Fan}},\
  }\bibfield  {title} {\bibinfo {title} {Lorentz--drude dipoles in the
  radiative limit and their modeling in finite-difference time-domain
  methods},\ }\href@noop {} {\bibfield  {journal} {\bibinfo  {journal} {Annalen
  der Physik}\ ,\ \bibinfo {pages} {e00156}} (\bibinfo {year}
  {2025})}\BibitemShut {NoStop}%
\bibitem [{\citenamefont {Hammer}\ and\ \citenamefont
  {Ivanova}(2009)}]{hammer2009eim}%
  \BibitemOpen
  \bibfield  {author} {\bibinfo {author} {\bibfnamefont {M.}~\bibnamefont
  {Hammer}}\ and\ \bibinfo {author} {\bibfnamefont {O.~V.}\ \bibnamefont
  {Ivanova}},\ }\bibfield  {title} {\bibinfo {title} {Effective index
  approximations of photonic crystal slabs: a 2-to-1-d assessment},\ }\href
  {https://doi.org/10.1007/s11082-009-9349-3} {\bibfield  {journal} {\bibinfo
  {journal} {Optical and Quantum Electronics}\ }\textbf {\bibinfo {volume}
  {41}},\ \bibinfo {pages} {267} (\bibinfo {year} {2009})}\BibitemShut
  {NoStop}%
\bibitem [{\citenamefont {Nikkhah}\ \emph {et~al.}(2024)\citenamefont
  {Nikkhah}, \citenamefont {Pirmoradi}, \citenamefont {Ashtiani}, \citenamefont
  {Edwards}, \citenamefont {Aflatouni},\ and\ \citenamefont
  {Engheta}}]{nikkhah2024_2D}%
  \BibitemOpen
  \bibfield  {author} {\bibinfo {author} {\bibfnamefont {V.}~\bibnamefont
  {Nikkhah}}, \bibinfo {author} {\bibfnamefont {A.}~\bibnamefont {Pirmoradi}},
  \bibinfo {author} {\bibfnamefont {F.}~\bibnamefont {Ashtiani}}, \bibinfo
  {author} {\bibfnamefont {B.}~\bibnamefont {Edwards}}, \bibinfo {author}
  {\bibfnamefont {F.}~\bibnamefont {Aflatouni}},\ and\ \bibinfo {author}
  {\bibfnamefont {N.}~\bibnamefont {Engheta}},\ }\bibfield  {title} {\bibinfo
  {title} {Inverse-designed low-index-contrast structures on a silicon
  photonics platform for vector--matrix multiplication},\ }\href@noop {}
  {\bibfield  {journal} {\bibinfo  {journal} {Nature Photonics}\ }\textbf
  {\bibinfo {volume} {18}},\ \bibinfo {pages} {501} (\bibinfo {year}
  {2024})}\BibitemShut {NoStop}%
\bibitem [{\citenamefont {Silver}\ \emph {et~al.}(2016)\citenamefont {Silver},
  \citenamefont {Huang}, \citenamefont {Maddison}, \citenamefont {Guez},
  \citenamefont {Sifre}, \citenamefont {Van Den~Driessche}, \citenamefont
  {Schrittwieser}, \citenamefont {Antonoglou}, \citenamefont {Panneershelvam},
  \citenamefont {Lanctot} \emph {et~al.}}]{silver2016alphago}%
  \BibitemOpen
  \bibfield  {author} {\bibinfo {author} {\bibfnamefont {D.}~\bibnamefont
  {Silver}}, \bibinfo {author} {\bibfnamefont {A.}~\bibnamefont {Huang}},
  \bibinfo {author} {\bibfnamefont {C.~J.}\ \bibnamefont {Maddison}}, \bibinfo
  {author} {\bibfnamefont {A.}~\bibnamefont {Guez}}, \bibinfo {author}
  {\bibfnamefont {L.}~\bibnamefont {Sifre}}, \bibinfo {author} {\bibfnamefont
  {G.}~\bibnamefont {Van Den~Driessche}}, \bibinfo {author} {\bibfnamefont
  {J.}~\bibnamefont {Schrittwieser}}, \bibinfo {author} {\bibfnamefont
  {I.}~\bibnamefont {Antonoglou}}, \bibinfo {author} {\bibfnamefont
  {V.}~\bibnamefont {Panneershelvam}}, \bibinfo {author} {\bibfnamefont
  {M.}~\bibnamefont {Lanctot}}, \emph {et~al.},\ }\bibfield  {title} {\bibinfo
  {title} {Mastering the game of go with deep neural networks and tree
  search},\ }\href@noop {} {\bibfield  {journal} {\bibinfo  {journal} {nature}\
  }\textbf {\bibinfo {volume} {529}},\ \bibinfo {pages} {484} (\bibinfo {year}
  {2016})}\BibitemShut {NoStop}%
\bibitem [{\citenamefont {Hwangbo}\ \emph {et~al.}(2019)\citenamefont
  {Hwangbo}, \citenamefont {Lee}, \citenamefont {Dosovitskiy}, \citenamefont
  {Bellicoso}, \citenamefont {Tsounis}, \citenamefont {Koltun},\ and\
  \citenamefont {Hutter}}]{hwangbo2019}%
  \BibitemOpen
  \bibfield  {author} {\bibinfo {author} {\bibfnamefont {J.}~\bibnamefont
  {Hwangbo}}, \bibinfo {author} {\bibfnamefont {J.}~\bibnamefont {Lee}},
  \bibinfo {author} {\bibfnamefont {A.}~\bibnamefont {Dosovitskiy}}, \bibinfo
  {author} {\bibfnamefont {D.}~\bibnamefont {Bellicoso}}, \bibinfo {author}
  {\bibfnamefont {V.}~\bibnamefont {Tsounis}}, \bibinfo {author} {\bibfnamefont
  {V.}~\bibnamefont {Koltun}},\ and\ \bibinfo {author} {\bibfnamefont
  {M.}~\bibnamefont {Hutter}},\ }\bibfield  {title} {\bibinfo {title} {Learning
  agile and dynamic motor skills for legged robots},\ }\href@noop {} {\bibfield
   {journal} {\bibinfo  {journal} {Science Robotics}\ }\textbf {\bibinfo
  {volume} {4}},\ \bibinfo {pages} {eaau5872} (\bibinfo {year}
  {2019})}\BibitemShut {NoStop}%
\bibitem [{\citenamefont {Xiang}\ \emph {et~al.}(2025)\citenamefont {Xiang},
  \citenamefont {Chen}, \citenamefont {Zhao}, \citenamefont {Shi},
  \citenamefont {Zeng}, \citenamefont {Zhang}, \citenamefont {Guo},
  \citenamefont {Han}, \citenamefont {Shi},\ and\ \citenamefont
  {Hao}}]{xiang2025spiking_rl}%
  \BibitemOpen
  \bibfield  {author} {\bibinfo {author} {\bibfnamefont {S.}~\bibnamefont
  {Xiang}}, \bibinfo {author} {\bibfnamefont {Y.}~\bibnamefont {Chen}},
  \bibinfo {author} {\bibfnamefont {H.}~\bibnamefont {Zhao}}, \bibinfo {author}
  {\bibfnamefont {S.}~\bibnamefont {Shi}}, \bibinfo {author} {\bibfnamefont
  {X.}~\bibnamefont {Zeng}}, \bibinfo {author} {\bibfnamefont {Y.}~\bibnamefont
  {Zhang}}, \bibinfo {author} {\bibfnamefont {X.}~\bibnamefont {Guo}}, \bibinfo
  {author} {\bibfnamefont {Y.}~\bibnamefont {Han}}, \bibinfo {author}
  {\bibfnamefont {Y.}~\bibnamefont {Shi}},\ and\ \bibinfo {author}
  {\bibfnamefont {Y.}~\bibnamefont {Hao}},\ }\bibfield  {title} {\bibinfo
  {title} {Nonlinear photonic neuromorphic chips for spiking reinforcement
  learning},\ }\href {https://doi.org/10.1364/OPTICA.578687} {\bibfield
  {journal} {\bibinfo  {journal} {Optica}\ }\textbf {\bibinfo {volume} {13}},\
  \bibinfo {pages} {457} (\bibinfo {year} {2025})}\BibitemShut {NoStop}%
\bibitem [{\citenamefont {Towers}\ \emph {et~al.}(2024)\citenamefont {Towers},
  \citenamefont {Kwiatkowski}, \citenamefont {Terry}, \citenamefont {Balis},
  \citenamefont {De~Cola}, \citenamefont {Deleu}, \citenamefont {Goul{\~a}o},
  \citenamefont {Kallinteris}, \citenamefont {Krimmel}, \citenamefont {KG}
  \emph {et~al.}}]{towers2024gymnasium}%
  \BibitemOpen
  \bibfield  {author} {\bibinfo {author} {\bibfnamefont {M.}~\bibnamefont
  {Towers}}, \bibinfo {author} {\bibfnamefont {A.}~\bibnamefont {Kwiatkowski}},
  \bibinfo {author} {\bibfnamefont {J.}~\bibnamefont {Terry}}, \bibinfo
  {author} {\bibfnamefont {J.~U.}\ \bibnamefont {Balis}}, \bibinfo {author}
  {\bibfnamefont {G.}~\bibnamefont {De~Cola}}, \bibinfo {author} {\bibfnamefont
  {T.}~\bibnamefont {Deleu}}, \bibinfo {author} {\bibfnamefont
  {M.}~\bibnamefont {Goul{\~a}o}}, \bibinfo {author} {\bibfnamefont
  {A.}~\bibnamefont {Kallinteris}}, \bibinfo {author} {\bibfnamefont
  {M.}~\bibnamefont {Krimmel}}, \bibinfo {author} {\bibfnamefont
  {A.}~\bibnamefont {KG}}, \emph {et~al.},\ }\bibfield  {title} {\bibinfo
  {title} {Gymnasium: A standard interface for reinforcement learning
  environments},\ }\href@noop {} {\bibfield  {journal} {\bibinfo  {journal}
  {arXiv preprint arXiv:2407.17032}\ } (\bibinfo {year} {2024})}\BibitemShut
  {NoStop}%
\bibitem [{\citenamefont {Schulman}\ \emph {et~al.}(2017)\citenamefont
  {Schulman}, \citenamefont {Wolski}, \citenamefont {Dhariwal}, \citenamefont
  {Radford},\ and\ \citenamefont {Klimov}}]{schulman2017ppo}%
  \BibitemOpen
  \bibfield  {author} {\bibinfo {author} {\bibfnamefont {J.}~\bibnamefont
  {Schulman}}, \bibinfo {author} {\bibfnamefont {F.}~\bibnamefont {Wolski}},
  \bibinfo {author} {\bibfnamefont {P.}~\bibnamefont {Dhariwal}}, \bibinfo
  {author} {\bibfnamefont {A.}~\bibnamefont {Radford}},\ and\ \bibinfo {author}
  {\bibfnamefont {O.}~\bibnamefont {Klimov}},\ }\bibfield  {title} {\bibinfo
  {title} {Proximal policy optimization algorithms},\ }\href@noop {} {\bibfield
   {journal} {\bibinfo  {journal} {arXiv preprint arXiv:1707.06347}\ }
  (\bibinfo {year} {2017})}\BibitemShut {NoStop}%
\bibitem [{\citenamefont {Haarnoja}\ \emph {et~al.}(2018)\citenamefont
  {Haarnoja}, \citenamefont {Zhou}, \citenamefont {Abbeel},\ and\ \citenamefont
  {Levine}}]{haarnoja2018sac}%
  \BibitemOpen
  \bibfield  {author} {\bibinfo {author} {\bibfnamefont {T.}~\bibnamefont
  {Haarnoja}}, \bibinfo {author} {\bibfnamefont {A.}~\bibnamefont {Zhou}},
  \bibinfo {author} {\bibfnamefont {P.}~\bibnamefont {Abbeel}},\ and\ \bibinfo
  {author} {\bibfnamefont {S.}~\bibnamefont {Levine}},\ }\bibfield  {title}
  {\bibinfo {title} {Soft actor-critic: Off-policy maximum entropy deep
  reinforcement learning with a stochastic actor},\ }in\ \href@noop {} {\emph
  {\bibinfo {booktitle} {International conference on machine learning}}}\
  (\bibinfo {organization} {Pmlr},\ \bibinfo {year} {2018})\ pp.\ \bibinfo
  {pages} {1861--1870}\BibitemShut {NoStop}%
\bibitem [{\citenamefont {Dinu}\ \emph {et~al.}(2003)\citenamefont {Dinu},
  \citenamefont {Quochi},\ and\ \citenamefont {Garcia}}]{dinu2003third}%
  \BibitemOpen
  \bibfield  {author} {\bibinfo {author} {\bibfnamefont {M.}~\bibnamefont
  {Dinu}}, \bibinfo {author} {\bibfnamefont {F.}~\bibnamefont {Quochi}},\ and\
  \bibinfo {author} {\bibfnamefont {H.}~\bibnamefont {Garcia}},\ }\bibfield
  {title} {\bibinfo {title} {Third-order nonlinearities in silicon at telecom
  wavelengths},\ }\href@noop {} {\bibfield  {journal} {\bibinfo  {journal}
  {Applied physics letters}\ }\textbf {\bibinfo {volume} {82}},\ \bibinfo
  {pages} {2954} (\bibinfo {year} {2003})}\BibitemShut {NoStop}%
\bibitem [{\citenamefont {Dulkeith}\ \emph {et~al.}(2006)\citenamefont
  {Dulkeith}, \citenamefont {Vlasov}, \citenamefont {Chen}, \citenamefont
  {Panoiu},\ and\ \citenamefont {Osgood~Jr}}]{dulkeith2006self}%
  \BibitemOpen
  \bibfield  {author} {\bibinfo {author} {\bibfnamefont {E.}~\bibnamefont
  {Dulkeith}}, \bibinfo {author} {\bibfnamefont {Y.~A.}\ \bibnamefont
  {Vlasov}}, \bibinfo {author} {\bibfnamefont {X.}~\bibnamefont {Chen}},
  \bibinfo {author} {\bibfnamefont {N.~C.}\ \bibnamefont {Panoiu}},\ and\
  \bibinfo {author} {\bibfnamefont {R.~M.}\ \bibnamefont {Osgood~Jr}},\
  }\bibfield  {title} {\bibinfo {title} {Self-phase-modulation in submicron
  silicon-on-insulator photonic wires},\ }\href@noop {} {\bibfield  {journal}
  {\bibinfo  {journal} {Optics express}\ }\textbf {\bibinfo {volume} {14}},\
  \bibinfo {pages} {5524} (\bibinfo {year} {2006})}\BibitemShut {NoStop}%
\bibitem [{\citenamefont {Lau}\ \emph {et~al.}(2022)\citenamefont {Lau},
  \citenamefont {Liu},\ and\ \citenamefont {Qiu}}]{lau2022comparison}%
  \BibitemOpen
  \bibfield  {author} {\bibinfo {author} {\bibfnamefont {K.~Y.}\ \bibnamefont
  {Lau}}, \bibinfo {author} {\bibfnamefont {X.}~\bibnamefont {Liu}},\ and\
  \bibinfo {author} {\bibfnamefont {J.}~\bibnamefont {Qiu}},\ }\bibfield
  {title} {\bibinfo {title} {A comparison for saturable absorbers: Carbon
  nanotube versus graphene},\ }\href@noop {} {\bibfield  {journal} {\bibinfo
  {journal} {Advanced Photonics Research}\ }\textbf {\bibinfo {volume} {3}},\
  \bibinfo {pages} {2200023} (\bibinfo {year} {2022})}\BibitemShut {NoStop}%
\bibitem [{\citenamefont {Radford}\ \emph {et~al.}(2019)\citenamefont
  {Radford}, \citenamefont {Wu}, \citenamefont {Child}, \citenamefont {Luan},
  \citenamefont {Amodei}, \citenamefont {Sutskever} \emph
  {et~al.}}]{radford2019language}%
  \BibitemOpen
  \bibfield  {author} {\bibinfo {author} {\bibfnamefont {A.}~\bibnamefont
  {Radford}}, \bibinfo {author} {\bibfnamefont {J.}~\bibnamefont {Wu}},
  \bibinfo {author} {\bibfnamefont {R.}~\bibnamefont {Child}}, \bibinfo
  {author} {\bibfnamefont {D.}~\bibnamefont {Luan}}, \bibinfo {author}
  {\bibfnamefont {D.}~\bibnamefont {Amodei}}, \bibinfo {author} {\bibfnamefont
  {I.}~\bibnamefont {Sutskever}}, \emph {et~al.},\ }\bibfield  {title}
  {\bibinfo {title} {Language models are unsupervised multitask learners},\
  }\href@noop {} {\bibfield  {journal} {\bibinfo  {journal} {OpenAI blog}\
  }\textbf {\bibinfo {volume} {1}},\ \bibinfo {pages} {9} (\bibinfo {year}
  {2019})}\BibitemShut {NoStop}%
\bibitem [{\citenamefont {Brown}\ \emph {et~al.}(2020)\citenamefont {Brown},
  \citenamefont {Mann}, \citenamefont {Ryder}, \citenamefont {Subbiah},
  \citenamefont {Kaplan}, \citenamefont {Dhariwal}, \citenamefont
  {Neelakantan}, \citenamefont {Shyam}, \citenamefont {Sastry}, \citenamefont
  {Askell} \emph {et~al.}}]{brown2020gpt3}%
  \BibitemOpen
  \bibfield  {author} {\bibinfo {author} {\bibfnamefont {T.}~\bibnamefont
  {Brown}}, \bibinfo {author} {\bibfnamefont {B.}~\bibnamefont {Mann}},
  \bibinfo {author} {\bibfnamefont {N.}~\bibnamefont {Ryder}}, \bibinfo
  {author} {\bibfnamefont {M.}~\bibnamefont {Subbiah}}, \bibinfo {author}
  {\bibfnamefont {J.~D.}\ \bibnamefont {Kaplan}}, \bibinfo {author}
  {\bibfnamefont {P.}~\bibnamefont {Dhariwal}}, \bibinfo {author}
  {\bibfnamefont {A.}~\bibnamefont {Neelakantan}}, \bibinfo {author}
  {\bibfnamefont {P.}~\bibnamefont {Shyam}}, \bibinfo {author} {\bibfnamefont
  {G.}~\bibnamefont {Sastry}}, \bibinfo {author} {\bibfnamefont
  {A.}~\bibnamefont {Askell}}, \emph {et~al.},\ }\bibfield  {title} {\bibinfo
  {title} {Language models are few-shot learners},\ }\href@noop {} {\bibfield
  {journal} {\bibinfo  {journal} {Advances in neural information processing
  systems}\ }\textbf {\bibinfo {volume} {33}},\ \bibinfo {pages} {1877}
  (\bibinfo {year} {2020})}\BibitemShut {NoStop}%
\bibitem [{\citenamefont {Touvron}\ \emph
  {et~al.}(2023{\natexlab{a}})\citenamefont {Touvron}, \citenamefont {Lavril},
  \citenamefont {Izacard}, \citenamefont {Martinet}, \citenamefont {Lachaux},
  \citenamefont {Lacroix}, \citenamefont {Rozi{\`e}re}, \citenamefont {Goyal},
  \citenamefont {Hambro}, \citenamefont {Azhar} \emph
  {et~al.}}]{touvron2023llama1}%
  \BibitemOpen
  \bibfield  {author} {\bibinfo {author} {\bibfnamefont {H.}~\bibnamefont
  {Touvron}}, \bibinfo {author} {\bibfnamefont {T.}~\bibnamefont {Lavril}},
  \bibinfo {author} {\bibfnamefont {G.}~\bibnamefont {Izacard}}, \bibinfo
  {author} {\bibfnamefont {X.}~\bibnamefont {Martinet}}, \bibinfo {author}
  {\bibfnamefont {M.-A.}\ \bibnamefont {Lachaux}}, \bibinfo {author}
  {\bibfnamefont {T.}~\bibnamefont {Lacroix}}, \bibinfo {author} {\bibfnamefont
  {B.}~\bibnamefont {Rozi{\`e}re}}, \bibinfo {author} {\bibfnamefont
  {N.}~\bibnamefont {Goyal}}, \bibinfo {author} {\bibfnamefont
  {E.}~\bibnamefont {Hambro}}, \bibinfo {author} {\bibfnamefont
  {F.}~\bibnamefont {Azhar}}, \emph {et~al.},\ }\bibfield  {title} {\bibinfo
  {title} {Llama: Open and efficient foundation language models},\ }\href@noop
  {} {\bibfield  {journal} {\bibinfo  {journal} {arXiv preprint
  arXiv:2302.13971}\ } (\bibinfo {year} {2023}{\natexlab{a}})}\BibitemShut
  {NoStop}%
\bibitem [{\citenamefont {Touvron}\ \emph
  {et~al.}(2023{\natexlab{b}})\citenamefont {Touvron}, \citenamefont {Martin},
  \citenamefont {Stone}, \citenamefont {Albert}, \citenamefont {Almahairi},
  \citenamefont {Babaei}, \citenamefont {Bashlykov}, \citenamefont {Batra},
  \citenamefont {Bhargava}, \citenamefont {Bhosale} \emph
  {et~al.}}]{touvron2023llama2}%
  \BibitemOpen
  \bibfield  {author} {\bibinfo {author} {\bibfnamefont {H.}~\bibnamefont
  {Touvron}}, \bibinfo {author} {\bibfnamefont {L.}~\bibnamefont {Martin}},
  \bibinfo {author} {\bibfnamefont {K.}~\bibnamefont {Stone}}, \bibinfo
  {author} {\bibfnamefont {P.}~\bibnamefont {Albert}}, \bibinfo {author}
  {\bibfnamefont {A.}~\bibnamefont {Almahairi}}, \bibinfo {author}
  {\bibfnamefont {Y.}~\bibnamefont {Babaei}}, \bibinfo {author} {\bibfnamefont
  {N.}~\bibnamefont {Bashlykov}}, \bibinfo {author} {\bibfnamefont
  {S.}~\bibnamefont {Batra}}, \bibinfo {author} {\bibfnamefont
  {P.}~\bibnamefont {Bhargava}}, \bibinfo {author} {\bibfnamefont
  {S.}~\bibnamefont {Bhosale}}, \emph {et~al.},\ }\bibfield  {title} {\bibinfo
  {title} {Llama 2: Open foundation and fine-tuned chat models},\ }\href@noop
  {} {\bibfield  {journal} {\bibinfo  {journal} {arXiv preprint
  arXiv:2307.09288}\ } (\bibinfo {year} {2023}{\natexlab{b}})}\BibitemShut
  {NoStop}%
\bibitem [{\citenamefont {Dubey}\ \emph {et~al.}(2024)\citenamefont {Dubey},
  \citenamefont {Jauhri}, \citenamefont {Pandey}, \citenamefont {Kadian},
  \citenamefont {Al-Dahle}, \citenamefont {Letman}, \citenamefont {Mathur},
  \citenamefont {Schelten}, \citenamefont {Yang}, \citenamefont {Fan} \emph
  {et~al.}}]{dubey2024llama3}%
  \BibitemOpen
  \bibfield  {author} {\bibinfo {author} {\bibfnamefont {A.}~\bibnamefont
  {Dubey}}, \bibinfo {author} {\bibfnamefont {A.}~\bibnamefont {Jauhri}},
  \bibinfo {author} {\bibfnamefont {A.}~\bibnamefont {Pandey}}, \bibinfo
  {author} {\bibfnamefont {A.}~\bibnamefont {Kadian}}, \bibinfo {author}
  {\bibfnamefont {A.}~\bibnamefont {Al-Dahle}}, \bibinfo {author}
  {\bibfnamefont {A.}~\bibnamefont {Letman}}, \bibinfo {author} {\bibfnamefont
  {A.}~\bibnamefont {Mathur}}, \bibinfo {author} {\bibfnamefont
  {A.}~\bibnamefont {Schelten}}, \bibinfo {author} {\bibfnamefont
  {A.}~\bibnamefont {Yang}}, \bibinfo {author} {\bibfnamefont {A.}~\bibnamefont
  {Fan}}, \emph {et~al.},\ }\bibfield  {title} {\bibinfo {title} {The llama 3
  herd of models},\ }\href@noop {} {\bibfield  {journal} {\bibinfo  {journal}
  {arXiv e-prints}\ ,\ \bibinfo {pages} {arXiv}} (\bibinfo {year}
  {2024})}\BibitemShut {NoStop}%
\bibitem [{\citenamefont {Bi}\ \emph {et~al.}(2024)\citenamefont {Bi},
  \citenamefont {Chen}, \citenamefont {Chen}, \citenamefont {Chen},
  \citenamefont {Dai}, \citenamefont {Deng}, \citenamefont {Ding},
  \citenamefont {Dong}, \citenamefont {Du}, \citenamefont {Fu} \emph
  {et~al.}}]{bi2024deepseek}%
  \BibitemOpen
  \bibfield  {author} {\bibinfo {author} {\bibfnamefont {X.}~\bibnamefont
  {Bi}}, \bibinfo {author} {\bibfnamefont {D.}~\bibnamefont {Chen}}, \bibinfo
  {author} {\bibfnamefont {G.}~\bibnamefont {Chen}}, \bibinfo {author}
  {\bibfnamefont {S.}~\bibnamefont {Chen}}, \bibinfo {author} {\bibfnamefont
  {D.}~\bibnamefont {Dai}}, \bibinfo {author} {\bibfnamefont {C.}~\bibnamefont
  {Deng}}, \bibinfo {author} {\bibfnamefont {H.}~\bibnamefont {Ding}}, \bibinfo
  {author} {\bibfnamefont {K.}~\bibnamefont {Dong}}, \bibinfo {author}
  {\bibfnamefont {Q.}~\bibnamefont {Du}}, \bibinfo {author} {\bibfnamefont
  {Z.}~\bibnamefont {Fu}}, \emph {et~al.},\ }\bibfield  {title} {\bibinfo
  {title} {Deepseek llm: Scaling open-source language models with
  longtermism},\ }\href@noop {} {\bibfield  {journal} {\bibinfo  {journal}
  {arXiv preprint arXiv:2401.02954}\ } (\bibinfo {year} {2024})}\BibitemShut
  {NoStop}%
\bibitem [{\citenamefont {Liu}\ \emph {et~al.}(2024{\natexlab{a}})\citenamefont
  {Liu}, \citenamefont {Feng}, \citenamefont {Wang}, \citenamefont {Wang},
  \citenamefont {Liu}, \citenamefont {Zhao}, \citenamefont {Dengr},
  \citenamefont {Ruan}, \citenamefont {Dai}, \citenamefont {Guo} \emph
  {et~al.}}]{liu2024deepseek-v2}%
  \BibitemOpen
  \bibfield  {author} {\bibinfo {author} {\bibfnamefont {A.}~\bibnamefont
  {Liu}}, \bibinfo {author} {\bibfnamefont {B.}~\bibnamefont {Feng}}, \bibinfo
  {author} {\bibfnamefont {B.}~\bibnamefont {Wang}}, \bibinfo {author}
  {\bibfnamefont {B.}~\bibnamefont {Wang}}, \bibinfo {author} {\bibfnamefont
  {B.}~\bibnamefont {Liu}}, \bibinfo {author} {\bibfnamefont {C.}~\bibnamefont
  {Zhao}}, \bibinfo {author} {\bibfnamefont {C.}~\bibnamefont {Dengr}},
  \bibinfo {author} {\bibfnamefont {C.}~\bibnamefont {Ruan}}, \bibinfo {author}
  {\bibfnamefont {D.}~\bibnamefont {Dai}}, \bibinfo {author} {\bibfnamefont
  {D.}~\bibnamefont {Guo}}, \emph {et~al.},\ }\bibfield  {title} {\bibinfo
  {title} {Deepseek-v2: A strong, economical, and efficient mixture-of-experts
  language model},\ }\href@noop {} {\bibfield  {journal} {\bibinfo  {journal}
  {arXiv preprint arXiv:2405.04434}\ } (\bibinfo {year}
  {2024}{\natexlab{a}})}\BibitemShut {NoStop}%
\bibitem [{\citenamefont {Liu}\ \emph {et~al.}(2024{\natexlab{b}})\citenamefont
  {Liu}, \citenamefont {Feng}, \citenamefont {Xue}, \citenamefont {Wang},
  \citenamefont {Wu}, \citenamefont {Lu}, \citenamefont {Zhao}, \citenamefont
  {Deng}, \citenamefont {Zhang}, \citenamefont {Ruan} \emph
  {et~al.}}]{liu2024deepseek-v3}%
  \BibitemOpen
  \bibfield  {author} {\bibinfo {author} {\bibfnamefont {A.}~\bibnamefont
  {Liu}}, \bibinfo {author} {\bibfnamefont {B.}~\bibnamefont {Feng}}, \bibinfo
  {author} {\bibfnamefont {B.}~\bibnamefont {Xue}}, \bibinfo {author}
  {\bibfnamefont {B.}~\bibnamefont {Wang}}, \bibinfo {author} {\bibfnamefont
  {B.}~\bibnamefont {Wu}}, \bibinfo {author} {\bibfnamefont {C.}~\bibnamefont
  {Lu}}, \bibinfo {author} {\bibfnamefont {C.}~\bibnamefont {Zhao}}, \bibinfo
  {author} {\bibfnamefont {C.}~\bibnamefont {Deng}}, \bibinfo {author}
  {\bibfnamefont {C.}~\bibnamefont {Zhang}}, \bibinfo {author} {\bibfnamefont
  {C.}~\bibnamefont {Ruan}}, \emph {et~al.},\ }\bibfield  {title} {\bibinfo
  {title} {Deepseek-v3 technical report},\ }\href@noop {} {\bibfield  {journal}
  {\bibinfo  {journal} {arXiv preprint arXiv:2412.19437}\ } (\bibinfo {year}
  {2024}{\natexlab{b}})}\BibitemShut {NoStop}%
\bibitem [{\citenamefont {Anderson}\ \emph {et~al.}(2023)\citenamefont
  {Anderson}, \citenamefont {Ma}, \citenamefont {Wang}, \citenamefont
  {Wright},\ and\ \citenamefont {McMahon}}]{anderson2023optical}%
  \BibitemOpen
  \bibfield  {author} {\bibinfo {author} {\bibfnamefont {M.}~\bibnamefont
  {Anderson}}, \bibinfo {author} {\bibfnamefont {S.-Y.}\ \bibnamefont {Ma}},
  \bibinfo {author} {\bibfnamefont {T.}~\bibnamefont {Wang}}, \bibinfo {author}
  {\bibfnamefont {L.}~\bibnamefont {Wright}},\ and\ \bibinfo {author}
  {\bibfnamefont {P.}~\bibnamefont {McMahon}},\ }\bibfield  {title} {\bibinfo
  {title} {Optical transformers},\ }\href@noop {} {\bibfield  {journal}
  {\bibinfo  {journal} {Transactions on Machine Learning Research}\ } (\bibinfo
  {year} {2023})}\BibitemShut {NoStop}%
\bibitem [{\citenamefont {Hua}\ \emph {et~al.}(2025)\citenamefont {Hua},
  \citenamefont {Divita}, \citenamefont {Yu}, \citenamefont {Peng},
  \citenamefont {Roques-Carmes} \emph {et~al.}}]{lightelligence2025pace}%
  \BibitemOpen
  \bibfield  {author} {\bibinfo {author} {\bibfnamefont {S.}~\bibnamefont
  {Hua}}, \bibinfo {author} {\bibfnamefont {E.}~\bibnamefont {Divita}},
  \bibinfo {author} {\bibfnamefont {S.}~\bibnamefont {Yu}}, \bibinfo {author}
  {\bibfnamefont {B.}~\bibnamefont {Peng}}, \bibinfo {author} {\bibfnamefont
  {C.}~\bibnamefont {Roques-Carmes}}, \emph {et~al.},\ }\bibfield  {title}
  {\bibinfo {title} {An integrated large-scale photonic accelerator with
  ultralow latency},\ }\href {https://doi.org/10.1038/s41586-025-08786-6}
  {\bibfield  {journal} {\bibinfo  {journal} {Nature}\ }\textbf {\bibinfo
  {volume} {640}},\ \bibinfo {pages} {361} (\bibinfo {year}
  {2025})}\BibitemShut {NoStop}%
\bibitem [{\citenamefont {Ahmed}\ \emph {et~al.}(2025)\citenamefont {Ahmed},
  \citenamefont {Baghdadi}, \citenamefont {Bernadskiy} \emph
  {et~al.}}]{ahmed2025upaia}%
  \BibitemOpen
  \bibfield  {author} {\bibinfo {author} {\bibfnamefont {S.~R.}\ \bibnamefont
  {Ahmed}}, \bibinfo {author} {\bibfnamefont {R.}~\bibnamefont {Baghdadi}},
  \bibinfo {author} {\bibfnamefont {M.}~\bibnamefont {Bernadskiy}}, \emph
  {et~al.},\ }\bibfield  {title} {\bibinfo {title} {Universal photonic
  artificial intelligence acceleration},\ }\href
  {https://doi.org/10.1038/s41586-025-08854-x} {\bibfield  {journal} {\bibinfo
  {journal} {Nature}\ }\textbf {\bibinfo {volume} {640}},\ \bibinfo {pages}
  {368} (\bibinfo {year} {2025})}\BibitemShut {NoStop}%
\bibitem [{\citenamefont {Englund}\ \emph {et~al.}(2005)\citenamefont
  {Englund}, \citenamefont {Fattal}, \citenamefont {Waks}, \citenamefont
  {Solomon}, \citenamefont {Zhang}, \citenamefont {Nakaoka}, \citenamefont
  {Arakawa}, \citenamefont {Yamamoto},\ and\ \citenamefont
  {Vu{\v{c}}kovi{\'c}}}]{englund2005PhC}%
  \BibitemOpen
  \bibfield  {author} {\bibinfo {author} {\bibfnamefont {D.}~\bibnamefont
  {Englund}}, \bibinfo {author} {\bibfnamefont {D.}~\bibnamefont {Fattal}},
  \bibinfo {author} {\bibfnamefont {E.}~\bibnamefont {Waks}}, \bibinfo {author}
  {\bibfnamefont {G.}~\bibnamefont {Solomon}}, \bibinfo {author} {\bibfnamefont
  {B.}~\bibnamefont {Zhang}}, \bibinfo {author} {\bibfnamefont
  {T.}~\bibnamefont {Nakaoka}}, \bibinfo {author} {\bibfnamefont
  {Y.}~\bibnamefont {Arakawa}}, \bibinfo {author} {\bibfnamefont
  {Y.}~\bibnamefont {Yamamoto}},\ and\ \bibinfo {author} {\bibfnamefont
  {J.}~\bibnamefont {Vu{\v{c}}kovi{\'c}}},\ }\bibfield  {title} {\bibinfo
  {title} {Controlling the spontaneous emission rate of single quantum dots in
  a two-dimensional photonic crystal},\ }\href@noop {} {\bibfield  {journal}
  {\bibinfo  {journal} {Physical review letters}\ }\textbf {\bibinfo {volume}
  {95}},\ \bibinfo {pages} {013904} (\bibinfo {year} {2005})}\BibitemShut
  {NoStop}%
\bibitem [{\citenamefont {Evans}\ \emph {et~al.}(2018)\citenamefont {Evans},
  \citenamefont {Bhaskar}, \citenamefont {Sukachev}, \citenamefont {Nguyen},
  \citenamefont {Sipahigil}, \citenamefont {Burek}, \citenamefont {Machielse},
  \citenamefont {Zhang}, \citenamefont {Zibrov}, \citenamefont {Bielejec},
  \citenamefont {Park}, \citenamefont {Lon{\v c}ar},\ and\ \citenamefont
  {Lukin}}]{evans2018}%
  \BibitemOpen
  \bibfield  {author} {\bibinfo {author} {\bibfnamefont {R.~E.}\ \bibnamefont
  {Evans}}, \bibinfo {author} {\bibfnamefont {M.~K.}\ \bibnamefont {Bhaskar}},
  \bibinfo {author} {\bibfnamefont {D.~D.}\ \bibnamefont {Sukachev}}, \bibinfo
  {author} {\bibfnamefont {C.~T.}\ \bibnamefont {Nguyen}}, \bibinfo {author}
  {\bibfnamefont {A.}~\bibnamefont {Sipahigil}}, \bibinfo {author}
  {\bibfnamefont {M.~J.}\ \bibnamefont {Burek}}, \bibinfo {author}
  {\bibfnamefont {B.}~\bibnamefont {Machielse}}, \bibinfo {author}
  {\bibfnamefont {G.~H.}\ \bibnamefont {Zhang}}, \bibinfo {author}
  {\bibfnamefont {A.~S.}\ \bibnamefont {Zibrov}}, \bibinfo {author}
  {\bibfnamefont {E.}~\bibnamefont {Bielejec}}, \bibinfo {author}
  {\bibfnamefont {H.}~\bibnamefont {Park}}, \bibinfo {author} {\bibfnamefont
  {M.}~\bibnamefont {Lon{\v c}ar}},\ and\ \bibinfo {author} {\bibfnamefont
  {M.~D.}\ \bibnamefont {Lukin}},\ }\bibfield  {title} {\bibinfo {title}
  {Photon-mediated interactions between quantum emitters in a diamond
  nanocavity},\ }\href {https://doi.org/10.1126/science.aau4691} {\bibfield
  {journal} {\bibinfo  {journal} {Science}\ }\textbf {\bibinfo {volume}
  {362}},\ \bibinfo {pages} {662} (\bibinfo {year} {2018})}\BibitemShut
  {NoStop}%
\bibitem [{\citenamefont {Laucht}\ \emph {et~al.}(2010)\citenamefont {Laucht},
  \citenamefont {Villas-B{\^o}as}, \citenamefont {Stobbe}, \citenamefont
  {Hauke}, \citenamefont {Hofbauer}, \citenamefont {B{\"o}hm}, \citenamefont
  {Lodahl}, \citenamefont {Amann}, \citenamefont {Kaniber},\ and\ \citenamefont
  {Finley}}]{laucht2010stark}%
  \BibitemOpen
  \bibfield  {author} {\bibinfo {author} {\bibfnamefont {A.}~\bibnamefont
  {Laucht}}, \bibinfo {author} {\bibfnamefont {J.}~\bibnamefont
  {Villas-B{\^o}as}}, \bibinfo {author} {\bibfnamefont {S.}~\bibnamefont
  {Stobbe}}, \bibinfo {author} {\bibfnamefont {N.}~\bibnamefont {Hauke}},
  \bibinfo {author} {\bibfnamefont {F.}~\bibnamefont {Hofbauer}}, \bibinfo
  {author} {\bibfnamefont {G.}~\bibnamefont {B{\"o}hm}}, \bibinfo {author}
  {\bibfnamefont {P.}~\bibnamefont {Lodahl}}, \bibinfo {author} {\bibfnamefont
  {M.-C.}\ \bibnamefont {Amann}}, \bibinfo {author} {\bibfnamefont
  {M.}~\bibnamefont {Kaniber}},\ and\ \bibinfo {author} {\bibfnamefont
  {J.}~\bibnamefont {Finley}},\ }\bibfield  {title} {\bibinfo {title} {Mutual
  coupling of two semiconductor quantum dots via an optical nanocavity},\
  }\href@noop {} {\bibfield  {journal} {\bibinfo  {journal} {Physical Review
  B—Condensed Matter and Materials Physics}\ }\textbf {\bibinfo {volume}
  {82}},\ \bibinfo {pages} {075305} (\bibinfo {year} {2010})}\BibitemShut
  {NoStop}%
\bibitem [{\citenamefont {Chen}\ \emph {et~al.}(2017)\citenamefont {Chen},
  \citenamefont {Salter}, \citenamefont {Knauer}, \citenamefont {Weng},
  \citenamefont {Frangeskou}, \citenamefont {Stephen}, \citenamefont {Ishmael},
  \citenamefont {Dolan}, \citenamefont {Johnson}, \citenamefont {Green} \emph
  {et~al.}}]{chen2017laser}%
  \BibitemOpen
  \bibfield  {author} {\bibinfo {author} {\bibfnamefont {Y.-C.}\ \bibnamefont
  {Chen}}, \bibinfo {author} {\bibfnamefont {P.~S.}\ \bibnamefont {Salter}},
  \bibinfo {author} {\bibfnamefont {S.}~\bibnamefont {Knauer}}, \bibinfo
  {author} {\bibfnamefont {L.}~\bibnamefont {Weng}}, \bibinfo {author}
  {\bibfnamefont {A.~C.}\ \bibnamefont {Frangeskou}}, \bibinfo {author}
  {\bibfnamefont {C.~J.}\ \bibnamefont {Stephen}}, \bibinfo {author}
  {\bibfnamefont {S.~N.}\ \bibnamefont {Ishmael}}, \bibinfo {author}
  {\bibfnamefont {P.~R.}\ \bibnamefont {Dolan}}, \bibinfo {author}
  {\bibfnamefont {S.}~\bibnamefont {Johnson}}, \bibinfo {author} {\bibfnamefont
  {B.~L.}\ \bibnamefont {Green}}, \emph {et~al.},\ }\bibfield  {title}
  {\bibinfo {title} {Laser writing of coherent colour centres in diamond},\
  }\href@noop {} {\bibfield  {journal} {\bibinfo  {journal} {Nature Photonics}\
  }\textbf {\bibinfo {volume} {11}},\ \bibinfo {pages} {77} (\bibinfo {year}
  {2017})}\BibitemShut {NoStop}%
\bibitem [{\citenamefont {Laferri{\`e}re}\ \emph {et~al.}(2022)\citenamefont
  {Laferri{\`e}re}, \citenamefont {Yeung}, \citenamefont {Miron}, \citenamefont
  {Northeast}, \citenamefont {Haffouz}, \citenamefont {Lapointe}, \citenamefont
  {Korkusinski}, \citenamefont {Poole}, \citenamefont {Williams},\ and\
  \citenamefont {Dalacu}}]{laferriere2022QD}%
  \BibitemOpen
  \bibfield  {author} {\bibinfo {author} {\bibfnamefont {P.}~\bibnamefont
  {Laferri{\`e}re}}, \bibinfo {author} {\bibfnamefont {E.}~\bibnamefont
  {Yeung}}, \bibinfo {author} {\bibfnamefont {I.}~\bibnamefont {Miron}},
  \bibinfo {author} {\bibfnamefont {D.~B.}\ \bibnamefont {Northeast}}, \bibinfo
  {author} {\bibfnamefont {S.}~\bibnamefont {Haffouz}}, \bibinfo {author}
  {\bibfnamefont {J.}~\bibnamefont {Lapointe}}, \bibinfo {author}
  {\bibfnamefont {M.}~\bibnamefont {Korkusinski}}, \bibinfo {author}
  {\bibfnamefont {P.~J.}\ \bibnamefont {Poole}}, \bibinfo {author}
  {\bibfnamefont {R.~L.}\ \bibnamefont {Williams}},\ and\ \bibinfo {author}
  {\bibfnamefont {D.}~\bibnamefont {Dalacu}},\ }\bibfield  {title} {\bibinfo
  {title} {Unity yield of deterministically positioned quantum dot single
  photon sources},\ }\href@noop {} {\bibfield  {journal} {\bibinfo  {journal}
  {Scientific Reports}\ }\textbf {\bibinfo {volume} {12}},\ \bibinfo {pages}
  {6376} (\bibinfo {year} {2022})}\BibitemShut {NoStop}%
\bibitem [{\citenamefont {Sipahigil}\ \emph {et~al.}(2014)\citenamefont
  {Sipahigil}, \citenamefont {Jahnke}, \citenamefont {Rogers}, \citenamefont
  {Teraji}, \citenamefont {Isoya}, \citenamefont {Zibrov}, \citenamefont
  {Jelezko},\ and\ \citenamefont {Lukin}}]{Sipahigil2014}%
  \BibitemOpen
  \bibfield  {author} {\bibinfo {author} {\bibfnamefont {A.}~\bibnamefont
  {Sipahigil}}, \bibinfo {author} {\bibfnamefont {K.~D.}\ \bibnamefont
  {Jahnke}}, \bibinfo {author} {\bibfnamefont {L.~J.}\ \bibnamefont {Rogers}},
  \bibinfo {author} {\bibfnamefont {T.}~\bibnamefont {Teraji}}, \bibinfo
  {author} {\bibfnamefont {J.}~\bibnamefont {Isoya}}, \bibinfo {author}
  {\bibfnamefont {A.~S.}\ \bibnamefont {Zibrov}}, \bibinfo {author}
  {\bibfnamefont {F.}~\bibnamefont {Jelezko}},\ and\ \bibinfo {author}
  {\bibfnamefont {M.~D.}\ \bibnamefont {Lukin}},\ }\bibfield  {title} {\bibinfo
  {title} {Indistinguishable photons from separated silicon-vacancy centers in
  diamond},\ }\href {https://doi.org/10.1103/PhysRevLett.113.113602} {\bibfield
   {journal} {\bibinfo  {journal} {Physical Review Letters}\ }\textbf {\bibinfo
  {volume} {113}},\ \bibinfo {pages} {113602} (\bibinfo {year}
  {2014})}\BibitemShut {NoStop}%
\bibitem [{\citenamefont {Rogers}\ \emph {et~al.}(2014)\citenamefont {Rogers},
  \citenamefont {Jahnke}, \citenamefont {Teraji}, \citenamefont {Marseglia},
  \citenamefont {M\"uller}, \citenamefont {Naydenov}, \citenamefont
  {Schauffert}, \citenamefont {Kranz}, \citenamefont {Isoya}, \citenamefont
  {McGuinness},\ and\ \citenamefont {Jelezko}}]{Rogers2014}%
  \BibitemOpen
  \bibfield  {author} {\bibinfo {author} {\bibfnamefont {L.~J.}\ \bibnamefont
  {Rogers}}, \bibinfo {author} {\bibfnamefont {K.~D.}\ \bibnamefont {Jahnke}},
  \bibinfo {author} {\bibfnamefont {T.}~\bibnamefont {Teraji}}, \bibinfo
  {author} {\bibfnamefont {L.}~\bibnamefont {Marseglia}}, \bibinfo {author}
  {\bibfnamefont {C.}~\bibnamefont {M\"uller}}, \bibinfo {author}
  {\bibfnamefont {B.}~\bibnamefont {Naydenov}}, \bibinfo {author}
  {\bibfnamefont {H.}~\bibnamefont {Schauffert}}, \bibinfo {author}
  {\bibfnamefont {C.}~\bibnamefont {Kranz}}, \bibinfo {author} {\bibfnamefont
  {J.}~\bibnamefont {Isoya}}, \bibinfo {author} {\bibfnamefont {L.~P.}\
  \bibnamefont {McGuinness}},\ and\ \bibinfo {author} {\bibfnamefont
  {F.}~\bibnamefont {Jelezko}},\ }\bibfield  {title} {\bibinfo {title}
  {Multiple intrinsically identical single-photon emitters in the solid
  state},\ }\href {https://doi.org/10.1038/ncomms5739} {\bibfield  {journal}
  {\bibinfo  {journal} {Nature Communications}\ }\textbf {\bibinfo {volume}
  {5}},\ \bibinfo {pages} {4739} (\bibinfo {year} {2014})}\BibitemShut
  {NoStop}%
\end{thebibliography}

\begin{thebibliography}{70}%
\makeatletter
\providecommand \@ifxundefined [1]{%
 \@ifx{#1\undefined}
}%
\providecommand \@ifnum [1]{%
 \ifnum #1\expandafter \@firstoftwo
 \else \expandafter \@secondoftwo
 \fi
}%
\providecommand \@ifx [1]{%
 \ifx #1\expandafter \@firstoftwo
 \else \expandafter \@secondoftwo
 \fi
}%
\providecommand \natexlab [1]{#1}%
\providecommand \enquote  [1]{``#1''}%
\providecommand \bibnamefont  [1]{#1}%
\providecommand \bibfnamefont [1]{#1}%
\providecommand \citenamefont [1]{#1}%
\providecommand \href@noop [0]{\@secondoftwo}%
\providecommand \href [0]{\begingroup \@sanitize@url \@href}%
\providecommand \@href[1]{\@@startlink{#1}\@@href}%
\providecommand \@@href[1]{\endgroup#1\@@endlink}%
\providecommand \@sanitize@url [0]{\catcode `\\12\catcode `\$12\catcode
  `\&12\catcode `\#12\catcode `\^12\catcode `\_12\catcode `\%12\relax}%
\providecommand \@@startlink[1]{}%
\providecommand \@@endlink[0]{}%
\providecommand \url  [0]{\begingroup\@sanitize@url \@url }%
\providecommand \@url [1]{\endgroup\@href {#1}{\urlprefix }}%
\providecommand \urlprefix  [0]{URL }%
\providecommand \Eprint [0]{\href }%
\providecommand \doibase [0]{https://doi.org/}%
\providecommand \selectlanguage [0]{\@gobble}%
\providecommand \bibinfo  [0]{\@secondoftwo}%
\providecommand \bibfield  [0]{\@secondoftwo}%
\providecommand \translation [1]{[#1]}%
\providecommand \BibitemOpen [0]{}%
\providecommand \bibitemStop [0]{}%
\providecommand \bibitemNoStop [0]{.\EOS\space}%
\providecommand \EOS [0]{\spacefactor3000\relax}%
\providecommand \BibitemShut  [1]{\csname bibitem#1\endcsname}%
\let\auto@bib@innerbib\@empty
\bibitem [{\citenamefont {Hughes}\ \emph {et~al.}(2018)\citenamefont {Hughes},
  \citenamefont {Minkov}, \citenamefont {Williamson},\ and\ \citenamefont
  {Fan}}]{si:hughes2018}%
  \BibitemOpen
  \bibfield  {author} {\bibinfo {author} {\bibfnamefont {T.~W.}\ \bibnamefont
  {Hughes}}, \bibinfo {author} {\bibfnamefont {M.}~\bibnamefont {Minkov}},
  \bibinfo {author} {\bibfnamefont {I.~A.}\ \bibnamefont {Williamson}},\ and\
  \bibinfo {author} {\bibfnamefont {S.}~\bibnamefont {Fan}},\ }\bibfield
  {title} {\bibinfo {title} {Adjoint method and inverse design for nonlinear
  nanophotonic devices},\ }\href@noop {} {\bibfield  {journal} {\bibinfo
  {journal} {ACS Photonics}\ }\textbf {\bibinfo {volume} {5}},\ \bibinfo
  {pages} {4781} (\bibinfo {year} {2018})}\BibitemShut {NoStop}%
\bibitem [{\citenamefont {Feldmann}\ \emph {et~al.}(2019)\citenamefont
  {Feldmann}, \citenamefont {Youngblood}, \citenamefont {Wright}, \citenamefont
  {Bhaskaran},\ and\ \citenamefont {Pernice}}]{si:feldmann2019}%
  \BibitemOpen
  \bibfield  {author} {\bibinfo {author} {\bibfnamefont {J.}~\bibnamefont
  {Feldmann}}, \bibinfo {author} {\bibfnamefont {N.}~\bibnamefont
  {Youngblood}}, \bibinfo {author} {\bibfnamefont {C.~D.}\ \bibnamefont
  {Wright}}, \bibinfo {author} {\bibfnamefont {H.}~\bibnamefont {Bhaskaran}},\
  and\ \bibinfo {author} {\bibfnamefont {W.~H.}\ \bibnamefont {Pernice}},\
  }\bibfield  {title} {\bibinfo {title} {All-optical spiking neurosynaptic
  networks with self-learning capabilities},\ }\href@noop {} {\bibfield
  {journal} {\bibinfo  {journal} {Nature}\ }\textbf {\bibinfo {volume} {569}},\
  \bibinfo {pages} {208} (\bibinfo {year} {2019})}\BibitemShut {NoStop}%
\bibitem [{\citenamefont {Hughes}\ \emph {et~al.}(2019)\citenamefont {Hughes},
  \citenamefont {Williamson}, \citenamefont {Minkov},\ and\ \citenamefont
  {Fan}}]{si:hughes2019}%
  \BibitemOpen
  \bibfield  {author} {\bibinfo {author} {\bibfnamefont {T.~W.}\ \bibnamefont
  {Hughes}}, \bibinfo {author} {\bibfnamefont {I.~A.}\ \bibnamefont
  {Williamson}}, \bibinfo {author} {\bibfnamefont {M.}~\bibnamefont {Minkov}},\
  and\ \bibinfo {author} {\bibfnamefont {S.}~\bibnamefont {Fan}},\ }\bibfield
  {title} {\bibinfo {title} {Wave physics as an analog recurrent neural
  network},\ }\href@noop {} {\bibfield  {journal} {\bibinfo  {journal} {Science
  advances}\ }\textbf {\bibinfo {volume} {5}},\ \bibinfo {pages} {eaay6946}
  (\bibinfo {year} {2019})}\BibitemShut {NoStop}%
\bibitem [{\citenamefont {Yu}\ \emph {et~al.}(2022)\citenamefont {Yu},
  \citenamefont {Zheng}, \citenamefont {Zhao}, \citenamefont {Wang},\ and\
  \citenamefont {Zhang}}]{si:yu2022reconfigurable}%
  \BibitemOpen
  \bibfield  {author} {\bibinfo {author} {\bibfnamefont {W.}~\bibnamefont
  {Yu}}, \bibinfo {author} {\bibfnamefont {S.}~\bibnamefont {Zheng}}, \bibinfo
  {author} {\bibfnamefont {Z.}~\bibnamefont {Zhao}}, \bibinfo {author}
  {\bibfnamefont {B.}~\bibnamefont {Wang}},\ and\ \bibinfo {author}
  {\bibfnamefont {W.}~\bibnamefont {Zhang}},\ }\bibfield  {title} {\bibinfo
  {title} {Reconfigurable low-threshold all-optical nonlinear activation
  functions based on an add-drop silicon microring resonator},\ }\href@noop {}
  {\bibfield  {journal} {\bibinfo  {journal} {IEEE Photonics Journal}\ }\textbf
  {\bibinfo {volume} {14}},\ \bibinfo {pages} {1} (\bibinfo {year}
  {2022})}\BibitemShut {NoStop}%
\bibitem [{\citenamefont {Shi}\ \emph {et~al.}(2022)\citenamefont {Shi},
  \citenamefont {Ren}, \citenamefont {Chen}, \citenamefont {Liu}, \citenamefont
  {Jin}, \citenamefont {Guo}, \citenamefont {Yu},\ and\ \citenamefont
  {Zhang}}]{si:shi2022nonlinear}%
  \BibitemOpen
  \bibfield  {author} {\bibinfo {author} {\bibfnamefont {Y.}~\bibnamefont
  {Shi}}, \bibinfo {author} {\bibfnamefont {J.}~\bibnamefont {Ren}}, \bibinfo
  {author} {\bibfnamefont {G.}~\bibnamefont {Chen}}, \bibinfo {author}
  {\bibfnamefont {W.}~\bibnamefont {Liu}}, \bibinfo {author} {\bibfnamefont
  {C.}~\bibnamefont {Jin}}, \bibinfo {author} {\bibfnamefont {X.}~\bibnamefont
  {Guo}}, \bibinfo {author} {\bibfnamefont {Y.}~\bibnamefont {Yu}},\ and\
  \bibinfo {author} {\bibfnamefont {X.}~\bibnamefont {Zhang}},\ }\bibfield
  {title} {\bibinfo {title} {Nonlinear germanium-silicon photodiode for
  activation and monitoring in photonic neuromorphic networks},\ }\href@noop {}
  {\bibfield  {journal} {\bibinfo  {journal} {Nature Communications}\ }\textbf
  {\bibinfo {volume} {13}},\ \bibinfo {pages} {6048} (\bibinfo {year}
  {2022})}\BibitemShut {NoStop}%
\bibitem [{\citenamefont {Wu}\ \emph {et~al.}(2022)\citenamefont {Wu},
  \citenamefont {Li}, \citenamefont {Tong}, \citenamefont {Dong},\ and\
  \citenamefont {Zhang}}]{si:wu2022low}%
  \BibitemOpen
  \bibfield  {author} {\bibinfo {author} {\bibfnamefont {B.}~\bibnamefont
  {Wu}}, \bibinfo {author} {\bibfnamefont {H.}~\bibnamefont {Li}}, \bibinfo
  {author} {\bibfnamefont {W.}~\bibnamefont {Tong}}, \bibinfo {author}
  {\bibfnamefont {J.}~\bibnamefont {Dong}},\ and\ \bibinfo {author}
  {\bibfnamefont {X.}~\bibnamefont {Zhang}},\ }\bibfield  {title} {\bibinfo
  {title} {Low-threshold all-optical nonlinear activation function based on a
  ge/si hybrid structure in a microring resonator},\ }\href@noop {} {\bibfield
  {journal} {\bibinfo  {journal} {Optical Materials Express}\ }\textbf
  {\bibinfo {volume} {12}},\ \bibinfo {pages} {970} (\bibinfo {year}
  {2022})}\BibitemShut {NoStop}%
\bibitem [{\citenamefont {Chen}\ \emph {et~al.}(2024)\citenamefont {Chen},
  \citenamefont {Yang}, \citenamefont {Wang}, \citenamefont {Wang},
  \citenamefont {Gao}, \citenamefont {Wu}, \citenamefont {Wang}, \citenamefont
  {Qiu},\ and\ \citenamefont {Tan}}]{si:chen2024ultra}%
  \BibitemOpen
  \bibfield  {author} {\bibinfo {author} {\bibfnamefont {C.}~\bibnamefont
  {Chen}}, \bibinfo {author} {\bibfnamefont {Z.}~\bibnamefont {Yang}}, \bibinfo
  {author} {\bibfnamefont {T.}~\bibnamefont {Wang}}, \bibinfo {author}
  {\bibfnamefont {Y.}~\bibnamefont {Wang}}, \bibinfo {author} {\bibfnamefont
  {K.}~\bibnamefont {Gao}}, \bibinfo {author} {\bibfnamefont {J.}~\bibnamefont
  {Wu}}, \bibinfo {author} {\bibfnamefont {J.}~\bibnamefont {Wang}}, \bibinfo
  {author} {\bibfnamefont {J.}~\bibnamefont {Qiu}},\ and\ \bibinfo {author}
  {\bibfnamefont {D.}~\bibnamefont {Tan}},\ }\bibfield  {title} {\bibinfo
  {title} {Ultra-broadband all-optical nonlinear activation function enabled by
  mote2/optical waveguide integrated devices},\ }\href@noop {} {\bibfield
  {journal} {\bibinfo  {journal} {Nature Communications}\ }\textbf {\bibinfo
  {volume} {15}},\ \bibinfo {pages} {9047} (\bibinfo {year}
  {2024})}\BibitemShut {NoStop}%
\bibitem [{\citenamefont {Yang}\ \emph {et~al.}(2024)\citenamefont {Yang},
  \citenamefont {He}, \citenamefont {Yan}, \citenamefont {Hu}, \citenamefont
  {Li}, \citenamefont {Dong},\ and\ \citenamefont {Wang}}]{si:yang2024inverse}%
  \BibitemOpen
  \bibfield  {author} {\bibinfo {author} {\bibfnamefont {Z.}~\bibnamefont
  {Yang}}, \bibinfo {author} {\bibfnamefont {J.}~\bibnamefont {He}}, \bibinfo
  {author} {\bibfnamefont {Z.}~\bibnamefont {Yan}}, \bibinfo {author}
  {\bibfnamefont {Y.}~\bibnamefont {Hu}}, \bibinfo {author} {\bibfnamefont
  {X.}~\bibnamefont {Li}}, \bibinfo {author} {\bibfnamefont {N.}~\bibnamefont
  {Dong}},\ and\ \bibinfo {author} {\bibfnamefont {J.}~\bibnamefont {Wang}},\
  }\bibfield  {title} {\bibinfo {title} {Inverse-designed integrated
  all-optical nonlinear activators for optical computing},\ }\href@noop {}
  {\bibfield  {journal} {\bibinfo  {journal} {Optics Express}\ }\textbf
  {\bibinfo {volume} {32}},\ \bibinfo {pages} {34001} (\bibinfo {year}
  {2024})}\BibitemShut {NoStop}%
\bibitem [{\citenamefont {Zhao}\ \emph {et~al.}(2025)\citenamefont {Zhao},
  \citenamefont {Lin}, \citenamefont {Samuel},\ and\ \citenamefont
  {Lawrence}}]{si:zhao2025high}%
  \BibitemOpen
  \bibfield  {author} {\bibinfo {author} {\bibfnamefont {B.}~\bibnamefont
  {Zhao}}, \bibinfo {author} {\bibfnamefont {L.}~\bibnamefont {Lin}}, \bibinfo
  {author} {\bibfnamefont {A.}~\bibnamefont {Samuel}},\ and\ \bibinfo {author}
  {\bibfnamefont {M.}~\bibnamefont {Lawrence}},\ }\bibfield  {title} {\bibinfo
  {title} {High-resolution and ultra-low power nonlinear image processing with
  passive high-quality factor metasurfaces},\ }\href@noop {} {\bibfield
  {journal} {\bibinfo  {journal} {arXiv preprint arXiv:2504.02981}\ } (\bibinfo
  {year} {2025})}\BibitemShut {NoStop}%
\bibitem [{\citenamefont {Huang}\ \emph {et~al.}(2019)\citenamefont {Huang},
  \citenamefont {De~Lima}, \citenamefont {Jha}, \citenamefont {Abbaslou},
  \citenamefont {Tait}, \citenamefont {Shastri},\ and\ \citenamefont
  {Prucnal}}]{si:huang2019programmable}%
  \BibitemOpen
  \bibfield  {author} {\bibinfo {author} {\bibfnamefont {C.}~\bibnamefont
  {Huang}}, \bibinfo {author} {\bibfnamefont {T.~F.}\ \bibnamefont {De~Lima}},
  \bibinfo {author} {\bibfnamefont {A.}~\bibnamefont {Jha}}, \bibinfo {author}
  {\bibfnamefont {S.}~\bibnamefont {Abbaslou}}, \bibinfo {author}
  {\bibfnamefont {A.~N.}\ \bibnamefont {Tait}}, \bibinfo {author}
  {\bibfnamefont {B.~J.}\ \bibnamefont {Shastri}},\ and\ \bibinfo {author}
  {\bibfnamefont {P.~R.}\ \bibnamefont {Prucnal}},\ }\bibfield  {title}
  {\bibinfo {title} {Programmable silicon photonic optical thresholder},\
  }\href@noop {} {\bibfield  {journal} {\bibinfo  {journal} {IEEE Photonics
  Technology Letters}\ }\textbf {\bibinfo {volume} {31}},\ \bibinfo {pages}
  {1834} (\bibinfo {year} {2019})}\BibitemShut {NoStop}%
\bibitem [{\citenamefont {Jha}\ \emph {et~al.}(2020)\citenamefont {Jha},
  \citenamefont {Huang},\ and\ \citenamefont
  {Prucnal}}]{si:jha2020reconfigurable}%
  \BibitemOpen
  \bibfield  {author} {\bibinfo {author} {\bibfnamefont {A.}~\bibnamefont
  {Jha}}, \bibinfo {author} {\bibfnamefont {C.}~\bibnamefont {Huang}},\ and\
  \bibinfo {author} {\bibfnamefont {P.~R.}\ \bibnamefont {Prucnal}},\
  }\bibfield  {title} {\bibinfo {title} {Reconfigurable all-optical nonlinear
  activation functions for neuromorphic photonics},\ }\href@noop {} {\bibfield
  {journal} {\bibinfo  {journal} {Optics letters}\ }\textbf {\bibinfo {volume}
  {45}},\ \bibinfo {pages} {4819} (\bibinfo {year} {2020})}\BibitemShut
  {NoStop}%
\bibitem [{\citenamefont {Wu}\ \emph {et~al.}(2025)\citenamefont {Wu},
  \citenamefont {Li}, \citenamefont {Ge},\ and\ \citenamefont
  {Feng}}]{si:wu2025field}%
  \BibitemOpen
  \bibfield  {author} {\bibinfo {author} {\bibfnamefont {T.}~\bibnamefont
  {Wu}}, \bibinfo {author} {\bibfnamefont {Y.}~\bibnamefont {Li}}, \bibinfo
  {author} {\bibfnamefont {L.}~\bibnamefont {Ge}},\ and\ \bibinfo {author}
  {\bibfnamefont {L.}~\bibnamefont {Feng}},\ }\bibfield  {title} {\bibinfo
  {title} {Field-programmable photonic nonlinearity},\ }\href@noop {}
  {\bibfield  {journal} {\bibinfo  {journal} {Nature Photonics}\ }\textbf
  {\bibinfo {volume} {19}},\ \bibinfo {pages} {725} (\bibinfo {year}
  {2025})}\BibitemShut {NoStop}%
\bibitem [{\citenamefont {Li}\ \emph {et~al.}(2023)\citenamefont {Li},
  \citenamefont {Sekine}, \citenamefont {Nehra}, \citenamefont {Gray},
  \citenamefont {Ledezma}, \citenamefont {Guo},\ and\ \citenamefont
  {Marandi}}]{si:li2023all}%
  \BibitemOpen
  \bibfield  {author} {\bibinfo {author} {\bibfnamefont {G.~H.}\ \bibnamefont
  {Li}}, \bibinfo {author} {\bibfnamefont {R.}~\bibnamefont {Sekine}}, \bibinfo
  {author} {\bibfnamefont {R.}~\bibnamefont {Nehra}}, \bibinfo {author}
  {\bibfnamefont {R.~M.}\ \bibnamefont {Gray}}, \bibinfo {author}
  {\bibfnamefont {L.}~\bibnamefont {Ledezma}}, \bibinfo {author} {\bibfnamefont
  {Q.}~\bibnamefont {Guo}},\ and\ \bibinfo {author} {\bibfnamefont
  {A.}~\bibnamefont {Marandi}},\ }\bibfield  {title} {\bibinfo {title}
  {All-optical ultrafast relu function for energy-efficient nanophotonic deep
  learning},\ }\href@noop {} {\bibfield  {journal} {\bibinfo  {journal}
  {Nanophotonics}\ }\textbf {\bibinfo {volume} {12}},\ \bibinfo {pages} {847}
  (\bibinfo {year} {2023})}\BibitemShut {NoStop}%
\bibitem [{\citenamefont {Pour~Fard}\ \emph {et~al.}(2020)\citenamefont
  {Pour~Fard}, \citenamefont {Williamson}, \citenamefont {Edwards},
  \citenamefont {Liu}, \citenamefont {Pai}, \citenamefont {Bartlett},
  \citenamefont {Minkov}, \citenamefont {Hughes}, \citenamefont {Fan},\ and\
  \citenamefont {Nguyen}}]{si:pour2020}%
  \BibitemOpen
  \bibfield  {author} {\bibinfo {author} {\bibfnamefont {M.~M.}\ \bibnamefont
  {Pour~Fard}}, \bibinfo {author} {\bibfnamefont {I.~A.}\ \bibnamefont
  {Williamson}}, \bibinfo {author} {\bibfnamefont {M.}~\bibnamefont {Edwards}},
  \bibinfo {author} {\bibfnamefont {K.}~\bibnamefont {Liu}}, \bibinfo {author}
  {\bibfnamefont {S.}~\bibnamefont {Pai}}, \bibinfo {author} {\bibfnamefont
  {B.}~\bibnamefont {Bartlett}}, \bibinfo {author} {\bibfnamefont
  {M.}~\bibnamefont {Minkov}}, \bibinfo {author} {\bibfnamefont {T.~W.}\
  \bibnamefont {Hughes}}, \bibinfo {author} {\bibfnamefont {S.}~\bibnamefont
  {Fan}},\ and\ \bibinfo {author} {\bibfnamefont {T.-A.}\ \bibnamefont
  {Nguyen}},\ }\bibfield  {title} {\bibinfo {title} {Experimental realization
  of arbitrary activation functions for optical neural networks},\ }\href@noop
  {} {\bibfield  {journal} {\bibinfo  {journal} {Optics Express}\ }\textbf
  {\bibinfo {volume} {28}},\ \bibinfo {pages} {12138} (\bibinfo {year}
  {2020})}\BibitemShut {NoStop}%
\bibitem [{\citenamefont {Ashtiani}\ \emph {et~al.}(2022)\citenamefont
  {Ashtiani}, \citenamefont {Geers},\ and\ \citenamefont
  {Aflatouni}}]{si:ashtiani2022}%
  \BibitemOpen
  \bibfield  {author} {\bibinfo {author} {\bibfnamefont {F.}~\bibnamefont
  {Ashtiani}}, \bibinfo {author} {\bibfnamefont {A.~J.}\ \bibnamefont
  {Geers}},\ and\ \bibinfo {author} {\bibfnamefont {F.}~\bibnamefont
  {Aflatouni}},\ }\bibfield  {title} {\bibinfo {title} {An on-chip photonic
  deep neural network for image classification},\ }\href@noop {} {\bibfield
  {journal} {\bibinfo  {journal} {Nature}\ }\textbf {\bibinfo {volume} {606}},\
  \bibinfo {pages} {501} (\bibinfo {year} {2022})}\BibitemShut {NoStop}%
\bibitem [{\citenamefont {Zhong}\ \emph {et~al.}(2023)\citenamefont {Zhong},
  \citenamefont {Liao}, \citenamefont {Dai}, \citenamefont {Wei}, \citenamefont
  {Ma}, \citenamefont {Wu}, \citenamefont {Zhang}, \citenamefont {Ye},
  \citenamefont {Luo}, \citenamefont {Chen} \emph
  {et~al.}}]{si:zhong2023graphene}%
  \BibitemOpen
  \bibfield  {author} {\bibinfo {author} {\bibfnamefont {C.}~\bibnamefont
  {Zhong}}, \bibinfo {author} {\bibfnamefont {K.}~\bibnamefont {Liao}},
  \bibinfo {author} {\bibfnamefont {T.}~\bibnamefont {Dai}}, \bibinfo {author}
  {\bibfnamefont {M.}~\bibnamefont {Wei}}, \bibinfo {author} {\bibfnamefont
  {H.}~\bibnamefont {Ma}}, \bibinfo {author} {\bibfnamefont {J.}~\bibnamefont
  {Wu}}, \bibinfo {author} {\bibfnamefont {Z.}~\bibnamefont {Zhang}}, \bibinfo
  {author} {\bibfnamefont {Y.}~\bibnamefont {Ye}}, \bibinfo {author}
  {\bibfnamefont {Y.}~\bibnamefont {Luo}}, \bibinfo {author} {\bibfnamefont
  {Z.}~\bibnamefont {Chen}}, \emph {et~al.},\ }\bibfield  {title} {\bibinfo
  {title} {Graphene/silicon heterojunction for reconfigurable phase-relevant
  activation function in coherent optical neural networks},\ }\href@noop {}
  {\bibfield  {journal} {\bibinfo  {journal} {Nature Communications}\ }\textbf
  {\bibinfo {volume} {14}},\ \bibinfo {pages} {6939} (\bibinfo {year}
  {2023})}\BibitemShut {NoStop}%
\bibitem [{\citenamefont {Feng}\ \emph {et~al.}(2025)\citenamefont {Feng},
  \citenamefont {Uzundal}, \citenamefont {Guo}, \citenamefont {Sanborn},
  \citenamefont {Qi}, \citenamefont {Xie}, \citenamefont {Zhang}, \citenamefont
  {Wu},\ and\ \citenamefont {Wang}}]{si:feng2025femtojoule}%
  \BibitemOpen
  \bibfield  {author} {\bibinfo {author} {\bibfnamefont {Q.}~\bibnamefont
  {Feng}}, \bibinfo {author} {\bibfnamefont {C.~B.}\ \bibnamefont {Uzundal}},
  \bibinfo {author} {\bibfnamefont {R.}~\bibnamefont {Guo}}, \bibinfo {author}
  {\bibfnamefont {C.}~\bibnamefont {Sanborn}}, \bibinfo {author} {\bibfnamefont
  {R.}~\bibnamefont {Qi}}, \bibinfo {author} {\bibfnamefont {J.}~\bibnamefont
  {Xie}}, \bibinfo {author} {\bibfnamefont {J.}~\bibnamefont {Zhang}}, \bibinfo
  {author} {\bibfnamefont {J.}~\bibnamefont {Wu}},\ and\ \bibinfo {author}
  {\bibfnamefont {F.}~\bibnamefont {Wang}},\ }\bibfield  {title} {\bibinfo
  {title} {Femtojoule optical nonlinearity for deep learning with incoherent
  illumination},\ }\href@noop {} {\bibfield  {journal} {\bibinfo  {journal}
  {Science Advances}\ }\textbf {\bibinfo {volume} {11}},\ \bibinfo {pages}
  {eads4224} (\bibinfo {year} {2025})}\BibitemShut {NoStop}%
\bibitem [{\citenamefont {Dulkeith}\ \emph {et~al.}(2006)\citenamefont
  {Dulkeith}, \citenamefont {Vlasov}, \citenamefont {Chen}, \citenamefont
  {Panoiu},\ and\ \citenamefont {Osgood~Jr}}]{si:dulkeith2006self}%
  \BibitemOpen
  \bibfield  {author} {\bibinfo {author} {\bibfnamefont {E.}~\bibnamefont
  {Dulkeith}}, \bibinfo {author} {\bibfnamefont {Y.~A.}\ \bibnamefont
  {Vlasov}}, \bibinfo {author} {\bibfnamefont {X.}~\bibnamefont {Chen}},
  \bibinfo {author} {\bibfnamefont {N.~C.}\ \bibnamefont {Panoiu}},\ and\
  \bibinfo {author} {\bibfnamefont {R.~M.}\ \bibnamefont {Osgood~Jr}},\
  }\bibfield  {title} {\bibinfo {title} {Self-phase-modulation in submicron
  silicon-on-insulator photonic wires},\ }\href@noop {} {\bibfield  {journal}
  {\bibinfo  {journal} {Optics express}\ }\textbf {\bibinfo {volume} {14}},\
  \bibinfo {pages} {5524} (\bibinfo {year} {2006})}\BibitemShut {NoStop}%
\bibitem [{\citenamefont {Kumar}\ \emph {et~al.}(2013)\citenamefont {Kumar},
  \citenamefont {Najmaei}, \citenamefont {Cui}, \citenamefont {Ceballos},
  \citenamefont {Ajayan}, \citenamefont {Lou},\ and\ \citenamefont
  {Zhao}}]{si:kumar2013second}%
  \BibitemOpen
  \bibfield  {author} {\bibinfo {author} {\bibfnamefont {N.}~\bibnamefont
  {Kumar}}, \bibinfo {author} {\bibfnamefont {S.}~\bibnamefont {Najmaei}},
  \bibinfo {author} {\bibfnamefont {Q.}~\bibnamefont {Cui}}, \bibinfo {author}
  {\bibfnamefont {F.}~\bibnamefont {Ceballos}}, \bibinfo {author}
  {\bibfnamefont {P.~M.}\ \bibnamefont {Ajayan}}, \bibinfo {author}
  {\bibfnamefont {J.}~\bibnamefont {Lou}},\ and\ \bibinfo {author}
  {\bibfnamefont {H.}~\bibnamefont {Zhao}},\ }\bibfield  {title} {\bibinfo
  {title} {Second harmonic microscopy of monolayer mos 2},\ }\href@noop {}
  {\bibfield  {journal} {\bibinfo  {journal} {Physical Review B—Condensed
  Matter and Materials Physics}\ }\textbf {\bibinfo {volume} {87}},\ \bibinfo
  {pages} {161403} (\bibinfo {year} {2013})}\BibitemShut {NoStop}%
\bibitem [{\citenamefont {Liu}\ \emph {et~al.}(2025)\citenamefont {Liu},
  \citenamefont {Liang}, \citenamefont {Zhou}, \citenamefont {Khan},
  \citenamefont {Lu}, \citenamefont {Yildirim}, \citenamefont {Sun},
  \citenamefont {Rahman}, \citenamefont {Liu}, \citenamefont {Yu} \emph
  {et~al.}}]{si:liu2025Te}%
  \BibitemOpen
  \bibfield  {author} {\bibinfo {author} {\bibfnamefont {B.}~\bibnamefont
  {Liu}}, \bibinfo {author} {\bibfnamefont {K.}~\bibnamefont {Liang}}, \bibinfo
  {author} {\bibfnamefont {Q.}~\bibnamefont {Zhou}}, \bibinfo {author}
  {\bibfnamefont {A.~R.}\ \bibnamefont {Khan}}, \bibinfo {author}
  {\bibfnamefont {Z.}~\bibnamefont {Lu}}, \bibinfo {author} {\bibfnamefont
  {T.}~\bibnamefont {Yildirim}}, \bibinfo {author} {\bibfnamefont
  {X.}~\bibnamefont {Sun}}, \bibinfo {author} {\bibfnamefont {S.}~\bibnamefont
  {Rahman}}, \bibinfo {author} {\bibfnamefont {Y.}~\bibnamefont {Liu}},
  \bibinfo {author} {\bibfnamefont {Z.}~\bibnamefont {Yu}}, \emph {et~al.},\
  }\bibfield  {title} {\bibinfo {title} {Giant second harmonic generation in
  two-dimensional tellurene with synthesis and thickness engineering},\
  }\href@noop {} {\bibfield  {journal} {\bibinfo  {journal} {Applied physics
  reviews}\ }\textbf {\bibinfo {volume} {12}} (\bibinfo {year}
  {2025})}\BibitemShut {NoStop}%
\bibitem [{\citenamefont {Bao}\ \emph {et~al.}(2009)\citenamefont {Bao},
  \citenamefont {Zhang}, \citenamefont {Wang}, \citenamefont {Ni},
  \citenamefont {Yan}, \citenamefont {Shen}, \citenamefont {Loh},\ and\
  \citenamefont {Tang}}]{si:bao2009atomic}%
  \BibitemOpen
  \bibfield  {author} {\bibinfo {author} {\bibfnamefont {Q.}~\bibnamefont
  {Bao}}, \bibinfo {author} {\bibfnamefont {H.}~\bibnamefont {Zhang}}, \bibinfo
  {author} {\bibfnamefont {Y.}~\bibnamefont {Wang}}, \bibinfo {author}
  {\bibfnamefont {Z.}~\bibnamefont {Ni}}, \bibinfo {author} {\bibfnamefont
  {Y.}~\bibnamefont {Yan}}, \bibinfo {author} {\bibfnamefont {Z.~X.}\
  \bibnamefont {Shen}}, \bibinfo {author} {\bibfnamefont {K.~P.}\ \bibnamefont
  {Loh}},\ and\ \bibinfo {author} {\bibfnamefont {D.~Y.}\ \bibnamefont
  {Tang}},\ }\bibfield  {title} {\bibinfo {title} {Atomic-layer graphene as a
  saturable absorber for ultrafast pulsed lasers},\ }\href@noop {} {\bibfield
  {journal} {\bibinfo  {journal} {Advanced Functional Materials}\ }\textbf
  {\bibinfo {volume} {19}},\ \bibinfo {pages} {3077} (\bibinfo {year}
  {2009})}\BibitemShut {NoStop}%
\bibitem [{\citenamefont {Bao}\ \emph {et~al.}(2011)\citenamefont {Bao},
  \citenamefont {Zhang}, \citenamefont {Ni}, \citenamefont {Wang},
  \citenamefont {Polavarapu}, \citenamefont {Shen}, \citenamefont {Xu},
  \citenamefont {Tang},\ and\ \citenamefont {Loh}}]{si:bao2011monolayer}%
  \BibitemOpen
  \bibfield  {author} {\bibinfo {author} {\bibfnamefont {Q.}~\bibnamefont
  {Bao}}, \bibinfo {author} {\bibfnamefont {H.}~\bibnamefont {Zhang}}, \bibinfo
  {author} {\bibfnamefont {Z.}~\bibnamefont {Ni}}, \bibinfo {author}
  {\bibfnamefont {Y.}~\bibnamefont {Wang}}, \bibinfo {author} {\bibfnamefont
  {L.}~\bibnamefont {Polavarapu}}, \bibinfo {author} {\bibfnamefont
  {Z.}~\bibnamefont {Shen}}, \bibinfo {author} {\bibfnamefont {Q.-H.}\
  \bibnamefont {Xu}}, \bibinfo {author} {\bibfnamefont {D.}~\bibnamefont
  {Tang}},\ and\ \bibinfo {author} {\bibfnamefont {K.~P.}\ \bibnamefont
  {Loh}},\ }\bibfield  {title} {\bibinfo {title} {Monolayer graphene as a
  saturable absorber in a mode-locked laser},\ }\href@noop {} {\bibfield
  {journal} {\bibinfo  {journal} {Nano Research}\ }\textbf {\bibinfo {volume}
  {4}},\ \bibinfo {pages} {297} (\bibinfo {year} {2011})}\BibitemShut {NoStop}%
\bibitem [{\citenamefont {Lamont}\ \emph {et~al.}(2008)\citenamefont {Lamont},
  \citenamefont {Luther-Davies}, \citenamefont {Choi}, \citenamefont {Madden},\
  and\ \citenamefont {Eggleton}}]{si:lamont2008supercontinuum}%
  \BibitemOpen
  \bibfield  {author} {\bibinfo {author} {\bibfnamefont {M.~R.}\ \bibnamefont
  {Lamont}}, \bibinfo {author} {\bibfnamefont {B.}~\bibnamefont
  {Luther-Davies}}, \bibinfo {author} {\bibfnamefont {D.-Y.}\ \bibnamefont
  {Choi}}, \bibinfo {author} {\bibfnamefont {S.}~\bibnamefont {Madden}},\ and\
  \bibinfo {author} {\bibfnamefont {B.~J.}\ \bibnamefont {Eggleton}},\
  }\bibfield  {title} {\bibinfo {title} {Supercontinuum generation in
  dispersion engineered highly nonlinear ($\gamma = 10$ /w/m) as$_2$s$_3$
  chalcogenide planar waveguide},\ }\href
  {https://doi.org/10.1364/OE.16.014938} {\bibfield  {journal} {\bibinfo
  {journal} {Optics Express}\ }\textbf {\bibinfo {volume} {16}},\ \bibinfo
  {pages} {14938} (\bibinfo {year} {2008})}\BibitemShut {NoStop}%
\bibitem [{\citenamefont {Pu}\ \emph {et~al.}(2016)\citenamefont {Pu},
  \citenamefont {Ottaviano}, \citenamefont {Semenova},\ and\ \citenamefont
  {Yvind}}]{si:pu2016efficient}%
  \BibitemOpen
  \bibfield  {author} {\bibinfo {author} {\bibfnamefont {M.}~\bibnamefont
  {Pu}}, \bibinfo {author} {\bibfnamefont {L.}~\bibnamefont {Ottaviano}},
  \bibinfo {author} {\bibfnamefont {E.}~\bibnamefont {Semenova}},\ and\
  \bibinfo {author} {\bibfnamefont {K.}~\bibnamefont {Yvind}},\ }\bibfield
  {title} {\bibinfo {title} {Efficient frequency comb generation in
  algaas-on-insulator},\ }\href {https://doi.org/10.1364/OPTICA.3.000823}
  {\bibfield  {journal} {\bibinfo  {journal} {Optica}\ }\textbf {\bibinfo
  {volume} {3}},\ \bibinfo {pages} {823} (\bibinfo {year} {2016})}\BibitemShut
  {NoStop}%
\bibitem [{\citenamefont {Wang}\ \emph {et~al.}(2017)\citenamefont {Wang},
  \citenamefont {Xiong}, \citenamefont {Andrade}, \citenamefont {Venkataraman},
  \citenamefont {Ren}, \citenamefont {Guo},\ and\ \citenamefont {Lon{\v
  c}ar}}]{si:wang2017linbo3shg}%
  \BibitemOpen
  \bibfield  {author} {\bibinfo {author} {\bibfnamefont {C.}~\bibnamefont
  {Wang}}, \bibinfo {author} {\bibfnamefont {X.}~\bibnamefont {Xiong}},
  \bibinfo {author} {\bibfnamefont {N.}~\bibnamefont {Andrade}}, \bibinfo
  {author} {\bibfnamefont {V.}~\bibnamefont {Venkataraman}}, \bibinfo {author}
  {\bibfnamefont {X.-F.}\ \bibnamefont {Ren}}, \bibinfo {author} {\bibfnamefont
  {G.-C.}\ \bibnamefont {Guo}},\ and\ \bibinfo {author} {\bibfnamefont
  {M.}~\bibnamefont {Lon{\v c}ar}},\ }\bibfield  {title} {\bibinfo {title}
  {Second harmonic generation in nano-structured thin-film lithium niobate
  waveguides},\ }\href {https://doi.org/10.1364/OE.25.006963} {\bibfield
  {journal} {\bibinfo  {journal} {Optics Express}\ }\textbf {\bibinfo {volume}
  {25}},\ \bibinfo {pages} {6963} (\bibinfo {year} {2017})}\BibitemShut
  {NoStop}%
\bibitem [{\citenamefont {Kopp}\ and\ \citenamefont
  {Lean}(2011)}]{si:kopp2011solar}%
  \BibitemOpen
  \bibfield  {author} {\bibinfo {author} {\bibfnamefont {G.}~\bibnamefont
  {Kopp}}\ and\ \bibinfo {author} {\bibfnamefont {J.~L.}\ \bibnamefont
  {Lean}},\ }\bibfield  {title} {\bibinfo {title} {A new, lower value of total
  solar irradiance: Evidence and climate significance},\ }\href@noop {}
  {\bibfield  {journal} {\bibinfo  {journal} {Geophysical Research Letters}\
  }\textbf {\bibinfo {volume} {38}} (\bibinfo {year} {2011})}\BibitemShut
  {NoStop}%
\bibitem [{\citenamefont {Zhou}\ \emph {et~al.}(2024)\citenamefont {Zhou},
  \citenamefont {Gangaraj}, \citenamefont {Zhou},\ and\ \citenamefont
  {Yu}}]{si:zhou2024fdtd}%
  \BibitemOpen
  \bibfield  {author} {\bibinfo {author} {\bibfnamefont {Q.}~\bibnamefont
  {Zhou}}, \bibinfo {author} {\bibfnamefont {S.}~\bibnamefont {Gangaraj}},
  \bibinfo {author} {\bibfnamefont {M.}~\bibnamefont {Zhou}},\ and\ \bibinfo
  {author} {\bibfnamefont {Z.}~\bibnamefont {Yu}},\ }\bibfield  {title}
  {\bibinfo {title} {Simulating quantum emitters in arbitrary photonic
  environments using fdtd: beyond the semi-classical regime},\ }\href@noop {}
  {\bibfield  {journal} {\bibinfo  {journal} {arXiv preprint arXiv:2410.16118}\
  } (\bibinfo {year} {2024})}\BibitemShut {NoStop}%
\bibitem [{\citenamefont {Wang}\ and\ \citenamefont
  {Fan}(2025)}]{si:wang2025lorentz}%
  \BibitemOpen
  \bibfield  {author} {\bibinfo {author} {\bibfnamefont {H.}~\bibnamefont
  {Wang}}\ and\ \bibinfo {author} {\bibfnamefont {S.}~\bibnamefont {Fan}},\
  }\bibfield  {title} {\bibinfo {title} {Lorentz--drude dipoles in the
  radiative limit and their modeling in finite-difference time-domain
  methods},\ }\href@noop {} {\bibfield  {journal} {\bibinfo  {journal} {Annalen
  der Physik}\ ,\ \bibinfo {pages} {e00156}} (\bibinfo {year}
  {2025})}\BibitemShut {NoStop}%
\bibitem [{\citenamefont {Nocedal}(2006)}]{si:nocedal2006numerical}%
  \BibitemOpen
  \bibfield  {author} {\bibinfo {author} {\bibfnamefont {J.}~\bibnamefont
  {Nocedal}},\ }\href@noop {} {\bibinfo {title} {Numerical optimization}}
  (\bibinfo {year} {2006})\BibitemShut {NoStop}%
\bibitem [{\citenamefont {Wirtinger}(1927)}]{si:wirtinger1927}%
  \BibitemOpen
  \bibfield  {author} {\bibinfo {author} {\bibfnamefont {W.}~\bibnamefont
  {Wirtinger}},\ }\bibfield  {title} {\bibinfo {title} {Zur formalen theorie
  der funktionen von mehr komplexen ver{\"a}nderlichen},\ }\href@noop {}
  {\bibfield  {journal} {\bibinfo  {journal} {Mathematische Annalen}\ }\textbf
  {\bibinfo {volume} {97}},\ \bibinfo {pages} {357} (\bibinfo {year}
  {1927})}\BibitemShut {NoStop}%
\bibitem [{\citenamefont {Poole}\ \emph {et~al.}(2016)\citenamefont {Poole},
  \citenamefont {Lahiri}, \citenamefont {Raghu}, \citenamefont
  {Sohl-Dickstein},\ and\ \citenamefont {Ganguli}}]{si:poole2016}%
  \BibitemOpen
  \bibfield  {author} {\bibinfo {author} {\bibfnamefont {B.}~\bibnamefont
  {Poole}}, \bibinfo {author} {\bibfnamefont {S.}~\bibnamefont {Lahiri}},
  \bibinfo {author} {\bibfnamefont {M.}~\bibnamefont {Raghu}}, \bibinfo
  {author} {\bibfnamefont {J.}~\bibnamefont {Sohl-Dickstein}},\ and\ \bibinfo
  {author} {\bibfnamefont {S.}~\bibnamefont {Ganguli}},\ }\bibfield  {title}
  {\bibinfo {title} {Exponential expressivity in deep neural networks through
  transient chaos},\ }\href@noop {} {\bibfield  {journal} {\bibinfo  {journal}
  {Advances in neural information processing systems}\ }\textbf {\bibinfo
  {volume} {29}} (\bibinfo {year} {2016})}\BibitemShut {NoStop}%
\bibitem [{\citenamefont {Raghu}\ \emph {et~al.}(2017)\citenamefont {Raghu},
  \citenamefont {Poole}, \citenamefont {Kleinberg}, \citenamefont {Ganguli},\
  and\ \citenamefont {Sohl-Dickstein}}]{si:raghu2017}%
  \BibitemOpen
  \bibfield  {author} {\bibinfo {author} {\bibfnamefont {M.}~\bibnamefont
  {Raghu}}, \bibinfo {author} {\bibfnamefont {B.}~\bibnamefont {Poole}},
  \bibinfo {author} {\bibfnamefont {J.}~\bibnamefont {Kleinberg}}, \bibinfo
  {author} {\bibfnamefont {S.}~\bibnamefont {Ganguli}},\ and\ \bibinfo {author}
  {\bibfnamefont {J.}~\bibnamefont {Sohl-Dickstein}},\ }\bibfield  {title}
  {\bibinfo {title} {On the expressive power of deep neural networks},\ }in\
  \href@noop {} {\emph {\bibinfo {booktitle} {international conference on
  machine learning}}}\ (\bibinfo {organization} {PMLR},\ \bibinfo {year}
  {2017})\ pp.\ \bibinfo {pages} {2847--2854}\BibitemShut {NoStop}%
\bibitem [{\citenamefont {Hepp}\ \emph {et~al.}(2014)\citenamefont {Hepp},
  \citenamefont {M{\"u}ller}, \citenamefont {Waselowski}, \citenamefont
  {Becker}, \citenamefont {Pingault}, \citenamefont {Sternschulte},
  \citenamefont {Steinm{\"u}ller-Nethl}, \citenamefont {Gali}, \citenamefont
  {Maze}, \citenamefont {Atat{\"u}re},\ and\ \citenamefont
  {Becher}}]{si:hepp2014}%
  \BibitemOpen
  \bibfield  {author} {\bibinfo {author} {\bibfnamefont {C.}~\bibnamefont
  {Hepp}}, \bibinfo {author} {\bibfnamefont {T.}~\bibnamefont {M{\"u}ller}},
  \bibinfo {author} {\bibfnamefont {V.}~\bibnamefont {Waselowski}}, \bibinfo
  {author} {\bibfnamefont {J.~N.}\ \bibnamefont {Becker}}, \bibinfo {author}
  {\bibfnamefont {B.}~\bibnamefont {Pingault}}, \bibinfo {author}
  {\bibfnamefont {H.}~\bibnamefont {Sternschulte}}, \bibinfo {author}
  {\bibfnamefont {D.}~\bibnamefont {Steinm{\"u}ller-Nethl}}, \bibinfo {author}
  {\bibfnamefont {A.}~\bibnamefont {Gali}}, \bibinfo {author} {\bibfnamefont
  {J.~R.}\ \bibnamefont {Maze}}, \bibinfo {author} {\bibfnamefont
  {M.}~\bibnamefont {Atat{\"u}re}},\ and\ \bibinfo {author} {\bibfnamefont
  {C.}~\bibnamefont {Becher}},\ }\bibfield  {title} {\bibinfo {title}
  {Electronic structure of the silicon vacancy color center in diamond},\
  }\href {https://doi.org/10.1103/PhysRevLett.112.036405} {\bibfield  {journal}
  {\bibinfo  {journal} {Physical Review Letters}\ }\textbf {\bibinfo {volume}
  {112}},\ \bibinfo {pages} {036405} (\bibinfo {year} {2014})}\BibitemShut
  {NoStop}%
\bibitem [{\citenamefont {Ding}\ \emph {et~al.}(2024)\citenamefont {Ding},
  \citenamefont {Haas}, \citenamefont {Guo}, \citenamefont {Kuruma},
  \citenamefont {Jin}, \citenamefont {Li}, \citenamefont {Awschalom},
  \citenamefont {Delegan}, \citenamefont {Heremans}, \citenamefont {High},\
  and\ \citenamefont {Lon{\v c}ar}}]{si:ding2024}%
  \BibitemOpen
  \bibfield  {author} {\bibinfo {author} {\bibfnamefont {S.~W.}\ \bibnamefont
  {Ding}}, \bibinfo {author} {\bibfnamefont {M.}~\bibnamefont {Haas}}, \bibinfo
  {author} {\bibfnamefont {X.}~\bibnamefont {Guo}}, \bibinfo {author}
  {\bibfnamefont {K.}~\bibnamefont {Kuruma}}, \bibinfo {author} {\bibfnamefont
  {C.}~\bibnamefont {Jin}}, \bibinfo {author} {\bibfnamefont {Z.}~\bibnamefont
  {Li}}, \bibinfo {author} {\bibfnamefont {D.~D.}\ \bibnamefont {Awschalom}},
  \bibinfo {author} {\bibfnamefont {N.}~\bibnamefont {Delegan}}, \bibinfo
  {author} {\bibfnamefont {F.~J.}\ \bibnamefont {Heremans}}, \bibinfo {author}
  {\bibfnamefont {A.~A.}\ \bibnamefont {High}},\ and\ \bibinfo {author}
  {\bibfnamefont {M.}~\bibnamefont {Lon{\v c}ar}},\ }\bibfield  {title}
  {\bibinfo {title} {High-q cavity interface for color centers in thin film
  diamond},\ }\href {https://doi.org/10.1038/s41467-024-50667-5} {\bibfield
  {journal} {\bibinfo  {journal} {Nat. Commun.}\ }\textbf {\bibinfo {volume}
  {15}},\ \bibinfo {pages} {6358} (\bibinfo {year} {2024})}\BibitemShut
  {NoStop}%
\bibitem [{\citenamefont {Chakravarthi}\ \emph {et~al.}(2023)\citenamefont
  {Chakravarthi}, \citenamefont {Yama}, \citenamefont {Abulnaga}, \citenamefont
  {Huang}, \citenamefont {Pederson}, \citenamefont {Hestroffer}, \citenamefont
  {Hatami}, \citenamefont {de~Leon},\ and\ \citenamefont
  {Fu}}]{si:chakravarthi2023}%
  \BibitemOpen
  \bibfield  {author} {\bibinfo {author} {\bibfnamefont {S.}~\bibnamefont
  {Chakravarthi}}, \bibinfo {author} {\bibfnamefont {N.~S.}\ \bibnamefont
  {Yama}}, \bibinfo {author} {\bibfnamefont {A.}~\bibnamefont {Abulnaga}},
  \bibinfo {author} {\bibfnamefont {D.}~\bibnamefont {Huang}}, \bibinfo
  {author} {\bibfnamefont {C.}~\bibnamefont {Pederson}}, \bibinfo {author}
  {\bibfnamefont {K.}~\bibnamefont {Hestroffer}}, \bibinfo {author}
  {\bibfnamefont {F.}~\bibnamefont {Hatami}}, \bibinfo {author} {\bibfnamefont
  {N.~P.}\ \bibnamefont {de~Leon}},\ and\ \bibinfo {author} {\bibfnamefont
  {K.-M.~C.}\ \bibnamefont {Fu}},\ }\bibfield  {title} {\bibinfo {title}
  {Hybrid integration of {GaP} photonic crystal cavities with silicon-vacancy
  centers in diamond by stamp-transfer},\ }\href
  {https://doi.org/10.1021/acs.nanolett.2c04890} {\bibfield  {journal}
  {\bibinfo  {journal} {Nano Lett.}\ }\textbf {\bibinfo {volume} {23}},\
  \bibinfo {pages} {3708} (\bibinfo {year} {2023})}\BibitemShut {NoStop}%
\bibitem [{\citenamefont {Yama}\ \emph {et~al.}(2026)\citenamefont {Yama},
  \citenamefont {Wu}, \citenamefont {Hatami},\ and\ \citenamefont
  {Fu}}]{si:yama2026}%
  \BibitemOpen
  \bibfield  {author} {\bibinfo {author} {\bibfnamefont {N.~S.}\ \bibnamefont
  {Yama}}, \bibinfo {author} {\bibfnamefont {C.-C.}\ \bibnamefont {Wu}},
  \bibinfo {author} {\bibfnamefont {F.}~\bibnamefont {Hatami}},\ and\ \bibinfo
  {author} {\bibfnamefont {K.-M.~C.}\ \bibnamefont {Fu}},\ }\bibfield  {title}
  {\bibinfo {title} {A scalable gallium-phosphide-on-diamond spin-photon
  interface},\ }\href@noop {} {\bibfield  {journal} {\bibinfo  {journal} {arXiv
  preprint arXiv:2601.04733}\ } (\bibinfo {year} {2026})}\BibitemShut {NoStop}%
\bibitem [{\citenamefont {Schr{\"o}der}\ \emph {et~al.}(2017)\citenamefont
  {Schr{\"o}der}, \citenamefont {Trusheim}, \citenamefont {Walsh},
  \citenamefont {Li}, \citenamefont {Zheng}, \citenamefont {Schukraft},
  \citenamefont {Sipahigil}, \citenamefont {Evans}, \citenamefont {Sukachev},
  \citenamefont {Nguyen}, \citenamefont {Pacheco}, \citenamefont {Camacho},
  \citenamefont {Bielejec}, \citenamefont {Lukin},\ and\ \citenamefont
  {Englund}}]{si:schroder2017}%
  \BibitemOpen
  \bibfield  {author} {\bibinfo {author} {\bibfnamefont {T.}~\bibnamefont
  {Schr{\"o}der}}, \bibinfo {author} {\bibfnamefont {M.~E.}\ \bibnamefont
  {Trusheim}}, \bibinfo {author} {\bibfnamefont {M.}~\bibnamefont {Walsh}},
  \bibinfo {author} {\bibfnamefont {L.}~\bibnamefont {Li}}, \bibinfo {author}
  {\bibfnamefont {J.}~\bibnamefont {Zheng}}, \bibinfo {author} {\bibfnamefont
  {M.}~\bibnamefont {Schukraft}}, \bibinfo {author} {\bibfnamefont
  {A.}~\bibnamefont {Sipahigil}}, \bibinfo {author} {\bibfnamefont {R.~E.}\
  \bibnamefont {Evans}}, \bibinfo {author} {\bibfnamefont {D.~D.}\ \bibnamefont
  {Sukachev}}, \bibinfo {author} {\bibfnamefont {C.~T.}\ \bibnamefont
  {Nguyen}}, \bibinfo {author} {\bibfnamefont {J.~L.}\ \bibnamefont {Pacheco}},
  \bibinfo {author} {\bibfnamefont {R.~M.}\ \bibnamefont {Camacho}}, \bibinfo
  {author} {\bibfnamefont {E.~S.}\ \bibnamefont {Bielejec}}, \bibinfo {author}
  {\bibfnamefont {M.~D.}\ \bibnamefont {Lukin}},\ and\ \bibinfo {author}
  {\bibfnamefont {D.}~\bibnamefont {Englund}},\ }\bibfield  {title} {\bibinfo
  {title} {Scalable focused ion beam creation of nearly lifetime-limited single
  quantum emitters in diamond nanostructures},\ }\href
  {https://doi.org/10.1038/ncomms15376} {\bibfield  {journal} {\bibinfo
  {journal} {Nature Communications}\ }\textbf {\bibinfo {volume} {8}},\
  \bibinfo {pages} {15376} (\bibinfo {year} {2017})}\BibitemShut {NoStop}%
\bibitem [{\citenamefont {Titze}\ \emph {et~al.}(2022)\citenamefont {Titze},
  \citenamefont {Byeon}, \citenamefont {Flores}, \citenamefont {Henshaw},
  \citenamefont {Harris}, \citenamefont {Mounce},\ and\ \citenamefont
  {Bielejec}}]{si:titze2022}%
  \BibitemOpen
  \bibfield  {author} {\bibinfo {author} {\bibfnamefont {M.}~\bibnamefont
  {Titze}}, \bibinfo {author} {\bibfnamefont {H.}~\bibnamefont {Byeon}},
  \bibinfo {author} {\bibfnamefont {A.}~\bibnamefont {Flores}}, \bibinfo
  {author} {\bibfnamefont {J.}~\bibnamefont {Henshaw}}, \bibinfo {author}
  {\bibfnamefont {C.~T.}\ \bibnamefont {Harris}}, \bibinfo {author}
  {\bibfnamefont {A.~M.}\ \bibnamefont {Mounce}},\ and\ \bibinfo {author}
  {\bibfnamefont {E.~S.}\ \bibnamefont {Bielejec}},\ }\bibfield  {title}
  {\bibinfo {title} {In situ ion counting for improved implanted ion error rate
  and silicon vacancy yield uncertainty},\ }\href
  {https://doi.org/10.1021/acs.nanolett.1c04646} {\bibfield  {journal}
  {\bibinfo  {journal} {Nano Lett.}\ }\textbf {\bibinfo {volume} {22}},\
  \bibinfo {pages} {3212} (\bibinfo {year} {2022})}\BibitemShut {NoStop}%
\bibitem [{\citenamefont {Jahnke}\ \emph {et~al.}(2015)\citenamefont {Jahnke},
  \citenamefont {Sipahigil}, \citenamefont {Binder}, \citenamefont {Doherty},
  \citenamefont {Metsch}, \citenamefont {Rogers}, \citenamefont {Manson},
  \citenamefont {Lukin},\ and\ \citenamefont {Jelezko}}]{si:jahnke2015}%
  \BibitemOpen
  \bibfield  {author} {\bibinfo {author} {\bibfnamefont {K.~D.}\ \bibnamefont
  {Jahnke}}, \bibinfo {author} {\bibfnamefont {A.}~\bibnamefont {Sipahigil}},
  \bibinfo {author} {\bibfnamefont {J.~M.}\ \bibnamefont {Binder}}, \bibinfo
  {author} {\bibfnamefont {M.~W.}\ \bibnamefont {Doherty}}, \bibinfo {author}
  {\bibfnamefont {M.}~\bibnamefont {Metsch}}, \bibinfo {author} {\bibfnamefont
  {L.~J.}\ \bibnamefont {Rogers}}, \bibinfo {author} {\bibfnamefont {N.~B.}\
  \bibnamefont {Manson}}, \bibinfo {author} {\bibfnamefont {M.~D.}\
  \bibnamefont {Lukin}},\ and\ \bibinfo {author} {\bibfnamefont
  {F.}~\bibnamefont {Jelezko}},\ }\bibfield  {title} {\bibinfo {title}
  {Electron--phonon processes of the silicon-vacancy centre in diamond},\
  }\href {https://doi.org/10.1088/1367-2630/17/4/043011} {\bibfield  {journal}
  {\bibinfo  {journal} {New Journal of Physics}\ }\textbf {\bibinfo {volume}
  {17}},\ \bibinfo {pages} {043011} (\bibinfo {year} {2015})}\BibitemShut
  {NoStop}%
\bibitem [{\citenamefont {Fan}\ \emph {et~al.}(2003)\citenamefont {Fan},
  \citenamefont {Suh},\ and\ \citenamefont {Joannopoulos}}]{si:fan2003temporal}%
  \BibitemOpen
  \bibfield  {author} {\bibinfo {author} {\bibfnamefont {S.}~\bibnamefont
  {Fan}}, \bibinfo {author} {\bibfnamefont {W.}~\bibnamefont {Suh}},\ and\
  \bibinfo {author} {\bibfnamefont {J.~D.}\ \bibnamefont {Joannopoulos}},\
  }\bibfield  {title} {\bibinfo {title} {Temporal coupled-mode theory for the
  fano resonance in optical resonators},\ }\href@noop {} {\bibfield  {journal}
  {\bibinfo  {journal} {Journal of the Optical Society of America A}\ }\textbf
  {\bibinfo {volume} {20}},\ \bibinfo {pages} {569} (\bibinfo {year}
  {2003})}\BibitemShut {NoStop}%
\bibitem [{\citenamefont {Hammer}\ and\ \citenamefont
  {Ivanova}(2009)}]{si:hammer2009eim}%
  \BibitemOpen
  \bibfield  {author} {\bibinfo {author} {\bibfnamefont {M.}~\bibnamefont
  {Hammer}}\ and\ \bibinfo {author} {\bibfnamefont {O.~V.}\ \bibnamefont
  {Ivanova}},\ }\bibfield  {title} {\bibinfo {title} {Effective index
  approximations of photonic crystal slabs: a 2-to-1-d assessment},\ }\href
  {https://doi.org/10.1007/s11082-009-9349-3} {\bibfield  {journal} {\bibinfo
  {journal} {Optical and Quantum Electronics}\ }\textbf {\bibinfo {volume}
  {41}},\ \bibinfo {pages} {267} (\bibinfo {year} {2009})}\BibitemShut
  {NoStop}%
\bibitem [{\citenamefont {Nikkhah}\ \emph {et~al.}(2024)\citenamefont
  {Nikkhah}, \citenamefont {Pirmoradi}, \citenamefont {Ashtiani}, \citenamefont
  {Edwards}, \citenamefont {Aflatouni},\ and\ \citenamefont
  {Engheta}}]{si:nikkhah2024_2D}%
  \BibitemOpen
  \bibfield  {author} {\bibinfo {author} {\bibfnamefont {V.}~\bibnamefont
  {Nikkhah}}, \bibinfo {author} {\bibfnamefont {A.}~\bibnamefont {Pirmoradi}},
  \bibinfo {author} {\bibfnamefont {F.}~\bibnamefont {Ashtiani}}, \bibinfo
  {author} {\bibfnamefont {B.}~\bibnamefont {Edwards}}, \bibinfo {author}
  {\bibfnamefont {F.}~\bibnamefont {Aflatouni}},\ and\ \bibinfo {author}
  {\bibfnamefont {N.}~\bibnamefont {Engheta}},\ }\bibfield  {title} {\bibinfo
  {title} {Inverse-designed low-index-contrast structures on a silicon
  photonics platform for vector--matrix multiplication},\ }\href@noop {}
  {\bibfield  {journal} {\bibinfo  {journal} {Nature Photonics}\ }\textbf
  {\bibinfo {volume} {18}},\ \bibinfo {pages} {501} (\bibinfo {year}
  {2024})}\BibitemShut {NoStop}%
\bibitem [{\citenamefont {B{\'e}zard}\ \emph {et~al.}(2024)\citenamefont
  {B{\'e}zard}, \citenamefont {Mindarava}, \citenamefont {Blinder},
  \citenamefont {Trebbia}, \citenamefont {Tamarat}, \citenamefont {Jelezko},\
  and\ \citenamefont {Lounis}}]{si:bezard2024}%
  \BibitemOpen
  \bibfield  {author} {\bibinfo {author} {\bibfnamefont {M.}~\bibnamefont
  {B{\'e}zard}}, \bibinfo {author} {\bibfnamefont {Y.}~\bibnamefont
  {Mindarava}}, \bibinfo {author} {\bibfnamefont {R.}~\bibnamefont {Blinder}},
  \bibinfo {author} {\bibfnamefont {J.-B.}\ \bibnamefont {Trebbia}}, \bibinfo
  {author} {\bibfnamefont {P.}~\bibnamefont {Tamarat}}, \bibinfo {author}
  {\bibfnamefont {F.}~\bibnamefont {Jelezko}},\ and\ \bibinfo {author}
  {\bibfnamefont {B.}~\bibnamefont {Lounis}},\ }\bibfield  {title} {\bibinfo
  {title} {Unveiling the high quantum efficiency of single silicon-vacancy
  centers through dielectric tuning of their local environment},\ }\href
  {https://doi.org/10.1116/5.0216709} {\bibfield  {journal} {\bibinfo
  {journal} {AVS Quantum Science}\ }\textbf {\bibinfo {volume} {6}},\ \bibinfo
  {pages} {031401} (\bibinfo {year} {2024})}\BibitemShut {NoStop}%
\bibitem [{\citenamefont {LeCun}(1998)}]{si:lecun1998mnist}%
  \BibitemOpen
  \bibfield  {author} {\bibinfo {author} {\bibfnamefont {Y.}~\bibnamefont
  {LeCun}},\ }\bibfield  {title} {\bibinfo {title} {The mnist database of
  handwritten digits},\ }\href@noop {} {\bibfield  {journal} {\bibinfo
  {journal} {http://yann. lecun. com/exdb/mnist/}\ } (\bibinfo {year}
  {1998})}\BibitemShut {NoStop}%
\bibitem [{\citenamefont {Xiao}\ \emph {et~al.}(2017)\citenamefont {Xiao},
  \citenamefont {Rasul},\ and\ \citenamefont
  {Vollgraf}}]{si:xiao2017fashionmnist}%
  \BibitemOpen
  \bibfield  {author} {\bibinfo {author} {\bibfnamefont {H.}~\bibnamefont
  {Xiao}}, \bibinfo {author} {\bibfnamefont {K.}~\bibnamefont {Rasul}},\ and\
  \bibinfo {author} {\bibfnamefont {R.}~\bibnamefont {Vollgraf}},\ }\bibfield
  {title} {\bibinfo {title} {Fashion-mnist: a novel image dataset for
  benchmarking machine learning algorithms},\ }\href@noop {} {\bibfield
  {journal} {\bibinfo  {journal} {arXiv preprint arXiv:1708.07747}\ } (\bibinfo
  {year} {2017})}\BibitemShut {NoStop}%
\bibitem [{\citenamefont {Lin}\ \emph {et~al.}(2018)\citenamefont {Lin},
  \citenamefont {Rivenson}, \citenamefont {Yardimci}, \citenamefont {Veli},
  \citenamefont {Luo}, \citenamefont {Jarrahi},\ and\ \citenamefont
  {Ozcan}}]{si:lin2018diffractive}%
  \BibitemOpen
  \bibfield  {author} {\bibinfo {author} {\bibfnamefont {X.}~\bibnamefont
  {Lin}}, \bibinfo {author} {\bibfnamefont {Y.}~\bibnamefont {Rivenson}},
  \bibinfo {author} {\bibfnamefont {N.~T.}\ \bibnamefont {Yardimci}}, \bibinfo
  {author} {\bibfnamefont {M.}~\bibnamefont {Veli}}, \bibinfo {author}
  {\bibfnamefont {Y.}~\bibnamefont {Luo}}, \bibinfo {author} {\bibfnamefont
  {M.}~\bibnamefont {Jarrahi}},\ and\ \bibinfo {author} {\bibfnamefont
  {A.}~\bibnamefont {Ozcan}},\ }\bibfield  {title} {\bibinfo {title}
  {All-optical machine learning using diffractive deep neural networks},\
  }\href@noop {} {\bibfield  {journal} {\bibinfo  {journal} {Science}\ }\textbf
  {\bibinfo {volume} {361}},\ \bibinfo {pages} {1004} (\bibinfo {year}
  {2018})}\BibitemShut {NoStop}%
\bibitem [{\citenamefont {Wu}\ \emph {et~al.}(2019)\citenamefont {Wu},
  \citenamefont {Zhou}, \citenamefont {Khoram}, \citenamefont {Liu},\ and\
  \citenamefont {Yu}}]{si:wu2019neuromorphic}%
  \BibitemOpen
  \bibfield  {author} {\bibinfo {author} {\bibfnamefont {Z.}~\bibnamefont
  {Wu}}, \bibinfo {author} {\bibfnamefont {M.}~\bibnamefont {Zhou}}, \bibinfo
  {author} {\bibfnamefont {E.}~\bibnamefont {Khoram}}, \bibinfo {author}
  {\bibfnamefont {B.}~\bibnamefont {Liu}},\ and\ \bibinfo {author}
  {\bibfnamefont {Z.}~\bibnamefont {Yu}},\ }\bibfield  {title} {\bibinfo
  {title} {Neuromorphic metasurface},\ }\href@noop {} {\bibfield  {journal}
  {\bibinfo  {journal} {Photonics Research}\ }\textbf {\bibinfo {volume} {8}},\
  \bibinfo {pages} {46} (\bibinfo {year} {2019})}\BibitemShut {NoStop}%
\bibitem [{\citenamefont {Towers}\ \emph {et~al.}(2024)\citenamefont {Towers},
  \citenamefont {Kwiatkowski}, \citenamefont {Terry}, \citenamefont {Balis},
  \citenamefont {De~Cola}, \citenamefont {Deleu}, \citenamefont {Goul{\~a}o},
  \citenamefont {Kallinteris}, \citenamefont {Krimmel}, \citenamefont {KG}
  \emph {et~al.}}]{si:towers2024gymnasium}%
  \BibitemOpen
  \bibfield  {author} {\bibinfo {author} {\bibfnamefont {M.}~\bibnamefont
  {Towers}}, \bibinfo {author} {\bibfnamefont {A.}~\bibnamefont {Kwiatkowski}},
  \bibinfo {author} {\bibfnamefont {J.}~\bibnamefont {Terry}}, \bibinfo
  {author} {\bibfnamefont {J.~U.}\ \bibnamefont {Balis}}, \bibinfo {author}
  {\bibfnamefont {G.}~\bibnamefont {De~Cola}}, \bibinfo {author} {\bibfnamefont
  {T.}~\bibnamefont {Deleu}}, \bibinfo {author} {\bibfnamefont
  {M.}~\bibnamefont {Goul{\~a}o}}, \bibinfo {author} {\bibfnamefont
  {A.}~\bibnamefont {Kallinteris}}, \bibinfo {author} {\bibfnamefont
  {M.}~\bibnamefont {Krimmel}}, \bibinfo {author} {\bibfnamefont
  {A.}~\bibnamefont {KG}}, \emph {et~al.},\ }\bibfield  {title} {\bibinfo
  {title} {Gymnasium: A standard interface for reinforcement learning
  environments},\ }\href@noop {} {\bibfield  {journal} {\bibinfo  {journal}
  {arXiv preprint arXiv:2407.17032}\ } (\bibinfo {year} {2024})}\BibitemShut
  {NoStop}%
\bibitem [{\citenamefont {Mnih}\ \emph {et~al.}(2015)\citenamefont {Mnih},
  \citenamefont {Kavukcuoglu}, \citenamefont {Silver}, \citenamefont {Rusu},
  \citenamefont {Veness}, \citenamefont {Bellemare}, \citenamefont {Graves},
  \citenamefont {Riedmiller}, \citenamefont {Fidjeland}, \citenamefont
  {Ostrovski} \emph {et~al.}}]{si:mnih2015atari}%
  \BibitemOpen
  \bibfield  {author} {\bibinfo {author} {\bibfnamefont {V.}~\bibnamefont
  {Mnih}}, \bibinfo {author} {\bibfnamefont {K.}~\bibnamefont {Kavukcuoglu}},
  \bibinfo {author} {\bibfnamefont {D.}~\bibnamefont {Silver}}, \bibinfo
  {author} {\bibfnamefont {A.~A.}\ \bibnamefont {Rusu}}, \bibinfo {author}
  {\bibfnamefont {J.}~\bibnamefont {Veness}}, \bibinfo {author} {\bibfnamefont
  {M.~G.}\ \bibnamefont {Bellemare}}, \bibinfo {author} {\bibfnamefont
  {A.}~\bibnamefont {Graves}}, \bibinfo {author} {\bibfnamefont
  {M.}~\bibnamefont {Riedmiller}}, \bibinfo {author} {\bibfnamefont {A.~K.}\
  \bibnamefont {Fidjeland}}, \bibinfo {author} {\bibfnamefont {G.}~\bibnamefont
  {Ostrovski}}, \emph {et~al.},\ }\bibfield  {title} {\bibinfo {title}
  {Human-level control through deep reinforcement learning},\ }\href@noop {}
  {\bibfield  {journal} {\bibinfo  {journal} {nature}\ }\textbf {\bibinfo
  {volume} {518}},\ \bibinfo {pages} {529} (\bibinfo {year}
  {2015})}\BibitemShut {NoStop}%
\bibitem [{\citenamefont {Schulman}\ \emph {et~al.}(2017)\citenamefont
  {Schulman}, \citenamefont {Wolski}, \citenamefont {Dhariwal}, \citenamefont
  {Radford},\ and\ \citenamefont {Klimov}}]{si:schulman2017ppo}%
  \BibitemOpen
  \bibfield  {author} {\bibinfo {author} {\bibfnamefont {J.}~\bibnamefont
  {Schulman}}, \bibinfo {author} {\bibfnamefont {F.}~\bibnamefont {Wolski}},
  \bibinfo {author} {\bibfnamefont {P.}~\bibnamefont {Dhariwal}}, \bibinfo
  {author} {\bibfnamefont {A.}~\bibnamefont {Radford}},\ and\ \bibinfo {author}
  {\bibfnamefont {O.}~\bibnamefont {Klimov}},\ }\bibfield  {title} {\bibinfo
  {title} {Proximal policy optimization algorithms},\ }\href@noop {} {\bibfield
   {journal} {\bibinfo  {journal} {arXiv preprint arXiv:1707.06347}\ }
  (\bibinfo {year} {2017})}\BibitemShut {NoStop}%
\bibitem [{\citenamefont {Raffin}\ \emph {et~al.}(2021)\citenamefont {Raffin},
  \citenamefont {Hill}, \citenamefont {Gleave}, \citenamefont {Kanervisto},
  \citenamefont {Ernestus},\ and\ \citenamefont {Dormann}}]{si:raffin2021stable}%
  \BibitemOpen
  \bibfield  {author} {\bibinfo {author} {\bibfnamefont {A.}~\bibnamefont
  {Raffin}}, \bibinfo {author} {\bibfnamefont {A.}~\bibnamefont {Hill}},
  \bibinfo {author} {\bibfnamefont {A.}~\bibnamefont {Gleave}}, \bibinfo
  {author} {\bibfnamefont {A.}~\bibnamefont {Kanervisto}}, \bibinfo {author}
  {\bibfnamefont {M.}~\bibnamefont {Ernestus}},\ and\ \bibinfo {author}
  {\bibfnamefont {N.}~\bibnamefont {Dormann}},\ }\bibfield  {title} {\bibinfo
  {title} {Stable-baselines3: Reliable reinforcement learning
  implementations},\ }\href@noop {} {\bibfield  {journal} {\bibinfo  {journal}
  {Journal of machine learning research}\ }\textbf {\bibinfo {volume} {22}},\
  \bibinfo {pages} {1} (\bibinfo {year} {2021})}\BibitemShut {NoStop}%
\bibitem [{\citenamefont {Haarnoja}\ \emph {et~al.}(2018)\citenamefont
  {Haarnoja}, \citenamefont {Zhou}, \citenamefont {Abbeel},\ and\ \citenamefont
  {Levine}}]{si:haarnoja2018sac}%
  \BibitemOpen
  \bibfield  {author} {\bibinfo {author} {\bibfnamefont {T.}~\bibnamefont
  {Haarnoja}}, \bibinfo {author} {\bibfnamefont {A.}~\bibnamefont {Zhou}},
  \bibinfo {author} {\bibfnamefont {P.}~\bibnamefont {Abbeel}},\ and\ \bibinfo
  {author} {\bibfnamefont {S.}~\bibnamefont {Levine}},\ }\bibfield  {title}
  {\bibinfo {title} {Soft actor-critic: Off-policy maximum entropy deep
  reinforcement learning with a stochastic actor},\ }in\ \href@noop {} {\emph
  {\bibinfo {booktitle} {International conference on machine learning}}}\
  (\bibinfo {organization} {Pmlr},\ \bibinfo {year} {2018})\ pp.\ \bibinfo
  {pages} {1861--1870}\BibitemShut {NoStop}%
\bibitem [{\citenamefont {Dinu}\ \emph {et~al.}(2003)\citenamefont {Dinu},
  \citenamefont {Quochi},\ and\ \citenamefont {Garcia}}]{si:dinu2003third}%
  \BibitemOpen
  \bibfield  {author} {\bibinfo {author} {\bibfnamefont {M.}~\bibnamefont
  {Dinu}}, \bibinfo {author} {\bibfnamefont {F.}~\bibnamefont {Quochi}},\ and\
  \bibinfo {author} {\bibfnamefont {H.}~\bibnamefont {Garcia}},\ }\bibfield
  {title} {\bibinfo {title} {Third-order nonlinearities in silicon at telecom
  wavelengths},\ }\href@noop {} {\bibfield  {journal} {\bibinfo  {journal}
  {Applied physics letters}\ }\textbf {\bibinfo {volume} {82}},\ \bibinfo
  {pages} {2954} (\bibinfo {year} {2003})}\BibitemShut {NoStop}%
\bibitem [{\citenamefont {Bristow}\ \emph {et~al.}(2007)\citenamefont
  {Bristow}, \citenamefont {Rotenberg},\ and\ \citenamefont
  {Van~Driel}}]{si:bristow2007two}%
  \BibitemOpen
  \bibfield  {author} {\bibinfo {author} {\bibfnamefont {A.~D.}\ \bibnamefont
  {Bristow}}, \bibinfo {author} {\bibfnamefont {N.}~\bibnamefont {Rotenberg}},\
  and\ \bibinfo {author} {\bibfnamefont {H.~M.}\ \bibnamefont {Van~Driel}},\
  }\bibfield  {title} {\bibinfo {title} {Two-photon absorption and kerr
  coefficients of silicon for 850--2200nm},\ }\href@noop {} {\bibfield
  {journal} {\bibinfo  {journal} {Applied physics letters}\ }\textbf {\bibinfo
  {volume} {90}} (\bibinfo {year} {2007})}\BibitemShut {NoStop}%
\bibitem [{\citenamefont {Lau}\ \emph {et~al.}(2022)\citenamefont {Lau},
  \citenamefont {Liu},\ and\ \citenamefont {Qiu}}]{si:lau2022comparison}%
  \BibitemOpen
  \bibfield  {author} {\bibinfo {author} {\bibfnamefont {K.~Y.}\ \bibnamefont
  {Lau}}, \bibinfo {author} {\bibfnamefont {X.}~\bibnamefont {Liu}},\ and\
  \bibinfo {author} {\bibfnamefont {J.}~\bibnamefont {Qiu}},\ }\bibfield
  {title} {\bibinfo {title} {A comparison for saturable absorbers: Carbon
  nanotube versus graphene},\ }\href@noop {} {\bibfield  {journal} {\bibinfo
  {journal} {Advanced Photonics Research}\ }\textbf {\bibinfo {volume} {3}},\
  \bibinfo {pages} {2200023} (\bibinfo {year} {2022})}\BibitemShut {NoStop}%
\bibitem [{\citenamefont {Vaswani}\ \emph {et~al.}(2017)\citenamefont
  {Vaswani}, \citenamefont {Shazeer}, \citenamefont {Parmar}, \citenamefont
  {Uszkoreit}, \citenamefont {Jones}, \citenamefont {Gomez}, \citenamefont
  {Kaiser},\ and\ \citenamefont {Polosukhin}}]{si:vaswani2017transformer}%
  \BibitemOpen
  \bibfield  {author} {\bibinfo {author} {\bibfnamefont {A.}~\bibnamefont
  {Vaswani}}, \bibinfo {author} {\bibfnamefont {N.}~\bibnamefont {Shazeer}},
  \bibinfo {author} {\bibfnamefont {N.}~\bibnamefont {Parmar}}, \bibinfo
  {author} {\bibfnamefont {J.}~\bibnamefont {Uszkoreit}}, \bibinfo {author}
  {\bibfnamefont {L.}~\bibnamefont {Jones}}, \bibinfo {author} {\bibfnamefont
  {A.~N.}\ \bibnamefont {Gomez}}, \bibinfo {author} {\bibfnamefont
  {{\L}.}~\bibnamefont {Kaiser}},\ and\ \bibinfo {author} {\bibfnamefont
  {I.}~\bibnamefont {Polosukhin}},\ }\bibfield  {title} {\bibinfo {title}
  {Attention is all you need},\ }\href@noop {} {\bibfield  {journal} {\bibinfo
  {journal} {Advances in neural information processing systems}\ }\textbf
  {\bibinfo {volume} {30}} (\bibinfo {year} {2017})}\BibitemShut {NoStop}%
\bibitem [{\citenamefont {Radford}\ \emph {et~al.}(2019)\citenamefont
  {Radford}, \citenamefont {Wu}, \citenamefont {Child}, \citenamefont {Luan},
  \citenamefont {Amodei}, \citenamefont {Sutskever} \emph
  {et~al.}}]{si:radford2019language}%
  \BibitemOpen
  \bibfield  {author} {\bibinfo {author} {\bibfnamefont {A.}~\bibnamefont
  {Radford}}, \bibinfo {author} {\bibfnamefont {J.}~\bibnamefont {Wu}},
  \bibinfo {author} {\bibfnamefont {R.}~\bibnamefont {Child}}, \bibinfo
  {author} {\bibfnamefont {D.}~\bibnamefont {Luan}}, \bibinfo {author}
  {\bibfnamefont {D.}~\bibnamefont {Amodei}}, \bibinfo {author} {\bibfnamefont
  {I.}~\bibnamefont {Sutskever}}, \emph {et~al.},\ }\bibfield  {title}
  {\bibinfo {title} {Language models are unsupervised multitask learners},\
  }\href@noop {} {\bibfield  {journal} {\bibinfo  {journal} {OpenAI blog}\
  }\textbf {\bibinfo {volume} {1}},\ \bibinfo {pages} {9} (\bibinfo {year}
  {2019})}\BibitemShut {NoStop}%
\bibitem [{\citenamefont {Brown}\ \emph {et~al.}(2020)\citenamefont {Brown},
  \citenamefont {Mann}, \citenamefont {Ryder}, \citenamefont {Subbiah},
  \citenamefont {Kaplan}, \citenamefont {Dhariwal}, \citenamefont
  {Neelakantan}, \citenamefont {Shyam}, \citenamefont {Sastry}, \citenamefont
  {Askell} \emph {et~al.}}]{si:brown2020gpt3}%
  \BibitemOpen
  \bibfield  {author} {\bibinfo {author} {\bibfnamefont {T.}~\bibnamefont
  {Brown}}, \bibinfo {author} {\bibfnamefont {B.}~\bibnamefont {Mann}},
  \bibinfo {author} {\bibfnamefont {N.}~\bibnamefont {Ryder}}, \bibinfo
  {author} {\bibfnamefont {M.}~\bibnamefont {Subbiah}}, \bibinfo {author}
  {\bibfnamefont {J.~D.}\ \bibnamefont {Kaplan}}, \bibinfo {author}
  {\bibfnamefont {P.}~\bibnamefont {Dhariwal}}, \bibinfo {author}
  {\bibfnamefont {A.}~\bibnamefont {Neelakantan}}, \bibinfo {author}
  {\bibfnamefont {P.}~\bibnamefont {Shyam}}, \bibinfo {author} {\bibfnamefont
  {G.}~\bibnamefont {Sastry}}, \bibinfo {author} {\bibfnamefont
  {A.}~\bibnamefont {Askell}}, \emph {et~al.},\ }\bibfield  {title} {\bibinfo
  {title} {Language models are few-shot learners},\ }\href@noop {} {\bibfield
  {journal} {\bibinfo  {journal} {Advances in neural information processing
  systems}\ }\textbf {\bibinfo {volume} {33}},\ \bibinfo {pages} {1877}
  (\bibinfo {year} {2020})}\BibitemShut {NoStop}%
\bibitem [{\citenamefont {Touvron}\ \emph
  {et~al.}(2023{\natexlab{a}})\citenamefont {Touvron}, \citenamefont {Lavril},
  \citenamefont {Izacard}, \citenamefont {Martinet}, \citenamefont {Lachaux},
  \citenamefont {Lacroix}, \citenamefont {Rozi{\`e}re}, \citenamefont {Goyal},
  \citenamefont {Hambro}, \citenamefont {Azhar} \emph
  {et~al.}}]{si:touvron2023llama1}%
  \BibitemOpen
  \bibfield  {author} {\bibinfo {author} {\bibfnamefont {H.}~\bibnamefont
  {Touvron}}, \bibinfo {author} {\bibfnamefont {T.}~\bibnamefont {Lavril}},
  \bibinfo {author} {\bibfnamefont {G.}~\bibnamefont {Izacard}}, \bibinfo
  {author} {\bibfnamefont {X.}~\bibnamefont {Martinet}}, \bibinfo {author}
  {\bibfnamefont {M.-A.}\ \bibnamefont {Lachaux}}, \bibinfo {author}
  {\bibfnamefont {T.}~\bibnamefont {Lacroix}}, \bibinfo {author} {\bibfnamefont
  {B.}~\bibnamefont {Rozi{\`e}re}}, \bibinfo {author} {\bibfnamefont
  {N.}~\bibnamefont {Goyal}}, \bibinfo {author} {\bibfnamefont
  {E.}~\bibnamefont {Hambro}}, \bibinfo {author} {\bibfnamefont
  {F.}~\bibnamefont {Azhar}}, \emph {et~al.},\ }\bibfield  {title} {\bibinfo
  {title} {Llama: Open and efficient foundation language models},\ }\href@noop
  {} {\bibfield  {journal} {\bibinfo  {journal} {arXiv preprint
  arXiv:2302.13971}\ } (\bibinfo {year} {2023}{\natexlab{a}})}\BibitemShut
  {NoStop}%
\bibitem [{\citenamefont {Touvron}\ \emph
  {et~al.}(2023{\natexlab{b}})\citenamefont {Touvron}, \citenamefont {Martin},
  \citenamefont {Stone}, \citenamefont {Albert}, \citenamefont {Almahairi},
  \citenamefont {Babaei}, \citenamefont {Bashlykov}, \citenamefont {Batra},
  \citenamefont {Bhargava}, \citenamefont {Bhosale} \emph
  {et~al.}}]{si:touvron2023llama2}%
  \BibitemOpen
  \bibfield  {author} {\bibinfo {author} {\bibfnamefont {H.}~\bibnamefont
  {Touvron}}, \bibinfo {author} {\bibfnamefont {L.}~\bibnamefont {Martin}},
  \bibinfo {author} {\bibfnamefont {K.}~\bibnamefont {Stone}}, \bibinfo
  {author} {\bibfnamefont {P.}~\bibnamefont {Albert}}, \bibinfo {author}
  {\bibfnamefont {A.}~\bibnamefont {Almahairi}}, \bibinfo {author}
  {\bibfnamefont {Y.}~\bibnamefont {Babaei}}, \bibinfo {author} {\bibfnamefont
  {N.}~\bibnamefont {Bashlykov}}, \bibinfo {author} {\bibfnamefont
  {S.}~\bibnamefont {Batra}}, \bibinfo {author} {\bibfnamefont
  {P.}~\bibnamefont {Bhargava}}, \bibinfo {author} {\bibfnamefont
  {S.}~\bibnamefont {Bhosale}}, \emph {et~al.},\ }\bibfield  {title} {\bibinfo
  {title} {Llama 2: Open foundation and fine-tuned chat models},\ }\href@noop
  {} {\bibfield  {journal} {\bibinfo  {journal} {arXiv preprint
  arXiv:2307.09288}\ } (\bibinfo {year} {2023}{\natexlab{b}})}\BibitemShut
  {NoStop}%
\bibitem [{\citenamefont {Dubey}\ \emph {et~al.}(2024)\citenamefont {Dubey},
  \citenamefont {Jauhri}, \citenamefont {Pandey}, \citenamefont {Kadian},
  \citenamefont {Al-Dahle}, \citenamefont {Letman}, \citenamefont {Mathur},
  \citenamefont {Schelten}, \citenamefont {Yang}, \citenamefont {Fan} \emph
  {et~al.}}]{si:dubey2024llama3}%
  \BibitemOpen
  \bibfield  {author} {\bibinfo {author} {\bibfnamefont {A.}~\bibnamefont
  {Dubey}}, \bibinfo {author} {\bibfnamefont {A.}~\bibnamefont {Jauhri}},
  \bibinfo {author} {\bibfnamefont {A.}~\bibnamefont {Pandey}}, \bibinfo
  {author} {\bibfnamefont {A.}~\bibnamefont {Kadian}}, \bibinfo {author}
  {\bibfnamefont {A.}~\bibnamefont {Al-Dahle}}, \bibinfo {author}
  {\bibfnamefont {A.}~\bibnamefont {Letman}}, \bibinfo {author} {\bibfnamefont
  {A.}~\bibnamefont {Mathur}}, \bibinfo {author} {\bibfnamefont
  {A.}~\bibnamefont {Schelten}}, \bibinfo {author} {\bibfnamefont
  {A.}~\bibnamefont {Yang}}, \bibinfo {author} {\bibfnamefont {A.}~\bibnamefont
  {Fan}}, \emph {et~al.},\ }\bibfield  {title} {\bibinfo {title} {The llama 3
  herd of models},\ }\href@noop {} {\bibfield  {journal} {\bibinfo  {journal}
  {arXiv e-prints}\ ,\ \bibinfo {pages} {arXiv}} (\bibinfo {year}
  {2024})}\BibitemShut {NoStop}%
\bibitem [{\citenamefont {Bi}\ \emph {et~al.}(2024)\citenamefont {Bi},
  \citenamefont {Chen}, \citenamefont {Chen}, \citenamefont {Chen},
  \citenamefont {Dai}, \citenamefont {Deng}, \citenamefont {Ding},
  \citenamefont {Dong}, \citenamefont {Du}, \citenamefont {Fu} \emph
  {et~al.}}]{si:bi2024deepseek}%
  \BibitemOpen
  \bibfield  {author} {\bibinfo {author} {\bibfnamefont {X.}~\bibnamefont
  {Bi}}, \bibinfo {author} {\bibfnamefont {D.}~\bibnamefont {Chen}}, \bibinfo
  {author} {\bibfnamefont {G.}~\bibnamefont {Chen}}, \bibinfo {author}
  {\bibfnamefont {S.}~\bibnamefont {Chen}}, \bibinfo {author} {\bibfnamefont
  {D.}~\bibnamefont {Dai}}, \bibinfo {author} {\bibfnamefont {C.}~\bibnamefont
  {Deng}}, \bibinfo {author} {\bibfnamefont {H.}~\bibnamefont {Ding}}, \bibinfo
  {author} {\bibfnamefont {K.}~\bibnamefont {Dong}}, \bibinfo {author}
  {\bibfnamefont {Q.}~\bibnamefont {Du}}, \bibinfo {author} {\bibfnamefont
  {Z.}~\bibnamefont {Fu}}, \emph {et~al.},\ }\bibfield  {title} {\bibinfo
  {title} {Deepseek llm: Scaling open-source language models with
  longtermism},\ }\href@noop {} {\bibfield  {journal} {\bibinfo  {journal}
  {arXiv preprint arXiv:2401.02954}\ } (\bibinfo {year} {2024})}\BibitemShut
  {NoStop}%
\bibitem [{\citenamefont {Liu}\ \emph {et~al.}(2024{\natexlab{a}})\citenamefont
  {Liu}, \citenamefont {Feng}, \citenamefont {Wang}, \citenamefont {Wang},
  \citenamefont {Liu}, \citenamefont {Zhao}, \citenamefont {Dengr},
  \citenamefont {Ruan}, \citenamefont {Dai}, \citenamefont {Guo} \emph
  {et~al.}}]{si:liu2024deepseek-v2}%
  \BibitemOpen
  \bibfield  {author} {\bibinfo {author} {\bibfnamefont {A.}~\bibnamefont
  {Liu}}, \bibinfo {author} {\bibfnamefont {B.}~\bibnamefont {Feng}}, \bibinfo
  {author} {\bibfnamefont {B.}~\bibnamefont {Wang}}, \bibinfo {author}
  {\bibfnamefont {B.}~\bibnamefont {Wang}}, \bibinfo {author} {\bibfnamefont
  {B.}~\bibnamefont {Liu}}, \bibinfo {author} {\bibfnamefont {C.}~\bibnamefont
  {Zhao}}, \bibinfo {author} {\bibfnamefont {C.}~\bibnamefont {Dengr}},
  \bibinfo {author} {\bibfnamefont {C.}~\bibnamefont {Ruan}}, \bibinfo {author}
  {\bibfnamefont {D.}~\bibnamefont {Dai}}, \bibinfo {author} {\bibfnamefont
  {D.}~\bibnamefont {Guo}}, \emph {et~al.},\ }\bibfield  {title} {\bibinfo
  {title} {Deepseek-v2: A strong, economical, and efficient mixture-of-experts
  language model},\ }\href@noop {} {\bibfield  {journal} {\bibinfo  {journal}
  {arXiv preprint arXiv:2405.04434}\ } (\bibinfo {year}
  {2024}{\natexlab{a}})}\BibitemShut {NoStop}%
\bibitem [{\citenamefont {Liu}\ \emph {et~al.}(2024{\natexlab{b}})\citenamefont
  {Liu}, \citenamefont {Feng}, \citenamefont {Xue}, \citenamefont {Wang},
  \citenamefont {Wu}, \citenamefont {Lu}, \citenamefont {Zhao}, \citenamefont
  {Deng}, \citenamefont {Zhang}, \citenamefont {Ruan} \emph
  {et~al.}}]{si:liu2024deepseek-v3}%
  \BibitemOpen
  \bibfield  {author} {\bibinfo {author} {\bibfnamefont {A.}~\bibnamefont
  {Liu}}, \bibinfo {author} {\bibfnamefont {B.}~\bibnamefont {Feng}}, \bibinfo
  {author} {\bibfnamefont {B.}~\bibnamefont {Xue}}, \bibinfo {author}
  {\bibfnamefont {B.}~\bibnamefont {Wang}}, \bibinfo {author} {\bibfnamefont
  {B.}~\bibnamefont {Wu}}, \bibinfo {author} {\bibfnamefont {C.}~\bibnamefont
  {Lu}}, \bibinfo {author} {\bibfnamefont {C.}~\bibnamefont {Zhao}}, \bibinfo
  {author} {\bibfnamefont {C.}~\bibnamefont {Deng}}, \bibinfo {author}
  {\bibfnamefont {C.}~\bibnamefont {Zhang}}, \bibinfo {author} {\bibfnamefont
  {C.}~\bibnamefont {Ruan}}, \emph {et~al.},\ }\bibfield  {title} {\bibinfo
  {title} {Deepseek-v3 technical report},\ }\href@noop {} {\bibfield  {journal}
  {\bibinfo  {journal} {arXiv preprint arXiv:2412.19437}\ } (\bibinfo {year}
  {2024}{\natexlab{b}})}\BibitemShut {NoStop}%
\bibitem [{\citenamefont {Shen}\ \emph {et~al.}(2017)\citenamefont {Shen},
  \citenamefont {Harris}, \citenamefont {Skirlo}, \citenamefont {Prabhu},
  \citenamefont {Baehr-Jones}, \citenamefont {Hochberg}, \citenamefont {Sun},
  \citenamefont {Zhao}, \citenamefont {Larochelle}, \citenamefont {Englund}
  \emph {et~al.}}]{si:shen2017}%
  \BibitemOpen
  \bibfield  {author} {\bibinfo {author} {\bibfnamefont {Y.}~\bibnamefont
  {Shen}}, \bibinfo {author} {\bibfnamefont {N.~C.}\ \bibnamefont {Harris}},
  \bibinfo {author} {\bibfnamefont {S.}~\bibnamefont {Skirlo}}, \bibinfo
  {author} {\bibfnamefont {M.}~\bibnamefont {Prabhu}}, \bibinfo {author}
  {\bibfnamefont {T.}~\bibnamefont {Baehr-Jones}}, \bibinfo {author}
  {\bibfnamefont {M.}~\bibnamefont {Hochberg}}, \bibinfo {author}
  {\bibfnamefont {X.}~\bibnamefont {Sun}}, \bibinfo {author} {\bibfnamefont
  {S.}~\bibnamefont {Zhao}}, \bibinfo {author} {\bibfnamefont {H.}~\bibnamefont
  {Larochelle}}, \bibinfo {author} {\bibfnamefont {D.}~\bibnamefont {Englund}},
  \emph {et~al.},\ }\bibfield  {title} {\bibinfo {title} {Deep learning with
  coherent nanophotonic circuits},\ }\href@noop {} {\bibfield  {journal}
  {\bibinfo  {journal} {Nature photonics}\ }\textbf {\bibinfo {volume} {11}},\
  \bibinfo {pages} {441} (\bibinfo {year} {2017})}\BibitemShut {NoStop}%
\bibitem [{\citenamefont {Anderson}\ \emph {et~al.}(2023)\citenamefont
  {Anderson}, \citenamefont {Ma}, \citenamefont {Wang}, \citenamefont
  {Wright},\ and\ \citenamefont {McMahon}}]{si:anderson2023optical}%
  \BibitemOpen
  \bibfield  {author} {\bibinfo {author} {\bibfnamefont {M.}~\bibnamefont
  {Anderson}}, \bibinfo {author} {\bibfnamefont {S.-Y.}\ \bibnamefont {Ma}},
  \bibinfo {author} {\bibfnamefont {T.}~\bibnamefont {Wang}}, \bibinfo {author}
  {\bibfnamefont {L.}~\bibnamefont {Wright}},\ and\ \bibinfo {author}
  {\bibfnamefont {P.}~\bibnamefont {McMahon}},\ }\bibfield  {title} {\bibinfo
  {title} {Optical transformers},\ }\href@noop {} {\bibfield  {journal}
  {\bibinfo  {journal} {Transactions on Machine Learning Research}\ } (\bibinfo
  {year} {2023})}\BibitemShut {NoStop}%
\bibitem [{\citenamefont {Hua}\ \emph {et~al.}(2025)\citenamefont {Hua},
  \citenamefont {Divita}, \citenamefont {Yu}, \citenamefont {Peng},
  \citenamefont {Roques-Carmes} \emph {et~al.}}]{si:lightelligence2025pace}%
  \BibitemOpen
  \bibfield  {author} {\bibinfo {author} {\bibfnamefont {S.}~\bibnamefont
  {Hua}}, \bibinfo {author} {\bibfnamefont {E.}~\bibnamefont {Divita}},
  \bibinfo {author} {\bibfnamefont {S.}~\bibnamefont {Yu}}, \bibinfo {author}
  {\bibfnamefont {B.}~\bibnamefont {Peng}}, \bibinfo {author} {\bibfnamefont
  {C.}~\bibnamefont {Roques-Carmes}}, \emph {et~al.},\ }\bibfield  {title}
  {\bibinfo {title} {An integrated large-scale photonic accelerator with
  ultralow latency},\ }\href {https://doi.org/10.1038/s41586-025-08786-6}
  {\bibfield  {journal} {\bibinfo  {journal} {Nature}\ }\textbf {\bibinfo
  {volume} {640}},\ \bibinfo {pages} {361} (\bibinfo {year}
  {2025})}\BibitemShut {NoStop}%
\bibitem [{\citenamefont {Ahmed}\ \emph {et~al.}(2025)\citenamefont {Ahmed},
  \citenamefont {Baghdadi}, \citenamefont {Bernadskiy} \emph
  {et~al.}}]{si:ahmed2025upaia}%
  \BibitemOpen
  \bibfield  {author} {\bibinfo {author} {\bibfnamefont {S.~R.}\ \bibnamefont
  {Ahmed}}, \bibinfo {author} {\bibfnamefont {R.}~\bibnamefont {Baghdadi}},
  \bibinfo {author} {\bibfnamefont {M.}~\bibnamefont {Bernadskiy}}, \emph
  {et~al.},\ }\bibfield  {title} {\bibinfo {title} {Universal photonic
  artificial intelligence acceleration},\ }\href
  {https://doi.org/10.1038/s41586-025-08854-x} {\bibfield  {journal} {\bibinfo
  {journal} {Nature}\ }\textbf {\bibinfo {volume} {640}},\ \bibinfo {pages}
  {368} (\bibinfo {year} {2025})}\BibitemShut {NoStop}%
\bibitem [{\citenamefont {Novotny}\ and\ \citenamefont
  {Hecht}(2012)}]{si:novotny2012principles}%
  \BibitemOpen
  \bibfield  {author} {\bibinfo {author} {\bibfnamefont {L.}~\bibnamefont
  {Novotny}}\ and\ \bibinfo {author} {\bibfnamefont {B.}~\bibnamefont
  {Hecht}},\ }\href@noop {} {\emph {\bibinfo {title} {Principles of
  nano-optics}}}\ (\bibinfo  {publisher} {Cambridge University Press},\
  \bibinfo {year} {2012})\BibitemShut {NoStop}%
\bibitem [{\citenamefont {Laucht}\ \emph {et~al.}(2010)\citenamefont {Laucht},
  \citenamefont {Villas-B{\^o}as}, \citenamefont {Stobbe}, \citenamefont
  {Hauke}, \citenamefont {Hofbauer}, \citenamefont {B{\"o}hm}, \citenamefont
  {Lodahl}, \citenamefont {Amann}, \citenamefont {Kaniber},\ and\ \citenamefont
  {Finley}}]{si:laucht2010stark}%
  \BibitemOpen
  \bibfield  {author} {\bibinfo {author} {\bibfnamefont {A.}~\bibnamefont
  {Laucht}}, \bibinfo {author} {\bibfnamefont {J.}~\bibnamefont
  {Villas-B{\^o}as}}, \bibinfo {author} {\bibfnamefont {S.}~\bibnamefont
  {Stobbe}}, \bibinfo {author} {\bibfnamefont {N.}~\bibnamefont {Hauke}},
  \bibinfo {author} {\bibfnamefont {F.}~\bibnamefont {Hofbauer}}, \bibinfo
  {author} {\bibfnamefont {G.}~\bibnamefont {B{\"o}hm}}, \bibinfo {author}
  {\bibfnamefont {P.}~\bibnamefont {Lodahl}}, \bibinfo {author} {\bibfnamefont
  {M.-C.}\ \bibnamefont {Amann}}, \bibinfo {author} {\bibfnamefont
  {M.}~\bibnamefont {Kaniber}},\ and\ \bibinfo {author} {\bibfnamefont
  {J.}~\bibnamefont {Finley}},\ }\bibfield  {title} {\bibinfo {title} {Mutual
  coupling of two semiconductor quantum dots via an optical nanocavity},\
  }\href@noop {} {\bibfield  {journal} {\bibinfo  {journal} {Physical Review
  B—Condensed Matter and Materials Physics}\ }\textbf {\bibinfo {volume}
  {82}},\ \bibinfo {pages} {075305} (\bibinfo {year} {2010})}\BibitemShut
  {NoStop}%
\end{thebibliography}
\end{document}